%% file: main.tex
\documentclass[10pt,journal,compsoc]{IEEEtran}
\usepackage{cite}
\usepackage{url}
\usepackage{subfig}
\usepackage[table]{xcolor}
\usepackage[utf8]{inputenc} % Eingabekodierung: UTF-8
\usepackage{fixltx2e} % Schickere Ausgabe
\usepackage[T1]{fontenc} % ordentliche Trennung
\usepackage{lmodern} % ordentliche Schriften
\usepackage{setspace,graphicx,tikz,tabularx} % für Elemente der Titelseite
\usepackage{amsmath}
\usepackage{amsfonts}
\usetikzlibrary{decorations.pathreplacing,chains,fit,shapes,calc,arrows,plotmarks,arrows.meta,shadows,positioning,shapes.arrows,shapes.geometric,shapes.multipart,calligraphy}
\usepackage{textcomp}
\usepackage{pgfplots}
\usepackage{svg}
\usepackage[binary-units=true,per-mode=symbol]{siunitx}
\usepackage[nolist]{acronym}
\usepackage{makecell}
\usepackage{booktabs}
\usepackage{algorithm}
\usepackage{algpseudocode}
\usepackage{listings}
\usepackage{csquotes}
\usepackage{stmaryrd}
\usepackage{caption}
\usepackage{fancyvrb}
\usepackage{fontawesome}
\usepackage{multirow,multicol,csvsimple}
\usepackage[inline]{enumitem}
\usepackage{amsthm}
\usepackage{pgfplotstable}
\usepackage{afterpage}
\usepackage{csvsimple}
\usepackage[nottoc]{tocbibind}
\usepackage{stfloats}

\usepackage[textsize=tiny]{todonotes}
\usepackage{longtable}
\usepackage{pdflscape}
\usepackage{multicol}\usepackage{adjustbox}
\usepackage{xspace}
\usepackage{makecell,rotating}
\usepackage{multirow}
\usepackage{soul}
\usepackage{makecell}

\usepackage{array}
\usepackage{verbatim}
\usepackage{float}

\usepackage{threeparttable}

\usepackage{breakurl}
\usepackage[unicode=true,breaklinks]{hyperref}
\usepackage[nameinlink,noabbrev,capitalize]{cleveref}
\creflabelformat{equation}{#2#1#3}

\newcolumntype{R}[2]{%
    >{\adjustbox{angle=#1,lap=\width-(#2)}\bgroup}%
    l%
    <{\egroup}%
}
\newcommand*\rot{\multicolumn{1}{R{45}{1mm}}}% no optional argument here, please!

\newcommand*\xor{\,\oplus\,}

\newcommand{\cf}{cf.\@\xspace}

\newcommand{\etc}{etc.\@\xspace}

\newcommand{\wrt}{w.r.t.\@\xspace}
\newcommand{\etal}{\textit{et al.}\@\xspace}
\newcommand{\eg}{\textit{e.g.}\@\xspace}

\newcommand{\ie}{\textit{i.e.}\@\xspace}

\algdef{SE}[FUNCTION]{Function}{EndFunction}%
   [2]{\algorithmicfunction\ \textproc{#1}\ifthenelse{\equal{#2}{}}{}{(#2)}}%
   {\algorithmicend\ \algorithmicfunction}%

\algnewcommand\algorithmicforeach{\textbf{for each}}
\algdef{S}[FOR]{ForEach}[1]{\algorithmicforeach\ #1\ \algorithmicdo}

\begin{document}
\input{acronyms.tex}
\title{A Thorough Investigation of Content-Defined Chunking Algorithms for Data Deduplication}

\author{Marcel Gregoriadis,
        Leonhard Balduf,
        Björn Scheuermann and~Johan~Pouwelse% <-this % stops a space
\IEEEcompsocitemizethanks{\IEEEcompsocthanksitem M. Gregoriadis and J. Pouwelse are with the Data-Intensive Systems Lab at Delft University of Technology, The Netherlands.\protect\\
E-mail: m.gregoriadis@tudelft.nl; j.a.pouwelse@tudelft.nl
\IEEEcompsocthanksitem L. Balduf and B. Scheuermann are with the Communication Networks Lab at Darmstadt University of Technology, Germany.
\protect\\
leonhard.balduf@tu-darmstadt.de; scheuermann@tu-darmstadt.de%
}% <-this % stops an unwanted space
}


\markboth{Submitted to IEEE Transactions on Cloud Computing}%
{Gregoriadis \MakeLowercase{\textit{et al.}}: A Thorough Investigation of Content-Defined Chunking Algorithms for Data Deduplication}

% \IEEEpubid{\makebox[\columnwidth]{\hfill 0000--0000/00/\$00.00~\copyright~2015 IEEE}%
% \hspace{\columnsep}\makebox[\columnwidth]{Published by the IEEE Computer Society\hfill}}


\IEEEtitleabstractindextext{%
\begin{abstract}
    \input{sections/abstract}
\end{abstract}


\noindent \textit{This work has been submitted to the IEEE for possible publication. Copyright may be transferred without notice, after which this version may no longer be accessible.}

}

\maketitle

\IEEEdisplaynontitleabstractindextext

% \fi
\IEEEpeerreviewmaketitle

\section{Introduction}\label{sec:intro}
\input{sections/introduction}

\section{Background}\label{sec:background}
\input{sections/background}

\section{Related Work}\label{sec:relatedWork}
\input{sections/related-work}

\section{Chunking Algorithms}\label{sec:algorithms}
\input{sections/algorithms}

\section{Experiment Setup}\label{sec:setup}
\input{sections/setup.tex}

\section{Computational Efficiency}\label{sec:efficiency}
\input{sections/efficiency.tex}

\section{Chunk Size Distribution}\label{sec:distribution}
\input{sections/csd.tex}

\section{Deduplication Ratio}\label{sec:dedup}
\input{sections/dedup.tex}

\section{Discussion}\label{sec:discussion}
\input{sections/discussion.tex}
\section{Conclusion}\label{sec:conclusion}
\input{sections/conclusion}

\ifCLASSOPTIONcompsoc
  \section*{Acknowledgments}
\else
  \section*{Acknowledgment}
\fi

This work was supported by the Dutch national NWO/TKI science
grant BLOCK.2019.004, as well as the German Research Foundation
(DFG) within the Collaborative Research Center (CRC) SFB 1053:
MAKI.

\ifCLASSOPTIONcaptionsoff
  \newpage
\fi



%\bibliographystyle{IEEEtran}
\bibliographystyle{IEEEtran}
\bibliography{IEEEabrv,ref}


%\newpage

\newpage
\begin{IEEEbiography}[{\includegraphics[width=1in,height=1.25in,clip,keepaspectratio]{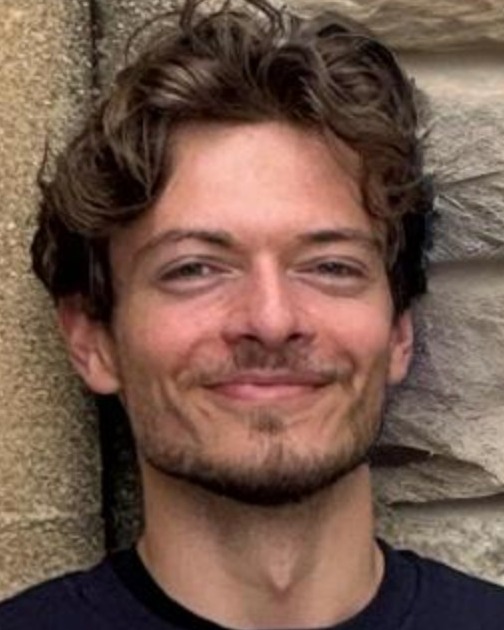}}]{Marcel Gregoriadis}
received the B.Sc. degree from Stuttgart Media University in 2020 and the M.Sc. degree from Humboldt University of Berlin in 2023, both in computer science.
He is currently working toward a Ph.D. in the Data-Intensive Systems group at Delft University of Technology,
where he is conducting research on decentralized information retrieval with intersections to artificial intelligence.
\end{IEEEbiography}
\vspace{-1.5em}
\begin{IEEEbiography}[{\includegraphics[width=1in,height=1.25in,clip,keepaspectratio]{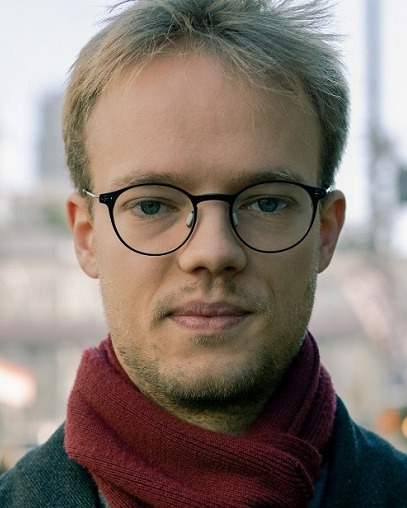}}]{Leonhard Balduf}
is a Ph.D. candidate at the Communication Networks Lab at TU Darmstadt in his third year.
He received his B.Sc. from Munich University of Applied Sciences and the M.Sc. from Humboldt University of Berlin in 2021.
His research interests are in measurement studies of distributed and peer-to-peer systems.
\end{IEEEbiography}
\vspace{-1.5em}
\begin{IEEEbiography}[{\includegraphics[width=1in,height=1.25in,clip,keepaspectratio]{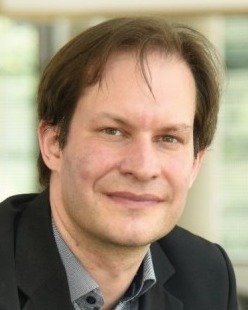}}]{Björn Scheuermann}
is a professor of Communication Networks at TU Darmstadt, Germany. He obtained his Ph.D. from the University of Düsseldorf, Germany, in 2007. After professorships in Düsseldorf, Würzburg, Bonn and at Humboldt University of Berlin he joined TU Darmstadt in 2021. His research interests include network protocol design and analysis, networked systems security and network hardware engineering.
\end{IEEEbiography}
\vspace{-1.5em}
\begin{IEEEbiography}[{\includegraphics[width=1in,height=1.25in,clip,keepaspectratio]{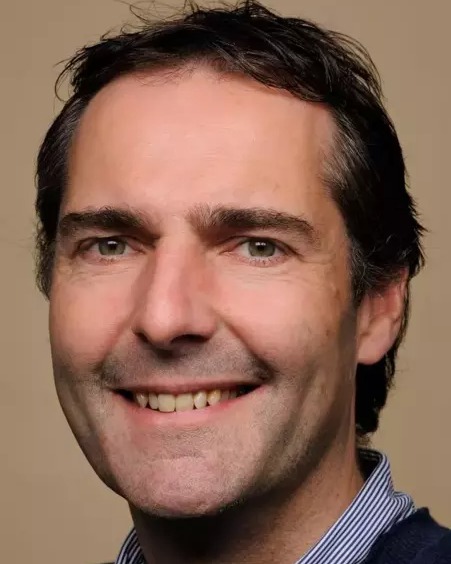}}]{Johan Pouwelse}
is an associate professor at Delft University of Technology, 
specialized in large-scale cooperative systems. 
During his Ph.D., he created the first system for cooperative resource management. The resulting driver got accepted into the Linux kernel and is still used by every Android and iOS device.
Also, he conducted the first resource usage measurements for IEEE 802.11b, known now as wifi. 
After receiving his Ph.D., he conducted one of the largest measurements of the BitTorrent P2P network. 
He founded the Tribler video-on-demand client in 2005, which has been installed by 1.8 million people over the past decade.
\end{IEEEbiography}

\clearpage

\begin{appendices}
\input{sections/appendix}
\end{appendices}


%\vfill


\end{document}

%% file: acronyms.tex
\begin{acronym}[derp]
        \acro{AE}[AE]{Asymmetric Extremum}
        \acro{RAM}[RAM]{Rapid Asymmetric Extremum}
        \acro{LMC}[LMC]{Local Maximum Chunking}
        \acro{FBC}[FBC]{Frequency-Based Chunking}
        \acro{BFBC}[BFBC]{Bytes-Frequency-Based Chunking}
        \acro{MII}[MII]{Minimal Incremental Interval}
        \acro{PCI}[PCI]{Parity Check of Interval}
        \acro{MUCH}[MUCH]{Multithreaded Content-Based File Chunking}
        \acro{CDC}[CDC]{Content-Defined Chunking}
        \acro{FSC}[FSC]{Fixed-Size Chunking}
        \acro{BSW}[BSW]{Basic Sliding Window}
        \acro{ipc}[IPC]{instructions per cycle}
        \acro{iqr}[IQR]{interquartile range}
        \acro{simd}[SIMD]{single instruction multiple data}
        \acro{rtt}[RTT]{round-trip time}
        \acro{NC}[NC]{Normalized Chunking}
        \acro{LMC}[LMC]{Local Maximum Chunking}
\end{acronym}

%% file: sections/abstract.tex
Data deduplication emerged as a powerful solution for reducing storage and bandwidth costs in cloud settings by eliminating redundancies at the level of chunks.
This has spurred the development of numerous Content-Defined Chunking (CDC) algorithms over the past two decades.
Despite advancements, the current state-of-the-art remains obscure, as a thorough and impartial analysis and comparison is lacking.
We conduct a rigorous theoretical analysis and impartial experimental comparison of several leading CDC algorithms.
Using four realistic datasets, we evaluate these algorithms against four key metrics:
throughput, deduplication ratio, average chunk size, and chunk-size variance.
Our analyses, in many instances, extend the findings of their original publications by reporting new results and putting existing ones into context.
Moreover, we highlight limitations that have previously gone unnoticed.
Our findings provide valuable insights that inform the selection and optimization of CDC algorithms for practical applications in data deduplication.

%% file: sections/introduction.tex
\IEEEPARstart{I}{n} the era of Big Data, cloud storage systems have become indispensable for managing the explosive growth of digital information~\cite{rydning2018digitization}. 
As storage is costly, these systems require efficient data reduction techniques. 
Concurrently, the advent of the Internet of Things has underscored the importance of minimizing data transfers between edge devices and central servers, often located in the cloud.
In large-scale systems, as data accumulates, it is typical for content to appear redundantly.
This results in an inefficient utilization of both bandwidth and storage.

Data deduplication emerges as a solution to this issue.
The strategy is to eliminate redundant content at a chunk-level, for instance, in blocks of \qty{8}{\kilo\byte}.
To this end, files are split into chunks and each chunk is indexed and identified by its cryptographic fingerprint.
Thereafter, a file is described as a sequence of such fingerprints.
Hence, duplicated blocks of data need to be stored or transferred only once,
and instances of the blocks can be referred to by their fingerprint.
As the size of a fingerprint %(typically 20 or 32 bytes)
is much smaller than the content it represents, this results in effective deduplication on systems where redundant content is prevalent.
Large-scale studies by Microsoft~\cite{meyer2012study,el2012primary} and EMC~\cite{wallace2012characteristics} report space savings of up to \qty{83}{\%} using this technique.
%We illustrate this process in \cref{fig:dedupProcess}.

\begin{figure}
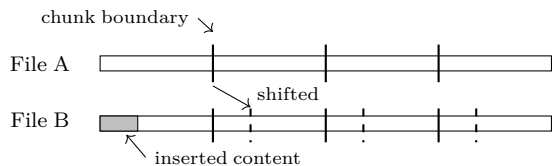

	\centering
	\include{figures/boundary-shift-problem.tex}
  \vspace{-2em}
  \caption{Boundary-shift problem.}
  \label{fig:boundaryShiftProblem}
\end{figure}

The algorithm by which the files are chunked has an important effect on deduplication.
The most straightforward solution is \ac{FSC}, where files are split into equal-sized chunks.
This strategy, however, suffers from the \emph{boundary-shift problem}.
It describes the situation that two (or more) files share similar content,
but the misalignment of their chunk boundaries hinders the detection of the existing redundancies (\cf \Cref{fig:boundaryShiftProblem}).
This problem is addressed by \ac{CDC} algorithms, which yield variable-sized chunks based on \emph{content} rather than \emph{position}.
To do this, CDC algorithms often rely on rolling hash functions, 
whose first application was the Rabin fingerprinting scheme~\cite{rabin,rabinkarp,rabinApplications}.

Over the years, numerous algorithms have been proposed,
claiming better efficiency (\ie, higher throughput), lower chunk-size variance, or better deduplication efficacy~\cite{fastcdc2,maxp,ram,pci,bfbc}.
However, a comprehensive and unbiased evaluation of these methods remains elusive. 
Each study typically presents its algorithm as the superior solution, often using curated datasets and assumptions that favor their approach. 
This fragmented landscape obscures a clear understanding of the true state-of-the-art in \ac{CDC}.
In our study, we select a set of \ac{CDC} algorithms for rigorous evaluation,
including Rabin~\cite{rabinApplications}, Buzhash~\cite{buzhash}, Gear~\cite{ddelta}, AE~\cite{ae2}, RAM~\cite{ram}, MII~\cite{mii}, PCI~\cite{pci}, and BFBC~\cite{bfbc}, and the \ac{NC} technique proposed for FastCDC~\cite{fastcdc2}. 
We reimplement these algorithms efficiently and compare their performance on four realistic datasets.
Our evaluation encompasses throughput, average chunk size and variance, and deduplication ratio.
We report new results and contrast them with existing literature.
In addition, we derive new theoretical insights, including novel formulas relating algorithm parameters to the expected average chunk size for AE, RAM, MII, and BFBC.
Moreover, we improve upon the existing formula for AE.
In summary, our research provides a comprehensive and unbiased evaluation, shedding new light on the capabilities and limitations of these algorithms.

The remainder of this article is structured as follows:
In \cref{sec:background},
we give an overview of the relevant algorithms and key techniques
that shaped the field of \ac{CDC} and define today's state-of-the-art.
Following this, in \Cref{sec:relatedWork} we outline related works in the field
of empirical measurements of \ac{CDC} algorithms.
%In addition, we elaborate on the significance of various key metrics,
%such as the chunk-size variance.
%After that, we briefly summarize related works in \cref{sec:relatedWork}.
In \cref{sec:algorithms}, we provide a detailed exposition of all the chunking algorithms 
that are subject to our in-depth analysis and comparison.
This includes remarks on their expected behavior and performance, such as the effect of their parameters on the expected chunk size.
Following this, we commence the experimental evaluation of the selected algorithms,
its detailed procedure described in \cref{sec:setup}.
The subsequent three sections present and analyze the results of our experiments 
with respect to our key metrics:
throughput (\cref{sec:efficiency}),
chunk size distribution (\cref{sec:distribution}),
and deduplication (\cref{sec:dedup}).
In \cref{sec:discussion},
we interpret interrelations between the results
and contemplate their implications in a broader context.
Finally, in \cref{sec:conclusion},
we arrive at our conclusions.

%% file: figures/boundary-shift-problem.tex
\begin{tikzpicture}

    % Define the label node and the target node
    \node[label] (cbl) at (0.2,1.7) {\scriptsize{chunk boundary}};
    \node (cb) at (1.5, 1.4) {};  % Dummy node to use as a reference for the arrow
    
    \node[label] (shifted) at (2.5, 0.7) {\scriptsize{shifted}};
    \node (cbShifted) at (2, 0.4) {};  % Dummy node to use as a reference for the arrow
    
    \node[label] (inserted) at (1.7, -0.15) {\scriptsize{inserted content}};

    % Draw an arrow from the label node to the target node
    \draw[->] (cbl.east) -- (cb.north);
    \draw[->] (1.5, 0.8) -- (cbShifted.north);
    \draw[->] (inserted.west) -- (0.35, 0.15);

    % File A
    \node at (-0.8,1.1) {\footnotesize{File A}};
    \draw[thick] (1.5, 1.35) -- (1.5, 0.85);
    \draw[thick] (3, 1.35) -- (3, 0.85);
    \draw[thick] (4.5, 1.35) -- (4.5, 0.85);
    \draw[draw=black] (0, 1) rectangle ++(6, 0.2);

    % File B
    \node at (-0.8,0.35) {\footnotesize{File B}};
    \draw[thick] (1.5, 0.5) -- (1.5, 0.05);
    \draw[thick] (3, 0.5) -- (3, 0.05);
    \draw[thick] (4.5, 0.5) -- (4.5, 0.05);
    \draw[thick,dashed] (2, 0.5) -- (2, 0.05);
    \draw[thick,dashed] (3.5, 0.5) -- (3.5, 0.05);
    \draw[thick,dashed] (5, 0.5) -- (5, 0.05);
    \draw[draw=black] (0, 0.2) rectangle ++(6, 0.2);

    % Draw a light gray rectangle
    \draw[draw=black,fill=lightgray] (0, 0.2) rectangle ++(0.5, 0.2);

\end{tikzpicture}

%% file: sections/background.tex
In the face of exponential data growth, data deduplication has emerged as a pivotal strategy for efficient data management~\cite{xia2016comprehensive}.
The algorithm by which chunking is performed poses the crucial feature by which efficacy and efficiency of the data deduplication process is determined.
This section provides an overview of the evolution of \ac{CDC} and the seminal innovations that shaped it.
%It also highlights the key challenges and considerations inherent to this technique.

%\subsection{Boundary-Shift Problem}
%\input{sections/background/boundary-shift-problem.tex}
%
%\subsection{Impact of Chunk Size}
%\input{sections/background/chunk-size.tex}
%
\subsection{Inception of CDC}
\input{sections/background/bsw.tex}

\subsection{Chunk-Size Variance}\label{sec:background:csVariance}
\input{sections/background/cs-variance.tex}
%
%\subsection{Normalized Chunking}\label{sec:background:normalized_chunking}
%\input{sections/background/normalized-chunking.tex}
%
%\subsection{Leap-Based Acceleration Techniques}
%\input{sections/background/leap-based.tex}

\subsection{Modern CDC Algorithms}
\input{sections/background/modern.tex}

%\subsection{Incremental Data Synchronization}
%\label{sec:background:incremental}
%\input{sections/background/incremental-sync.tex}
%
%\subsection{Statistical Approaches}
%\input{sections/background/rechunking-stateful.tex}
%
%\subsection{Rechunking and Stateful Algorithms}\label{sec:background:rechunking}
%\input{sections/background/rechunking-stateful.tex}

%\subsection{Parallelization}
%\input{sections/background/parallelization.tex}

%% file: sections/background/bsw.tex
\begin{figure}[h]
  \centering
  \input{figures/rabin.tex}
  \caption{Window sliding over a file byte-by-byte.}
  \label{fig:slidingWindow}
\end{figure}
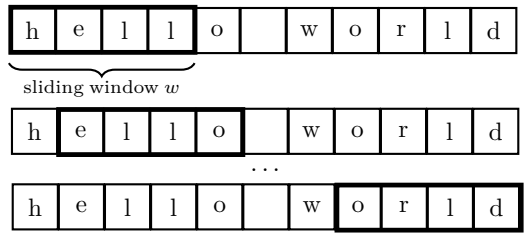

CDC algorithms avoid boundary shifting by setting chunk cut-points based not on position but \emph{content}.
Traditionally, this has been the result of hash-based comparisons on a sliding window that is iterated over a file byte-by-byte (\cf \Cref{fig:slidingWindow}).
The fingerprint on each iteration of the sliding window is compared against a bitmask to determine new chunk boundaries.
Since hash functions are deterministic,
this results in chunk boundaries that are set in a content-dependent manner.
This idea, which we refer to as \ac{BSW}, marks the advent of data deduplication.
It can be attributed to two pioneering studies from the early 2000s~\cite{lbfs,quinlan2002venti}.

%% file: figures/rabin.tex
\begin{tikzpicture}
\tikzstyle{every path}=[thick]

\edef\sizetape{0.6cm}
\tikzstyle{tmtape}=[draw,minimum size=\sizetape]
\tikzstyle{tmhead}=[arrow box,draw,minimum size=.5cm,arrow box
arrows={east:.25cm, west:0.25cm}]

%% FIRST TAPE

\begin{scope}[start chain=1 going right,node distance=-0.15mm,local bounding box=tape1]
    \draw[draw=black,line width=1.8pt] (-0.29,-0.29) rectangle ++(2.45,0.6);
    \node [on chain=1,tmtape] (start) {h};
    \node [on chain=1,tmtape] {e};
    \node [on chain=1,tmtape] {l};
    \node [on chain=1,tmtape] (end) {l};
    \node [on chain=1,tmtape] {o};
    \node [on chain=1,tmtape] {};
    \node [on chain=1,tmtape] {w};
    \node [on chain=1,tmtape] {o};
    \node [on chain=1,tmtape] {r};
    \node [on chain=1,tmtape] {l};
    \node [on chain=1,tmtape] {d};
    \draw [decorate,decoration={brace,amplitude=5pt,mirror,raise=3ex}]
        (start.west) -- (end.east) node[midway,yshift=-2.3em]{\scriptsize{sliding window $w$}};
\end{scope}
  
%% SECOND TAPE

\begin{scope}[start chain=1 going right,node distance=-0.15mm,local bounding box=tape2, shift={($(tape1.west)-(-0.34,1cm)$)}]
    \draw[draw=black,line width=1.8pt] (0.32,-0.31) rectangle ++(2.45,0.6);
    \node [on chain=1,tmtape] {h};
    \node [on chain=1,tmtape] {e};
    \node [on chain=1,tmtape] {l};
    \node [on chain=1,tmtape] {l};
    \node [on chain=1,tmtape] {o};
    \node [on chain=1,tmtape] {};
    \node [on chain=1,tmtape] {w};
    \node [on chain=1,tmtape] {o};
    \node [on chain=1,tmtape] {r};
    \node [on chain=1,tmtape] {l};
    \node [on chain=1,tmtape] {d};
\end{scope}
  
%% LAST TAPE

\begin{scope}[start chain=1 going right,node distance=-0.15mm, shift={($(tape2.west)-(-0.31,1cm)$)}]
    \node [draw=none,yshift=14,xshift=87] {$\ldots$};
    \draw[draw=black,line width=1.8pt] (4.01,-0.31) rectangle ++(2.45,0.6);
    \node [on chain=1,tmtape] {h};
    \node [on chain=1,tmtape] {e};
    \node [on chain=1,tmtape] {l};
    \node [on chain=1,tmtape] {l};
    \node [on chain=1,tmtape] {o};
    \node [on chain=1,tmtape] {};
    \node [on chain=1,tmtape] {w};
    \node [on chain=1,tmtape] {o};
    \node [on chain=1,tmtape] {r};
    \node [on chain=1,tmtape] {l};
    \node [on chain=1,tmtape] {d};
\end{scope}

\end{tikzpicture}

%% file: sections/background/cs-variance.tex
Early on, BSW-based CDC was criticized for two shortcomings:
low throughput and high chunk-size variance.
While the problem of low throughput was ameliorated by more efficient hash algorithms~\cite{buzhash,ddelta},
high chunk-size variance remained a problem inherent to the \ac{BSW} approach.
High chunk-size variance gives rise to the issue of pathological chunk sizes.
%, \ie,
%chunks that are either extremely small or very large relative to the target chunk size.
%% Problem with large chunks
Very large chunks can be the product of a recurring pattern in the byte sequence
which happens to not meet the criteria for setting a chunk boundary~\cite{schleimer2003winnowing}.
%This is a typical phenomenon in real data
%and a crucial weak point of the BSW algorithm.
These chunks are undesirable
because they impair deduplication:
First, large chunks are generally more difficult to deduplicate
as the chances for smaller chunks of data to be redundant
% in any dataset
are higher.
% than for larger chunks.
Second, when an existing file is modified,
this modification is more likely to affect a larger chunk than it is to affect a small chunk,
%This intuition was formalized in~\cite{tttd}.
%Larger affected chunks
which in turn leads to a higher number of of bytes 
affected by the modification,
hence a negative effect on deduplication.
%% Problem with small chunks
Pathologically small chunks are undesirable as well
because they produce more metadata, more computation overhead, 
and, in distributed settings, greater overhead due to round trip times.

%% Problem with min/max settings
%The BSW algorithm is therefore often complemented by setting minimum and maximum chunk size constraints~\cite{lbfs,rabinPrimary,rabinNetwork,lu2019empirical}.
%That is, no chunk boundary is formed before reaching the minimum size,
%and a chunk boundary is forced after the chunk reaches the maximum size.
%This solves pathological chunk sizes at the cost of content-dependence
%and therefore deduplication performance.
%The chunk-size distribution of BSW, considering uniformly distributed random data input,
%exhibits a logarithmic decline in the frequency of growing chunk sizes:
%The probability for the first byte in a new chunk to become a boundary is 
%$\mathrm{P}(B_1)=2^{-b}$, 
%therefore the probability for the second byte to become a boundary is $\mathrm{P}(B_2)=2^{-b}-\mathrm{P}(B_1)$; 
%for the $i$th byte it is $\mathrm{P}(B_i)=2^{-b} (1-2^{-b})^{i-1}$.
%Considering the long-tailed chunk size distribution,
%and its accumulation at both the minimum and maximum chunk size boundary,
%the number of affected chunks can be significant.
%This is because the first possible byte is the most likely to be selected a chunk boundary,
%and the last possible byte (the maximum) accumulates the likelihood for all succeeding position bytes.

%% file: sections/background/modern.tex
High chunk-size variance and low throughput led to the emergence of an alternative approach.
Starting in 2009, researchers started proposing CDC algorithms
based on the identification of local extrema in the input data~\cite{maxp,lmc,ae,ram}.
By using byte comparisons rather than hash functions, these algorithms claim to achieve higher throughput than BSW algorithms.
Furthermore, they are attributed with a significantly lower chunk-size variance~\cite{lmc,ae}.

Later on, researchers focused on the specific application of CDC for incremental synchronization, as for example in rsync~\cite{rsync}.
This use case does not consider data reduction in storage systems, but the incremental synchronization of data between machines.
Files are chunked on both ends to determine the segments of data that need to be transmitted.
However, the produced chunks are never stored.
This condition puts a relaxation on the constraint for low chunk-size variance.
We find this focus particularly in the works of Zhang \etal in 2019~\cite{mii} and 2020~\cite{pci}.
Those algorithms do not fundamentally differ, however, from the BSW or extremum-based approaches.

Lastly, the third and most recent approach is chunking based on a dynamically predetermined set of divisors, and emerged in 2020~\cite{bfbc}.
The technique relies on a statistical analysis of the expected dataset.
Specifically, the algorithms will conduct a statistical frequency estimation of byte pairs~\cite{bfbc,jehlol2022big} 
(or triplets~\cite{jehlol2023enhancing}),
and, based on that, determine a set of byte pair/triplet divisors.
The matching condition is thus reduced to a simple table lookup, promising superior throughput compared to classical CDC algorithms.

%% file: sections/related-work.tex
Previous works have studied the state-of-the-art of data deduplication and chunking algorithms~\cite{xia2016comprehensive,goel2023detailed}.
However, these studies primarily focus on theoretical discussions based on the original works that introduced the algorithms.
Few researchers have attempted to reproduce the results or investigate the state-of-the-art through experimental means.
Most experimental evaluations and comparisons of chunking algorithms are part of works introducing new algorithms~\cite{ae,ram,pci,ellappan2021dynamic}.
We find that there is no consistent set of datasets and comparable evaluation methodologies to judge these algorithms based on their original publications.

In their work~\cite{ellappan2021dynamic}, Ellapan \etal present the superior throughput of their own algorithm.
In their measurements, however, they did not isolate the effects of the chunking algorithm from the computationally expensive SHA-1 fingerprinting applied to the produced chunks.
Consequently, the throughput is heavily skewed in favor of algorithms that generate a smaller number of chunks.
Notably, their own algorithm produces the fewest chunks across all evaluated algorithms and datasets.

The authors of PCI~\cite{pci} compare their algorithm against Rabin, LMC, AE, RAM, and MII.
The datasets used in the experiments are artificial, based on sequences of zeros with random byte insertions or deletions in defined intervals.
%They argue that an evaluation based on randomly generated rather than real data is necessary
%to filter out any special effects individual algorithms might have on certain datasets, which could otherwise lead to inconsistent conclusions.
%Although their results are interesting,
%as they demonstrate algorithms' resistance to byte shifting,
%we believe that realistic datasets hold special value.
Ultimately, chunking algorithms are applied to real-world data, which is not reflected in the datasets used by the authors.
For instance, the issue of low-entropy strings is disregarded within this experiment.
%While this is less of an issue for data synchronization, \acp{rtt} remain relevant in this context.

We find one study that compares specifically the throughput of Rabin, LMC, AE, RAM, and PCI on random datasets~\cite{viji2021comparative}.
In their experiment, RAM significantly outperformed the other algorithms,
followed by AE and PCI;
results that our own experimental study confirms as well.
While useful, this work lacks \emph{realistic} datasets, making it difficult to derive actionable recommendations from it.

%% file: sections/algorithms.tex
\begin{figure}[tb]
	\centering
	\input{figures/algos-tree.tex}
	\vspace{-1em}
	\caption{Taxonomy of evaluated chunking algorithms.}
	\label{fig:algosTree}
\end{figure}
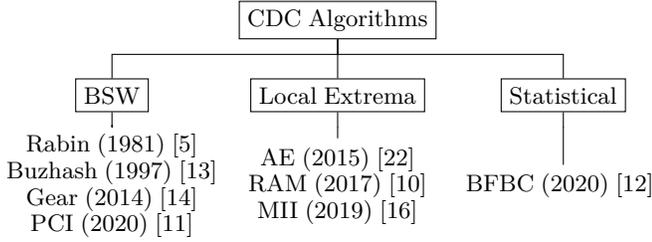

%In our in-depth literature review, we have delved into the field of data deduplication,
%specifically the history and evolution of \ac{CDC} algorithms.
We conducted a thorough investigation of related literature to identify a set of state-of-the-art algorithms to carry on for our theoretical analysis and experimental comparison.
This includes recent as well as traditional \ac{CDC} algorithms (\cf \Cref{fig:algosTree}).
%Notably, although the hash algorithms of Rabin and Buzhash are much older,
%we did not find them applied in the context of data deduplication prior to the introduction of LBFS~\cite{lbfs} in 2001.
In this section, we reintroduce these algorithms with detailed technical descriptions.
We examine the algorithms through the lens of theoretical behavior,
extending the descriptions and derivations made by the original authors where possible.
The pseudocode to our implementations of the chunking algorithms can be found in \hyperref[app:algos]{Appendix~A}.
%Our contribution further extends to stochastic analyses of algorithms, where these were either not provided in the original publication, or were done incorrectly.

\subsection{Basic Sliding Window (BSW)}\label{sec:algorithms:bsw}
\input{sections/algorithms/bsw.tex}

%\subsection{QuickCDC}\label{sec:algorithms:quickcdc}
%\input{sections/algorithms/quickcdc.tex}

\subsection{Asymmetric Extremum (AE)}\label{sec:algorithms:ae}
\input{sections/algorithms/ae.tex}

\subsection{Rapid Asymmetric Extremum (RAM)}\label{sec:algorithms:ram}
\input{sections/algorithms/ram.tex}

\subsection{Minimal Incremental Interval (MII)}\label{sec:algorithms:mii}
\input{sections/algorithms/mii.tex}

\subsection{Parity Check of Interval (PCI)}\label{sec:algorithms:pci}
\input{sections/algorithms/pci.tex}

\subsection{Bytes-Frequency--Based Chunking (BFBC)}\label{sec:algorithms:bfbc}
\input{sections/algorithms/bfbc.tex}

%% file: figures/algos-tree.tex
\small
\begin{tikzpicture}[
    node distance = 1cm,
    every node/.style = {draw, rectangle},
    level 1/.style={sibling distance=3cm},
    level 2/.style={sibling distance=3cm},
    level 3/.style={sibling distance=3cm},
    level distance = 1cm,
    edge from parent path={
        (\tikzparentnode.south)
        -- ++(0,-0.2cm)
        -| (\tikzchildnode.north)
    }
]
\node {CDC Algorithms}
    child { node {BSW}
                child { node[align=center,draw=none,yshift=-0.2cm] {
                    Rabin (1981) \cite{rabin}\\
                	Buzhash (1997) \cite{buzhash} \\
                	Gear (2014) \cite{ddelta}\\
                	PCI (2020) \cite{pci}
                }
        }
    }
    child { node {Local Extrema}
        child { node[align=center,draw=none,yshift=-0.2cm] {
%            MAXP (2009) \cite{maxp}\\
%            LMC (2010) \cite{lmc}\\
            %LeapCDC (2015) \cite{leapBased}\\
            AE (2015) \cite{ae}\\
            RAM (2017) \cite{ram}\\
            MII (2019) \cite{mii}
        } }
    }
    child { node {Statistical}
        child { node[align=center,draw=none,yshift=-0.2cm] {
%            SDM (2017) \cite{sdm}\\
%            RDM (2017) \cite{rdm}\\
            BFBC (2020) \cite{bfbc}
        } }
%        child { node {Rechunking}
%            child { node[align=center,draw=none,yshift=-1.2cm] {
%                Fingerdiff (2006) \cite{fingerdiff}\\
%                FBC (2010) \cite{fbc}\\
%                Bimodal (2010) \cite{bimodal}\\
%                Anchor-driven subchunk (2011) \cite{anchorDrivenSubchunk}\\
%                Heysteresis (2013) \cite{zhou2013hysteresis}
%            } }
%        }
    };
\end{tikzpicture}

%% file: sections/algorithms/bsw.tex
\acused{BSW}
\ac{BSW} algorithms operate by sliding a fixed-size window of size $w$ over the stream of input data,
deriving a fingerprint for the current window using a function $H$,
and emitting a chunk boundary if the calculated fingerprint fulfills a given condition.
%Generally, these algorithms implement a minimum chunk size equal to the window size $w$,
%since the window is usually pre-filled completely first, before fingerprints are derived.
The \ac{BSW} variants differ in the hash function producing the fingerprint, the function judging the fingerprint, and the choice of window size.
Typically, \emph{rolling hash} functions are used for their efficiency.
These functions can update their output in constant time when used in a sliding window.
Further, the judging function usually checks for a number $b$ of least-significant bits to be zero.
If $H$ is distributed uniformly at random, this can then be expected to occur for any window with a probability of $2^{-b}$.
%Generally, if $H$ is assumed to be distributed uniformly at random,
%it does not matter \emph{which} $b$ bits are tested to be zero.
Therefore, an average chunk size of $\mu$ can be aimed by setting $b=log_2(\mu)$.

% The general algorithm is shown in \Cref{alg:bsw},
% where $H$ denotes a function to derive a fingerprint from the contents of a window,
% and the judging function simply checks whether the $b$ least-significant bits of the fingerprint are zero.

%\begin{figure*}[tbp]
%\centering
%
%\subfloat[Rabin]{\scalebox{0.85}{\input{fig/hash_value_distribution_rabin}}}
%    \label{fig:hash_rabin}
%\hfil
%\subfloat[Buzhash]{\scalebox{0.85}{\input{fig/hash_value_distribution_buzhash}}}
%    \label{fig:hash_buzhash}
%\hfil
%\subfloat[Gear64]{\scalebox{0.85}{\input{fig/hash_value_distribution_gear64}}}
%    \label{fig:hash_gear64}
%\hfil
%\subfloat[Adler32]{\scalebox{0.85}{\input{fig/hash_value_distribution_adler32}}}
%    \label{fig:hash_adler32}
%
%\vspace{-1em}
%\input{fig/hash_value_distribution_legendonly}
%\vspace{-2em}
%
%\caption{Hash value distributions for different BSW algorithms. In (a) and (b), the plots are all overlapping as they yield similar results; In (c), we only show \qty{64}{B} as it is the only legal window size with Gear64.
%	\leo{this figure is nice, but not referenced :(}
%	\leo{also, weirdly, the subfloat captions use a different font}
%}
%\label{fig:hash}
%\end{figure*}

%\subsubsection{Variants}

BSW variants differ in the hash function they utilize.
Rabin-based chunking is the first prominent application of the BSW algorithm, and moreover of CDC in general \cite{lbfs,rabinNetwork,rabinPrimary}.
It is rooted in the fingerprinting schema presented in~\cite{rabin,rabinApplications}.
Rabin was often criticized for being slow~\cite{fastcdc,sun2015redundant}, which spurred the development of more efficient rolling hash functions.
Hashing by cyclic polynomials~\cite{buzhash}, 
or \textit{Buzhash},
presents a more efficient rolling hash function.
Another efficient implementation was presented as Gear~\cite{ddelta}.
Due to its shifting behavior,
the matching condition for Gear uses the \emph{most} significant bits of $H$.
%We postulate that it should be possible to reduce the effective window size by moving the position of the matching mask,
%but do not investigate this further.
In \cref{table:bswVariants}, we show how to compute both the first and consecutive hashes for a sliding window over a stream of bytes.
In the presented formulas, $x$ is a prime number, 
$\rho^b(x)$ denotes a binary rotation of $x$ by $b$ bits,
and $h: [0,255] \mapsto [0,2^{32}]$ denotes a predefined table.
We note that we experimentally verified uniform hash value distributions for these functions,
as this poses a vital criterion for effective CDC~\cite{chapuis2016throughput}.

\begin{table*}[t]
    \centering
    \caption{Rolling Hash Functions With Initial Computation ($H_\text{prev}$) and Update Method ($H_\text{next}$)}
    \label{table:bswVariants}
    \input{tables/bsw-variants.tex}
\end{table*}

%\paragraph{Parallel Processing}
%\label{sec:algorithms:bsw:simd}

Another optimization applicable to BSW algorithms is data-parallelism.
This is a common optimization technique with dedicated instructions on all modern instruction sets \cite{sscdc,xia2023design}.
It is generally not obvious how to parallelize CDC algorithms to operate on different parts of the input data,
due to the boundaries of \emph{previous} blocks affecting the \emph{current} state of the algorithm.
%In some cases, and with care, however, algorithms can be vectorized~\cite{amiri2020simd}.
%We will examine one such example in the case of $64$-bit Gear in the following.
\Ac{BSW} algorithms have a property that can be exploited in this regard:
They operate on a fixed window, \ie, the number of input bytes affecting their current state is limited.
%Notably, Gear also has this property although it does not explicitly operate on a window, because the shifting operations limit the window size implicitly to the width of the hash.
%For $64$-bit Gear, this limit is \qty{64}{\byte}.
Using this property, it is possible to write data-parallel versions of these algorithms using \ac{simd} instructions.
For Gear, in particular, this is made easy due to the simplicity of the algorithm itself,
and \ac{simd} implementations exist.\footnote{\eg, \url{https://crates.io/crates/gearhash}}
%We use an existing implementation
%\footnote{\url{https://crates.io/crates/gearhash}}
%of $64$-bit Gear using \ac{simd} instructions
%and compare it to the scalar variant.

%We show an overview of the achievable throughput of the scalar and SIMD variants on the random dataset in \Cref{fig:perf_gear_simd_throughput_comparison_chunk_sizes_random}
%and microarchitectural counters in \Cref{tab:perf_gear_simd_comparison_random}.
%In general, we can see that the SIMD implementation is much faster on our machine, with factors between $\approx 1.5$ to $\approx 2.8$.
%Furthermore, the difference between the two implementations becomes more pronounced for larger target chunk sizes.
%This is expected from the previous discussion, \ie, the code falls back to a scalar version less frequently for larger target chunk sizes
%since fewer chunk boundaries are found.
%In terms of microarchitectural performance,
%we can see that the larger the target chunk size, the fewer instructions per byte are utilized by the SIMD version,
%which again follows from the code falling back to scalar code less frequently.
%The same applies to branches.
%In conclusion, while not always possible, data-parallelism can offer a large increase in performance.

%% file: tables/bsw-variants.tex
% \small
\begin{tabular}{lll}
\toprule
Rolling Hash & $H_\text{prev} = H(B_1,\ldots,B_w)$ & $H_\text{next} = H(B_2,\ldots,B_{w+1})$ \\
\midrule
Rabin~\cite{rabin} & $B_1x^{w-1} + B_2x^{w-2} + \ldots + B_{w}$ & $(H_\text{prev} - B_1x^{w-1})x+B_{w+1}$ \\
Buzhash~\cite{buzhash} & $\rho^{w-1}(h(B_1)) \xor \rho^{w-2}(h(B_2)) \xor \ldots \xor h(B_w)$ & $\rho(H_\text{prev}) \xor \rho^w(h(B_1)) \xor h(B_{w+1})$ \\
Gear~\cite{ddelta} & $h(B_1)\cdot 2^{w-1} + h(B_2)\cdot 2^{w-2} + \ldots + h(B_w)$ & $(H_\text{prev} \ll 1) + h(B_{w+1}) \mod 2^w$ \\
\bottomrule
\end{tabular}

%% file: sections/algorithms/ae.tex
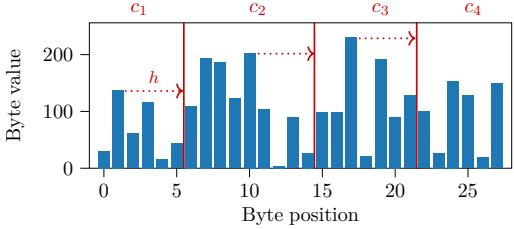
\begin{figure}[h]
  \centering
  \scalebox{0.8}{\input{figures/ae.tex}}
  \caption{Illustration of the AE chunking algorithm on a sequence of 27 bytes and a horizon $h=4$. The vertical red lines mark the cut points which then determine the resulting chunks $c_i$.
  }
  \label{fig:ae}
\end{figure}

%The authors of \ac{AE}~\cite{ae,ae2} noticed that the chunk throughput in \ac{LMC}~\cite{lmc}
%can be improved by employing an asymmetric instead of a symmetric window.
%This is because of the necessity to backtrack every value in $[i-h,i)$ 
%($h$ is growing dynamically)
%before declaring a new chunk boundary.
\acused{AE}
AE~\cite{ae,ae2} emerged from the alternative line of extremum-based approaches, \ie, it does not rely on hash functions.
It determines chunk boundaries based on an extreme value within an asymmetric window.
The window is comprised of a fixed-size horizon of length $h$ (to the right) and another dynamically sized horizon (to the left).
A chunk boundary is declared at index $i+h$ if the byte at $i$ is the local extremum from the previous chunk cut-point to $i+h$;
more precisely, if $B_i > \max\{B_j\}_{j=1}^{i-1} \land B_i \geq \max\{B_j\}_{j=i+1}^{i+h}$.
We illustrate the chunking mechanism in \cref{fig:ae}.% and present the algorithm in \cref{alg:ae}.

The average chunk size in AE largely depends on the parameter setting $h$. 
Larger values yield larger chunks.
In order to produce chunks of specific average size $\mu$,
understanding the relationship between $\mu$ and $h$ is crucial.
The authors suggest that the average chunk size on random data input is expected to be $(e-1)\cdot h$, therefore $h = \mu/(e-1)$.
This formula has been implemented in their open-source testing framework Destor\footnote{\url{https://github.com/fomy/destor}},
which was used in various works as the basis for the experimental evaluation of AE in comparison to other algorithms
\cite{ellappan2021dynamic,fastcdc2,jin2023accelerating,ellappan2023smart},
as well as the more recent benchmarking tool DedupBench~\cite{liu2023dedupbench}.
Our experimental as well as theoretical analysis, however, suggests that this formula cannot be used to accurately predict the average size of produced chunks.
When employing it to determine the parameter $h$,
AE yields means significantly lower than the target.
%AE persistently yielded chunk size means lower than expected.
%For example, with an expected average of \qty{8}{KiB} 
%(\ie, $h=\lfloor 8192/(e-1)\rceil=4768$),
%the actual mean turned out at \qty{5023}{B},
%a deviation of almost \qty{40}{\%}.
We observed these results on uniformly distributed random data,
motivating us to revisit the mechanism to determine $h$.
Hereby, we notice that the authors disregard
not only the discrete value range of the random variables in $[0,255]$,
but also the conditioning of the probabilities based on dependent events.
Thus, we conduct our own stochastic analysis and come to the following conclusions:
If $h$ is large, the extreme value is likely to be 255, and therefore the probability for an unknown random byte to match this value is $\frac{1}{256}$.
Based on this reasoning, we propose using the approximation $h\approx\mu -256$ where $\mu\geq\qty{2}{KiB}$.
For smaller target chunk sizes, we rely on empirical evaluations, the results which we list in \cref{table:aeH}.

\begin{table}[h]
    \centering
    \caption{Empirical Results for Parameter $h$ in AE for Target Chunk Sizes $\mu<\qty{2}{KiB}$}
    \label{table:aeH}
    \input{tables/ae-h.tex} 
\end{table}

%% file: figures/ae.tex
% This file was created with tikzplotlib v0.10.1.
\begin{tikzpicture}

\tikzstyle{line} = [draw, latex'-latex']

\definecolor{darkgray176}{RGB}{176,176,176}
\definecolor{steelblue31119180}{RGB}{31,119,180}

%\node[draw=none,color=steelblue31119180] at (0.52,1.55) {\footnotesize $i$};
\draw (0.83,2.65) node[draw=none,color=red!80!black,align=center] {$c_1$};
\path [->,draw=red!80!black,dotted,thick] (0.59,1.285) -- node [text width=1.2cm,color=red!80!black,midway,above,align=center] {\footnotesize $h$} (1.56,1.285);
\path [draw=red!80!black,thick] (1.57,0) -- node [] {} (1.57,2.42);

\draw (2.805,2.65) node[draw=none,color=red!80!black,align=center] {$c_2$};
\path [->,draw=red!80!black,dotted,thick] (2.8,1.905) -- node [text width=1.2cm,color=red!80!black,midway,above,align=center] {} (3.74,1.905);
\path [draw=red!80!black,thick] (3.74,0) -- node [] {} (3.74,2.42);

\draw (4.845,2.65) node[draw=none,color=red!80!black,align=center] {$c_3$};
\path [->,draw=red!80!black,dotted,thick] (4.48,2.16) -- node [text width=1.2cm,color=red!80!black,midway,above,align=center] {} (5.42,2.16);
\path [draw=red!80!black,thick] (5.44,0) -- node [] {} (5.44,2.42);

\draw (6.38,2.65) node[draw=none,color=red!80!black,align=center] {$c_4$};
        
\begin{axis}[
tick align=outside,
tick pos=left,
xlabel = {Byte position},
ylabel = {Byte value},
x grid style={darkgray176},
xmin=-1, xmax=28,
xtick style={color=black},
y grid style={darkgray176},
ymin=0, ymax=256,
ytick style={color=black},
height=4cm,width=8.6cm
]
\draw[draw=none,fill=steelblue31119180] (axis cs:-0.4,0) rectangle (axis cs:0.4,30);
\draw[draw=none,fill=steelblue31119180] (axis cs:0.6,0) rectangle (axis cs:1.4,138);
\draw[draw=none,fill=steelblue31119180] (axis cs:1.6,0) rectangle (axis cs:2.4,61);
\draw[draw=none,fill=steelblue31119180] (axis cs:2.6,0) rectangle (axis cs:3.4,117);
\draw[draw=none,fill=steelblue31119180] (axis cs:3.6,0) rectangle (axis cs:4.4,16);
\draw[draw=none,fill=steelblue31119180] (axis cs:4.6,0) rectangle (axis cs:5.4,44);
\draw[draw=none,fill=steelblue31119180] (axis cs:5.6,0) rectangle (axis cs:6.4,110);
\draw[draw=none,fill=steelblue31119180] (axis cs:6.6,0) rectangle (axis cs:7.4,194);
\draw[draw=none,fill=steelblue31119180] (axis cs:7.6,0) rectangle (axis cs:8.4,186);
\draw[draw=none,fill=steelblue31119180] (axis cs:8.6,0) rectangle (axis cs:9.4,123);
\draw[draw=none,fill=steelblue31119180] (axis cs:9.6,0) rectangle (axis cs:10.4,203);
\draw[draw=none,fill=steelblue31119180] (axis cs:10.6,0) rectangle (axis cs:11.4,104);
\draw[draw=none,fill=steelblue31119180] (axis cs:11.6,0) rectangle (axis cs:12.4,4);
\draw[draw=none,fill=steelblue31119180] (axis cs:12.6,0) rectangle (axis cs:13.4,89);
\draw[draw=none,fill=steelblue31119180] (axis cs:13.6,0) rectangle (axis cs:14.4,27);
\draw[draw=none,fill=steelblue31119180] (axis cs:14.6,0) rectangle (axis cs:15.4,98);
\draw[draw=none,fill=steelblue31119180] (axis cs:15.6,0) rectangle (axis cs:16.4,98);
\draw[draw=none,fill=steelblue31119180] (axis cs:16.6,0) rectangle (axis cs:17.4,230);
\draw[draw=none,fill=steelblue31119180] (axis cs:17.6,0) rectangle (axis cs:18.4,21);
\draw[draw=none,fill=steelblue31119180] (axis cs:18.6,0) rectangle (axis cs:19.4,192);
\draw[draw=none,fill=steelblue31119180] (axis cs:19.6,0) rectangle (axis cs:20.4,90);
\draw[draw=none,fill=steelblue31119180] (axis cs:20.6,0) rectangle (axis cs:21.4,128);
\draw[draw=none,fill=steelblue31119180] (axis cs:21.6,0) rectangle (axis cs:22.4,100);
\draw[draw=none,fill=steelblue31119180] (axis cs:22.6,0) rectangle (axis cs:23.4,27);
\draw[draw=none,fill=steelblue31119180] (axis cs:23.6,0) rectangle (axis cs:24.4,154);
\draw[draw=none,fill=steelblue31119180] (axis cs:24.6,0) rectangle (axis cs:25.4,128);
\draw[draw=none,fill=steelblue31119180] (axis cs:25.6,0) rectangle (axis cs:26.4,20);
\draw[draw=none,fill=steelblue31119180] (axis cs:26.6,0) rectangle (axis cs:27.4,149);
\end{axis}

\end{tikzpicture}

%% file: tables/ae-h.tex
% \small
\begin{tabular}{rrrrr}
\toprule
$\mu$ & $h$ \\
\midrule
512 & 348 \\
770 & 563 \\
1024 & 793 \\
\bottomrule
\end{tabular}

%% file: sections/algorithms/ram.tex
\acused{RAM}
The throughput of AE has been improved even further in \ac{RAM}~\cite{ram}.
This algorithm essentially swaps the order of the fixed and dynamically sized windows.
It first employs a fixed-sized window, in which it finds the maximum value $x$;
it then determines the first byte that is larger or equal $x$ as the next cut-point.
This requires fewer comparison operations than AE and is stated to be $\approx\qty{25}{\%}$ faster.

The authors note that the performance of RAM can suffer when the given data has low entropy.
This is apparent for large $x$, as it becomes ever more unlikely to find any byte $\geq x$.
%This can happen, for example, when in a text-based file the fixed window contains the special character `€' (whose encoding has a high byte value),
%and is then followed by a long sequence of only ASCII-characters (whose byte range is lower).
%In that case, the chunk could even grow infinitely.
To counteract this behavior, the authors recommend setting a maximum chunk size,
trading off deduplication efficacy.
As our objective is to measure the inherent characteristics of the algorithms,
and give them a fair comparison,
we deliberately do not impose any such limit on our implementation.% (\cf \cref{alg:ram}).

The authors do not explicitly state how to tune the algorithm's parameter $h$
for desired average chunk sizes.
We analytically derive the relationship between $\mu$ and $h$ as expressed in \cref{eq:ramH}.
By solving for $h$ numerically, this formula can be used to tune RAM for different target chunk sizes.
Our experimental evaluation supports this derivation with near-perfect empirical means on uniformly distributed random data.
%We expand on our extensive calculations in \cref{app:ram_probs}.

\begin{equation}
	\mu=h + \left(1-\frac{\sum_{m=0}^{255} \left(m \left(\left(\frac{m+1}{256}\right)^h - \left(\frac{m}{256}\right)^h\right)\right)}{256}\right)^{-1}
\label{eq:ramH}
\end{equation}

%% file: sections/algorithms/mii.tex
\acused{MII}
\ac{MII}~\cite{mii} is another instance of extremum-based algorithms, with less emphasis on the achievement of low chunk-size variance.
It applies a fixed window $w$ and determines a chunk cut-point after each byte position $i$ 
for which the predicate $B_i>B_{i-1}>\ldots>B_{i-w}$ is true.
In simpler words, \ac{MII} sets a chunk boundary after an incremental interval of length $w$.

Larger intervals, \ie, higher $w$, will accordingly lead to larger chunks.
An exact formula is not provided by the authors.
For the purpose of our own experiments,
we propose \cref{eq:miiAvg}.
The rationale for this formula is that in every window of $w$ bytes,
there exist $256^w$ possible combinations.
For the above-stated predicate to be true,
every byte in the window must occur uniquely.
Furthermore, for every possible combination of $w$ distinct values,
there exists exactly one order in which they are ascending.
The total number of this possibility is captured in $\binom{256}{w}$.

\begin{equation}\label{eq:miiAvg}
	\mu(w) = \left(\frac{\binom{256}{w}}{256^w}\right)^{-1}+w
\end{equation}

This equation shows, as also the authors have mentioned, the factorial growth of the average chunk size with growing $w$
and the weak control over it that comes with that.
As our experiments reveal (\cf \cref{tab:csd_means_sd_full}),
this formula predicts the average chunk size with an offset of $\approx -\qty{15}{\%}$.
We speculate this is due to dependent probabilities of not matching the previous window,
which we have not regarded in our formula.
%We speculate that this is due to the conditioning of the expectation about the values in any given window
%based on the failure of preceding windows to match the predicate.
%This is something we have disregarded in our formula.

%% file: sections/algorithms/pci.tex
\acused{PCI}
The authors of \ac{MII} later proposed \ac{PCI}~\cite{pci} as an improvement to \ac{MII},
which they critiqued for having weak control over the average chunk size.
%As we have shown, tuning $w$ has a factorial growth effect on the expected average chunk size.
\ac{PCI} works schematically similar to the \ac{BSW} algorithm,
but uses the popcount, \ie, the number of 1-bits in the sliding window,
instead of a hash function.
A chunk boundary is set if the popcount exceeds a specific threshold~$\theta$.
It shares properties of rolling hash algorithms,
as subsequent iterations require only the removal of the popcount of the leftmost byte 
and the addition of the popcount of the rightmost byte.
We note that the pseudocode in the original publication omits this optimization.
%we present the optimized version in \cref{alg:pciOpt}.

Contrary to the statement in the original paper,
the popcount is not subject to a discrete \textit{uniform} 
but rather a discrete \textit{binomial} distribution 
for uniformly distributed random byte sequences.
Specifically, this leads to a probability of $\binom{8w}{\theta} \cdot 2^{-8w}$ 
for every popcount $\theta\in[0,8w]$ in a window of $w$ bytes.

The average chunk size is determined by the ratio between $\theta$ and $w$,
rather than their absolute value.
Because of the binomial probability distribution 
for $\theta$ bits in $8w$ bits to be 1-bits,
the average chunk size grows superlinearly with increasing $\frac{\theta}{8w}$,
if $\frac{\theta}{8w} > 0.5$.
However, if $w$ can be chosen freely,
any granularity of $\frac{\theta}{8w}$, and thereby virtually any average chunk size, can be targeted.
Note, $w$ simultaneously sets an implicit lower bound on the chunk size.

Since the expectation of the popcount in a sliding window is influenced by the validation of the matching condition on preceding bytes,
the aforementioned formula does not conclusively inform about the frequency of chunk cut-points given $w$ and $\theta$.
%As we conclude in \cref{app:pci_probs},
%those parameters are best determined empirically.
As an exact solution is out of scope of our work, we determine the parameters empirically for our experiments.
We run a simulation of the algorithm on a sequence of \qty{10}{MB} of data for all possible parameters in the range $w=[32,64]$.
The results are shown in \cref{table:pciParams}.
In the last column, we show the ratio between popcount-threshold and window size.
This leads to an interesting observation:
The naive assumption is that the average chunk size is subordinate to the ratio $\frac{\theta}{w}$, rather than their absolute values.
However, one must also acknowledge the role of $w$ as an implicit lower bound on the chunk size.
Ultimately, both $w$ itself as well as the ratio $\frac{\theta}{w}$ influence the produced chunk size.

\begin{table}[t]
    \centering
    \caption{Empirical Results for PCI Parameters $w,\theta$ to Approximate Specific Targets $\mu$ as Obtained in Simulation}
    \label{table:pciParams}
    \input{tables/pci-params.tex} 
\end{table}

%% file: tables/pci-params.tex
\newcommand{\pciParamRow}[2]{\num{#1} & \num[evaluate-expression]{#1*8} & \num{#2} & \num[evaluate-expression, round-mode=places, round-precision=3]{#2/(8*#1)}}
% \small
\begin{tabularx}{0.36\textwidth}{rrrrr}
\toprule
$\mu$ & $w$ bytes & $8w$ bits & $\theta$ & $\frac{\theta}{8w}$ \\
\midrule
512 & \pciParamRow{58}{253} \\
770 & \pciParamRow{40}{181} \\
1024 & \pciParamRow{34}{157} \\
2048 & \pciParamRow{61}{273} \\
4096 & \pciParamRow{39}{183} \\
5482 & \pciParamRow{56}{256} \\
8192 & \pciParamRow{57}{262} \\
\bottomrule
\end{tabularx}

%% file: sections/algorithms/bfbc.tex
\acused{BFBC}
\ac{BFBC}~\cite{bfbc} operates differently than the other algorithms in that it is tailored to the dataset.
The initialization is composed of a statistical frequency analysis.
This analysis identifies the top-$k$ frequent byte pairs, which are then used as divisors in the chunking process, alongside minimum and maximum chunk size thresholds. 
This approach aims to be faster but also achieve superior deduplication compared to traditional CDC algorithms.

As the original publication does not indicate implementation details,
the data structure to hold the set of divisors presents an interesting design choice.
In our implementation we utilize a \qty{8}{\kibi\byte} bitset,
which results in constant-time lookups, regardless of the number of divisors.

Meeting a desired chunk size is challenging and depends
on the distribution of byte pairs and therefore the content of the dataset itself.
In our experiments, we noticed that the most frequent byte pairs in realistic datasets tend to occur excessively
(\eg, \texttt{NULL-NULL}, or \texttt{/>} in HTML files).
Using such a byte pair as a divisor leads to chunk boundaries often created a few bytes after the minimum threshold.
As we require comparable average chunk sizes for our analysis,
and using the minimum chunk size as a means to control it negatively impacts the deduplication ratio,
we design an algorithm to select a set of divisors that would result in the desired average chunk size.
In our experiments, we run this modification of BFBC as an additional algorithm, denoted BFBC*.
It only differs in the procedure which determines its divisors, explained in the following.

\subsubsection{Determining BFBC* Divisors}\label{sec:bfbc_star}

We propose an algorithm that finds a set of divisors
in the list of frequent byte pairs that match a given target chunk size $\mu$
\wrt a minimum chunk length $\lambda_\text{min}$.
For our explanations, we will denote $F$ as an ordered list of only the frequencies of the most-frequently occurring byte pairs in descending order,
\eg, $F=(10112, 8435, 8003, \ldots)$.
Furthermore, $D$ represents the set of indices of $F$ used as divisors.

We can further say, without consideration of $\lambda_\text{min}$, 
that the subjected file will be split in as many chunks
as the accumulated frequencies over all divisors, plus $1$.
Knowing the file’s length $l$, we can determine the definitive average chunk size.
When considering $\lambda_\text{min}>0$,
we can set up the assumption that the divisors are uniformly distributed across the file,
and then simply subtract the number of hypothetical chunks multiplied with $\lambda_\text{min}$
to get an adjusted average chunk size.
Finally, this leaves us with \cref{eq:bfbcAvg} 
for the calculation of the expected average chunk size.

\begin{equation}\label{eq:bfbcAvg}
	\mu(D) = \frac{l}{1+\sum_{i\in D} F_i}+\lambda_\text{min}
\end{equation}

%We present the algorithm as pseudocode in \cref{alg:bfbcParams},
%where $\mu: D \subset \mathbb{N} \mapsto \mathbb{N}$ returns the hypothetical mean chunk size given the divisors $D$.
%This value can be derived as the length of the data divided by the total number of divisor occurences. That is, $l / \sum_{i \in D} F_i$.

Our algorithm uses \cref{eq:bfbcAvg} while iterating over potential divisors to make predictions about the outcome of the average chunk size.
The algorithm is greedy, minimizing the difference between target mean and achieved mean.

%For illustration, it follows the intuition of paying a bill using cash:
%We take the highest-valued coin or bank note that is lower than the bill amount,
%and repeat this until we have reached or surpassed the bill amount.
%The difference is that while paying a bill aims for an amount $\geq$ the target,
%our goal is to minimize the absolute difference between the target and the given amount.
%This is achieved \via lines 7--11 in the algorithm.
%For efficiency reasons, we break the loop when the target chunk size is reached with a tolerance range of \qty{1}{\%}.

%\begin{algorithm}
%	\input{algorithms/bfbc-params.tex}
%	\label{alg:bfbcParams}
%\end{algorithm}

%% file: sections/setup.tex
We implement all algorithms from \cref{sec:algorithms} in a standardized environment, tuned for performance and efficiency.
In addition, we created a test framework that analyzes the throughput, average chunk size and chunk size distribution, and deduplication ratio
for a range of target chunk sizes and datasets.
We have been careful to isolate the performance of the algorithms from factors that are not determined by the characteristics of the chunking algorithms themselves and would therefore lead to false conclusions or unfair comparisons.
Our analysis excludes additional overhead from the processes of fingerprinting and disk I/O as much as possible.
We detail the measures taken to ensure reproducibility.

In all our experiments, we use target chunk sizes from \qty{512}{B} to \qty{8}{KiB} in exponential steps
as values in that range are common in literature and practice~\cite{rabinPrimary,ae,fastcdc,mii}.
\ac{MII} represents a special case among the set of evaluated algorithms,
as adjustments of its parameter $w$ result in target chunk sizes of $\ldots, 130, 770, 5482, 45037, \ldots$ bytes.
%Varying its paramter $w$ has a factorial growth effect on the expected average chunk size.
This makes aligning $w$ to our range of target chunk sizes impossible.
%According to our formula, incremental steps of $w$ yield possible target chunk sizes of $\ldots, 130, 770, 5482, 45037, \ldots$ bytes.
Therefore, we use $w$ where $\mu(w)$ is closest to $\mu$
in our analysis of throughput,
where the chunk sizes do not have a notable effect on the performance.
Further, in order to be able to make a fair assessment of \ac{MII} in the other experiments,
we add target chunk sizes \qty{770}{B} and \qty{5482}{B}.
Note, however, that those targets cannot be met by BSW algorithms.

Our testing framework\footnote{\url{https://github.com/mrd0ll4r/cdc-algorithm-tester}},
as well as our implementations of the algorithms\footnote{\url{https://github.com/mrd0ll4r/cdchunking-rs}},
are published on GitHub under permissive licenses.

\subsection{Parameters Settings}
\label{sec:setup:parameters}
The parameters in each algorithm ultimately adjust the target chunk size.
All parameter settings in our experiments are informed by either the literature, our own stochastic analysis, or, in some cases, empirical findings.
They align with the algorithm descriptions provided in \cref{sec:algorithms}.
An overview is shown in \cref{table:algoParams}.

In order to determine optimal window sizes for Rabin and Buzhash,
we run an experiment to measure their deduplication performance on randomly distributed data.
The best results were achieved with $w=32$ and $w=64$, respectively (\cf \cref{fig:dedup_window_sizes}).
Recall that Gear has an implicit window size equal to the width of its hash, \eg, \qty{32}{\byte} for 32-bit Gear.

Similar to MII and PCI, finding parameter values for BFBC is not trivial.
%BFBC uses the top-$k$ most frequent byte pairs in a dataset.
%The question about the correct value for $k$ cannot be answered generally,
%\ie, without adjusting to the respective dataset.
%Moreover, already the top-1 most frequent byte pair occurs frequently enough in some datasets
%that it becomes impossible to fine-tune it for specific targets.
We follow the recommendation of the authors:
We employ a minimum chunk size $\lambda_\text{min}$ shortly before the target and set $k=3$,
aiming for chunks being created within a short interval after $\lambda_\text{min}$.
In addition, we extend the set of algorithms by BFBC*,
by which we refer to BFBC using our improved algorithm for determining divisors (\cf \cref{sec:bfbc_star}).
BFBC* does not enforce a minimum chunk length.

\begin{table}[ht]
    \centering
    \caption{Algorithmic Parameter Settings}
    \label{table:algoParams}
    \input{tables/algo-params.tex}
\end{table}

\subsection{Datasets}\label{sec:setup:datasets}

As the deduplication ratio and the chunk size distribution highly depend on the given dataset,
we have collected multiple real-world datasets
in addition to one artificially generated dataset with maximal entropy.
Our choice of datasets is inspired by the kind that is typically used when evaluating \ac{CDC} algorithms~\cite{fastcdc,ram,bimodal,quickcdc,bfbc}.
In our repository, we include the exact scripts used to craft the datasets.
Those encompass:
\begin{itemize}
	\item \textbf{LNX:} A selection of 14 Linux ISO images representing various distributions, as well as different versions of the same distribution.
	Note that ISO is an uncompressed file format.
	\item \textbf{PDF:} Over 2000 PDF files (scientific articles) retrieved from arXiv\footnote{\url{https://info.arxiv.org/help/bulk_data}}.
	PDF is a complex format and includes both textual content and binary content (\eg, to represent images).
	\item \textbf{WEB:} Daily snapshots of the website \href{https://nytimes.com}{nytimes.com} for the entire year of 2022,
downloaded from the Internet Archive and recursively crawling three levels of links.
As such, this dataset is a mix of textual (HTML, CSS, and JavaScript)
as well as binary files (images, videos, font files, \etc).
The latter makes up \qty{89}{\%} of the content.
\item \textbf{CODE:} Source code distributions of various releases of the open-source projects GCC, GDB, and Emacs (94 versions in total).
The content in this dataset is \qty{97}{\%} textual and thereby the most likely to benefit from CDC.
\end{itemize}

The datasets were intentionally chosen to be suitable beneficiaries for CDC
because of textual file format or the high intra-correlation 
given by series of consecutive versions and their incremental changes.
%Other file types, such as multimedia files, executables, or compressed archive formats
%are considered poor candidates for CDC because of their internal compression and arbitrary correlation.
%For these cases, \ac{FSC} may be a superior choice for performance reasons.
We still chose to include an artificial dataset of random data, \textbf{RAND},
which corresponds to the theoretical considerations about the expected behavior of the algorithms as laid out in \cref{sec:algorithms}.
All datasets are approximately \qty{10}{\gibi\byte} in size, which is important for comparable results.
An overview of the datasets and their characteristics is given in \Cref{table:datasets}.
The entropy of a dataset is represented as its size after GZIP compression relative to its original size.

\begin{table}[ht]
    \centering
    \caption{Experimental Datasets}
    \label{table:datasets}
    \input{tables/datasets.tex}
\end{table}

Apart from RAND, each dataset represents a collection of multiple files.
However, CDC algorithms operate on only one data stream.
In order to make valid and comparable claims and observations,
it is necessary to collect each dataset into one single file.
We do this through simple concatenation.
This comes with a caveat when interpreting the results:
They might not represent the performance that would be found for the same datasets in real storage systems,
as those systems usually deduplicate on the level of individual files.
This can be advantageous because similar (or related) files will have their first chunks' starting position aligned at byte index 0.
With our method, chunks can be formed starting in one file and ending in another.
This means that even files that are identical might not be detected as duplicates,
especially when the target chunk size is large relative to the file size.
In order to get comparable results on various target chunk sizes, datasets, and algorithms,
we have to contemplate the datasets as streams of data of a specific type (\eg, source code).
For the same reasons we omit the last chunk produced for every dataset in the evaluation of our experiments.
%The intention here again is to not have the results distorted by chunk boundaries
%that exist because a file ends rather than by the actual effects of CDC.

%However, there is a different case to be made for our analysis on deduplication performance.
%For our experiment on deduplication performance, we bundle the dataset as (uncompressed) TAR archives.
%TAR archives have the special feature that content is maintained in blocks of \qty{512}{B}.
%The intention here is to not have the results distorted by chunk boundaries 
%that exist because a file ends rather than by the actual effects of CDC.
%For the same reason, %in the evaluation of the chunk size distribution and deduplication performance, 
%we ignore the very last chunk produced for every dataset.
%However, a feature of TAR archives is that content is maintained in blocks of \qty{512}{B}.
%This has the consequence that small files ($\leq \qty{512}{B}$)
%can be altered without shifting boundaries to the subsequent files.
%This would yield false conclusions about the deduplication performance of \ac{FSC}.
%Thus, we furthermore present an overview of the distribution of file sizes in each dataset,
%shown in \cref{fig:dataset_file_sizes}.
%As can be seen, a significant amount of files in CODE are about or smaller than a single TAR block.

%\begin{figure}
%	\centering%
%	\input{plots/dataset-file-sizes.tex}\unskip%
%	\caption[Dataset File Sizes]{
%		Cumulative distribution of file sizes in every dataset.	}%
%	\label{fig:dataset_file_sizes}
%\end{figure}

\subsection{Benchmark Program}\label{sec:setup:benchmark}
In order to measure the algorithms, both in terms of throughput as well as the chunks produced,
we present a framework and implementation of a benchmark program, written in the Rust programming language,
with a focus on performance and efficiency.
The framework makes it easy to implement new chunking algorithms and test their performance.
It consists of a \emph{driver}, which reads the input file in large blocks and uses them to drive a selected algorithm.
The algorithms are presented with consecutive blocks of file data,
on which they are to find a boundary, advancing their internal state as they ingest the blocks.
This allows for a performant, real-world oriented implementation.
Ultimately, the algorithms operate on sequences of bytes,
closely following the psueudocode laid out in their descriptions.
The algorithms are collected into a single benchmark program, which is compiled with optimizations enabled,
targeted at the executing machine.
The benchmark program operates as follows:
The selected input file is read,
fed into the selected algorithm,
produced chunks are fingerprinted,
and the fingerprint and size of the chunks is output.

When evaluating the throughput of an algorithm, however,
the resulting chunks are not fingerprinted.
In this case, the benchmark program tracks and outputs a single value, the sum of all chunks' sizes,
to prevent compiler optimizations from removing the chunking code altogether.
The entire file is served from a RAMdisk,
ensuring that the speed of reading the file is not a limiting factor,
which we verify by implementing and evaluation FSC using the same framework. 
We ensure ample memory remains for program execution.
Apart from our benchmark program, the system is in an idle state.
All benchmarks are evaluated sequentially, in order not to influence each other.
We execute the the benchmark program for each dataset, algorithm, and target chunk size a total of $n=10$ times.
In addition, we \enquote{warm up} the system once per dataset/algorithm combination.

We execute our benchmarks on a machine running Ubuntu 22.04 with an Intel Xeon Gold 6154 CPU at \qty{3.00}{GHz},
capture performance counters using {\ttfamily perf},
and report on statistics derived from these results.
While we execute our benchmarks on one specific system,
we believe that general trends are transferable to other systems.
Properties of the algorithms,
such as cache utilization or ease of branch prediction,
are influential on all modern systems.
Note also that the results for deduplication performance and chunk size distribution are independent of the executing machine,
as all algorithms are deterministic.

%% file: tables/algo-params.tex
% \small
\begin{tabularx}{\columnwidth}{ll}
\toprule
Alg. & Parameter Settings \\
\midrule
BSW & $w=32$, $b=log_2(\mu-w)$ \\
AE & \(h=\begin{cases}
\mu-256 & \text{if } \mu<\qty{2}{KiB}\text{,}\\
\text{\cf\cref{table:aeH}} & \text{otherwise}
\end{cases}\) \\
RAM & \makecell[l]{$h$ such that\\
	$h + \left(1-\frac{\sum_{m=0}^{255} \left(m \left(\left(\frac{m+1}{256}\right)^h - \left(\frac{m}{256}\right)^h\right)\right)}{256}\right)^{-1} = \mu$} \\
MII & $w$ such that $\left(\binom{256}{w}\cdot256^{-w}\right)^{-1}\approx\mu$ \\
PCI & \cf\cref{table:pciParams} \\
BFBC & $k=3$ and $\lambda_\text{min}=\mu - 128$ \\
BFBC* & $D$ such that $\frac{l}{1+\sum_{i\in D} F_i}+\lambda_\text{min}\approx\mu$ \\
\bottomrule
\end{tabularx}

%% file: tables/datasets.tex
% \small
\begin{tabularx}{\columnwidth}{lrX}
\toprule
Name & Entropy & Description \\
\midrule
RAND & \qty{100.0}{\%} & Randomly generated binary. \\
LNX & \qty{98.6}{\%} & Linux ISO images. \\
PDF & \qty{87.5}{\%} & Collection of research papers. \\
WEB & \qty{72.8}{\%} & Daily website snapshots. \\
CODE & \qty{22.1}{\%} & Consecutive versions from code repositories. \\
\bottomrule
\end{tabularx}

%% file: sections/efficiency.tex
Much of the discussion and development around chunking algorithms has been fueled by the need to achieve higher throughput
while maintaining good deduplication.
Rabin, one of the oldest \ac{CDC} algorithms, is widely known to be slow.
This has spurred the development of newer, faster \ac{CDC} algorithms.
\begin{comment}
Some of them, such as \eg Buzhash or Gear, still utilize the \ac{BSW} approach:
They move a fixed-size window over the stream of data and apply a rolling hash function to it.
As such, the innovations in these cases are the hash functions, in particular.
On the other hand, algorithms utilizing extrema, such as AE or RAM, use a different approach to find chunk boundaries.
Of the two, RAM aims to be an improvement upon AE in terms of throughput.
Yet other approaches exist:
MII inspects the sequence of input bytes itself for a streak of increasing values.
BFBC operates using a list of popular byte pairs.
This requires either reading the input twice, or operating on a generic distribution of byte pair frequencies that \enquote{fits} the dataset well enough.
\end{comment}
We dedicate this section to investigate the performance of the algorithms in terms of throughput, or computational efficiency.
We report on overall achievable throughput of the algorithms,
as well as microarchitectural details to explain certain behaviors.

%%%%%%%%%%%%%%%%%%%%%%%%%%%%%%%%%%%%%%%%%%%%%%%%%%%%%%%%%%%%%%%%%%
\subsection{Setup and Methodology}
\label{sec:perf:setup}

\begin{comment}
We implement two \emph{drivers}:
The standard driver, which sequentially feeds \emph{all} input file data into the inner algorithm,
and a QuickCDC driver, which supports caching and jumping, as described in \Cref{sec:algorithms:quickcdc}.
Recall how QuickCDC introduces two novel ideas for chunking:
A selection of the degree of normalization applied to a given CDC algorithm to dynamically adjust expected chunk size,
and a chunk caching and jumping algorithm, to avoid re-chunking identical data.
We extract the latter into the QuickCDC driver and can thus evaluate it on different wrapped CDC algorithms.
%\leo{this is a noteworthy contribution by us, because I don't think anyone else realized this.}
\end{comment}

We execute our benchmark program as described in \Cref{sec:setup:benchmark} and collect both the execution time as well as {\ttfamily perf} counters.
We normalize results by the size of the dataset where applicable.
We expect different distributions for the recorded metrics:
\begin{enumerate*}
	\item For all microarchitectural performance counters, such as \ac{ipc}, number of instructions, number of branches, \etc, we assume a normal distribution.
	As such, we derive the mean value and corresponding standard error for these metrics.
	We expect very little spread in most of these metrics through multiple runs, as we compile our program just once, and all algorithms are deterministic.
	The data confirms these expectations, with a standard error for \emph{all} reported metrics of $\leq 0.01$, which we thus omit.
	%Nevertheless, small fluctuations, especially \wrt caching, remain and are reported.
	
	\item For the runtime of the benchmarks, and conversely the throughput as a measure of input processed per execution time,
	we assume a skewed distribution in accordance with \cite{lemire2023}.
	As such, we report the median and \ac{iqr}, calculated as $Q_3 - Q_1$.
\end{enumerate*}

\begin{comment}
Notably, the throughput of different algorithms is \emph{not} independent of the dataset or the target chunk size.
This is mainly for three reasons:
\begin{enumerate*}
	\item Smaller target chunk sizes lead to lower expected throughput as the driver code is executed more frequently.
	\marcel{since \emph{driver} isn't introduced anymore with standard/quickcdc driver, I find the term confusing}
	\leo{The term is now introduced in the previous section :) }
	We control for this by using infrequent reads from the input file and generally make the driver as efficient as possible.

	\item The expected chunk size of an algorithm is theoretically derived for uniformly random input data (\cf \Cref{sec:algorithms}).
	While desirable, we will show in \Cref{sec:distribution} that this does not always transfer to other datasets.
	These different distributions of chunk sizes can lead to the same influences of the driver code as discussed previously.

	\item Lastly, some algorithms, BFBC in our case, are specifically made to be sensitive to the input dataset.
\end{enumerate*}
\end{comment}

We also evaluate the SIMD optimization for BSW algorithms discussed in \cref{sec:algorithms:bsw}.
For this, we use an existing implementation\footnote{\url{https://crates.io/crates/gearhash}} of the 64-bit Gear algorithm with manual vectorization, which we refer to as Gear64+.
The instruction set utilized by the implementation depends on the support of the current machine.
On our system, the algorithm uses AVX2 instructions.
Note that the original Gear algorithm uses 32-bit.
For transparency about what contributed to the difference in results, we therefore additionally implement 64-bit Gear without the use of manual vectorization, denoted Gear64.

%%%%%%%%%%%%%%%%%%%%%%%%%%%%%%%%%%%%%%%%%%%%%%%%%%%%%%%%%%%%%%%%%%
\subsection{Overview on Synthetic Dataset}
\label{sec:perf:overview}

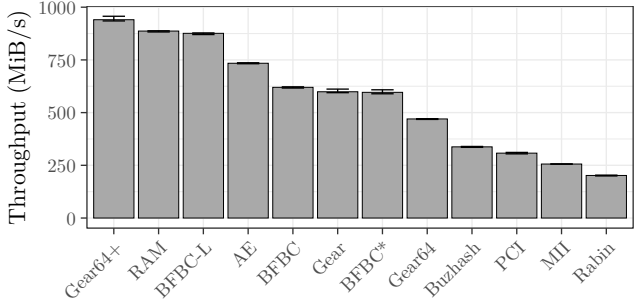
\begin{figure}[bt]
	\centering%
	\vspace{-0.2cm}%
	\scalebox{0.85}{
		%\tiny
		\input{fig/perf_overview_throughput_random_2kib}}\unskip%
%	\vspace{-0.3cm}%
	\caption{
		CDC throughput, median values and quartiles, $\mu=\qty{2}{\kibi\byte}$, RAND dataset. BFBC-L indicates BFBC on the CODE dataset.
	}%
	\label{fig:perf_overview_throughput_random_2kib}
\end{figure}

We first provide an overview of the achievable throughput using the various algorithms.
To that end, we evaluate each of them on the RAND dataset with a target chunk size of \qty{2}{KiB} (\Cref{fig:perf_overview_throughput_random_2kib}).
Due to the specific content-dependence of BFBC with regard to its efficiency, we additionally present measurements for BFBC on the CODE dataset, referred to as BFBC-L.
Evaluating on the random dataset corresponds to the same theoretical considerations on expected chunk size as given in \cref{sec:algorithms}.
%A target chunk size of \qty{2}{KiB} presents a realistic choice.
To better understand these results, we furthermore provide relevant microarchitectural counters in \Cref{tab:perf_overview}.
We investigate the results in detail in the following.

\begin{table}[bt]
	\centering
	\caption{
		Computational Performance of Chunking Algorithms ($\mu=\qty{2}{\kibi\byte}$, RAND Dataset;
		BFBC-L Indicates BFBC on the CODE Dataset;
		%$\mu$-arch. performance counters indicate means; all standard errors $\leq 0.01$ (not shown).
		Br./B = Branches/Input Byte. BM-\% = Branch Misprediction Percentage)
	}
	\label{tab:perf_overview}
	{
		\setlength\tabcolsep{5pt}
		% \small
		\input{tab/perf_overview_random_2kib}\unskip%
	}
\end{table}

\subsubsection{BSW Algorithms}
The BSW algorithms (Rabin, Buzhash, Gear, and PCI) generally make up the lower end of the performance scale, with the exception of Gear.
Although these algorithms all operate in $\mathcal{O}(1)$ per input byte, we can see that they differ substantially in constant complexity:
Rabin and PCI place past 30 instructions per input byte, Buzhash places just shy of that, which results the highest throughput of the three at $\approx \qty{340}{MiB/s}$.
Gear uses only $\approx 13$ instructions per byte, leading to a much higher throughput of $\approx \qty{600}{MiB/s}$.

\begin{figure}[tb]
	\centering%
	\scalebox{0.85}{
		\input{fig/perf_gear_simd_throughput_comparison_chunk_sizes_random}\unskip%
	}
	\caption{
		Throughput of scalar and SIMD implementations of Gear64, median and quartiles, RAND dataset.
	}%
	\label{fig:perf_gear_simd_throughput_comparison_chunk_sizes_random}
\end{figure}
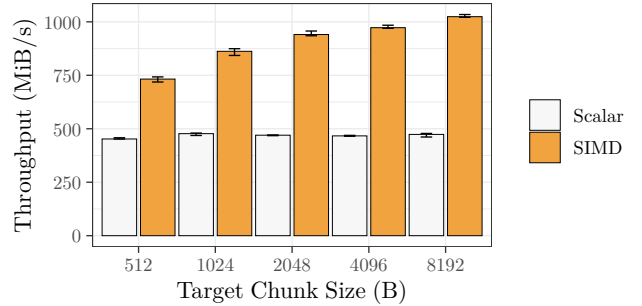

The SIMD implementation (Gear64+) uses multiple \enquote{heads}, spaced out by a number of bytes dependent on the instructions supported on the target machine.
Each head then performs the Gear algorithm as usual, although all heads execute in parallel using \ac{simd} instructions.
Once any of the heads finds a boundary,
the code falls back to a scalar variant in order to ensure none of the \emph{previous} heads detects a chunk point in any of the yet-unprocessed data.
The SIMD implementation is around twice as fast as the scalar variant (Gear64).
This difference becomes more pronounced the larger the target chunk sizes (\cf \Cref{fig:perf_gear_simd_throughput_comparison_chunk_sizes_random}).
This is expected from the implementation, \ie, the code needs to fall back to a scalar version less frequently for larger target chunk sizes
since fewer chunk boundaries are found.
In terms of microarchitectural performance (\Cref{tab:perf_gear_simd_comparison_random}),
we can see that the larger the target chunk size, the fewer instructions per byte are utilized by the SIMD version,
which again follows from the implementation falling back to scalar code less frequently.
The same applies to branches.
Finally, by comparing Gear with Gear64, we can also see that there is an expected efficiency drawback that comes with generating larger hashes.
Note that it should also be possible to apply manual vectorization to the 32-bit variant, as well as to other BSW algorithms,
but this is not the focus of our work.
In conclusion, while not always possible, data-parallelism can offer a large increase in performance.

\begin{table}[tb]
	\centering
	\caption{
		Performance of Scalar and SIMD Implementations of Gear64 on the RAND Dataset
		%$\mu$-arch. performance counters indicate means, all standard errors $\leq 0.01$ (not shown).
		(Br. = Branches)
	}
	\label{tab:perf_gear_simd_comparison_random}
	{
		% \small
		\input{tab/perf_gear_simd_comparison_random}\unskip%
	}
\end{table}

\subsubsection{Extremum-Based Algorithms}
The algorithms utilizing local extrema (AE, RAM, and MII) generally perform very well, with the exception of MII.
Of the three, RAM performs the best at $\approx \qty{870}{MiB/s}$, markedly better than AE at $\approx \qty{730}{MiB/s}$.
We thus conclude that RAM achieves its goal of being a faster AE~\cite{ram}.
%\leo{about MII: were these the people that said "look, our algorithm only has two instructions, so it's fast"?}
MII, although algorithmically simple, performs poorly at only $\approx \qty{260}{MiB/s}$.
It requires only slightly more instructions per input byte than AE and RAM.
%($\approx \qty{12}{IPB}$ vs. $\approx \qty{11}{IPB}$ and $\approx \qty{8}{IPB}$, respectively).
However, we can see that it utilizes the CPU poorly at only $\approx \qty{1}{IPC}$.
A closer look shows that this is due to the poor predictability of the branch comparing the current input byte to the previous one.
We find that \qty{15}{\percent} of branches are mispredicted, the worst behavior of all algorithms examined.

\subsubsection{BFBC}
As the results in \cref{sec:distribution} will demonstrate, the original BFBC algorithm is not well-suited for high-entropy datasets.
We compare BFBC on RAND to BFBC on CODE (denoted BFBC-L).
BFBC struggles on datasets like RAND because the top-3 most frequent byte pairs occur as frequently as any other.
This causes the algorithm to skip the minimum chunk size, yet fail to find a boundary quickly.
This is evident in the higher number of instructions and branches per input byte.
On low-entropy datasets like CODE the top-3 most frequent byte pairs are usually encountered shortly after skipping the minimum chunk size.
We can also compare BFBC to BFBC* to judge the effects of skipping.
Recall that BFBC* uses a different set of divisors and does not skip, but otherwise functions the same as standard BFBC.
In particular, comparison against the chosen byte pairs happens in constant time in our implementation,
\ie, the number of selected byte pairs should have no effect on throughput.
The absence of skipping results in a slight elevation in number of instructions and branches per input byte, which leads to a correspondent decrease in throughput.

\subsubsection{Fixed-size Chunking}
Finally, if content-defined chunking is not a concern, fixed-size chunking unsurprisingly outperforms all CDC contenders at more than \qty{2.5}{GiB/s} of processed input.
These results also show that our benchmarking system is not limited by I/O.

%%%%%%%%%%%%%%%%%%%%%%%%%%%%%%%%%%%%%%%%%%%%%%%%%%%%
\subsection{Key Takeaways}
\label{sec:perf:conclusion}

In this section, we focused on performance in terms of throughput as a measure of computational efficiency for a range of algorithms.
In summary, we can derive the following conclusions:

\subsubsection{Complexity in Pseudocode versus Performance on Real-world Systems}
Modern CPUs are complex.
It is not always obvious how an algorithm, given in pseudocode, performs on a real system.
We were able to show that various microarchitectural factors, in particular caching and branch predictability, play an important role in achieving higher throughput.
Some algorithms (\eg, MII) are algorithmically simple yet difficult for the machine to execute.
Others (\eg, Gearhash) are both simple and fast to execute.

\subsubsection{BFBC Variants}
We found that the BFBC variants are relatively fast, with differences stemming from skipping over parts of a new chunk and the dataset.
We did not evaluate the time it takes or the feasibility of deriving byte pair frequencies before chunking.
In general, while the algorithms perform well, their application is potentially limited by this requirement.

\subsubsection{BSW Algorithms}
Of all the \ac{BSW} algorithms, Gearhash presents itself with low algorithmic complexity and good real-world performance.
If supported on the machine, and an implementation is feasible, vectorization of these algorithms, in particular of Gear, achieves very high throughput.
In our evaluation, \ac{simd}-accelerated Gear outperforms \emph{all} other \ac{CDC} algorithms.
We postulate whether it would be more fruitful in the future to develop algorithms with this in mind,
instead of developing new \ac{CDC} algorithms.

\subsubsection{Algorithms Using Local Extrema}
Of the algorithms utilizing extrema (MII, RAM, AE), we find that MII performs poorly due to the difficulty of predicting its branch.
Both AE and RAM perform very well overall.
Between the two, we find that RAM indeed outperforms AE.
%In fact, if we do not consider \ac{simd}-accelerated Gearhash, RAM showed the best throughput overall.

%% file: fig/perf_overview_throughput_random_2kib.tex
% Created by tikzDevice version 0.12.6 on 2024-08-04 13:01:37
% !TEX encoding = UTF-8 Unicode
\begin{tikzpicture}[x=1pt,y=1pt]
\definecolor{fillColor}{RGB}{255,255,255}
\path[use as bounding box,fill=fillColor,fill opacity=0.00] (0,0) rectangle (289.08,144.54);
\begin{scope}
\path[clip] (  0.00,  0.00) rectangle (289.08,144.54);
\definecolor{drawColor}{RGB}{255,255,255}
\definecolor{fillColor}{RGB}{255,255,255}

\path[draw=drawColor,line width= 0.6pt,line join=round,line cap=round,fill=fillColor] (  0.00,  0.00) rectangle (289.08,144.54);
\end{scope}
\begin{scope}
\path[clip] ( 40.51, 40.20) rectangle (283.58,139.04);
\definecolor{fillColor}{RGB}{255,255,255}

\path[fill=fillColor] ( 40.51, 40.20) rectangle (283.58,139.04);
\definecolor{drawColor}{gray}{0.92}

\path[draw=drawColor,line width= 0.3pt,line join=round] ( 40.51, 56.43) --
	(283.58, 56.43);

\path[draw=drawColor,line width= 0.3pt,line join=round] ( 40.51, 79.90) --
	(283.58, 79.90);

\path[draw=drawColor,line width= 0.3pt,line join=round] ( 40.51,103.38) --
	(283.58,103.38);

\path[draw=drawColor,line width= 0.3pt,line join=round] ( 40.51,126.85) --
	(283.58,126.85);

\path[draw=drawColor,line width= 0.6pt,line join=round] ( 40.51, 44.69) --
	(283.58, 44.69);

\path[draw=drawColor,line width= 0.6pt,line join=round] ( 40.51, 68.17) --
	(283.58, 68.17);

\path[draw=drawColor,line width= 0.6pt,line join=round] ( 40.51, 91.64) --
	(283.58, 91.64);

\path[draw=drawColor,line width= 0.6pt,line join=round] ( 40.51,115.11) --
	(283.58,115.11);

\path[draw=drawColor,line width= 0.6pt,line join=round] ( 40.51,138.59) --
	(283.58,138.59);

\path[draw=drawColor,line width= 0.6pt,line join=round] ( 52.46, 40.20) --
	( 52.46,139.04);

\path[draw=drawColor,line width= 0.6pt,line join=round] ( 72.39, 40.20) --
	( 72.39,139.04);

\path[draw=drawColor,line width= 0.6pt,line join=round] ( 92.31, 40.20) --
	( 92.31,139.04);

\path[draw=drawColor,line width= 0.6pt,line join=round] (112.24, 40.20) --
	(112.24,139.04);

\path[draw=drawColor,line width= 0.6pt,line join=round] (132.16, 40.20) --
	(132.16,139.04);

\path[draw=drawColor,line width= 0.6pt,line join=round] (152.08, 40.20) --
	(152.08,139.04);

\path[draw=drawColor,line width= 0.6pt,line join=round] (172.01, 40.20) --
	(172.01,139.04);

\path[draw=drawColor,line width= 0.6pt,line join=round] (191.93, 40.20) --
	(191.93,139.04);

\path[draw=drawColor,line width= 0.6pt,line join=round] (211.85, 40.20) --
	(211.85,139.04);

\path[draw=drawColor,line width= 0.6pt,line join=round] (231.78, 40.20) --
	(231.78,139.04);

\path[draw=drawColor,line width= 0.6pt,line join=round] (251.70, 40.20) --
	(251.70,139.04);

\path[draw=drawColor,line width= 0.6pt,line join=round] (271.63, 40.20) --
	(271.63,139.04);
\definecolor{drawColor}{RGB}{0,0,0}
\definecolor{fillColor}{RGB}{169,169,169}

\path[draw=drawColor,line width= 0.3pt,fill=fillColor] (103.27, 44.69) rectangle (121.20,113.62);

\path[draw=drawColor,line width= 0.3pt,fill=fillColor] (123.19, 44.69) rectangle (141.13,102.90);

\path[draw=drawColor,line width= 0.3pt,fill=fillColor] ( 83.35, 44.69) rectangle (101.28,126.95);

\path[draw=drawColor,line width= 0.3pt,fill=fillColor] (163.04, 44.69) rectangle (180.97,100.67);

\path[draw=drawColor,line width= 0.3pt,fill=fillColor] (202.89, 44.69) rectangle (220.82, 76.39);

\path[draw=drawColor,line width= 0.3pt,fill=fillColor] (143.12, 44.69) rectangle (161.05,100.93);

\path[draw=drawColor,line width= 0.3pt,fill=fillColor] (182.97, 44.69) rectangle (200.90, 88.82);

\path[draw=drawColor,line width= 0.3pt,fill=fillColor] ( 43.50, 44.69) rectangle ( 61.43,133.01);

\path[draw=drawColor,line width= 0.3pt,fill=fillColor] (242.74, 44.69) rectangle (260.67, 68.74);

\path[draw=drawColor,line width= 0.3pt,fill=fillColor] (222.81, 44.69) rectangle (240.74, 73.59);

\path[draw=drawColor,line width= 0.3pt,fill=fillColor] (262.66, 44.69) rectangle (280.59, 63.65);

\path[draw=drawColor,line width= 0.3pt,fill=fillColor] ( 63.42, 44.69) rectangle ( 81.35,127.96);

\path[draw=drawColor,line width= 0.6pt,line join=round] (107.25,113.84) --
	(117.22,113.84);

\path[draw=drawColor,line width= 0.6pt,line join=round] (112.24,113.84) --
	(112.24,113.41);

\path[draw=drawColor,line width= 0.6pt,line join=round] (107.25,113.41) --
	(117.22,113.41);

\path[draw=drawColor,line width= 0.6pt,line join=round] (127.18,103.13) --
	(137.14,103.13);

\path[draw=drawColor,line width= 0.6pt,line join=round] (132.16,103.13) --
	(132.16,102.58);

\path[draw=drawColor,line width= 0.6pt,line join=round] (127.18,102.58) --
	(137.14,102.58);

\path[draw=drawColor,line width= 0.6pt,line join=round] ( 87.33,127.12) --
	( 97.29,127.12);

\path[draw=drawColor,line width= 0.6pt,line join=round] ( 92.31,127.12) --
	( 92.31,126.41);

\path[draw=drawColor,line width= 0.6pt,line join=round] ( 87.33,126.41) --
	( 97.29,126.41);

\path[draw=drawColor,line width= 0.6pt,line join=round] (167.03,101.80) --
	(176.99,101.80);

\path[draw=drawColor,line width= 0.6pt,line join=round] (172.01,101.80) --
	(172.01,100.03);

\path[draw=drawColor,line width= 0.6pt,line join=round] (167.03,100.03) --
	(176.99,100.03);

\path[draw=drawColor,line width= 0.6pt,line join=round] (206.87, 76.57) --
	(216.84, 76.57);

\path[draw=drawColor,line width= 0.6pt,line join=round] (211.85, 76.57) --
	(211.85, 76.17);

\path[draw=drawColor,line width= 0.6pt,line join=round] (206.87, 76.17) --
	(216.84, 76.17);

\path[draw=drawColor,line width= 0.6pt,line join=round] (147.10,102.11) --
	(157.06,102.11);

\path[draw=drawColor,line width= 0.6pt,line join=round] (152.08,102.11) --
	(152.08,100.52);

\path[draw=drawColor,line width= 0.6pt,line join=round] (147.10,100.52) --
	(157.06,100.52);

\path[draw=drawColor,line width= 0.6pt,line join=round] (186.95, 88.93) --
	(196.91, 88.93);

\path[draw=drawColor,line width= 0.6pt,line join=round] (191.93, 88.93) --
	(191.93, 88.61);

\path[draw=drawColor,line width= 0.6pt,line join=round] (186.95, 88.61) --
	(196.91, 88.61);

\path[draw=drawColor,line width= 0.6pt,line join=round] ( 47.48,134.55) --
	( 57.45,134.55);

\path[draw=drawColor,line width= 0.6pt,line join=round] ( 52.46,134.55) --
	( 52.46,132.47);

\path[draw=drawColor,line width= 0.6pt,line join=round] ( 47.48,132.47) --
	( 57.45,132.47);

\path[draw=drawColor,line width= 0.6pt,line join=round] (246.72, 68.78) --
	(256.68, 68.78);

\path[draw=drawColor,line width= 0.6pt,line join=round] (251.70, 68.78) --
	(251.70, 68.69);

\path[draw=drawColor,line width= 0.6pt,line join=round] (246.72, 68.69) --
	(256.68, 68.69);

\path[draw=drawColor,line width= 0.6pt,line join=round] (226.80, 73.91) --
	(236.76, 73.91);

\path[draw=drawColor,line width= 0.6pt,line join=round] (231.78, 73.91) --
	(231.78, 73.27);

\path[draw=drawColor,line width= 0.6pt,line join=round] (226.80, 73.27) --
	(236.76, 73.27);

\path[draw=drawColor,line width= 0.6pt,line join=round] (266.64, 63.79) --
	(276.61, 63.79);

\path[draw=drawColor,line width= 0.6pt,line join=round] (271.63, 63.79) --
	(271.63, 63.44);

\path[draw=drawColor,line width= 0.6pt,line join=round] (266.64, 63.44) --
	(276.61, 63.44);

\path[draw=drawColor,line width= 0.6pt,line join=round] ( 67.41,128.08) --
	( 77.37,128.08);

\path[draw=drawColor,line width= 0.6pt,line join=round] ( 72.39,128.08) --
	( 72.39,127.56);

\path[draw=drawColor,line width= 0.6pt,line join=round] ( 67.41,127.56) --
	( 77.37,127.56);
\definecolor{drawColor}{gray}{0.20}

\path[draw=drawColor,line width= 0.6pt,line join=round,line cap=round] ( 40.51, 40.20) rectangle (283.58,139.04);
\end{scope}
\begin{scope}
\path[clip] (  0.00,  0.00) rectangle (289.08,144.54);
\definecolor{drawColor}{gray}{0.30}

\node[text=drawColor,anchor=base east,inner sep=0pt, outer sep=0pt, scale=  0.88] at ( 35.56, 41.66) {0};

\node[text=drawColor,anchor=base east,inner sep=0pt, outer sep=0pt, scale=  0.88] at ( 35.56, 65.14) {250};

\node[text=drawColor,anchor=base east,inner sep=0pt, outer sep=0pt, scale=  0.88] at ( 35.56, 88.61) {500};

\node[text=drawColor,anchor=base east,inner sep=0pt, outer sep=0pt, scale=  0.88] at ( 35.56,112.08) {750};

\node[text=drawColor,anchor=base east,inner sep=0pt, outer sep=0pt, scale=  0.88] at ( 35.56,135.56) {1000};
\end{scope}
\begin{scope}
\path[clip] (  0.00,  0.00) rectangle (289.08,144.54);
\definecolor{drawColor}{gray}{0.20}

\path[draw=drawColor,line width= 0.6pt,line join=round] ( 37.76, 44.69) --
	( 40.51, 44.69);

\path[draw=drawColor,line width= 0.6pt,line join=round] ( 37.76, 68.17) --
	( 40.51, 68.17);

\path[draw=drawColor,line width= 0.6pt,line join=round] ( 37.76, 91.64) --
	( 40.51, 91.64);

\path[draw=drawColor,line width= 0.6pt,line join=round] ( 37.76,115.11) --
	( 40.51,115.11);

\path[draw=drawColor,line width= 0.6pt,line join=round] ( 37.76,138.59) --
	( 40.51,138.59);
\end{scope}
\begin{scope}
\path[clip] (  0.00,  0.00) rectangle (289.08,144.54);
\definecolor{drawColor}{gray}{0.20}

\path[draw=drawColor,line width= 0.6pt,line join=round] ( 52.46, 37.45) --
	( 52.46, 40.20);

\path[draw=drawColor,line width= 0.6pt,line join=round] ( 72.39, 37.45) --
	( 72.39, 40.20);

\path[draw=drawColor,line width= 0.6pt,line join=round] ( 92.31, 37.45) --
	( 92.31, 40.20);

\path[draw=drawColor,line width= 0.6pt,line join=round] (112.24, 37.45) --
	(112.24, 40.20);

\path[draw=drawColor,line width= 0.6pt,line join=round] (132.16, 37.45) --
	(132.16, 40.20);

\path[draw=drawColor,line width= 0.6pt,line join=round] (152.08, 37.45) --
	(152.08, 40.20);

\path[draw=drawColor,line width= 0.6pt,line join=round] (172.01, 37.45) --
	(172.01, 40.20);

\path[draw=drawColor,line width= 0.6pt,line join=round] (191.93, 37.45) --
	(191.93, 40.20);

\path[draw=drawColor,line width= 0.6pt,line join=round] (211.85, 37.45) --
	(211.85, 40.20);

\path[draw=drawColor,line width= 0.6pt,line join=round] (231.78, 37.45) --
	(231.78, 40.20);

\path[draw=drawColor,line width= 0.6pt,line join=round] (251.70, 37.45) --
	(251.70, 40.20);

\path[draw=drawColor,line width= 0.6pt,line join=round] (271.63, 37.45) --
	(271.63, 40.20);
\end{scope}
\begin{scope}
\path[clip] (  0.00,  0.00) rectangle (289.08,144.54);
\definecolor{drawColor}{gray}{0.30}

\node[text=drawColor,rotate= 45.00,anchor=base east,inner sep=0pt, outer sep=0pt, scale=  0.88] at ( 56.75, 30.96) {Gear64+};

\node[text=drawColor,rotate= 45.00,anchor=base east,inner sep=0pt, outer sep=0pt, scale=  0.88] at ( 76.67, 30.96) {RAM};

\node[text=drawColor,rotate= 45.00,anchor=base east,inner sep=0pt, outer sep=0pt, scale=  0.88] at ( 96.60, 30.96) {BFBC-L};

\node[text=drawColor,rotate= 45.00,anchor=base east,inner sep=0pt, outer sep=0pt, scale=  0.88] at (116.52, 30.96) {AE};

\node[text=drawColor,rotate= 45.00,anchor=base east,inner sep=0pt, outer sep=0pt, scale=  0.88] at (136.45, 30.96) {BFBC};

\node[text=drawColor,rotate= 45.00,anchor=base east,inner sep=0pt, outer sep=0pt, scale=  0.88] at (156.37, 30.96) {Gear};

\node[text=drawColor,rotate= 45.00,anchor=base east,inner sep=0pt, outer sep=0pt, scale=  0.88] at (176.29, 30.96) {BFBC*};

\node[text=drawColor,rotate= 45.00,anchor=base east,inner sep=0pt, outer sep=0pt, scale=  0.88] at (196.22, 30.96) {Gear64};

\node[text=drawColor,rotate= 45.00,anchor=base east,inner sep=0pt, outer sep=0pt, scale=  0.88] at (216.14, 30.96) {Buzhash};

\node[text=drawColor,rotate= 45.00,anchor=base east,inner sep=0pt, outer sep=0pt, scale=  0.88] at (236.06, 30.96) {PCI};

\node[text=drawColor,rotate= 45.00,anchor=base east,inner sep=0pt, outer sep=0pt, scale=  0.88] at (255.99, 30.96) {MII};

\node[text=drawColor,rotate= 45.00,anchor=base east,inner sep=0pt, outer sep=0pt, scale=  0.88] at (275.91, 30.96) {Rabin};
\end{scope}
\begin{scope}
\path[clip] (  0.00,  0.00) rectangle (289.08,144.54);
\definecolor{drawColor}{RGB}{0,0,0}

\node[text=drawColor,rotate= 90.00,anchor=base,inner sep=0pt, outer sep=0pt, scale=  1.10] at ( 13.08, 89.62) {Throughput (MiB/s)};
\end{scope}
\end{tikzpicture}

%% file: tab/perf_overview_random_2kib.tex
% latex table generated in R 4.3.3 by xtable 1.8-4 package
% Fri Aug 23 14:11:35 2024
\begin{tabular}{lrrrrrr}
  \toprule
  & \multicolumn{2}{c}{\makecell{Throughput\\(MiB/s)}} & & & & \\
\cmidrule(lr){2-3}
Alg. & Median & IQR & Inst./B & IPC & Br./B & BM-\% \\ \midrule
FSC & \num{2529} & \num{51} & \num{0.40} & \num{0.31} & \num{0.06} & \num{0.99} \\ 
  Gear64+ & \num{941} & \num{22} & \num{8.03} & \num{2.33} & \num{0.73} & \num{0.38} \\ 
  RAM & \num{887} & \num{6} & \num{8.45} & \num{2.40} & \num{1.75} & \num{0.46} \\ 
  BFBC-L & \num{876} & \num{8} & \num{10.93} & \num{3.05} & \num{1.79} & \num{0.11} \\ 
  AE & \num{734} & \num{5} & \num{11.42} & \num{2.56} & \num{2.56} & \num{0.19} \\ 
  BFBC & \num{620} & \num{6} & \num{16.78} & \num{3.24} & \num{2.49} & \num{0.05} \\ 
  Gear & \num{599} & \num{17} & \num{13.39} & \num{2.50} & \num{1.73} & \num{0.09} \\ 
  BFBC* & \num{596} & \num{19} & \num{17.38} & \num{3.24} & \num{2.56} & \num{0.06} \\ 
  Gear64 & \num{470} & \num{3} & \num{15.41} & \num{2.24} & \num{2.57} & \num{0.06} \\ 
  Buzhash & \num{338} & \num{4} & \num{29.34} & \num{3.06} & \num{5.02} & \num{0.10} \\ 
  PCI & \num{308} & \num{7} & \num{31.25} & \num{3.03} & \num{2.54} & \num{0.07} \\ 
  MII & \num{256} & \num{1} & \num{12.43} & \num{0.99} & \num{2.16} & \num{15.06} \\ 
  Rabin & \num{202} & \num{4} & \num{34.27} & \num{2.19} & \num{5.04} & \num{0.05} \\ 
   \bottomrule
\end{tabular}

%% file: tab/perf_gear_simd_comparison_random.tex
% latex table generated in R 4.3.3 by xtable 1.8-4 package
% Mon Jul 29 13:46:04 2024
\begin{tabular}{lrrrrrr}
  \toprule
  && \multicolumn{2}{c}{\makecell{Throughput\\(MiB/s)}} &&&\\
\cmidrule(lr){3-4}
Alg. & $\mu$ (B) & Median & IQR & Inst./B & IPC & \makecell{Br.\\($\times 10^9$)} \\ \midrule
Scalar & \num{512} & \num{452.2} & $6.1$ & \num{15.57} & \num{2.27} & \num{27.82} \\ 
  SIMD & \num{512} & \num{732.4} & $23.9$ & \num{11.28} & \num{2.57} & \num{11.25} \\ 
  Scalar & \num{8192} & \num{473.6} & $16.6$ & \num{15.38} & \num{2.23} & \num{27.50} \\ 
  SIMD & \num{8192} & \num{1024.1} & $13.2$ & \num{7.02} & \num{2.21} & \num{6.30} \\ 
   \bottomrule
\end{tabular}

%% file: fig/perf_gear_simd_throughput_comparison_chunk_sizes_random.tex
% Created by tikzDevice version 0.12.6 on 2024-07-29 12:02:17
% !TEX encoding = UTF-8 Unicode
\begin{tikzpicture}[x=1pt,y=1pt]
\definecolor{fillColor}{RGB}{255,255,255}
\path[use as bounding box,fill=fillColor,fill opacity=0.00] (0,0) rectangle (289.08,144.54);
\begin{scope}
\path[clip] (  0.00,  0.00) rectangle (289.08,144.54);
\definecolor{drawColor}{RGB}{255,255,255}
\definecolor{fillColor}{RGB}{255,255,255}

\path[draw=drawColor,line width= 0.6pt,line join=round,line cap=round,fill=fillColor] (  0.00,  0.00) rectangle (289.08,144.54);
\end{scope}
\begin{scope}
\path[clip] ( 40.51, 30.69) rectangle (218.14,139.04);
\definecolor{fillColor}{RGB}{255,255,255}

\path[fill=fillColor] ( 40.51, 30.69) rectangle (218.14,139.04);
\definecolor{drawColor}{gray}{0.92}

\path[draw=drawColor,line width= 0.3pt,line join=round] ( 40.51, 47.52) --
	(218.14, 47.52);

\path[draw=drawColor,line width= 0.3pt,line join=round] ( 40.51, 71.33) --
	(218.14, 71.33);

\path[draw=drawColor,line width= 0.3pt,line join=round] ( 40.51, 95.14) --
	(218.14, 95.14);

\path[draw=drawColor,line width= 0.3pt,line join=round] ( 40.51,118.95) --
	(218.14,118.95);

\path[draw=drawColor,line width= 0.6pt,line join=round] ( 40.51, 35.61) --
	(218.14, 35.61);

\path[draw=drawColor,line width= 0.6pt,line join=round] ( 40.51, 59.42) --
	(218.14, 59.42);

\path[draw=drawColor,line width= 0.6pt,line join=round] ( 40.51, 83.23) --
	(218.14, 83.23);

\path[draw=drawColor,line width= 0.6pt,line join=round] ( 40.51,107.04) --
	(218.14,107.04);

\path[draw=drawColor,line width= 0.6pt,line join=round] ( 40.51,130.86) --
	(218.14,130.86);

\path[draw=drawColor,line width= 0.6pt,line join=round] ( 61.01, 30.69) --
	( 61.01,139.04);

\path[draw=drawColor,line width= 0.6pt,line join=round] ( 95.17, 30.69) --
	( 95.17,139.04);

\path[draw=drawColor,line width= 0.6pt,line join=round] (129.33, 30.69) --
	(129.33,139.04);

\path[draw=drawColor,line width= 0.6pt,line join=round] (163.49, 30.69) --
	(163.49,139.04);

\path[draw=drawColor,line width= 0.6pt,line join=round] (197.64, 30.69) --
	(197.64,139.04);
\definecolor{drawColor}{RGB}{0,0,0}
\definecolor{fillColor}{gray}{0.97}

\path[draw=drawColor,line width= 0.3pt,fill=fillColor] ( 44.78, 35.61) rectangle ( 60.15, 78.68);

\path[draw=drawColor,line width= 0.3pt,fill=fillColor] ( 78.94, 35.61) rectangle ( 94.31, 81.05);

\path[draw=drawColor,line width= 0.3pt,fill=fillColor] (113.10, 35.61) rectangle (128.47, 80.37);

\path[draw=drawColor,line width= 0.3pt,fill=fillColor] (147.26, 35.61) rectangle (162.63, 80.06);

\path[draw=drawColor,line width= 0.3pt,fill=fillColor] (181.42, 35.61) rectangle (196.79, 80.72);
\definecolor{fillColor}{RGB}{241,163,64}

\path[draw=drawColor,line width= 0.3pt,fill=fillColor] ( 61.86, 35.61) rectangle ( 77.23,105.36);

\path[draw=drawColor,line width= 0.3pt,fill=fillColor] ( 96.02, 35.61) rectangle (111.39,117.73);

\path[draw=drawColor,line width= 0.3pt,fill=fillColor] (130.18, 35.61) rectangle (145.55,125.20);

\path[draw=drawColor,line width= 0.3pt,fill=fillColor] (164.34, 35.61) rectangle (179.71,128.25);

\path[draw=drawColor,line width= 0.3pt,fill=fillColor] (198.50, 35.61) rectangle (213.87,133.15);

\path[draw=drawColor,line width= 0.6pt,line join=round] ( 49.90, 79.23) --
	( 55.03, 79.23);

\path[draw=drawColor,line width= 0.6pt,line join=round] ( 52.47, 79.23) --
	( 52.47, 78.65);

\path[draw=drawColor,line width= 0.6pt,line join=round] ( 49.90, 78.65) --
	( 55.03, 78.65);

\path[draw=drawColor,line width= 0.6pt,line join=round] ( 84.06, 81.31) --
	( 89.19, 81.31);

\path[draw=drawColor,line width= 0.6pt,line join=round] ( 86.63, 81.31) --
	( 86.63, 80.24);

\path[draw=drawColor,line width= 0.6pt,line join=round] ( 84.06, 80.24) --
	( 89.19, 80.24);

\path[draw=drawColor,line width= 0.6pt,line join=round] (118.22, 80.48) --
	(123.35, 80.48);

\path[draw=drawColor,line width= 0.6pt,line join=round] (120.79, 80.48) --
	(120.79, 80.16);

\path[draw=drawColor,line width= 0.6pt,line join=round] (118.22, 80.16) --
	(123.35, 80.16);

\path[draw=drawColor,line width= 0.6pt,line join=round] (152.38, 80.29) --
	(157.51, 80.29);

\path[draw=drawColor,line width= 0.6pt,line join=round] (154.95, 80.29) --
	(154.95, 79.90);

\path[draw=drawColor,line width= 0.6pt,line join=round] (152.38, 79.90) --
	(157.51, 79.90);

\path[draw=drawColor,line width= 0.6pt,line join=round] (186.54, 81.15) --
	(191.67, 81.15);

\path[draw=drawColor,line width= 0.6pt,line join=round] (189.10, 81.15) --
	(189.10, 79.56);

\path[draw=drawColor,line width= 0.6pt,line join=round] (186.54, 79.56) --
	(191.67, 79.56);

\path[draw=drawColor,line width= 0.6pt,line join=round] ( 66.98,106.33) --
	( 72.11,106.33);

\path[draw=drawColor,line width= 0.6pt,line join=round] ( 69.55,106.33) --
	( 69.55,104.05);

\path[draw=drawColor,line width= 0.6pt,line join=round] ( 66.98,104.05) --
	( 72.11,104.05);

\path[draw=drawColor,line width= 0.6pt,line join=round] (101.14,118.88) --
	(106.27,118.88);

\path[draw=drawColor,line width= 0.6pt,line join=round] (103.71,118.88) --
	(103.71,115.88);

\path[draw=drawColor,line width= 0.6pt,line join=round] (101.14,115.88) --
	(106.27,115.88);

\path[draw=drawColor,line width= 0.6pt,line join=round] (135.30,126.76) --
	(140.43,126.76);

\path[draw=drawColor,line width= 0.6pt,line join=round] (137.87,126.76) --
	(137.87,124.65);

\path[draw=drawColor,line width= 0.6pt,line join=round] (135.30,124.65) --
	(140.43,124.65);

\path[draw=drawColor,line width= 0.6pt,line join=round] (169.46,129.34) --
	(174.59,129.34);

\path[draw=drawColor,line width= 0.6pt,line join=round] (172.02,129.34) --
	(172.02,127.96);

\path[draw=drawColor,line width= 0.6pt,line join=round] (169.46,127.96) --
	(174.59,127.96);

\path[draw=drawColor,line width= 0.6pt,line join=round] (203.62,134.11) --
	(208.75,134.11);

\path[draw=drawColor,line width= 0.6pt,line join=round] (206.18,134.11) --
	(206.18,132.86);

\path[draw=drawColor,line width= 0.6pt,line join=round] (203.62,132.86) --
	(208.75,132.86);
\definecolor{drawColor}{gray}{0.20}

\path[draw=drawColor,line width= 0.6pt,line join=round,line cap=round] ( 40.51, 30.69) rectangle (218.14,139.04);
\end{scope}
\begin{scope}
\path[clip] (  0.00,  0.00) rectangle (289.08,144.54);
\definecolor{drawColor}{gray}{0.30}

\node[text=drawColor,anchor=base east,inner sep=0pt, outer sep=0pt, scale=  0.88] at ( 35.56, 32.58) {0};

\node[text=drawColor,anchor=base east,inner sep=0pt, outer sep=0pt, scale=  0.88] at ( 35.56, 56.39) {250};

\node[text=drawColor,anchor=base east,inner sep=0pt, outer sep=0pt, scale=  0.88] at ( 35.56, 80.20) {500};

\node[text=drawColor,anchor=base east,inner sep=0pt, outer sep=0pt, scale=  0.88] at ( 35.56,104.01) {750};

\node[text=drawColor,anchor=base east,inner sep=0pt, outer sep=0pt, scale=  0.88] at ( 35.56,127.83) {1000};
\end{scope}
\begin{scope}
\path[clip] (  0.00,  0.00) rectangle (289.08,144.54);
\definecolor{drawColor}{gray}{0.20}

\path[draw=drawColor,line width= 0.6pt,line join=round] ( 37.76, 35.61) --
	( 40.51, 35.61);

\path[draw=drawColor,line width= 0.6pt,line join=round] ( 37.76, 59.42) --
	( 40.51, 59.42);

\path[draw=drawColor,line width= 0.6pt,line join=round] ( 37.76, 83.23) --
	( 40.51, 83.23);

\path[draw=drawColor,line width= 0.6pt,line join=round] ( 37.76,107.04) --
	( 40.51,107.04);

\path[draw=drawColor,line width= 0.6pt,line join=round] ( 37.76,130.86) --
	( 40.51,130.86);
\end{scope}
\begin{scope}
\path[clip] (  0.00,  0.00) rectangle (289.08,144.54);
\definecolor{drawColor}{gray}{0.20}

\path[draw=drawColor,line width= 0.6pt,line join=round] ( 61.01, 27.94) --
	( 61.01, 30.69);

\path[draw=drawColor,line width= 0.6pt,line join=round] ( 95.17, 27.94) --
	( 95.17, 30.69);

\path[draw=drawColor,line width= 0.6pt,line join=round] (129.33, 27.94) --
	(129.33, 30.69);

\path[draw=drawColor,line width= 0.6pt,line join=round] (163.49, 27.94) --
	(163.49, 30.69);

\path[draw=drawColor,line width= 0.6pt,line join=round] (197.64, 27.94) --
	(197.64, 30.69);
\end{scope}
\begin{scope}
\path[clip] (  0.00,  0.00) rectangle (289.08,144.54);
\definecolor{drawColor}{gray}{0.30}

\node[text=drawColor,anchor=base,inner sep=0pt, outer sep=0pt, scale=  0.88] at ( 61.01, 19.68) {512};

\node[text=drawColor,anchor=base,inner sep=0pt, outer sep=0pt, scale=  0.88] at ( 95.17, 19.68) {1024};

\node[text=drawColor,anchor=base,inner sep=0pt, outer sep=0pt, scale=  0.88] at (129.33, 19.68) {2048};

\node[text=drawColor,anchor=base,inner sep=0pt, outer sep=0pt, scale=  0.88] at (163.49, 19.68) {4096};

\node[text=drawColor,anchor=base,inner sep=0pt, outer sep=0pt, scale=  0.88] at (197.64, 19.68) {8192};
\end{scope}
\begin{scope}
\path[clip] (  0.00,  0.00) rectangle (289.08,144.54);
\definecolor{drawColor}{RGB}{0,0,0}

\node[text=drawColor,anchor=base,inner sep=0pt, outer sep=0pt, scale=  1.10] at (129.33,  7.64) {Target Chunk Size (B)};
\end{scope}
\begin{scope}
\path[clip] (  0.00,  0.00) rectangle (289.08,144.54);
\definecolor{drawColor}{RGB}{0,0,0}

\node[text=drawColor,rotate= 90.00,anchor=base,inner sep=0pt, outer sep=0pt, scale=  1.10] at ( 13.08, 84.86) {Throughput (MiB/s)};
\end{scope}
\begin{scope}
\path[clip] (  0.00,  0.00) rectangle (289.08,144.54);
\definecolor{fillColor}{RGB}{255,255,255}

\path[fill=fillColor] (229.14, 62.16) rectangle (283.58,107.57);
\end{scope}
\begin{scope}
\path[clip] (  0.00,  0.00) rectangle (289.08,144.54);
\definecolor{fillColor}{RGB}{255,255,255}

\path[fill=fillColor] (234.64, 82.11) rectangle (249.09, 96.57);
\end{scope}
\begin{scope}
\path[clip] (  0.00,  0.00) rectangle (289.08,144.54);
\definecolor{drawColor}{RGB}{0,0,0}
\definecolor{fillColor}{gray}{0.97}

\path[draw=drawColor,line width= 0.3pt,fill=fillColor] (235.07, 82.54) rectangle (248.67, 96.14);
\end{scope}
\begin{scope}
\path[clip] (  0.00,  0.00) rectangle (289.08,144.54);
\definecolor{fillColor}{RGB}{255,255,255}

\path[fill=fillColor] (234.64, 67.66) rectangle (249.09, 82.11);
\end{scope}
\begin{scope}
\path[clip] (  0.00,  0.00) rectangle (289.08,144.54);
\definecolor{drawColor}{RGB}{0,0,0}
\definecolor{fillColor}{RGB}{241,163,64}

\path[draw=drawColor,line width= 0.3pt,fill=fillColor] (235.07, 68.09) rectangle (248.67, 81.69);
\end{scope}
\begin{scope}
\path[clip] (  0.00,  0.00) rectangle (289.08,144.54);
\definecolor{drawColor}{RGB}{0,0,0}

\node[text=drawColor,anchor=base west,inner sep=0pt, outer sep=0pt, scale=  0.88] at (254.59, 86.31) {Scalar};
\end{scope}
\begin{scope}
\path[clip] (  0.00,  0.00) rectangle (289.08,144.54);
\definecolor{drawColor}{RGB}{0,0,0}

\node[text=drawColor,anchor=base west,inner sep=0pt, outer sep=0pt, scale=  0.88] at (254.59, 71.86) {SIMD};
\end{scope}
\end{tikzpicture}

%% file: sections/csd.tex
In this section, we explore the chunk size distribution of the selected algorithms.
Our analysis is based on empirical data collected from running these algorithms on diverse datasets, reflecting both high-entropy and low-entropy scenarios.
The chunk size distributions produced by any \ac{CDC} algorithm can be characterized by two relevant statistics:
\begin{enumerate*}
	\item The empirical mean chunk size $\bar{cs}$ produced, which should be close to the target $\mu$.
	Intuitively, this reflects how easy it is to configure an algorithm for a target and how predictable its behavior is.
	
	\item The spread of the distribution around the mean, calculated as the empirical standard deviation, $s$.
	As laid out in \Cref{sec:background:csVariance}, pathological chunk sizes are undesirable.
	On the other hand, an algorithm must show some flexibility in the size of chunks produced in order to effectively combat the boundary shift problem.
\end{enumerate*}
Additionally, it is helpful to not just evaluate an algorithm based on $\bar{cs}$ and $s$, but also examine the shape of the distribution function,
and understand the mechanics behind it.

\subsection{Distributions}

\begin{figure}[tb]
\centering
\subfloat[RAND]{\scalebox{0.85}{\input{fig/csd_random.tex}}%
    \label{fig:csd_rand}}
\hfil
\subfloat[CODE]{\scalebox{0.85}{\input{fig/csd_code.tex}}%
    \label{fig:csd_code}}
\vspace{-1em}
\makebox[\textwidth][l]{\hspace{-0.8cm}\input{fig/csd_legendonly.tex}}
\vspace{-2.5em}
\caption{Chunk size distributions for target chunk size $\mu=1024$ (in case of MII, $\mu=770$).}
\label{fig:csd_datasets}
\end{figure}
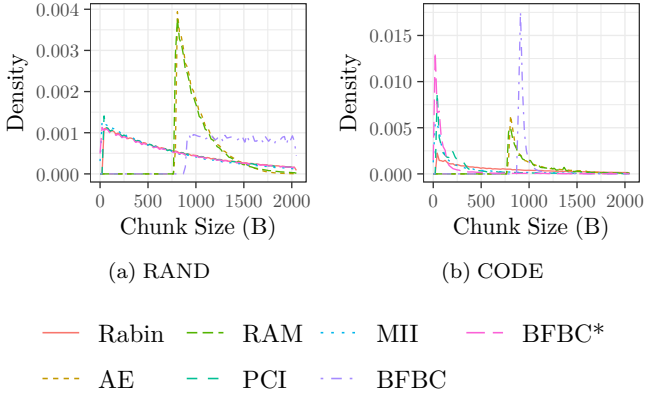

In \cref{fig:csd_datasets}, we show the distribution of the produced chunk sizes for a selection of algorithms.
For ease of interpretation, and because their distributions resemble Rabin's, we do not show Buzhash and Gear, although we discuss minor differences in \cref{sec:csd:quantitative}.
For a comprehensive overview of the distributions across all settings, and to better distinguish between individual algorithms, please visit \url{https://mrd0ll4r.github.io/cdc-algorithm-tester}, where we provide the data through interactive charts.

Strikingly, almost all algorithms exhibit a similar shape.
The reason is that they underlie the same stochastic property:
Each position of the data stream, looked upon independently, is equally likely to become a chunk cut-point (\cf \Cref{sec:background}).
However, each position's probability also depends on all previous positions not having fulfilled the same matching condition.
Therefore, we observe that almost all distributions peak at a minimal size, determined by the window size or other feature of the algorithm,
and then decay, forming a heavy-tailed distribution.
The BSW algorithms, utilizing small windows of typically \qty{32}{\byte} to \qty{256}{\byte},
form a large number of small chunks,
but then compensate by producing a long tail that shifts the mean closer to the target.
This inevitably results in a large chunk-size variance.
The local maxima-based algorithms, in contrast,
present themselves with a minimum chunk size, bound by the horizon size in AE and RAM, close to the target $\mu$. 
%Secondly, the probabilistic difficulty is decreased.
%\leo{what does this mean?}
Their distributions drop much more rapidly, with most of the chunk sizes forming within a smaller region around the target.
The only exception to this pattern is the distribution of BFBC on RAND.
Recall how BFBC operates on a fixed set of $k=3$ most popular byte pairs of the dataset (\cf \Cref{sec:algorithms:bfbc}), as well as skipping $\mu-\qty{128}{\byte}$.
This is a valid strategy for low-entropy datasets, as the most frequent byte pairs are expected to occur very frequently,
leading to chunk boundaries closely aligning with $\mu$.
Because the top frequent byte pairs in those datasets occur with such high frequency, 
chunks never grow much beyond the minimum chunk size.
For example, we find that in CODE the most frequent byte pair were two spaces, presumably for indentation; 
in LNX, PDF, and WEB, we find that it is two null bytes, presumably for padding.
On pseudorandom data, however,
the top most frequent byte pairs occur as frequently as any other.
This leads to an almost uniform distribution of chunk sizes after the skipped minimum size,
with single chunks sized up to \qty{300}{KB}.

\subsection{Quantitative Results}\label{sec:csd:quantitative}

We now move to present the mean chunk size $\bar{cs}$ and standard deviation $s$
for all datasets and algorithms evaluated.
To that end, we include an extensive table in the appendix (\Cref{tab:csd_means_sd_full}),
and an aggregated version here (\Cref{tab:csd_overview}).
The cell colors follow a continuous scale,
where a light color indicates values near the optimum.
For the mean $\bar{cs}$, this is $\mu$.
The color range reaches its maximum where the mean deviates $\pm\qty{100}{\%}$ from the target $\mu$.
For the standard deviation $s$,
the color range is relative to the empirical mean $\bar{cs}$, in a range $[0, 2\bar{cs}]$.
Recall (\cf \Cref{sec:algorithms:ram}) that we examine all algorithms without a limit on the chunk size.

\begin{comment}
We deliberately ran the algorithms without constraining the chunk size.
Our intention is to show each algorithm's inherent capability of finding chunk boundaries in a controlled and balanced way, on different datasets.
Furthermore, we want to ensure comparability with our results in \cref{sec:dedup},
where we show every algorithm's deduplication ratio uncorrupted by such constraints.
\end{comment}

\begin{table}[tb]
    \caption{Aggregated Overview of the Relative Performance of Chunk-size Variance and Mean}
%    	For Gear with NC, the targets 770 and \qty{5482}{B} were disregarded.
    \centering
    \input{tab/csd_overview}
    \label{tab:csd_overview}
\end{table}

%% Observations
We make the following observations:
Firstly, judging by results rendered for RAND,
we see our formulas for determining parameters for AE and RAM confirmed,
as means are observed very closely to the desired target.
The empirical means for MII, on the other hand, exceed the target by 13--15\,\%.
Additionally, we observe that, for almost all algorithms, chunk-size variance seemingly correlates inversely with dataset entropy,
\ie, datasets with low entropy tend to lead to higher variance in chunk sizes produced.
Often, with BSW algorithms, we furthermore observe an increase in variance when target chunk sizes are higher.
With respect to the mean, the performance varies with no obvious pattern.

% Observations: BSW and NC
\subsubsection{BSW Algorithms and Normalized Chunking}

\begin{figure}[t]
	\centering
	\subfloat[RAND]{\scalebox{0.85}{\input{fig/csd_gear_variants_random}}%
		\label{fig:csd_gear_variants_rand}}
	\hfill
	\subfloat[CODE]{\scalebox{0.85}{\input{fig/csd_gear_variants_code}}%
		\label{fig:csd_gear_variants_code}}
	
	\input{fig/csd_gear_variants_legendonly}
	\vspace{-2em}
	\caption{Effects of NC on the chunk size distribution of Gear, with target chunk size \qty{2}{KiB}.
		Dots beneath the x-axis mark the mean chunk size produced.}
	\label{fig:csd_gear_variants}
\end{figure}
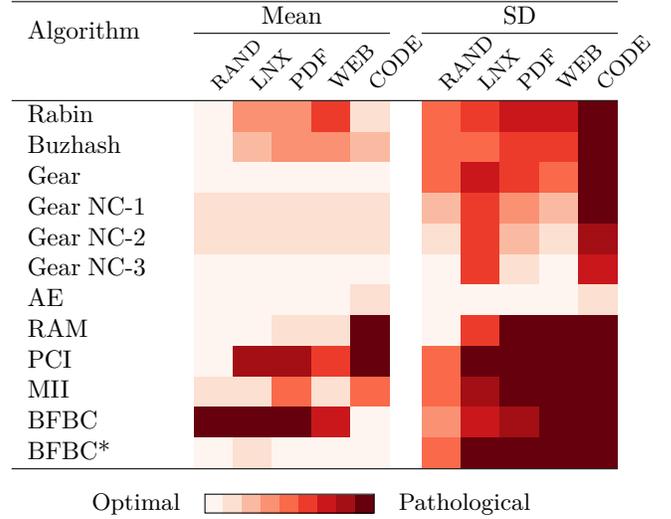
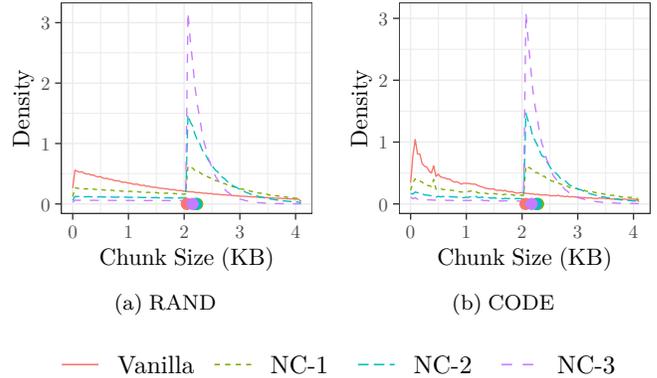

Different hash functions within BSW algorithms yield different distributions.
While Gear seems to be better at maintaining means close to the target,
all BSW algorithms notoriously struggle with chunk-size variance.
This effect is gradually reduced with increasing levels of NC, see~\Cref{fig:csd_gear_variants}.

% Observations: AE and RAM
\subsubsection{Local Extrema-based Algorithms}
The lowest levels of variance are produced by AE.
Historically, this was also the motivation behind local extrema approaches in general.
In contrast, RAM fulfills this promise only on the RAND dataset.
On the realistic datasets, the results are pathological as RAM fails to find chunk boundaries.
The authors warned against poor performance on files containing low-entropy.
In their paper~\cite{ram}, the experiments yielded very similar results in comparison with AE.
Surprisingly, we find also pathological performance on LNX and PDF, which still represent realistic datasets with fairly high entropy.

% Observations: PCI and MII
\subsubsection{Algorithms for Data Synchronization}
PCI exhibits a significant deviation from the target chunk size and high chunk-size variance across most settings.
Its predecessor, MII, although restrictive in the targets it can tune to, performs slightly better in this regard.
However, since chunk-size variance is not a critical concern in the application for data synchronization, this may not be a substantial drawback.

% Observations: BFBC(*)
\subsubsection{BFBC}
Finally, BFBC attains pathological means if datasets have high entropy, for reasons explained previously.
The BFBC* variant
%, with its algorithm for picking byte pairs specifically to meet a target chunk size, 
fixes this issue.
However, high chunk-size variance remains a problem as the uniform distribution of the divisors within the datasets is never given, neither with BFBC nor with BFBC*.

\subsection{Key Takeaways}
\label{sec:csd:takeaways}

In this section, we focused on the distribution of chunk sizes produced by CDC algorithms on both synthetic as well as real-world datasets.
In summary, we arrive at the following conclusions.

\subsubsection{Heavy-Tailed Distributions}
Almost all algorithms produce chunk sizes with a heavy-tailed distribution.
We find a large number of small chunks shortly after a minimum defined by the algorithm, \eg, the size of the window, which is always smaller than the target chunk size.
This is followed by a heavy tail, which moves the produced mean closer to the target.

\subsubsection{Chunk-Size Variance}
We find that aforementioned skew in the distributions is more pronounced for BSW algorithms, which operate on a relatively small window.
AE and RAM produce a distribution of similar shape, although forming much more closely around the target mean, leading to lower chunk-size variance.

\subsubsection{Normalized Chunking}
Gear with normalized chunking presents a variant with lower chunk-size variance than plain Gear through its use of two matching conditions.
This presents itself in the distribution as two pronounced peaks.

%% file: fig/csd_random.tex
% Created by tikzDevice version 0.12.6 on 2024-07-21 09:04:57
% !TEX encoding = UTF-8 Unicode
\begin{tikzpicture}[x=1pt,y=1pt]
\definecolor{fillColor}{RGB}{255,255,255}
\path[use as bounding box,fill=fillColor,fill opacity=0.00] (0,0) rectangle (144.54,115.63);
\begin{scope}
\path[clip] (  0.00,  0.00) rectangle (144.54,115.63);
\definecolor{drawColor}{RGB}{255,255,255}
\definecolor{fillColor}{RGB}{255,255,255}

\path[draw=drawColor,line width= 0.6pt,line join=round,line cap=round,fill=fillColor] (  0.00,  0.00) rectangle (144.54,115.63);
\end{scope}
\begin{scope}
\path[clip] ( 42.95, 30.69) rectangle (139.04,110.13);
\definecolor{fillColor}{RGB}{255,255,255}

\path[fill=fillColor] ( 42.95, 30.69) rectangle (139.04,110.13);
\definecolor{drawColor}{gray}{0.92}

\path[draw=drawColor,line width= 0.3pt,line join=round] ( 42.95, 43.46) --
	(139.04, 43.46);

\path[draw=drawColor,line width= 0.3pt,line join=round] ( 42.95, 61.79) --
	(139.04, 61.79);

\path[draw=drawColor,line width= 0.3pt,line join=round] ( 42.95, 80.13) --
	(139.04, 80.13);

\path[draw=drawColor,line width= 0.3pt,line join=round] ( 42.95, 98.46) --
	(139.04, 98.46);

\path[draw=drawColor,line width= 0.3pt,line join=round] ( 57.98, 30.69) --
	( 57.98,110.13);

\path[draw=drawColor,line width= 0.3pt,line join=round] ( 79.31, 30.69) --
	( 79.31,110.13);

\path[draw=drawColor,line width= 0.3pt,line join=round] (100.64, 30.69) --
	(100.64,110.13);

\path[draw=drawColor,line width= 0.3pt,line join=round] (121.96, 30.69) --
	(121.96,110.13);

\path[draw=drawColor,line width= 0.6pt,line join=round] ( 42.95, 34.30) --
	(139.04, 34.30);

\path[draw=drawColor,line width= 0.6pt,line join=round] ( 42.95, 52.63) --
	(139.04, 52.63);

\path[draw=drawColor,line width= 0.6pt,line join=round] ( 42.95, 70.96) --
	(139.04, 70.96);

\path[draw=drawColor,line width= 0.6pt,line join=round] ( 42.95, 89.29) --
	(139.04, 89.29);

\path[draw=drawColor,line width= 0.6pt,line join=round] ( 42.95,107.62) --
	(139.04,107.62);

\path[draw=drawColor,line width= 0.6pt,line join=round] ( 47.32, 30.69) --
	( 47.32,110.13);

\path[draw=drawColor,line width= 0.6pt,line join=round] ( 68.65, 30.69) --
	( 68.65,110.13);

\path[draw=drawColor,line width= 0.6pt,line join=round] ( 89.97, 30.69) --
	( 89.97,110.13);

\path[draw=drawColor,line width= 0.6pt,line join=round] (111.30, 30.69) --
	(111.30,110.13);

\path[draw=drawColor,line width= 0.6pt,line join=round] (132.63, 30.69) --
	(132.63,110.13);
\definecolor{drawColor}{RGB}{248,118,109}

\path[draw=drawColor,line width= 0.6pt,line join=round] ( 47.32, 34.30) --
	( 48.20, 34.30) --
	( 49.09, 54.24) --
	( 49.97, 54.79) --
	( 50.85, 54.39) --
	( 51.73, 53.00) --
	( 52.62, 53.56) --
	( 53.50, 53.18) --
	( 54.38, 51.93) --
	( 55.26, 52.46) --
	( 56.14, 52.08) --
	( 57.03, 50.86) --
	( 57.91, 51.47) --
	( 58.79, 51.00) --
	( 59.67, 49.96) --
	( 60.56, 50.33) --
	( 61.44, 50.12) --
	( 62.32, 49.72) --
	( 63.20, 48.65) --
	( 64.09, 49.13) --
	( 64.97, 48.92) --
	( 65.85, 47.94) --
	( 66.73, 48.23) --
	( 67.62, 48.01) --
	( 68.50, 47.10) --
	( 69.38, 47.46) --
	( 70.26, 47.15) --
	( 71.14, 46.33) --
	( 72.03, 46.69) --
	( 72.91, 46.37) --
	( 73.79, 45.63) --
	( 74.67, 45.97) --
	( 75.56, 45.75) --
	( 76.44, 45.48) --
	( 77.32, 44.75) --
	( 78.20, 45.05) --
	( 79.09, 44.77) --
	( 79.97, 44.16) --
	( 80.85, 44.48) --
	( 81.73, 44.22) --
	( 82.61, 43.50) --
	( 83.50, 43.85) --
	( 84.38, 43.66) --
	( 85.26, 43.03) --
	( 86.14, 43.24) --
	( 87.03, 43.05) --
	( 87.91, 42.46) --
	( 88.79, 42.69) --
	( 89.67, 42.53) --
	( 90.56, 42.41) --
	( 91.44, 41.83) --
	( 92.32, 42.05) --
	( 93.20, 41.89) --
	( 94.09, 41.38) --
	( 94.97, 41.59) --
	( 95.85, 41.44) --
	( 96.73, 40.97) --
	( 97.61, 41.18) --
	( 98.50, 41.10) --
	( 99.38, 40.65) --
	(100.26, 40.77) --
	(101.14, 40.66) --
	(102.03, 40.22) --
	(102.91, 40.37) --
	(103.79, 40.26) --
	(104.67, 39.87) --
	(105.56, 40.02) --
	(106.44, 39.93) --
	(107.32, 39.80) --
	(108.20, 39.46) --
	(109.08, 39.54) --
	(109.97, 39.50) --
	(110.85, 39.14) --
	(111.73, 39.27) --
	(112.61, 39.17) --
	(113.50, 38.87) --
	(114.38, 38.98) --
	(115.26, 38.91) --
	(116.14, 38.55) --
	(117.03, 38.69) --
	(117.91, 38.60) --
	(118.79, 38.33) --
	(119.67, 38.45) --
	(120.56, 38.38) --
	(121.44, 38.32) --
	(122.32, 38.07) --
	(123.20, 38.13) --
	(124.08, 38.04) --
	(124.97, 37.79) --
	(125.85, 37.91) --
	(126.73, 37.86) --
	(127.61, 37.59) --
	(128.50, 37.71) --
	(129.38, 37.62) --
	(130.26, 37.41) --
	(131.14, 37.46) --
	(132.03, 37.41) --
	(132.91, 37.22) --
	(133.79, 37.29) --
	(134.67, 35.86);
\definecolor{drawColor}{RGB}{196,154,0}

\path[draw=drawColor,line width= 0.6pt,dash pattern=on 2pt off 2pt ,line join=round] ( 47.32, 34.30) --
	( 48.20, 34.30) --
	( 49.09, 34.30) --
	( 49.97, 34.30) --
	( 50.85, 34.30) --
	( 51.73, 34.30) --
	( 52.62, 34.30) --
	( 53.50, 34.30) --
	( 54.38, 34.30) --
	( 55.26, 34.30) --
	( 56.14, 34.30) --
	( 57.03, 34.30) --
	( 57.91, 34.30) --
	( 58.79, 34.30) --
	( 59.67, 34.30) --
	( 60.56, 34.30) --
	( 61.44, 34.30) --
	( 62.32, 34.30) --
	( 63.20, 34.30) --
	( 64.09, 34.30) --
	( 64.97, 34.30) --
	( 65.85, 34.30) --
	( 66.73, 34.30) --
	( 67.62, 34.30) --
	( 68.50, 34.30) --
	( 69.38, 34.30) --
	( 70.26, 34.30) --
	( 71.14, 34.30) --
	( 72.03, 34.30) --
	( 72.91, 34.30) --
	( 73.79, 34.30) --
	( 74.67, 34.30) --
	( 75.56, 34.30) --
	( 76.44, 34.30) --
	( 77.32, 34.30) --
	( 78.20, 34.30) --
	( 79.09, 34.30) --
	( 79.97, 34.30) --
	( 80.85, 45.11) --
	( 81.73,106.52) --
	( 82.61, 97.59) --
	( 83.50, 95.28) --
	( 84.38, 90.36) --
	( 85.26, 83.54) --
	( 86.14, 81.90) --
	( 87.03, 77.95) --
	( 87.91, 72.61) --
	( 88.79, 71.42) --
	( 89.67, 68.44) --
	( 90.56, 65.68) --
	( 91.44, 61.77) --
	( 92.32, 60.96) --
	( 93.20, 58.78) --
	( 94.09, 55.79) --
	( 94.97, 55.10) --
	( 95.85, 53.50) --
	( 96.73, 51.05) --
	( 97.61, 50.60) --
	( 98.50, 49.32) --
	( 99.38, 47.47) --
	(100.26, 46.97) --
	(101.14, 46.05) --
	(102.03, 44.59) --
	(102.91, 44.29) --
	(103.79, 43.48) --
	(104.67, 42.36) --
	(105.56, 42.06) --
	(106.44, 41.51) --
	(107.32, 40.97) --
	(108.20, 40.11) --
	(109.08, 39.93) --
	(109.97, 39.52) --
	(110.85, 38.85) --
	(111.73, 38.71) --
	(112.61, 38.39) --
	(113.50, 37.88) --
	(114.38, 37.77) --
	(115.26, 37.40) --
	(116.14, 36.84) --
	(117.03, 36.54) --
	(117.91, 36.19) --
	(118.79, 35.81) --
	(119.67, 35.67) --
	(120.56, 35.43) --
	(121.44, 35.25) --
	(122.32, 35.07) --
	(123.20, 35.01) --
	(124.08, 34.89) --
	(124.97, 34.78) --
	(125.85, 34.72) --
	(126.73, 34.65) --
	(127.61, 34.59) --
	(128.50, 34.55) --
	(129.38, 34.52) --
	(130.26, 34.47) --
	(131.14, 34.45) --
	(132.03, 34.43) --
	(132.91, 34.39) --
	(133.79, 34.39) --
	(134.67, 34.34);
\definecolor{drawColor}{RGB}{83,180,0}

\path[draw=drawColor,line width= 0.6pt,dash pattern=on 4pt off 2pt ,line join=round] ( 47.32, 34.30) --
	( 48.20, 34.30) --
	( 49.09, 34.30) --
	( 49.97, 34.30) --
	( 50.85, 34.30) --
	( 51.73, 34.30) --
	( 52.62, 34.30) --
	( 53.50, 34.30) --
	( 54.38, 34.30) --
	( 55.26, 34.30) --
	( 56.14, 34.30) --
	( 57.03, 34.30) --
	( 57.91, 34.30) --
	( 58.79, 34.30) --
	( 59.67, 34.30) --
	( 60.56, 34.30) --
	( 61.44, 34.30) --
	( 62.32, 34.30) --
	( 63.20, 34.30) --
	( 64.09, 34.30) --
	( 64.97, 34.30) --
	( 65.85, 34.30) --
	( 66.73, 34.30) --
	( 67.62, 34.30) --
	( 68.50, 34.30) --
	( 69.38, 34.30) --
	( 70.26, 34.30) --
	( 71.14, 34.30) --
	( 72.03, 34.30) --
	( 72.91, 34.30) --
	( 73.79, 34.30) --
	( 74.67, 34.30) --
	( 75.56, 34.30) --
	( 76.44, 34.30) --
	( 77.32, 34.30) --
	( 78.20, 34.30) --
	( 79.09, 34.30) --
	( 79.97, 34.30) --
	( 80.85, 87.55) --
	( 81.73,103.38) --
	( 82.61, 94.50) --
	( 83.50, 92.21) --
	( 84.38, 87.41) --
	( 85.26, 80.82) --
	( 86.14, 79.13) --
	( 87.03, 75.32) --
	( 87.91, 70.21) --
	( 88.79, 69.06) --
	( 89.67, 66.17) --
	( 90.56, 63.57) --
	( 91.44, 59.89) --
	( 92.32, 59.11) --
	( 93.20, 56.98) --
	( 94.09, 54.23) --
	( 94.97, 53.60) --
	( 95.85, 52.09) --
	( 96.73, 49.81) --
	( 97.61, 49.32) --
	( 98.50, 48.17) --
	( 99.38, 46.46) --
	(100.26, 45.96) --
	(101.14, 45.09) --
	(102.03, 43.79) --
	(102.91, 43.49) --
	(103.79, 42.71) --
	(104.67, 41.73) --
	(105.56, 41.46) --
	(106.44, 40.90) --
	(107.32, 40.43) --
	(108.20, 39.64) --
	(109.08, 39.45) --
	(109.97, 39.09) --
	(110.85, 38.48) --
	(111.73, 38.34) --
	(112.61, 38.05) --
	(113.50, 37.59) --
	(114.38, 37.48) --
	(115.26, 37.22) --
	(116.14, 36.88) --
	(117.03, 36.77) --
	(117.91, 36.60) --
	(118.79, 36.31) --
	(119.67, 36.25) --
	(120.56, 36.09) --
	(121.44, 35.94) --
	(122.32, 35.74) --
	(123.20, 35.73) --
	(124.08, 35.60) --
	(124.97, 35.42) --
	(125.85, 35.41) --
	(126.73, 35.32) --
	(127.61, 35.20) --
	(128.50, 35.16) --
	(129.38, 35.11) --
	(130.26, 35.00) --
	(131.14, 34.98) --
	(132.03, 34.93) --
	(132.91, 34.84) --
	(133.79, 34.83) --
	(134.67, 34.57);
\definecolor{drawColor}{RGB}{0,192,148}

\path[draw=drawColor,line width= 0.6pt,dash pattern=on 4pt off 4pt ,line join=round] ( 47.32, 34.30) --
	( 48.20, 34.30) --
	( 49.09, 60.13) --
	( 49.97, 55.01) --
	( 50.85, 54.61) --
	( 51.73, 53.29) --
	( 52.62, 53.81) --
	( 53.50, 53.45) --
	( 54.38, 52.18) --
	( 55.26, 52.65) --
	( 56.14, 52.23) --
	( 57.03, 51.12) --
	( 57.91, 51.59) --
	( 58.79, 51.17) --
	( 59.67, 50.03) --
	( 60.56, 50.56) --
	( 61.44, 50.15) --
	( 62.32, 49.82) --
	( 63.20, 48.78) --
	( 64.09, 49.19) --
	( 64.97, 48.91) --
	( 65.85, 47.91) --
	( 66.73, 48.26) --
	( 67.62, 48.01) --
	( 68.50, 47.04) --
	( 69.38, 47.48) --
	( 70.26, 47.18) --
	( 71.14, 46.33) --
	( 72.03, 46.62) --
	( 72.91, 46.34) --
	( 73.79, 45.65) --
	( 74.67, 45.89) --
	( 75.56, 45.61) --
	( 76.44, 45.46) --
	( 77.32, 44.67) --
	( 78.20, 44.97) --
	( 79.09, 44.69) --
	( 79.97, 44.00) --
	( 80.85, 44.33) --
	( 81.73, 44.12) --
	( 82.61, 43.44) --
	( 83.50, 43.69) --
	( 84.38, 43.53) --
	( 85.26, 42.87) --
	( 86.14, 43.15) --
	( 87.03, 42.96) --
	( 87.91, 42.39) --
	( 88.79, 42.58) --
	( 89.67, 42.44) --
	( 90.56, 42.28) --
	( 91.44, 41.75) --
	( 92.32, 41.92) --
	( 93.20, 41.80) --
	( 94.09, 41.26) --
	( 94.97, 41.46) --
	( 95.85, 41.34) --
	( 96.73, 40.81) --
	( 97.61, 41.02) --
	( 98.50, 40.91) --
	( 99.38, 40.51) --
	(100.26, 40.66) --
	(101.14, 40.52) --
	(102.03, 40.10) --
	(102.91, 40.24) --
	(103.79, 40.11) --
	(104.67, 39.77) --
	(105.56, 39.91) --
	(106.44, 39.78) --
	(107.32, 39.69) --
	(108.20, 39.32) --
	(109.08, 39.44) --
	(109.97, 39.35) --
	(110.85, 38.98) --
	(111.73, 39.10) --
	(112.61, 39.02) --
	(113.50, 38.66) --
	(114.38, 38.86) --
	(115.26, 38.76) --
	(116.14, 38.44) --
	(117.03, 38.55) --
	(117.91, 38.46) --
	(118.79, 38.23) --
	(119.67, 38.31) --
	(120.56, 38.23) --
	(121.44, 38.11) --
	(122.32, 37.88) --
	(123.20, 37.96) --
	(124.08, 37.88) --
	(124.97, 37.68) --
	(125.85, 37.76) --
	(126.73, 37.68) --
	(127.61, 37.47) --
	(128.50, 37.53) --
	(129.38, 37.52) --
	(130.26, 37.29) --
	(131.14, 37.34) --
	(132.03, 37.30) --
	(132.91, 37.08) --
	(133.79, 37.15) --
	(134.67, 35.78);
\definecolor{drawColor}{RGB}{0,182,235}

\path[draw=drawColor,line width= 0.6pt,dash pattern=on 1pt off 3pt ,line join=round] ( 47.32, 40.00) --
	( 48.20, 57.30) --
	( 49.09, 55.67) --
	( 49.97, 56.22) --
	( 50.85, 55.68) --
	( 51.73, 54.29) --
	( 52.62, 54.81) --
	( 53.50, 54.32) --
	( 54.38, 52.96) --
	( 55.26, 53.38) --
	( 56.14, 52.84) --
	( 57.03, 51.60) --
	( 57.91, 52.09) --
	( 58.79, 51.65) --
	( 59.67, 50.43) --
	( 60.56, 50.83) --
	( 61.44, 50.51) --
	( 62.32, 50.11) --
	( 63.20, 49.00) --
	( 64.09, 49.41) --
	( 64.97, 49.07) --
	( 65.85, 47.96) --
	( 66.73, 48.36) --
	( 67.62, 48.04) --
	( 68.50, 47.09) --
	( 69.38, 47.38) --
	( 70.26, 47.08) --
	( 71.14, 46.22) --
	( 72.03, 46.56) --
	( 72.91, 46.24) --
	( 73.79, 45.39) --
	( 74.67, 45.72) --
	( 75.56, 45.39) --
	( 76.44, 45.10) --
	( 77.32, 44.40) --
	( 78.20, 44.66) --
	( 79.09, 44.44) --
	( 79.97, 43.73) --
	( 80.85, 43.96) --
	( 81.73, 43.75) --
	( 82.61, 43.03) --
	( 83.50, 43.28) --
	( 84.38, 43.12) --
	( 85.26, 42.47) --
	( 86.14, 42.69) --
	( 87.03, 42.45) --
	( 87.91, 41.90) --
	( 88.79, 42.11) --
	( 89.67, 41.93) --
	( 90.56, 41.75) --
	( 91.44, 41.22) --
	( 92.32, 41.42) --
	( 93.20, 41.24) --
	( 94.09, 40.76) --
	( 94.97, 40.96) --
	( 95.85, 40.81) --
	( 96.73, 40.33) --
	( 97.61, 40.51) --
	( 98.50, 40.35) --
	( 99.38, 39.92) --
	(100.26, 40.03) --
	(101.14, 39.86) --
	(102.03, 39.53) --
	(102.91, 39.67) --
	(103.79, 39.54) --
	(104.67, 39.19) --
	(105.56, 39.32) --
	(106.44, 39.19) --
	(107.32, 39.06) --
	(108.20, 38.73) --
	(109.08, 38.89) --
	(109.97, 38.76) --
	(110.85, 38.45) --
	(111.73, 38.54) --
	(112.61, 38.44) --
	(113.50, 38.15) --
	(114.38, 38.28) --
	(115.26, 38.13) --
	(116.14, 37.89) --
	(117.03, 38.01) --
	(117.91, 37.89) --
	(118.79, 37.64) --
	(119.67, 37.75) --
	(120.56, 37.67) --
	(121.44, 37.60) --
	(122.32, 37.35) --
	(123.20, 37.41) --
	(124.08, 37.36) --
	(124.97, 37.14) --
	(125.85, 37.21) --
	(126.73, 37.15) --
	(127.61, 36.94) --
	(128.50, 37.02) --
	(129.38, 36.95) --
	(130.26, 36.76) --
	(131.14, 36.82) --
	(132.03, 36.77) --
	(132.91, 36.59) --
	(133.79, 36.68) --
	(134.67, 35.53);
\definecolor{drawColor}{RGB}{165,138,255}

\path[draw=drawColor,line width= 0.6pt,dash pattern=on 1pt off 3pt on 4pt off 3pt ,line join=round] ( 47.32, 34.30) --
	( 48.20, 34.30) --
	( 49.09, 34.30) --
	( 49.97, 34.30) --
	( 50.85, 34.30) --
	( 51.73, 34.30) --
	( 52.62, 34.30) --
	( 53.50, 34.30) --
	( 54.38, 34.30) --
	( 55.26, 34.30) --
	( 56.14, 34.30) --
	( 57.03, 34.30) --
	( 57.91, 34.30) --
	( 58.79, 34.30) --
	( 59.67, 34.30) --
	( 60.56, 34.30) --
	( 61.44, 34.30) --
	( 62.32, 34.30) --
	( 63.20, 34.30) --
	( 64.09, 34.30) --
	( 64.97, 34.30) --
	( 65.85, 34.30) --
	( 66.73, 34.30) --
	( 67.62, 34.30) --
	( 68.50, 34.30) --
	( 69.38, 34.30) --
	( 70.26, 34.30) --
	( 71.14, 34.30) --
	( 72.03, 34.30) --
	( 72.91, 34.30) --
	( 73.79, 34.30) --
	( 74.67, 34.30) --
	( 75.56, 34.30) --
	( 76.44, 34.30) --
	( 77.32, 34.30) --
	( 78.20, 34.30) --
	( 79.09, 34.30) --
	( 79.97, 34.30) --
	( 80.85, 34.30) --
	( 81.73, 34.30) --
	( 82.61, 34.30) --
	( 83.50, 34.30) --
	( 84.38, 34.30) --
	( 85.26, 37.43) --
	( 86.14, 50.95) --
	( 87.03, 51.45) --
	( 87.91, 49.44) --
	( 88.79, 51.81) --
	( 89.67, 51.63) --
	( 90.56, 51.24) --
	( 91.44, 50.95) --
	( 92.32, 50.99) --
	( 93.20, 51.24) --
	( 94.09, 49.76) --
	( 94.97, 50.37) --
	( 95.85, 51.74) --
	( 96.73, 51.74) --
	( 97.61, 50.41) --
	( 98.50, 49.47) --
	( 99.38, 49.98) --
	(100.26, 51.31) --
	(101.14, 50.70) --
	(102.03, 48.61) --
	(102.91, 51.49) --
	(103.79, 50.26) --
	(104.67, 50.48) --
	(105.56, 50.73) --
	(106.44, 50.41) --
	(107.32, 49.40) --
	(108.20, 49.29) --
	(109.08, 49.98) --
	(109.97, 51.45) --
	(110.85, 50.34) --
	(111.73, 48.43) --
	(112.61, 51.63) --
	(113.50, 49.29) --
	(114.38, 50.88) --
	(115.26, 51.42) --
	(116.14, 48.86) --
	(117.03, 48.90) --
	(117.91, 50.19) --
	(118.79, 50.12) --
	(119.67, 48.61) --
	(120.56, 51.35) --
	(121.44, 50.05) --
	(122.32, 48.82) --
	(123.20, 49.94) --
	(124.08, 49.47) --
	(124.97, 47.85) --
	(125.85, 49.94) --
	(126.73, 49.47) --
	(127.61, 49.65) --
	(128.50, 50.66) --
	(129.38, 50.95) --
	(130.26, 47.35) --
	(131.14, 50.52) --
	(132.03, 49.22) --
	(132.91, 51.27) --
	(133.79, 48.79) --
	(134.67, 42.37);
\definecolor{drawColor}{RGB}{251,97,215}

\path[draw=drawColor,line width= 0.6pt,dash pattern=on 7pt off 3pt ,line join=round] ( 47.32, 43.33) --
	( 48.20, 55.03) --
	( 49.09, 53.64) --
	( 49.97, 54.23) --
	( 50.85, 53.79) --
	( 51.73, 52.43) --
	( 52.62, 53.11) --
	( 53.50, 52.63) --
	( 54.38, 51.44) --
	( 55.26, 51.93) --
	( 56.14, 51.54) --
	( 57.03, 50.40) --
	( 57.91, 50.83) --
	( 58.79, 50.55) --
	( 59.67, 49.55) --
	( 60.56, 49.87) --
	( 61.44, 49.61) --
	( 62.32, 49.33) --
	( 63.20, 48.31) --
	( 64.09, 48.72) --
	( 64.97, 48.34) --
	( 65.85, 47.34) --
	( 66.73, 47.81) --
	( 67.62, 47.54) --
	( 68.50, 46.67) --
	( 69.38, 47.04) --
	( 70.26, 46.74) --
	( 71.14, 45.96) --
	( 72.03, 46.25) --
	( 72.91, 46.02) --
	( 73.79, 45.22) --
	( 74.67, 45.58) --
	( 75.56, 45.29) --
	( 76.44, 45.12) --
	( 77.32, 44.38) --
	( 78.20, 44.72) --
	( 79.09, 44.48) --
	( 79.97, 43.81) --
	( 80.85, 44.07) --
	( 81.73, 43.91) --
	( 82.61, 43.23) --
	( 83.50, 43.47) --
	( 84.38, 43.26) --
	( 85.26, 42.68) --
	( 86.14, 42.94) --
	( 87.03, 42.70) --
	( 87.91, 42.20) --
	( 88.79, 42.45) --
	( 89.67, 42.27) --
	( 90.56, 42.12) --
	( 91.44, 41.61) --
	( 92.32, 41.78) --
	( 93.20, 41.63) --
	( 94.09, 41.19) --
	( 94.97, 41.32) --
	( 95.85, 41.18) --
	( 96.73, 40.70) --
	( 97.61, 40.95) --
	( 98.50, 40.82) --
	( 99.38, 40.34) --
	(100.26, 40.53) --
	(101.14, 40.44) --
	(102.03, 39.97) --
	(102.91, 40.19) --
	(103.79, 40.07) --
	(104.67, 39.70) --
	(105.56, 39.77) --
	(106.44, 39.69) --
	(107.32, 39.59) --
	(108.20, 39.22) --
	(109.08, 39.41) --
	(109.97, 39.30) --
	(110.85, 38.94) --
	(111.73, 39.07) --
	(112.61, 38.97) --
	(113.50, 38.64) --
	(114.38, 38.80) --
	(115.26, 38.70) --
	(116.14, 38.41) --
	(117.03, 38.54) --
	(117.91, 38.44) --
	(118.79, 38.16) --
	(119.67, 38.26) --
	(120.56, 38.18) --
	(121.44, 38.14) --
	(122.32, 37.87) --
	(123.20, 37.97) --
	(124.08, 37.89) --
	(124.97, 37.66) --
	(125.85, 37.75) --
	(126.73, 37.67) --
	(127.61, 37.45) --
	(128.50, 37.54) --
	(129.38, 37.50) --
	(130.26, 37.26) --
	(131.14, 37.36) --
	(132.03, 37.28) --
	(132.91, 37.11) --
	(133.79, 37.21) --
	(134.67, 35.78);
\definecolor{drawColor}{gray}{0.20}

\path[draw=drawColor,line width= 0.6pt,line join=round,line cap=round] ( 42.95, 30.69) rectangle (139.04,110.13);
\end{scope}
\begin{scope}
\path[clip] (  0.00,  0.00) rectangle (144.54,115.63);
\definecolor{drawColor}{gray}{0.30}

\node[text=drawColor,anchor=base east,inner sep=0pt, outer sep=0pt, scale=  0.88] at ( 38.00, 31.27) {0.000};

\node[text=drawColor,anchor=base east,inner sep=0pt, outer sep=0pt, scale=  0.88] at ( 38.00, 49.60) {0.001};

\node[text=drawColor,anchor=base east,inner sep=0pt, outer sep=0pt, scale=  0.88] at ( 38.00, 67.93) {0.002};

\node[text=drawColor,anchor=base east,inner sep=0pt, outer sep=0pt, scale=  0.88] at ( 38.00, 86.26) {0.003};

\node[text=drawColor,anchor=base east,inner sep=0pt, outer sep=0pt, scale=  0.88] at ( 38.00,104.59) {0.004};
\end{scope}
\begin{scope}
\path[clip] (  0.00,  0.00) rectangle (144.54,115.63);
\definecolor{drawColor}{gray}{0.20}

\path[draw=drawColor,line width= 0.6pt,line join=round] ( 40.20, 34.30) --
	( 42.95, 34.30);

\path[draw=drawColor,line width= 0.6pt,line join=round] ( 40.20, 52.63) --
	( 42.95, 52.63);

\path[draw=drawColor,line width= 0.6pt,line join=round] ( 40.20, 70.96) --
	( 42.95, 70.96);

\path[draw=drawColor,line width= 0.6pt,line join=round] ( 40.20, 89.29) --
	( 42.95, 89.29);

\path[draw=drawColor,line width= 0.6pt,line join=round] ( 40.20,107.62) --
	( 42.95,107.62);
\end{scope}
\begin{scope}
\path[clip] (  0.00,  0.00) rectangle (144.54,115.63);
\definecolor{drawColor}{gray}{0.20}

\path[draw=drawColor,line width= 0.6pt,line join=round] ( 47.32, 27.94) --
	( 47.32, 30.69);

\path[draw=drawColor,line width= 0.6pt,line join=round] ( 68.65, 27.94) --
	( 68.65, 30.69);

\path[draw=drawColor,line width= 0.6pt,line join=round] ( 89.97, 27.94) --
	( 89.97, 30.69);

\path[draw=drawColor,line width= 0.6pt,line join=round] (111.30, 27.94) --
	(111.30, 30.69);

\path[draw=drawColor,line width= 0.6pt,line join=round] (132.63, 27.94) --
	(132.63, 30.69);
\end{scope}
\begin{scope}
\path[clip] (  0.00,  0.00) rectangle (144.54,115.63);
\definecolor{drawColor}{gray}{0.30}

\node[text=drawColor,anchor=base,inner sep=0pt, outer sep=0pt, scale=  0.88] at ( 47.32, 19.68) {0};

\node[text=drawColor,anchor=base,inner sep=0pt, outer sep=0pt, scale=  0.88] at ( 68.65, 19.68) {500};

\node[text=drawColor,anchor=base,inner sep=0pt, outer sep=0pt, scale=  0.88] at ( 89.97, 19.68) {1000};

\node[text=drawColor,anchor=base,inner sep=0pt, outer sep=0pt, scale=  0.88] at (111.30, 19.68) {1500};

\node[text=drawColor,anchor=base,inner sep=0pt, outer sep=0pt, scale=  0.88] at (132.63, 19.68) {2000};
\end{scope}
\begin{scope}
\path[clip] (  0.00,  0.00) rectangle (144.54,115.63);
\definecolor{drawColor}{RGB}{0,0,0}

\node[text=drawColor,anchor=base,inner sep=0pt, outer sep=0pt, scale=  1.10] at ( 91.00,  7.64) {Chunk Size (B)};
\end{scope}
\begin{scope}
\path[clip] (  0.00,  0.00) rectangle (144.54,115.63);
\definecolor{drawColor}{RGB}{0,0,0}

\node[text=drawColor,rotate= 90.00,anchor=base,inner sep=0pt, outer sep=0pt, scale=  1.10] at ( 13.08, 70.41) {Density};
\end{scope}
\end{tikzpicture}

%% file: fig/csd_code.tex
% Created by tikzDevice version 0.12.6 on 2024-07-21 09:07:39
% !TEX encoding = UTF-8 Unicode
\begin{tikzpicture}[x=1pt,y=1pt]
\definecolor{fillColor}{RGB}{255,255,255}
\path[use as bounding box,fill=fillColor,fill opacity=0.00] (0,0) rectangle (144.54,115.63);
\begin{scope}
\path[clip] (  0.00,  0.00) rectangle (144.54,115.63);
\definecolor{drawColor}{RGB}{255,255,255}
\definecolor{fillColor}{RGB}{255,255,255}

\path[draw=drawColor,line width= 0.6pt,line join=round,line cap=round,fill=fillColor] (  0.00,  0.00) rectangle (144.54,115.63);
\end{scope}
\begin{scope}
\path[clip] ( 42.95, 30.69) rectangle (139.04,110.13);
\definecolor{fillColor}{RGB}{255,255,255}

\path[fill=fillColor] ( 42.95, 30.69) rectangle (139.04,110.13);
\definecolor{drawColor}{gray}{0.92}

\path[draw=drawColor,line width= 0.3pt,line join=round] ( 42.95, 44.57) --
	(139.04, 44.57);

\path[draw=drawColor,line width= 0.3pt,line join=round] ( 42.95, 65.13) --
	(139.04, 65.13);

\path[draw=drawColor,line width= 0.3pt,line join=round] ( 42.95, 85.68) --
	(139.04, 85.68);

\path[draw=drawColor,line width= 0.3pt,line join=round] ( 42.95,106.23) --
	(139.04,106.23);

\path[draw=drawColor,line width= 0.3pt,line join=round] ( 57.98, 30.69) --
	( 57.98,110.13);

\path[draw=drawColor,line width= 0.3pt,line join=round] ( 79.31, 30.69) --
	( 79.31,110.13);

\path[draw=drawColor,line width= 0.3pt,line join=round] (100.64, 30.69) --
	(100.64,110.13);

\path[draw=drawColor,line width= 0.3pt,line join=round] (121.96, 30.69) --
	(121.96,110.13);

\path[draw=drawColor,line width= 0.6pt,line join=round] ( 42.95, 34.30) --
	(139.04, 34.30);

\path[draw=drawColor,line width= 0.6pt,line join=round] ( 42.95, 54.85) --
	(139.04, 54.85);

\path[draw=drawColor,line width= 0.6pt,line join=round] ( 42.95, 75.40) --
	(139.04, 75.40);

\path[draw=drawColor,line width= 0.6pt,line join=round] ( 42.95, 95.95) --
	(139.04, 95.95);

\path[draw=drawColor,line width= 0.6pt,line join=round] ( 47.32, 30.69) --
	( 47.32,110.13);

\path[draw=drawColor,line width= 0.6pt,line join=round] ( 68.65, 30.69) --
	( 68.65,110.13);

\path[draw=drawColor,line width= 0.6pt,line join=round] ( 89.97, 30.69) --
	( 89.97,110.13);

\path[draw=drawColor,line width= 0.6pt,line join=round] (111.30, 30.69) --
	(111.30,110.13);

\path[draw=drawColor,line width= 0.6pt,line join=round] (132.63, 30.69) --
	(132.63,110.13);
\definecolor{drawColor}{RGB}{248,118,109}

\path[draw=drawColor,line width= 0.6pt,line join=round] ( 47.32, 34.30) --
	( 48.20, 34.30) --
	( 49.09, 43.56) --
	( 49.97, 40.28) --
	( 50.85, 40.18) --
	( 51.73, 39.84) --
	( 52.62, 40.35) --
	( 53.50, 39.45) --
	( 54.38, 38.83) --
	( 55.26, 38.76) --
	( 56.14, 38.81) --
	( 57.03, 38.13) --
	( 57.91, 38.45) --
	( 58.79, 38.22) --
	( 59.67, 37.88) --
	( 60.56, 37.86) --
	( 61.44, 37.80) --
	( 62.32, 37.76) --
	( 63.20, 37.45) --
	( 64.09, 37.39) --
	( 64.97, 37.39) --
	( 65.85, 37.19) --
	( 66.73, 37.21) --
	( 67.62, 37.09) --
	( 68.50, 36.79) --
	( 69.38, 37.00) --
	( 70.26, 36.84) --
	( 71.14, 36.67) --
	( 72.03, 36.77) --
	( 72.91, 36.74) --
	( 73.79, 36.47) --
	( 74.67, 36.59) --
	( 75.56, 36.53) --
	( 76.44, 36.46) --
	( 77.32, 36.38) --
	( 78.20, 36.59) --
	( 79.09, 36.48) --
	( 79.97, 36.39) --
	( 80.85, 36.41) --
	( 81.73, 36.40) --
	( 82.61, 36.12) --
	( 83.50, 36.21) --
	( 84.38, 36.11) --
	( 85.26, 36.13) --
	( 86.14, 36.06) --
	( 87.03, 35.97) --
	( 87.91, 35.90) --
	( 88.79, 35.88) --
	( 89.67, 35.92) --
	( 90.56, 35.89) --
	( 91.44, 35.85) --
	( 92.32, 35.82) --
	( 93.20, 35.87) --
	( 94.09, 35.81) --
	( 94.97, 35.80) --
	( 95.85, 35.79) --
	( 96.73, 35.63) --
	( 97.61, 35.63) --
	( 98.50, 35.66) --
	( 99.38, 35.61) --
	(100.26, 35.56) --
	(101.14, 35.54) --
	(102.03, 35.46) --
	(102.91, 35.52) --
	(103.79, 35.46) --
	(104.67, 35.38) --
	(105.56, 35.39) --
	(106.44, 35.50) --
	(107.32, 35.38) --
	(108.20, 35.29) --
	(109.08, 35.36) --
	(109.97, 35.27) --
	(110.85, 35.35) --
	(111.73, 35.31) --
	(112.61, 35.27) --
	(113.50, 35.20) --
	(114.38, 35.28) --
	(115.26, 35.19) --
	(116.14, 35.14) --
	(117.03, 35.14) --
	(117.91, 35.13) --
	(118.79, 35.13) --
	(119.67, 35.11) --
	(120.56, 35.16) --
	(121.44, 35.08) --
	(122.32, 35.06) --
	(123.20, 35.04) --
	(124.08, 35.04) --
	(124.97, 34.98) --
	(125.85, 35.00) --
	(126.73, 35.02) --
	(127.61, 34.95) --
	(128.50, 34.96) --
	(129.38, 34.96) --
	(130.26, 34.94) --
	(131.14, 34.90) --
	(132.03, 34.92) --
	(132.91, 34.88) --
	(133.79, 34.91) --
	(134.67, 34.62);
\definecolor{drawColor}{RGB}{196,154,0}

\path[draw=drawColor,line width= 0.6pt,dash pattern=on 2pt off 2pt ,line join=round] ( 47.32, 34.30) --
	( 48.20, 34.30) --
	( 49.09, 34.30) --
	( 49.97, 34.30) --
	( 50.85, 34.30) --
	( 51.73, 34.30) --
	( 52.62, 34.30) --
	( 53.50, 34.30) --
	( 54.38, 34.30) --
	( 55.26, 34.30) --
	( 56.14, 34.30) --
	( 57.03, 34.30) --
	( 57.91, 34.30) --
	( 58.79, 34.30) --
	( 59.67, 34.30) --
	( 60.56, 34.30) --
	( 61.44, 34.30) --
	( 62.32, 34.30) --
	( 63.20, 34.30) --
	( 64.09, 34.30) --
	( 64.97, 34.30) --
	( 65.85, 34.30) --
	( 66.73, 34.30) --
	( 67.62, 34.30) --
	( 68.50, 34.30) --
	( 69.38, 34.30) --
	( 70.26, 34.30) --
	( 71.14, 34.30) --
	( 72.03, 34.30) --
	( 72.91, 34.30) --
	( 73.79, 34.30) --
	( 74.67, 34.30) --
	( 75.56, 34.30) --
	( 76.44, 34.30) --
	( 77.32, 34.30) --
	( 78.20, 34.30) --
	( 79.09, 34.30) --
	( 79.97, 34.30) --
	( 80.85, 40.94) --
	( 81.73, 60.25) --
	( 82.61, 54.75) --
	( 83.50, 49.14) --
	( 84.38, 45.93) --
	( 85.26, 43.50) --
	( 86.14, 42.67) --
	( 87.03, 41.59) --
	( 87.91, 40.27) --
	( 88.79, 40.06) --
	( 89.67, 39.53) --
	( 90.56, 39.20) --
	( 91.44, 38.53) --
	( 92.32, 38.43) --
	( 93.20, 38.07) --
	( 94.09, 37.79) --
	( 94.97, 37.66) --
	( 95.85, 37.48) --
	( 96.73, 37.12) --
	( 97.61, 37.03) --
	( 98.50, 36.86) --
	( 99.38, 36.63) --
	(100.26, 36.66) --
	(101.14, 36.53) --
	(102.03, 36.35) --
	(102.91, 36.39) --
	(103.79, 36.27) --
	(104.67, 36.16) --
	(105.56, 36.09) --
	(106.44, 36.10) --
	(107.32, 36.01) --
	(108.20, 35.86) --
	(109.08, 35.86) --
	(109.97, 35.78) --
	(110.85, 35.67) --
	(111.73, 35.67) --
	(112.61, 35.70) --
	(113.50, 35.53) --
	(114.38, 35.68) --
	(115.26, 35.55) --
	(116.14, 35.24) --
	(117.03, 35.17) --
	(117.91, 35.08) --
	(118.79, 34.97) --
	(119.67, 34.95) --
	(120.56, 34.91) --
	(121.44, 34.82) --
	(122.32, 34.75) --
	(123.20, 34.78) --
	(124.08, 34.71) --
	(124.97, 34.66) --
	(125.85, 34.64) --
	(126.73, 34.61) --
	(127.61, 34.58) --
	(128.50, 34.58) --
	(129.38, 34.55) --
	(130.26, 34.52) --
	(131.14, 34.51) --
	(132.03, 34.49) --
	(132.91, 34.46) --
	(133.79, 34.48) --
	(134.67, 34.38);
\definecolor{drawColor}{RGB}{83,180,0}

\path[draw=drawColor,line width= 0.6pt,dash pattern=on 4pt off 2pt ,line join=round] ( 47.32, 34.30) --
	( 48.20, 34.30) --
	( 49.09, 34.30) --
	( 49.97, 34.30) --
	( 50.85, 34.30) --
	( 51.73, 34.30) --
	( 52.62, 34.30) --
	( 53.50, 34.30) --
	( 54.38, 34.30) --
	( 55.26, 34.30) --
	( 56.14, 34.30) --
	( 57.03, 34.30) --
	( 57.91, 34.30) --
	( 58.79, 34.30) --
	( 59.67, 34.30) --
	( 60.56, 34.30) --
	( 61.44, 34.30) --
	( 62.32, 34.30) --
	( 63.20, 34.30) --
	( 64.09, 34.30) --
	( 64.97, 34.30) --
	( 65.85, 34.30) --
	( 66.73, 34.30) --
	( 67.62, 34.30) --
	( 68.50, 34.30) --
	( 69.38, 34.30) --
	( 70.26, 34.30) --
	( 71.14, 34.30) --
	( 72.03, 34.30) --
	( 72.91, 34.30) --
	( 73.79, 34.30) --
	( 74.67, 34.30) --
	( 75.56, 34.30) --
	( 76.44, 34.30) --
	( 77.32, 34.30) --
	( 78.20, 34.30) --
	( 79.09, 34.30) --
	( 79.97, 34.30) --
	( 80.85, 53.00) --
	( 81.73, 54.83) --
	( 82.61, 50.30) --
	( 83.50, 47.32) --
	( 84.38, 45.51) --
	( 85.26, 42.64) --
	( 86.14, 42.59) --
	( 87.03, 40.94) --
	( 87.91, 40.55) --
	( 88.79, 40.09) --
	( 89.67, 40.32) --
	( 90.56, 40.26) --
	( 91.44, 38.93) --
	( 92.32, 37.95) --
	( 93.20, 38.25) --
	( 94.09, 38.47) --
	( 94.97, 37.23) --
	( 95.85, 37.08) --
	( 96.73, 37.06) --
	( 97.61, 36.87) --
	( 98.50, 37.02) --
	( 99.38, 37.43) --
	(100.26, 36.52) --
	(101.14, 36.81) --
	(102.03, 35.99) --
	(102.91, 36.09) --
	(103.79, 35.94) --
	(104.67, 35.81) --
	(105.56, 37.22) --
	(106.44, 35.83) --
	(107.32, 35.64) --
	(108.20, 35.54) --
	(109.08, 35.61) --
	(109.97, 35.33) --
	(110.85, 35.50) --
	(111.73, 35.44) --
	(112.61, 35.36) --
	(113.50, 35.25) --
	(114.38, 35.35) --
	(115.26, 35.17) --
	(116.14, 34.97) --
	(117.03, 35.06) --
	(117.91, 34.95) --
	(118.79, 35.04) --
	(119.67, 34.97) --
	(120.56, 35.17) --
	(121.44, 34.80) --
	(122.32, 34.76) --
	(123.20, 34.93) --
	(124.08, 34.74) --
	(124.97, 34.69) --
	(125.85, 35.00) --
	(126.73, 34.82) --
	(127.61, 34.68) --
	(128.50, 34.69) --
	(129.38, 34.80) --
	(130.26, 34.63) --
	(131.14, 34.67) --
	(132.03, 34.67) --
	(132.91, 34.71) --
	(133.79, 34.72) --
	(134.67, 34.57);
\definecolor{drawColor}{RGB}{0,192,148}

\path[draw=drawColor,line width= 0.6pt,dash pattern=on 4pt off 4pt ,line join=round] ( 47.32, 34.30) --
	( 48.20, 34.30) --
	( 49.09, 70.61) --
	( 49.97, 50.50) --
	( 50.85, 47.07) --
	( 51.73, 43.61) --
	( 52.62, 42.60) --
	( 53.50, 41.70) --
	( 54.38, 41.62) --
	( 55.26, 44.85) --
	( 56.14, 44.00) --
	( 57.03, 41.86) --
	( 57.91, 40.66) --
	( 58.79, 40.75) --
	( 59.67, 39.78) --
	( 60.56, 39.80) --
	( 61.44, 38.53) --
	( 62.32, 37.17) --
	( 63.20, 36.42) --
	( 64.09, 36.12) --
	( 64.97, 36.15) --
	( 65.85, 35.98) --
	( 66.73, 35.75) --
	( 67.62, 35.74) --
	( 68.50, 35.67) --
	( 69.38, 35.51) --
	( 70.26, 35.46) --
	( 71.14, 35.24) --
	( 72.03, 35.32) --
	( 72.91, 35.37) --
	( 73.79, 35.16) --
	( 74.67, 35.21) --
	( 75.56, 35.26) --
	( 76.44, 35.06) --
	( 77.32, 35.15) --
	( 78.20, 35.07) --
	( 79.09, 34.97) --
	( 79.97, 34.98) --
	( 80.85, 34.91) --
	( 81.73, 34.91) --
	( 82.61, 35.00) --
	( 83.50, 34.94) --
	( 84.38, 34.93) --
	( 85.26, 34.72) --
	( 86.14, 34.86) --
	( 87.03, 34.80) --
	( 87.91, 34.66) --
	( 88.79, 34.83) --
	( 89.67, 34.76) --
	( 90.56, 34.63) --
	( 91.44, 34.73) --
	( 92.32, 34.68) --
	( 93.20, 34.76) --
	( 94.09, 34.64) --
	( 94.97, 34.55) --
	( 95.85, 34.74) --
	( 96.73, 34.68) --
	( 97.61, 34.60) --
	( 98.50, 34.60) --
	( 99.38, 34.65) --
	(100.26, 34.62) --
	(101.14, 34.70) --
	(102.03, 34.59) --
	(102.91, 34.74) --
	(103.79, 34.60) --
	(104.67, 34.53) --
	(105.56, 34.55) --
	(106.44, 34.68) --
	(107.32, 34.59) --
	(108.20, 34.50) --
	(109.08, 34.52) --
	(109.97, 34.60) --
	(110.85, 34.58) --
	(111.73, 34.54) --
	(112.61, 34.50) --
	(113.50, 34.58) --
	(114.38, 34.50) --
	(115.26, 34.57) --
	(116.14, 34.49) --
	(117.03, 34.47) --
	(117.91, 34.45) --
	(118.79, 34.49) --
	(119.67, 34.48) --
	(120.56, 34.46) --
	(121.44, 34.49) --
	(122.32, 34.45) --
	(123.20, 34.48) --
	(124.08, 34.47) --
	(124.97, 34.49) --
	(125.85, 34.48) --
	(126.73, 34.46) --
	(127.61, 34.42) --
	(128.50, 34.46) --
	(129.38, 34.48) --
	(130.26, 34.42) --
	(131.14, 34.46) --
	(132.03, 34.42) --
	(132.91, 34.42) --
	(133.79, 34.41) --
	(134.67, 34.36);
\definecolor{drawColor}{RGB}{0,182,235}

\path[draw=drawColor,line width= 0.6pt,dash pattern=on 1pt off 3pt ,line join=round] ( 47.32, 39.43) --
	( 48.20, 57.39) --
	( 49.09, 56.33) --
	( 49.97, 53.52) --
	( 50.85, 47.53) --
	( 51.73, 44.53) --
	( 52.62, 42.65) --
	( 53.50, 41.05) --
	( 54.38, 40.14) --
	( 55.26, 39.81) --
	( 56.14, 39.16) --
	( 57.03, 38.37) --
	( 57.91, 38.03) --
	( 58.79, 37.74) --
	( 59.67, 37.33) --
	( 60.56, 37.25) --
	( 61.44, 36.95) --
	( 62.32, 36.73) --
	( 63.20, 36.52) --
	( 64.09, 36.66) --
	( 64.97, 36.52) --
	( 65.85, 36.13) --
	( 66.73, 36.07) --
	( 67.62, 35.99) --
	( 68.50, 35.78) --
	( 69.38, 35.81) --
	( 70.26, 35.70) --
	( 71.14, 35.58) --
	( 72.03, 35.59) --
	( 72.91, 35.62) --
	( 73.79, 35.43) --
	( 74.67, 35.45) --
	( 75.56, 35.37) --
	( 76.44, 35.28) --
	( 77.32, 35.21) --
	( 78.20, 35.35) --
	( 79.09, 35.18) --
	( 79.97, 35.12) --
	( 80.85, 35.14) --
	( 81.73, 35.11) --
	( 82.61, 35.04) --
	( 83.50, 35.08) --
	( 84.38, 35.03) --
	( 85.26, 34.95) --
	( 86.14, 34.99) --
	( 87.03, 34.94) --
	( 87.91, 34.92) --
	( 88.79, 34.92) --
	( 89.67, 34.95) --
	( 90.56, 34.86) --
	( 91.44, 34.80) --
	( 92.32, 34.85) --
	( 93.20, 34.83) --
	( 94.09, 34.80) --
	( 94.97, 34.80) --
	( 95.85, 34.82) --
	( 96.73, 34.77) --
	( 97.61, 34.73) --
	( 98.50, 34.74) --
	( 99.38, 34.71) --
	(100.26, 34.72) --
	(101.14, 34.69) --
	(102.03, 34.66) --
	(102.91, 34.67) --
	(103.79, 34.65) --
	(104.67, 34.63) --
	(105.56, 34.65) --
	(106.44, 34.64) --
	(107.32, 34.63) --
	(108.20, 34.62) --
	(109.08, 34.61) --
	(109.97, 34.63) --
	(110.85, 34.63) --
	(111.73, 34.61) --
	(112.61, 34.60) --
	(113.50, 34.60) --
	(114.38, 34.60) --
	(115.26, 34.57) --
	(116.14, 34.53) --
	(117.03, 34.55) --
	(117.91, 34.55) --
	(118.79, 34.53) --
	(119.67, 34.53) --
	(120.56, 34.53) --
	(121.44, 34.52) --
	(122.32, 34.50) --
	(123.20, 34.52) --
	(124.08, 34.51) --
	(124.97, 34.49) --
	(125.85, 34.48) --
	(126.73, 34.48) --
	(127.61, 34.47) --
	(128.50, 34.49) --
	(129.38, 34.49) --
	(130.26, 34.47) --
	(131.14, 34.47) --
	(132.03, 34.48) --
	(132.91, 34.46) --
	(133.79, 34.47) --
	(134.67, 34.38);
\definecolor{drawColor}{RGB}{165,138,255}

\path[draw=drawColor,line width= 0.6pt,dash pattern=on 1pt off 3pt on 4pt off 3pt ,line join=round] ( 47.32, 34.30) --
	( 48.20, 34.30) --
	( 49.09, 34.30) --
	( 49.97, 34.30) --
	( 50.85, 34.30) --
	( 51.73, 34.30) --
	( 52.62, 34.30) --
	( 53.50, 34.30) --
	( 54.38, 34.30) --
	( 55.26, 34.30) --
	( 56.14, 34.30) --
	( 57.03, 34.30) --
	( 57.91, 34.30) --
	( 58.79, 34.30) --
	( 59.67, 34.30) --
	( 60.56, 34.30) --
	( 61.44, 34.30) --
	( 62.32, 34.30) --
	( 63.20, 34.30) --
	( 64.09, 34.30) --
	( 64.97, 34.30) --
	( 65.85, 34.30) --
	( 66.73, 34.30) --
	( 67.62, 34.30) --
	( 68.50, 34.30) --
	( 69.38, 34.30) --
	( 70.26, 34.30) --
	( 71.14, 34.30) --
	( 72.03, 34.30) --
	( 72.91, 34.30) --
	( 73.79, 34.30) --
	( 74.67, 34.30) --
	( 75.56, 34.30) --
	( 76.44, 34.30) --
	( 77.32, 34.30) --
	( 78.20, 34.30) --
	( 79.09, 34.30) --
	( 79.97, 34.30) --
	( 80.85, 34.30) --
	( 81.73, 34.30) --
	( 82.61, 34.30) --
	( 83.50, 34.30) --
	( 84.38, 34.30) --
	( 85.26, 66.06) --
	( 86.14,106.52) --
	( 87.03, 72.29) --
	( 87.91, 52.53) --
	( 88.79, 44.54) --
	( 89.67, 40.77) --
	( 90.56, 38.76) --
	( 91.44, 37.41) --
	( 92.32, 36.76) --
	( 93.20, 36.09) --
	( 94.09, 35.71) --
	( 94.97, 35.44) --
	( 95.85, 35.28) --
	( 96.73, 35.05) --
	( 97.61, 34.91) --
	( 98.50, 34.87) --
	( 99.38, 34.74) --
	(100.26, 34.67) --
	(101.14, 34.68) --
	(102.03, 34.59) --
	(102.91, 34.55) --
	(103.79, 34.52) --
	(104.67, 34.49) --
	(105.56, 34.47) --
	(106.44, 34.46) --
	(107.32, 34.44) --
	(108.20, 34.43) --
	(109.08, 34.44) --
	(109.97, 34.42) --
	(110.85, 34.40) --
	(111.73, 34.39) --
	(112.61, 34.38) --
	(113.50, 34.39) --
	(114.38, 34.38) --
	(115.26, 34.37) --
	(116.14, 34.36) --
	(117.03, 34.37) --
	(117.91, 34.35) --
	(118.79, 34.34) --
	(119.67, 34.35) --
	(120.56, 34.35) --
	(121.44, 34.35) --
	(122.32, 34.34) --
	(123.20, 34.35) --
	(124.08, 34.34) --
	(124.97, 34.33) --
	(125.85, 34.33) --
	(126.73, 34.33) --
	(127.61, 34.33) --
	(128.50, 34.33) --
	(129.38, 34.32) --
	(130.26, 34.32) --
	(131.14, 34.32) --
	(132.03, 34.33) --
	(132.91, 34.32) --
	(133.79, 34.32) --
	(134.67, 34.31);
\definecolor{drawColor}{RGB}{251,97,215}

\path[draw=drawColor,line width= 0.6pt,dash pattern=on 7pt off 3pt ,line join=round] ( 47.32, 40.46) --
	( 48.20, 88.64) --
	( 49.09, 61.53) --
	( 49.97, 62.24) --
	( 50.85, 50.91) --
	( 51.73, 46.43) --
	( 52.62, 44.80) --
	( 53.50, 39.63) --
	( 54.38, 38.05) --
	( 55.26, 37.32) --
	( 56.14, 36.62) --
	( 57.03, 36.36) --
	( 57.91, 36.08) --
	( 58.79, 35.97) --
	( 59.67, 35.65) --
	( 60.56, 35.57) --
	( 61.44, 35.39) --
	( 62.32, 35.28) --
	( 63.20, 35.14) --
	( 64.09, 35.18) --
	( 64.97, 35.18) --
	( 65.85, 35.11) --
	( 66.73, 35.02) --
	( 67.62, 34.97) --
	( 68.50, 34.88) --
	( 69.38, 35.02) --
	( 70.26, 34.88) --
	( 71.14, 34.73) --
	( 72.03, 34.72) --
	( 72.91, 34.72) --
	( 73.79, 34.62) --
	( 74.67, 34.62) --
	( 75.56, 34.59) --
	( 76.44, 34.60) --
	( 77.32, 34.62) --
	( 78.20, 34.66) --
	( 79.09, 34.68) --
	( 79.97, 34.58) --
	( 80.85, 34.63) --
	( 81.73, 34.57) --
	( 82.61, 34.53) --
	( 83.50, 34.57) --
	( 84.38, 34.55) --
	( 85.26, 34.56) --
	( 86.14, 34.58) --
	( 87.03, 34.54) --
	( 87.91, 34.55) --
	( 88.79, 34.55) --
	( 89.67, 34.52) --
	( 90.56, 34.50) --
	( 91.44, 34.50) --
	( 92.32, 34.48) --
	( 93.20, 34.57) --
	( 94.09, 34.55) --
	( 94.97, 34.48) --
	( 95.85, 34.55) --
	( 96.73, 34.55) --
	( 97.61, 34.49) --
	( 98.50, 34.44) --
	( 99.38, 34.46) --
	(100.26, 34.43) --
	(101.14, 34.45) --
	(102.03, 34.42) --
	(102.91, 34.42) --
	(103.79, 34.42) --
	(104.67, 34.41) --
	(105.56, 34.42) --
	(106.44, 34.40) --
	(107.32, 34.40) --
	(108.20, 34.39) --
	(109.08, 34.39) --
	(109.97, 34.38) --
	(110.85, 34.37) --
	(111.73, 34.40) --
	(112.61, 34.37) --
	(113.50, 34.38) --
	(114.38, 34.39) --
	(115.26, 34.37) --
	(116.14, 34.38) --
	(117.03, 34.38) --
	(117.91, 34.39) --
	(118.79, 34.37) --
	(119.67, 34.36) --
	(120.56, 34.39) --
	(121.44, 34.37) --
	(122.32, 34.35) --
	(123.20, 34.36) --
	(124.08, 34.36) --
	(124.97, 34.37) --
	(125.85, 34.37) --
	(126.73, 34.38) --
	(127.61, 34.36) --
	(128.50, 34.35) --
	(129.38, 34.35) --
	(130.26, 34.35) --
	(131.14, 34.37) --
	(132.03, 34.37) --
	(132.91, 34.36) --
	(133.79, 34.34) --
	(134.67, 34.32);
\definecolor{drawColor}{gray}{0.20}

\path[draw=drawColor,line width= 0.6pt,line join=round,line cap=round] ( 42.95, 30.69) rectangle (139.04,110.13);
\end{scope}
\begin{scope}
\path[clip] (  0.00,  0.00) rectangle (144.54,115.63);
\definecolor{drawColor}{gray}{0.30}

\node[text=drawColor,anchor=base east,inner sep=0pt, outer sep=0pt, scale=  0.88] at ( 38.00, 31.27) {0.000};

\node[text=drawColor,anchor=base east,inner sep=0pt, outer sep=0pt, scale=  0.88] at ( 38.00, 51.82) {0.005};

\node[text=drawColor,anchor=base east,inner sep=0pt, outer sep=0pt, scale=  0.88] at ( 38.00, 72.37) {0.010};

\node[text=drawColor,anchor=base east,inner sep=0pt, outer sep=0pt, scale=  0.88] at ( 38.00, 92.92) {0.015};
\end{scope}
\begin{scope}
\path[clip] (  0.00,  0.00) rectangle (144.54,115.63);
\definecolor{drawColor}{gray}{0.20}

\path[draw=drawColor,line width= 0.6pt,line join=round] ( 40.20, 34.30) --
	( 42.95, 34.30);

\path[draw=drawColor,line width= 0.6pt,line join=round] ( 40.20, 54.85) --
	( 42.95, 54.85);

\path[draw=drawColor,line width= 0.6pt,line join=round] ( 40.20, 75.40) --
	( 42.95, 75.40);

\path[draw=drawColor,line width= 0.6pt,line join=round] ( 40.20, 95.95) --
	( 42.95, 95.95);
\end{scope}
\begin{scope}
\path[clip] (  0.00,  0.00) rectangle (144.54,115.63);
\definecolor{drawColor}{gray}{0.20}

\path[draw=drawColor,line width= 0.6pt,line join=round] ( 47.32, 27.94) --
	( 47.32, 30.69);

\path[draw=drawColor,line width= 0.6pt,line join=round] ( 68.65, 27.94) --
	( 68.65, 30.69);

\path[draw=drawColor,line width= 0.6pt,line join=round] ( 89.97, 27.94) --
	( 89.97, 30.69);

\path[draw=drawColor,line width= 0.6pt,line join=round] (111.30, 27.94) --
	(111.30, 30.69);

\path[draw=drawColor,line width= 0.6pt,line join=round] (132.63, 27.94) --
	(132.63, 30.69);
\end{scope}
\begin{scope}
\path[clip] (  0.00,  0.00) rectangle (144.54,115.63);
\definecolor{drawColor}{gray}{0.30}

\node[text=drawColor,anchor=base,inner sep=0pt, outer sep=0pt, scale=  0.88] at ( 47.32, 19.68) {0};

\node[text=drawColor,anchor=base,inner sep=0pt, outer sep=0pt, scale=  0.88] at ( 68.65, 19.68) {500};

\node[text=drawColor,anchor=base,inner sep=0pt, outer sep=0pt, scale=  0.88] at ( 89.97, 19.68) {1000};

\node[text=drawColor,anchor=base,inner sep=0pt, outer sep=0pt, scale=  0.88] at (111.30, 19.68) {1500};

\node[text=drawColor,anchor=base,inner sep=0pt, outer sep=0pt, scale=  0.88] at (132.63, 19.68) {2000};
\end{scope}
\begin{scope}
\path[clip] (  0.00,  0.00) rectangle (144.54,115.63);
\definecolor{drawColor}{RGB}{0,0,0}

\node[text=drawColor,anchor=base,inner sep=0pt, outer sep=0pt, scale=  1.10] at ( 91.00,  7.64) {Chunk Size (B)};
\end{scope}
\begin{scope}
\path[clip] (  0.00,  0.00) rectangle (144.54,115.63);
\definecolor{drawColor}{RGB}{0,0,0}

\node[text=drawColor,rotate= 90.00,anchor=base,inner sep=0pt, outer sep=0pt, scale=  1.10] at ( 13.08, 70.41) {Density};
\end{scope}
\end{tikzpicture}

%% file: fig/csd_legendonly.tex
% Created by tikzDevice version 0.12.6 on 2024-07-21 09:10:08
% !TEX encoding = UTF-8 Unicode
\begin{tikzpicture}[x=1pt,y=1pt]
\definecolor{fillColor}{RGB}{255,255,255}
\path[use as bounding box,fill=fillColor,fill opacity=0.00] (0,0) rectangle (289.08, 72.27);
\begin{scope}
\path[clip] (  0.00,  0.00) rectangle (289.08, 72.27);
\definecolor{drawColor}{RGB}{248,118,109}

\path[draw=drawColor,line width= 0.6pt,line join=round] ( 42.10, 44.81) -- ( 55.98, 44.81);
\end{scope}
\begin{scope}
\path[clip] (  0.00,  0.00) rectangle (289.08, 72.27);
\definecolor{drawColor}{RGB}{196,154,0}

\path[draw=drawColor,line width= 0.6pt,dash pattern=on 2pt off 2pt ,line join=round] ( 42.10, 27.46) -- ( 55.98, 27.46);
\end{scope}
\begin{scope}
\path[clip] (  0.00,  0.00) rectangle (289.08, 72.27);
\definecolor{drawColor}{RGB}{83,180,0}

\path[draw=drawColor,line width= 0.6pt,dash pattern=on 4pt off 2pt ,line join=round] ( 96.66, 44.81) -- (110.54, 44.81);
\end{scope}
\begin{scope}
\path[clip] (  0.00,  0.00) rectangle (289.08, 72.27);
\definecolor{drawColor}{RGB}{0,192,148}

\path[draw=drawColor,line width= 0.6pt,dash pattern=on 4pt off 4pt ,line join=round] ( 96.66, 27.46) -- (110.54, 27.46);
\end{scope}
\begin{scope}
\path[clip] (  0.00,  0.00) rectangle (289.08, 72.27);
\definecolor{drawColor}{RGB}{0,182,235}

\path[draw=drawColor,line width= 0.6pt,dash pattern=on 1pt off 3pt ,line join=round] (147.31, 44.81) -- (161.19, 44.81);
\end{scope}
\begin{scope}
\path[clip] (  0.00,  0.00) rectangle (289.08, 72.27);
\definecolor{drawColor}{RGB}{165,138,255}

\path[draw=drawColor,line width= 0.6pt,dash pattern=on 1pt off 3pt on 4pt off 3pt ,line join=round] (147.31, 27.46) -- (161.19, 27.46);
\end{scope}
\begin{scope}
\path[clip] (  0.00,  0.00) rectangle (289.08, 72.27);
\definecolor{drawColor}{RGB}{251,97,215}

\path[draw=drawColor,line width= 0.6pt,dash pattern=on 7pt off 3pt ,line join=round] (202.43, 44.81) -- (216.30, 44.81);
\end{scope}
\begin{scope}
\path[clip] (  0.00,  0.00) rectangle (289.08, 72.27);
\definecolor{drawColor}{RGB}{0,0,0}

\node[text=drawColor,anchor=base west,inner sep=0pt, outer sep=0pt, scale=  1.00] at ( 63.21, 41.36) {Rabin};
\end{scope}
\begin{scope}
\path[clip] (  0.00,  0.00) rectangle (289.08, 72.27);
\definecolor{drawColor}{RGB}{0,0,0}

\node[text=drawColor,anchor=base west,inner sep=0pt, outer sep=0pt, scale=  1.00] at ( 63.21, 24.02) {AE};
\end{scope}
\begin{scope}
\path[clip] (  0.00,  0.00) rectangle (289.08, 72.27);
\definecolor{drawColor}{RGB}{0,0,0}

\node[text=drawColor,anchor=base west,inner sep=0pt, outer sep=0pt, scale=  1.00] at (117.78, 41.36) {RAM};
\end{scope}
\begin{scope}
\path[clip] (  0.00,  0.00) rectangle (289.08, 72.27);
\definecolor{drawColor}{RGB}{0,0,0}

\node[text=drawColor,anchor=base west,inner sep=0pt, outer sep=0pt, scale=  1.00] at (117.78, 24.02) {PCI};
\end{scope}
\begin{scope}
\path[clip] (  0.00,  0.00) rectangle (289.08, 72.27);
\definecolor{drawColor}{RGB}{0,0,0}

\node[text=drawColor,anchor=base west,inner sep=0pt, outer sep=0pt, scale=  1.00] at (168.42, 41.36) {MII};
\end{scope}
\begin{scope}
\path[clip] (  0.00,  0.00) rectangle (289.08, 72.27);
\definecolor{drawColor}{RGB}{0,0,0}

\node[text=drawColor,anchor=base west,inner sep=0pt, outer sep=0pt, scale=  1.00] at (168.42, 24.02) {BFBC};
\end{scope}
\begin{scope}
\path[clip] (  0.00,  0.00) rectangle (289.08, 72.27);
\definecolor{drawColor}{RGB}{0,0,0}

\node[text=drawColor,anchor=base west,inner sep=0pt, outer sep=0pt, scale=  1.00] at (223.54, 41.36) {BFBC*};
\end{scope}
\end{tikzpicture}

%% file: tab/csd_overview.tex
\definecolor{tempcolor1}{rgb}{1,0.96,0.94} 
\definecolor{tempcolor2}{rgb}{0.99,0.88,0.82} 
\definecolor{tempcolor3}{rgb}{0.4,0,0.05} 
\definecolor{tempcolor4}{rgb}{0.98,0.57,0.45} 
\definecolor{tempcolor5}{rgb}{0.98,0.73,0.63} 
\definecolor{tempcolor6}{rgb}{0.64,0.06,0.08} 
\definecolor{tempcolor7}{rgb}{0.98,0.41,0.29} 
\definecolor{tempcolor8}{rgb}{0.93,0.23,0.17} 
\definecolor{tempcolor9}{rgb}{0.79,0.09,0.11} 
\definecolor{white}{rgb}{1.0, 1.0, 1.0} 
\definecolor{black}{rgb}{0.0, 0.0, 0.0}

\begin{center}
\begin{normalsize}
\color{black}
% \small
\begin{tabular}{lrrrrrcrrrrrc}
\hline
\multirow{2}{*}{Algorithm}&&\multicolumn{5}{c}{Mean}&&\multicolumn{5}{c}{SD}\\ 
\cline{3-7}\cline{9-13}
&&\rot{\scriptsize{RAND}}&\rot{\footnotesize{LNX}}&\rot{\footnotesize{PDF}}&\rot{\footnotesize{WEB}}&\rot{\footnotesize{CODE}}&&\rot{\footnotesize{RAND}}&\rot{\footnotesize{LNX}}&\rot{\footnotesize{PDF}}&\rot{\footnotesize{WEB}}&\rot{\footnotesize{CODE}}\\ 
\hline

Rabin&&\cellcolor{tempcolor1}&\cellcolor{tempcolor4}&\cellcolor{tempcolor4}&\cellcolor{tempcolor8}\color{white}&\cellcolor{tempcolor2}&&\cellcolor{tempcolor7}\color{white}&\cellcolor{tempcolor8}\color{white}&\cellcolor{tempcolor9}\color{white}&\cellcolor{tempcolor9}\color{white}&\cellcolor{tempcolor3}\color{white}\\ 
Buzhash&&\cellcolor{tempcolor1}&\cellcolor{tempcolor5}&\cellcolor{tempcolor4}&\cellcolor{tempcolor4}&\cellcolor{tempcolor5}&&\cellcolor{tempcolor7}\color{white}&\cellcolor{tempcolor7}\color{white}&\cellcolor{tempcolor8}\color{white}&\cellcolor{tempcolor8}\color{white}&\cellcolor{tempcolor3}\color{white}\\ 
Gear&&\cellcolor{tempcolor1}&\cellcolor{tempcolor1}&\cellcolor{tempcolor1}&\cellcolor{tempcolor1}&\cellcolor{tempcolor1}&&\cellcolor{tempcolor7}\color{white}&\cellcolor{tempcolor9}\color{white}&\cellcolor{tempcolor8}\color{white}&\cellcolor{tempcolor7}\color{white}&\cellcolor{tempcolor3}\color{white}\\ 
Gear NC-1&&\cellcolor{tempcolor2}&\cellcolor{tempcolor2}&\cellcolor{tempcolor2}&\cellcolor{tempcolor2}&\cellcolor{tempcolor2}&&\cellcolor{tempcolor5}&\cellcolor{tempcolor8}\color{white}&\cellcolor{tempcolor4}&\cellcolor{tempcolor5}&\cellcolor{tempcolor3}\color{white}\\ 
Gear NC-2&&\cellcolor{tempcolor2}&\cellcolor{tempcolor2}&\cellcolor{tempcolor2}&\cellcolor{tempcolor2}&\cellcolor{tempcolor2}&&\cellcolor{tempcolor2}&\cellcolor{tempcolor8}\color{white}&\cellcolor{tempcolor5}&\cellcolor{tempcolor2}&\cellcolor{tempcolor6}\color{white}\\ 
Gear NC-3&&\cellcolor{tempcolor1}&\cellcolor{tempcolor1}&\cellcolor{tempcolor1}&\cellcolor{tempcolor1}&\cellcolor{tempcolor1}&&\cellcolor{tempcolor1}&\cellcolor{tempcolor8}\color{white}&\cellcolor{tempcolor2}&\cellcolor{tempcolor1}&\cellcolor{tempcolor9}\color{white}\\ 
AE&&\cellcolor{tempcolor1}&\cellcolor{tempcolor1}&\cellcolor{tempcolor1}&\cellcolor{tempcolor1}&\cellcolor{tempcolor2}&&\cellcolor{tempcolor1}&\cellcolor{tempcolor1}&\cellcolor{tempcolor1}&\cellcolor{tempcolor1}&\cellcolor{tempcolor2}\\ 
RAM&&\cellcolor{tempcolor1}&\cellcolor{tempcolor1}&\cellcolor{tempcolor2}&\cellcolor{tempcolor2}&\cellcolor{tempcolor3}\color{white}&&\cellcolor{tempcolor1}&\cellcolor{tempcolor8}\color{white}&\cellcolor{tempcolor3}\color{white}&\cellcolor{tempcolor3}\color{white}&\cellcolor{tempcolor3}\color{white}\\ 
PCI&&\cellcolor{tempcolor1}&\cellcolor{tempcolor6}\color{white}&\cellcolor{tempcolor6}\color{white}&\cellcolor{tempcolor8}\color{white}&\cellcolor{tempcolor3}\color{white}&&\cellcolor{tempcolor7}\color{white}&\cellcolor{tempcolor3}\color{white}&\cellcolor{tempcolor3}\color{white}&\cellcolor{tempcolor3}\color{white}&\cellcolor{tempcolor3}\color{white}\\ 
MII&&\cellcolor{tempcolor2}&\cellcolor{tempcolor2}&\cellcolor{tempcolor7}\color{white}&\cellcolor{tempcolor2}&\cellcolor{tempcolor7}\color{white}&&\cellcolor{tempcolor7}\color{white}&\cellcolor{tempcolor6}\color{white}&\cellcolor{tempcolor3}\color{white}&\cellcolor{tempcolor3}\color{white}&\cellcolor{tempcolor3}\color{white}\\ 
BFBC&&\cellcolor{tempcolor3}\color{white}&\cellcolor{tempcolor3}\color{white}&\cellcolor{tempcolor3}\color{white}&\cellcolor{tempcolor9}\color{white}&\cellcolor{tempcolor1}&&\cellcolor{tempcolor4}&\cellcolor{tempcolor9}\color{white}&\cellcolor{tempcolor6}\color{white}&\cellcolor{tempcolor3}\color{white}&\cellcolor{tempcolor3}\color{white}\\ 
BFBC*&&\cellcolor{tempcolor1}&\cellcolor{tempcolor2}&\cellcolor{tempcolor1}&\cellcolor{tempcolor1}&\cellcolor{tempcolor1}&&\cellcolor{tempcolor7}\color{white}&\cellcolor{tempcolor3}\color{white}&\cellcolor{tempcolor3}\color{white}&\cellcolor{tempcolor3}\color{white}&\cellcolor{tempcolor3}\color{white}\\ 

\hline
\end{tabular}
\end{normalsize}
\end{center}

\color{black}

\begin{tikzpicture}

\begin{scope}[shift={(12,2)}]
	\foreach \x [count=\xi] in {tempcolor1, tempcolor2, tempcolor5, tempcolor4, tempcolor7, tempcolor8, tempcolor9, tempcolor6, tempcolor3} {
	  \fill[\x] (0.25*\xi,0) rectangle ++(0.25,0.25);
	}
	\node[draw, thin, minimum width=2.25cm, minimum height=0.25cm, anchor=south west, inner sep=0] (legend) at (0.25,0) {};

	\node[left=of legend, xshift=8mm] {\small{Optimal}};
	\node[right=of legend, xshift=-8mm] {\small{Pathological}};
\end{scope}

\end{tikzpicture}

%% file: fig/csd_gear_variants_random.tex
% Created by tikzDevice version 0.12.6 on 2024-04-19 15:33:03
% !TEX encoding = UTF-8 Unicode
\begin{tikzpicture}[x=1pt,y=1pt]
\definecolor{fillColor}{RGB}{255,255,255}
\path[use as bounding box,fill=fillColor,fill opacity=0.00] (0,0) rectangle (144.54,130.09);
\begin{scope}
\path[clip] (  0.00,  0.00) rectangle (144.54,130.09);
\definecolor{drawColor}{RGB}{255,255,255}
\definecolor{fillColor}{RGB}{255,255,255}

\path[draw=drawColor,line width= 0.6pt,line join=round,line cap=round,fill=fillColor] (  0.00,  0.00) rectangle (144.54,130.09);
\end{scope}
\begin{scope}
\path[clip] ( 27.31, 30.69) rectangle (139.04,124.59);
\definecolor{fillColor}{RGB}{255,255,255}

\path[fill=fillColor] ( 27.31, 30.69) rectangle (139.04,124.59);
\definecolor{drawColor}{gray}{0.92}

\path[draw=drawColor,line width= 0.3pt,line join=round] ( 27.31, 48.41) --
	(139.04, 48.41);

\path[draw=drawColor,line width= 0.3pt,line join=round] ( 27.31, 75.32) --
	(139.04, 75.32);

\path[draw=drawColor,line width= 0.3pt,line join=round] ( 27.31,102.23) --
	(139.04,102.23);

\path[draw=drawColor,line width= 0.3pt,line join=round] ( 44.79, 30.69) --
	( 44.79,124.59);

\path[draw=drawColor,line width= 0.3pt,line join=round] ( 69.59, 30.69) --
	( 69.59,124.59);

\path[draw=drawColor,line width= 0.3pt,line join=round] ( 94.39, 30.69) --
	( 94.39,124.59);

\path[draw=drawColor,line width= 0.3pt,line join=round] (119.18, 30.69) --
	(119.18,124.59);

\path[draw=drawColor,line width= 0.6pt,line join=round] ( 27.31, 34.95) --
	(139.04, 34.95);

\path[draw=drawColor,line width= 0.6pt,line join=round] ( 27.31, 61.86) --
	(139.04, 61.86);

\path[draw=drawColor,line width= 0.6pt,line join=round] ( 27.31, 88.77) --
	(139.04, 88.77);

\path[draw=drawColor,line width= 0.6pt,line join=round] ( 27.31,115.68) --
	(139.04,115.68);

\path[draw=drawColor,line width= 0.6pt,line join=round] ( 32.39, 30.69) --
	( 32.39,124.59);

\path[draw=drawColor,line width= 0.6pt,line join=round] ( 57.19, 30.69) --
	( 57.19,124.59);

\path[draw=drawColor,line width= 0.6pt,line join=round] ( 81.99, 30.69) --
	( 81.99,124.59);

\path[draw=drawColor,line width= 0.6pt,line join=round] (106.78, 30.69) --
	(106.78,124.59);

\path[draw=drawColor,line width= 0.6pt,line join=round] (131.58, 30.69) --
	(131.58,124.59);
\definecolor{drawColor}{RGB}{248,118,109}

\path[draw=drawColor,line width= 0.6pt,line join=round] ( 32.39, 42.03) --
	( 33.42, 50.02) --
	( 34.44, 49.42) --
	( 35.47, 49.15) --
	( 36.50, 49.26) --
	( 37.52, 48.63) --
	( 38.55, 48.32) --
	( 39.57, 48.33) --
	( 40.60, 47.80) --
	( 41.63, 47.85) --
	( 42.65, 47.27) --
	( 43.68, 46.99) --
	( 44.70, 47.13) --
	( 45.73, 46.54) --
	( 46.76, 46.28) --
	( 47.78, 46.34) --
	( 48.81, 45.88) --
	( 49.83, 45.89) --
	( 50.86, 45.37) --
	( 51.88, 45.23) --
	( 52.91, 45.27) --
	( 53.94, 44.77) --
	( 54.96, 44.59) --
	( 55.99, 44.66) --
	( 57.01, 44.21) --
	( 58.04, 44.24) --
	( 59.07, 43.79) --
	( 60.09, 43.71) --
	( 61.12, 43.81) --
	( 62.14, 43.34) --
	( 63.17, 43.20) --
	( 64.20, 43.28) --
	( 65.22, 42.82) --
	( 66.25, 42.89) --
	( 67.27, 42.51) --
	( 68.30, 42.36) --
	( 69.33, 42.42) --
	( 70.35, 42.11) --
	( 71.38, 41.91) --
	( 72.40, 41.92) --
	( 73.43, 41.67) --
	( 74.46, 41.68) --
	( 75.48, 41.40) --
	( 76.51, 41.27) --
	( 77.53, 41.32) --
	( 78.56, 41.03) --
	( 79.59, 40.91) --
	( 80.61, 40.96) --
	( 81.64, 40.66) --
	( 82.66, 40.72) --
	( 83.69, 40.43) --
	( 84.72, 40.36) --
	( 85.74, 40.36) --
	( 86.77, 40.12) --
	( 87.79, 39.97) --
	( 88.82, 40.05) --
	( 89.85, 39.84) --
	( 90.87, 39.70) --
	( 91.90, 39.74) --
	( 92.92, 39.53) --
	( 93.95, 39.58) --
	( 94.98, 39.33) --
	( 96.00, 39.25) --
	( 97.03, 39.30) --
	( 98.05, 39.09) --
	( 99.08, 39.00) --
	(100.10, 39.02) --
	(101.13, 38.85) --
	(102.16, 38.90) --
	(103.18, 38.72) --
	(104.21, 38.62) --
	(105.23, 38.62) --
	(106.26, 38.47) --
	(107.29, 38.38) --
	(108.31, 38.43) --
	(109.34, 38.27) --
	(110.36, 38.27) --
	(111.39, 38.17) --
	(112.42, 38.04) --
	(113.44, 38.13) --
	(114.47, 37.95) --
	(115.49, 37.88) --
	(116.52, 37.88) --
	(117.55, 37.78) --
	(118.57, 37.78) --
	(119.60, 37.68) --
	(120.62, 37.57) --
	(121.65, 37.62) --
	(122.68, 37.47) --
	(123.70, 37.43) --
	(124.73, 37.49) --
	(125.75, 37.35) --
	(126.78, 37.37) --
	(127.81, 37.23) --
	(128.83, 37.21) --
	(129.86, 37.24) --
	(130.88, 37.08) --
	(131.91, 37.05) --
	(132.94, 37.08) --
	(133.96, 35.99);
\definecolor{drawColor}{RGB}{124,174,0}

\path[draw=drawColor,line width= 0.6pt,dash pattern=on 2pt off 2pt ,line join=round] ( 32.39, 38.21) --
	( 33.42, 42.18) --
	( 34.44, 41.93) --
	( 35.47, 41.83) --
	( 36.50, 41.94) --
	( 37.52, 41.70) --
	( 38.55, 41.61) --
	( 39.57, 41.71) --
	( 40.60, 41.54) --
	( 41.63, 41.58) --
	( 42.65, 41.40) --
	( 43.68, 41.26) --
	( 44.70, 41.45) --
	( 45.73, 41.21) --
	( 46.76, 41.12) --
	( 47.78, 41.19) --
	( 48.81, 40.98) --
	( 49.83, 41.08) --
	( 50.86, 40.85) --
	( 51.88, 40.79) --
	( 52.91, 40.91) --
	( 53.94, 40.65) --
	( 54.96, 40.62) --
	( 55.99, 40.72) --
	( 57.01, 40.52) --
	( 58.04, 40.60) --
	( 59.07, 40.43) --
	( 60.09, 40.37) --
	( 61.12, 40.42) --
	( 62.14, 40.26) --
	( 63.17, 40.21) --
	( 64.20, 40.26) --
	( 65.22, 40.06) --
	( 66.25, 40.13) --
	( 67.27, 39.94) --
	( 68.30, 39.95) --
	( 69.33, 40.02) --
	( 70.35, 39.82) --
	( 71.38, 39.74) --
	( 72.40, 39.84) --
	( 73.43, 39.68) --
	( 74.46, 39.78) --
	( 75.48, 39.58) --
	( 76.51, 39.52) --
	( 77.53, 39.62) --
	( 78.56, 39.44) --
	( 79.59, 39.42) --
	( 80.61, 39.51) --
	( 81.64, 39.30) --
	( 82.66, 39.38) --
	( 83.69, 51.77) --
	( 84.72, 51.19) --
	( 85.74, 50.94) --
	( 86.77, 49.88) --
	( 87.79, 49.24) --
	( 88.82, 49.07) --
	( 89.85, 48.29) --
	( 90.87, 47.75) --
	( 91.90, 47.49) --
	( 92.92, 46.73) --
	( 93.95, 46.53) --
	( 94.98, 45.73) --
	( 96.00, 45.34) --
	( 97.03, 45.14) --
	( 98.05, 44.55) --
	( 99.08, 44.14) --
	(100.10, 43.99) --
	(101.13, 43.45) --
	(102.16, 43.35) --
	(103.18, 42.78) --
	(104.21, 42.47) --
	(105.23, 42.39) --
	(106.26, 41.89) --
	(107.29, 41.59) --
	(108.31, 41.52) --
	(109.34, 41.07) --
	(110.36, 40.98) --
	(111.39, 40.66) --
	(112.42, 40.41) --
	(113.44, 40.33) --
	(114.47, 39.96) --
	(115.49, 39.77) --
	(116.52, 39.69) --
	(117.55, 39.41) --
	(118.57, 39.37) --
	(119.60, 39.01) --
	(120.62, 38.90) --
	(121.65, 38.82) --
	(122.68, 38.62) --
	(123.70, 38.46) --
	(124.73, 38.38) --
	(125.75, 38.17) --
	(126.78, 38.11) --
	(127.81, 37.93) --
	(128.83, 37.79) --
	(129.86, 37.77) --
	(130.88, 37.59) --
	(131.91, 37.48) --
	(132.94, 37.48) --
	(133.96, 36.17);
\definecolor{drawColor}{RGB}{0,191,196}

\path[draw=drawColor,line width= 0.6pt,dash pattern=on 4pt off 2pt ,line join=round] ( 32.39, 36.47) --
	( 33.42, 38.33) --
	( 34.44, 38.23) --
	( 35.47, 38.23) --
	( 36.50, 38.32) --
	( 37.52, 38.17) --
	( 38.55, 38.15) --
	( 39.57, 38.23) --
	( 40.60, 38.13) --
	( 41.63, 38.20) --
	( 42.65, 38.09) --
	( 43.68, 38.04) --
	( 44.70, 38.17) --
	( 45.73, 38.05) --
	( 46.76, 38.04) --
	( 47.78, 38.05) --
	( 48.81, 38.02) --
	( 49.83, 38.05) --
	( 50.86, 37.94) --
	( 51.88, 37.95) --
	( 52.91, 38.04) --
	( 53.94, 37.90) --
	( 54.96, 37.92) --
	( 55.99, 37.99) --
	( 57.01, 37.87) --
	( 58.04, 37.90) --
	( 59.07, 37.87) --
	( 60.09, 37.82) --
	( 61.12, 37.89) --
	( 62.14, 37.85) --
	( 63.17, 37.81) --
	( 64.20, 37.83) --
	( 65.22, 37.78) --
	( 66.25, 37.82) --
	( 67.27, 37.75) --
	( 68.30, 37.75) --
	( 69.33, 37.81) --
	( 70.35, 37.70) --
	( 71.38, 37.67) --
	( 72.40, 37.73) --
	( 73.43, 37.65) --
	( 74.46, 37.71) --
	( 75.48, 37.61) --
	( 76.51, 37.60) --
	( 77.53, 37.66) --
	( 78.56, 37.61) --
	( 79.59, 37.58) --
	( 80.61, 37.64) --
	( 81.64, 37.53) --
	( 82.66, 37.61) --
	( 83.69, 74.52) --
	( 84.72, 71.46) --
	( 85.74, 69.67) --
	( 86.77, 66.03) --
	( 87.79, 63.58) --
	( 88.82, 62.12) --
	( 89.85, 59.38) --
	( 90.87, 57.50) --
	( 91.90, 56.15) --
	( 92.92, 54.10) --
	( 93.95, 53.11) --
	( 94.98, 51.14) --
	( 96.00, 49.93) --
	( 97.03, 49.11) --
	( 98.05, 47.72) --
	( 99.08, 46.75) --
	(100.10, 46.10) --
	(101.13, 44.97) --
	(102.16, 44.38) --
	(103.18, 43.38) --
	(104.21, 42.80) --
	(105.23, 42.39) --
	(106.26, 41.68) --
	(107.29, 41.11) --
	(108.31, 40.79) --
	(109.34, 40.16) --
	(110.36, 39.85) --
	(111.39, 39.39) --
	(112.42, 39.03) --
	(113.44, 38.83) --
	(114.47, 38.46) --
	(115.49, 38.19) --
	(116.52, 37.98) --
	(117.55, 37.72) --
	(118.57, 37.52) --
	(119.60, 37.25) --
	(120.62, 37.10) --
	(121.65, 36.99) --
	(122.68, 36.77) --
	(123.70, 36.63) --
	(124.73, 36.55) --
	(125.75, 36.35) --
	(126.78, 36.30) --
	(127.81, 36.16) --
	(128.83, 36.07) --
	(129.86, 36.02) --
	(130.88, 35.91) --
	(131.91, 35.84) --
	(132.94, 35.80) --
	(133.96, 35.34);
\definecolor{drawColor}{RGB}{199,124,255}

\path[draw=drawColor,line width= 0.6pt,dash pattern=on 4pt off 4pt ,line join=round] ( 32.39, 35.70) --
	( 33.42, 36.63) --
	( 34.44, 36.55) --
	( 35.47, 36.59) --
	( 36.50, 36.63) --
	( 37.52, 36.57) --
	( 38.55, 36.56) --
	( 39.57, 36.60) --
	( 40.60, 36.54) --
	( 41.63, 36.58) --
	( 42.65, 36.55) --
	( 43.68, 36.52) --
	( 44.70, 36.55) --
	( 45.73, 36.54) --
	( 46.76, 36.53) --
	( 47.78, 36.56) --
	( 48.81, 36.53) --
	( 49.83, 36.54) --
	( 50.86, 36.51) --
	( 51.88, 36.50) --
	( 52.91, 36.54) --
	( 53.94, 36.51) --
	( 54.96, 36.48) --
	( 55.99, 36.53) --
	( 57.01, 36.50) --
	( 58.04, 36.54) --
	( 59.07, 36.48) --
	( 60.09, 36.47) --
	( 61.12, 36.49) --
	( 62.14, 36.49) --
	( 63.17, 36.48) --
	( 64.20, 36.49) --
	( 65.22, 36.43) --
	( 66.25, 36.49) --
	( 67.27, 36.47) --
	( 68.30, 36.47) --
	( 69.33, 36.48) --
	( 70.35, 36.43) --
	( 71.38, 36.44) --
	( 72.40, 36.47) --
	( 73.43, 36.42) --
	( 74.46, 36.47) --
	( 75.48, 36.41) --
	( 76.51, 36.40) --
	( 77.53, 36.46) --
	( 78.56, 36.41) --
	( 79.59, 36.41) --
	( 80.61, 36.41) --
	( 81.64, 36.40) --
	( 82.66, 36.44) --
	( 83.69,120.32) --
	( 84.72,107.60) --
	( 85.74, 98.23) --
	( 86.77, 87.45) --
	( 87.79, 79.57) --
	( 88.82, 73.84) --
	( 89.85, 67.22) --
	( 90.87, 62.41) --
	( 91.90, 58.87) --
	( 92.92, 54.76) --
	( 93.95, 52.13) --
	( 94.98, 49.34) --
	( 96.00, 47.10) --
	( 97.03, 45.52) --
	( 98.05, 43.72) --
	( 99.08, 42.50) --
	(100.10, 41.45) --
	(101.13, 40.34) --
	(102.16, 39.68) --
	(103.18, 38.85) --
	(104.21, 38.30) --
	(105.23, 37.84) --
	(106.26, 37.34) --
	(107.29, 36.99) --
	(108.31, 36.74) --
	(109.34, 36.42) --
	(110.36, 36.21) --
	(111.39, 36.01) --
	(112.42, 35.86) --
	(113.44, 35.74) --
	(114.47, 35.61) --
	(115.49, 35.50) --
	(116.52, 35.45) --
	(117.55, 35.35) --
	(118.57, 35.29) --
	(119.60, 35.25) --
	(120.62, 35.20) --
	(121.65, 35.16) --
	(122.68, 35.13) --
	(123.70, 35.11) --
	(124.73, 35.08) --
	(125.75, 35.07) --
	(126.78, 35.05) --
	(127.81, 35.03) --
	(128.83, 35.02) --
	(129.86, 35.01) --
	(130.88, 35.00) --
	(131.91, 35.00) --
	(132.94, 34.99) --
	(133.96, 34.97);
\definecolor{drawColor}{RGB}{248,118,109}
\definecolor{fillColor}{RGB}{248,118,109}

\path[draw=drawColor,line width= 0.4pt,line join=round,line cap=round,fill=fillColor] ( 83.19, 34.95) circle (  2.50);
\definecolor{drawColor}{RGB}{124,174,0}
\definecolor{fillColor}{RGB}{124,174,0}

\path[draw=drawColor,line width= 0.4pt,line join=round,line cap=round,fill=fillColor] ( 87.77, 34.95) circle (  2.50);
\definecolor{drawColor}{RGB}{0,191,196}
\definecolor{fillColor}{RGB}{0,191,196}

\path[draw=drawColor,line width= 0.4pt,line join=round,line cap=round,fill=fillColor] ( 87.19, 34.95) circle (  2.50);
\definecolor{drawColor}{RGB}{199,124,255}
\definecolor{fillColor}{RGB}{199,124,255}

\path[draw=drawColor,line width= 0.4pt,line join=round,line cap=round,fill=fillColor] ( 85.71, 34.95) circle (  2.50);
\definecolor{drawColor}{gray}{0.20}

\path[draw=drawColor,line width= 0.6pt,line join=round,line cap=round] ( 27.31, 30.69) rectangle (139.04,124.59);
\end{scope}
\begin{scope}
\path[clip] (  0.00,  0.00) rectangle (144.54,130.09);
\definecolor{drawColor}{gray}{0.30}

\node[text=drawColor,anchor=base east,inner sep=0pt, outer sep=0pt, scale=  0.88] at ( 22.36, 31.92) {0};

\node[text=drawColor,anchor=base east,inner sep=0pt, outer sep=0pt, scale=  0.88] at ( 22.36, 58.83) {1};

\node[text=drawColor,anchor=base east,inner sep=0pt, outer sep=0pt, scale=  0.88] at ( 22.36, 85.74) {2};

\node[text=drawColor,anchor=base east,inner sep=0pt, outer sep=0pt, scale=  0.88] at ( 22.36,112.65) {3};
\end{scope}
\begin{scope}
\path[clip] (  0.00,  0.00) rectangle (144.54,130.09);
\definecolor{drawColor}{gray}{0.20}

\path[draw=drawColor,line width= 0.6pt,line join=round] ( 24.56, 34.95) --
	( 27.31, 34.95);

\path[draw=drawColor,line width= 0.6pt,line join=round] ( 24.56, 61.86) --
	( 27.31, 61.86);

\path[draw=drawColor,line width= 0.6pt,line join=round] ( 24.56, 88.77) --
	( 27.31, 88.77);

\path[draw=drawColor,line width= 0.6pt,line join=round] ( 24.56,115.68) --
	( 27.31,115.68);
\end{scope}
\begin{scope}
\path[clip] (  0.00,  0.00) rectangle (144.54,130.09);
\definecolor{drawColor}{gray}{0.20}

\path[draw=drawColor,line width= 0.6pt,line join=round] ( 32.39, 27.94) --
	( 32.39, 30.69);

\path[draw=drawColor,line width= 0.6pt,line join=round] ( 57.19, 27.94) --
	( 57.19, 30.69);

\path[draw=drawColor,line width= 0.6pt,line join=round] ( 81.99, 27.94) --
	( 81.99, 30.69);

\path[draw=drawColor,line width= 0.6pt,line join=round] (106.78, 27.94) --
	(106.78, 30.69);

\path[draw=drawColor,line width= 0.6pt,line join=round] (131.58, 27.94) --
	(131.58, 30.69);
\end{scope}
\begin{scope}
\path[clip] (  0.00,  0.00) rectangle (144.54,130.09);
\definecolor{drawColor}{gray}{0.30}

\node[text=drawColor,anchor=base,inner sep=0pt, outer sep=0pt, scale=  0.88] at ( 32.39, 19.68) {0};

\node[text=drawColor,anchor=base,inner sep=0pt, outer sep=0pt, scale=  0.88] at ( 57.19, 19.68) {1};

\node[text=drawColor,anchor=base,inner sep=0pt, outer sep=0pt, scale=  0.88] at ( 81.99, 19.68) {2};

\node[text=drawColor,anchor=base,inner sep=0pt, outer sep=0pt, scale=  0.88] at (106.78, 19.68) {3};

\node[text=drawColor,anchor=base,inner sep=0pt, outer sep=0pt, scale=  0.88] at (131.58, 19.68) {4};
\end{scope}
\begin{scope}
\path[clip] (  0.00,  0.00) rectangle (144.54,130.09);
\definecolor{drawColor}{RGB}{0,0,0}

\node[text=drawColor,anchor=base,inner sep=0pt, outer sep=0pt, scale=  1.10] at ( 83.18,  7.64) {Chunk Size (KB)};
\end{scope}
\begin{scope}
\path[clip] (  0.00,  0.00) rectangle (144.54,130.09);
\definecolor{drawColor}{RGB}{0,0,0}

\node[text=drawColor,rotate= 90.00,anchor=base,inner sep=0pt, outer sep=0pt, scale=  1.10] at ( 13.08, 77.64) {Density};
\end{scope}
\end{tikzpicture}

%% file: fig/csd_gear_variants_code.tex
% Created by tikzDevice version 0.12.6 on 2024-04-24 12:13:03
% !TEX encoding = UTF-8 Unicode
\begin{tikzpicture}[x=1pt,y=1pt]
\definecolor{fillColor}{RGB}{255,255,255}
\path[use as bounding box,fill=fillColor,fill opacity=0.00] (0,0) rectangle (144.54,130.09);
\begin{scope}
\path[clip] (  0.00,  0.00) rectangle (144.54,130.09);
\definecolor{drawColor}{RGB}{255,255,255}
\definecolor{fillColor}{RGB}{255,255,255}

\path[draw=drawColor,line width= 0.6pt,line join=round,line cap=round,fill=fillColor] (  0.00,  0.00) rectangle (144.54,130.09);
\end{scope}
\begin{scope}
\path[clip] ( 27.31, 30.69) rectangle (139.04,124.59);
\definecolor{fillColor}{RGB}{255,255,255}

\path[fill=fillColor] ( 27.31, 30.69) rectangle (139.04,124.59);
\definecolor{drawColor}{gray}{0.92}

\path[draw=drawColor,line width= 0.3pt,line join=round] ( 27.31, 48.73) --
	(139.04, 48.73);

\path[draw=drawColor,line width= 0.3pt,line join=round] ( 27.31, 76.28) --
	(139.04, 76.28);

\path[draw=drawColor,line width= 0.3pt,line join=round] ( 27.31,103.82) --
	(139.04,103.82);

\path[draw=drawColor,line width= 0.3pt,line join=round] ( 44.79, 30.69) --
	( 44.79,124.59);

\path[draw=drawColor,line width= 0.3pt,line join=round] ( 69.59, 30.69) --
	( 69.59,124.59);

\path[draw=drawColor,line width= 0.3pt,line join=round] ( 94.39, 30.69) --
	( 94.39,124.59);

\path[draw=drawColor,line width= 0.3pt,line join=round] (119.18, 30.69) --
	(119.18,124.59);

\path[draw=drawColor,line width= 0.6pt,line join=round] ( 27.31, 34.95) --
	(139.04, 34.95);

\path[draw=drawColor,line width= 0.6pt,line join=round] ( 27.31, 62.50) --
	(139.04, 62.50);

\path[draw=drawColor,line width= 0.6pt,line join=round] ( 27.31, 90.05) --
	(139.04, 90.05);

\path[draw=drawColor,line width= 0.6pt,line join=round] ( 27.31,117.60) --
	(139.04,117.60);

\path[draw=drawColor,line width= 0.6pt,line join=round] ( 32.39, 30.69) --
	( 32.39,124.59);

\path[draw=drawColor,line width= 0.6pt,line join=round] ( 57.19, 30.69) --
	( 57.19,124.59);

\path[draw=drawColor,line width= 0.6pt,line join=round] ( 81.99, 30.69) --
	( 81.99,124.59);

\path[draw=drawColor,line width= 0.6pt,line join=round] (106.78, 30.69) --
	(106.78,124.59);

\path[draw=drawColor,line width= 0.6pt,line join=round] (131.58, 30.69) --
	(131.58,124.59);
\definecolor{drawColor}{RGB}{248,118,109}

\path[draw=drawColor,line width= 0.6pt,line join=round] ( 32.39, 44.38) --
	( 33.42, 56.46) --
	( 34.44, 63.66) --
	( 35.47, 57.18) --
	( 36.50, 57.19) --
	( 37.52, 52.98) --
	( 38.55, 51.74) --
	( 39.57, 51.41) --
	( 40.60, 49.60) --
	( 41.63, 48.70) --
	( 42.65, 51.85) --
	( 43.68, 47.21) --
	( 44.70, 48.23) --
	( 45.73, 47.00) --
	( 46.76, 46.70) --
	( 47.78, 46.48) --
	( 48.81, 45.97) --
	( 49.83, 46.02) --
	( 50.86, 45.19) --
	( 51.88, 44.65) --
	( 52.91, 44.84) --
	( 53.94, 44.37) --
	( 54.96, 43.73) --
	( 55.99, 44.18) --
	( 57.01, 44.12) --
	( 58.04, 44.10) --
	( 59.07, 43.90) --
	( 60.09, 43.78) --
	( 61.12, 43.77) --
	( 62.14, 43.24) --
	( 63.17, 42.78) --
	( 64.20, 42.58) --
	( 65.22, 42.08) --
	( 66.25, 41.88) --
	( 67.27, 41.66) --
	( 68.30, 41.60) --
	( 69.33, 41.71) --
	( 70.35, 41.07) --
	( 71.38, 41.00) --
	( 72.40, 41.10) --
	( 73.43, 40.78) --
	( 74.46, 40.98) --
	( 75.48, 40.51) --
	( 76.51, 40.37) --
	( 77.53, 40.42) --
	( 78.56, 40.08) --
	( 79.59, 39.77) --
	( 80.61, 39.87) --
	( 81.64, 40.01) --
	( 82.66, 40.06) --
	( 83.69, 39.61) --
	( 84.72, 39.63) --
	( 85.74, 39.41) --
	( 86.77, 39.51) --
	( 87.79, 39.13) --
	( 88.82, 39.31) --
	( 89.85, 39.05) --
	( 90.87, 38.98) --
	( 91.90, 38.99) --
	( 92.92, 38.92) --
	( 93.95, 38.82) --
	( 94.98, 38.53) --
	( 96.00, 38.65) --
	( 97.03, 38.54) --
	( 98.05, 38.85) --
	( 99.08, 38.39) --
	(100.10, 38.49) --
	(101.13, 38.16) --
	(102.16, 38.41) --
	(103.18, 38.07) --
	(104.21, 38.20) --
	(105.23, 38.08) --
	(106.26, 38.04) --
	(107.29, 38.01) --
	(108.31, 37.75) --
	(109.34, 37.79) --
	(110.36, 37.72) --
	(111.39, 37.57) --
	(112.42, 37.71) --
	(113.44, 37.50) --
	(114.47, 37.38) --
	(115.49, 37.38) --
	(116.52, 37.63) --
	(117.55, 37.45) --
	(118.57, 37.47) --
	(119.60, 37.26) --
	(120.62, 37.30) --
	(121.65, 37.12) --
	(122.68, 37.16) --
	(123.70, 37.19) --
	(124.73, 37.14) --
	(125.75, 37.00) --
	(126.78, 37.08) --
	(127.81, 37.12) --
	(128.83, 36.96) --
	(129.86, 36.94) --
	(130.88, 37.09) --
	(131.91, 37.04) --
	(132.94, 36.92) --
	(133.96, 35.89);
\definecolor{drawColor}{RGB}{124,174,0}

\path[draw=drawColor,line width= 0.6pt,dash pattern=on 2pt off 2pt ,line join=round] ( 32.39, 41.02) --
	( 33.42, 44.37) --
	( 34.44, 46.20) --
	( 35.47, 45.18) --
	( 36.50, 45.09) --
	( 37.52, 44.73) --
	( 38.55, 43.47) --
	( 39.57, 43.19) --
	( 40.60, 43.10) --
	( 41.63, 42.73) --
	( 42.65, 45.86) --
	( 43.68, 41.50) --
	( 44.70, 41.90) --
	( 45.73, 41.67) --
	( 46.76, 41.41) --
	( 47.78, 41.32) --
	( 48.81, 41.15) --
	( 49.83, 41.05) --
	( 50.86, 41.03) --
	( 51.88, 40.84) --
	( 52.91, 40.74) --
	( 53.94, 40.37) --
	( 54.96, 40.29) --
	( 55.99, 40.21) --
	( 57.01, 40.28) --
	( 58.04, 40.47) --
	( 59.07, 40.65) --
	( 60.09, 41.05) --
	( 61.12, 40.78) --
	( 62.14, 39.97) --
	( 63.17, 39.86) --
	( 64.20, 40.01) --
	( 65.22, 39.76) --
	( 66.25, 39.61) --
	( 67.27, 39.42) --
	( 68.30, 39.23) --
	( 69.33, 39.53) --
	( 70.35, 39.46) --
	( 71.38, 39.21) --
	( 72.40, 39.32) --
	( 73.43, 39.06) --
	( 74.46, 39.26) --
	( 75.48, 39.09) --
	( 76.51, 39.21) --
	( 77.53, 39.06) --
	( 78.56, 38.81) --
	( 79.59, 38.85) --
	( 80.61, 38.93) --
	( 81.64, 38.86) --
	( 82.66, 38.77) --
	( 83.69, 51.89) --
	( 84.72, 51.25) --
	( 85.74, 50.88) --
	( 86.77, 49.75) --
	( 87.79, 48.87) --
	( 88.82, 48.90) --
	( 89.85, 48.16) --
	( 90.87, 47.18) --
	( 91.90, 47.35) --
	( 92.92, 46.24) --
	( 93.95, 46.06) --
	( 94.98, 45.22) --
	( 96.00, 44.98) --
	( 97.03, 44.96) --
	( 98.05, 44.32) --
	( 99.08, 43.76) --
	(100.10, 43.58) --
	(101.13, 43.11) --
	(102.16, 42.90) --
	(103.18, 42.24) --
	(104.21, 42.32) --
	(105.23, 42.02) --
	(106.26, 41.73) --
	(107.29, 41.54) --
	(108.31, 41.27) --
	(109.34, 41.22) --
	(110.36, 40.67) --
	(111.39, 40.54) --
	(112.42, 40.22) --
	(113.44, 40.33) --
	(114.47, 40.01) --
	(115.49, 39.95) --
	(116.52, 39.60) --
	(117.55, 39.55) --
	(118.57, 39.32) --
	(119.60, 39.04) --
	(120.62, 38.87) --
	(121.65, 38.79) --
	(122.68, 38.77) --
	(123.70, 38.54) --
	(124.73, 38.62) --
	(125.75, 38.10) --
	(126.78, 38.10) --
	(127.81, 38.02) --
	(128.83, 38.00) --
	(129.86, 37.79) --
	(130.88, 37.91) --
	(131.91, 37.74) --
	(132.94, 37.63) --
	(133.96, 36.47);
\definecolor{drawColor}{RGB}{0,191,196}

\path[draw=drawColor,line width= 0.6pt,dash pattern=on 4pt off 2pt ,line join=round] ( 32.39, 39.64) --
	( 33.42, 39.33) --
	( 34.44, 40.26) --
	( 35.47, 39.95) --
	( 36.50, 39.93) --
	( 37.52, 39.52) --
	( 38.55, 38.96) --
	( 39.57, 38.89) --
	( 40.60, 38.75) --
	( 41.63, 38.54) --
	( 42.65, 38.37) --
	( 43.68, 37.95) --
	( 44.70, 38.35) --
	( 45.73, 38.08) --
	( 46.76, 38.30) --
	( 47.78, 38.12) --
	( 48.81, 38.03) --
	( 49.83, 38.22) --
	( 50.86, 38.07) --
	( 51.88, 38.07) --
	( 52.91, 38.00) --
	( 53.94, 37.87) --
	( 54.96, 37.73) --
	( 55.99, 37.73) --
	( 57.01, 37.96) --
	( 58.04, 37.84) --
	( 59.07, 38.12) --
	( 60.09, 38.17) --
	( 61.12, 38.59) --
	( 62.14, 37.83) --
	( 63.17, 37.80) --
	( 64.20, 38.04) --
	( 65.22, 37.62) --
	( 66.25, 37.72) --
	( 67.27, 37.73) --
	( 68.30, 37.43) --
	( 69.33, 37.74) --
	( 70.35, 37.50) --
	( 71.38, 37.36) --
	( 72.40, 37.42) --
	( 73.43, 37.22) --
	( 74.46, 37.46) --
	( 75.48, 37.29) --
	( 76.51, 37.36) --
	( 77.53, 37.38) --
	( 78.56, 37.26) --
	( 79.59, 37.34) --
	( 80.61, 37.36) --
	( 81.64, 37.21) --
	( 82.66, 37.16) --
	( 83.69, 75.30) --
	( 84.72, 71.84) --
	( 85.74, 69.27) --
	( 86.77, 65.06) --
	( 87.79, 63.13) --
	( 88.82, 61.11) --
	( 89.85, 58.44) --
	( 90.87, 56.35) --
	( 91.90, 55.92) --
	( 92.92, 53.87) --
	( 93.95, 52.23) --
	( 94.98, 50.52) --
	( 96.00, 49.23) --
	( 97.03, 48.71) --
	( 98.05, 47.67) --
	( 99.08, 46.31) --
	(100.10, 45.74) --
	(101.13, 45.26) --
	(102.16, 44.33) --
	(103.18, 43.23) --
	(104.21, 42.99) --
	(105.23, 42.50) --
	(106.26, 41.84) --
	(107.29, 41.43) --
	(108.31, 41.22) --
	(109.34, 40.67) --
	(110.36, 40.30) --
	(111.39, 39.85) --
	(112.42, 39.42) --
	(113.44, 39.20) --
	(114.47, 38.91) --
	(115.49, 38.63) --
	(116.52, 38.58) --
	(117.55, 38.04) --
	(118.57, 38.14) --
	(119.60, 37.69) --
	(120.62, 37.67) --
	(121.65, 37.47) --
	(122.68, 37.14) --
	(123.70, 37.06) --
	(124.73, 37.10) --
	(125.75, 36.75) --
	(126.78, 36.78) --
	(127.81, 36.51) --
	(128.83, 36.41) --
	(129.86, 36.52) --
	(130.88, 36.28) --
	(131.91, 36.31) --
	(132.94, 36.22) --
	(133.96, 35.53);
\definecolor{drawColor}{RGB}{199,124,255}

\path[draw=drawColor,line width= 0.6pt,dash pattern=on 4pt off 4pt ,line join=round] ( 32.39, 38.05) --
	( 33.42, 37.24) --
	( 34.44, 37.71) --
	( 35.47, 37.04) --
	( 36.50, 37.11) --
	( 37.52, 37.20) --
	( 38.55, 36.76) --
	( 39.57, 36.73) --
	( 40.60, 36.76) --
	( 41.63, 36.71) --
	( 42.65, 36.65) --
	( 43.68, 36.63) --
	( 44.70, 36.76) --
	( 45.73, 36.62) --
	( 46.76, 36.49) --
	( 47.78, 36.62) --
	( 48.81, 36.50) --
	( 49.83, 36.57) --
	( 50.86, 36.49) --
	( 51.88, 36.47) --
	( 52.91, 36.41) --
	( 53.94, 36.35) --
	( 54.96, 36.47) --
	( 55.99, 36.48) --
	( 57.01, 36.52) --
	( 58.04, 36.49) --
	( 59.07, 36.50) --
	( 60.09, 36.64) --
	( 61.12, 36.76) --
	( 62.14, 36.61) --
	( 63.17, 36.57) --
	( 64.20, 36.67) --
	( 65.22, 36.44) --
	( 66.25, 36.47) --
	( 67.27, 36.33) --
	( 68.30, 36.29) --
	( 69.33, 36.33) --
	( 70.35, 36.26) --
	( 71.38, 36.29) --
	( 72.40, 36.30) --
	( 73.43, 36.25) --
	( 74.46, 36.36) --
	( 75.48, 36.30) --
	( 76.51, 36.14) --
	( 77.53, 36.24) --
	( 78.56, 36.23) --
	( 79.59, 36.28) --
	( 80.61, 36.28) --
	( 81.64, 36.23) --
	( 82.66, 36.25) --
	( 83.69,120.32) --
	( 84.72,107.72) --
	( 85.74, 96.60) --
	( 86.77, 85.67) --
	( 87.79, 78.27) --
	( 88.82, 72.90) --
	( 89.85, 66.40) --
	( 90.87, 61.71) --
	( 91.90, 59.06) --
	( 92.92, 54.76) --
	( 93.95, 52.52) --
	( 94.98, 49.58) --
	( 96.00, 47.34) --
	( 97.03, 46.16) --
	( 98.05, 44.23) --
	( 99.08, 43.10) --
	(100.10, 42.62) --
	(101.13, 41.29) --
	(102.16, 40.65) --
	(103.18, 39.61) --
	(104.21, 38.86) --
	(105.23, 38.54) --
	(106.26, 38.22) --
	(107.29, 37.58) --
	(108.31, 37.37) --
	(109.34, 37.09) --
	(110.36, 36.81) --
	(111.39, 36.66) --
	(112.42, 36.48) --
	(113.44, 36.19) --
	(114.47, 36.12) --
	(115.49, 35.88) --
	(116.52, 35.88) --
	(117.55, 35.70) --
	(118.57, 35.70) --
	(119.60, 35.48) --
	(120.62, 35.52) --
	(121.65, 35.43) --
	(122.68, 35.44) --
	(123.70, 35.29) --
	(124.73, 35.28) --
	(125.75, 35.22) --
	(126.78, 35.31) --
	(127.81, 35.16) --
	(128.83, 35.13) --
	(129.86, 35.14) --
	(130.88, 35.15) --
	(131.91, 35.08) --
	(132.94, 35.12) --
	(133.96, 35.02);
\definecolor{drawColor}{RGB}{248,118,109}
\definecolor{fillColor}{RGB}{248,118,109}

\path[draw=drawColor,line width= 0.4pt,line join=round,line cap=round,fill=fillColor] ( 83.54, 34.95) circle (  2.50);
\definecolor{drawColor}{RGB}{124,174,0}
\definecolor{fillColor}{RGB}{124,174,0}

\path[draw=drawColor,line width= 0.4pt,line join=round,line cap=round,fill=fillColor] ( 89.08, 34.95) circle (  2.50);
\definecolor{drawColor}{RGB}{0,191,196}
\definecolor{fillColor}{RGB}{0,191,196}

\path[draw=drawColor,line width= 0.4pt,line join=round,line cap=round,fill=fillColor] ( 88.25, 34.95) circle (  2.50);
\definecolor{drawColor}{RGB}{199,124,255}
\definecolor{fillColor}{RGB}{199,124,255}

\path[draw=drawColor,line width= 0.4pt,line join=round,line cap=round,fill=fillColor] ( 86.25, 34.95) circle (  2.50);
\definecolor{drawColor}{gray}{0.20}

\path[draw=drawColor,line width= 0.6pt,line join=round,line cap=round] ( 27.31, 30.69) rectangle (139.04,124.59);
\end{scope}
\begin{scope}
\path[clip] (  0.00,  0.00) rectangle (144.54,130.09);
\definecolor{drawColor}{gray}{0.30}

\node[text=drawColor,anchor=base east,inner sep=0pt, outer sep=0pt, scale=  0.88] at ( 22.36, 31.92) {0};

\node[text=drawColor,anchor=base east,inner sep=0pt, outer sep=0pt, scale=  0.88] at ( 22.36, 59.47) {1};

\node[text=drawColor,anchor=base east,inner sep=0pt, outer sep=0pt, scale=  0.88] at ( 22.36, 87.02) {2};

\node[text=drawColor,anchor=base east,inner sep=0pt, outer sep=0pt, scale=  0.88] at ( 22.36,114.57) {3};
\end{scope}
\begin{scope}
\path[clip] (  0.00,  0.00) rectangle (144.54,130.09);
\definecolor{drawColor}{gray}{0.20}

\path[draw=drawColor,line width= 0.6pt,line join=round] ( 24.56, 34.95) --
	( 27.31, 34.95);

\path[draw=drawColor,line width= 0.6pt,line join=round] ( 24.56, 62.50) --
	( 27.31, 62.50);

\path[draw=drawColor,line width= 0.6pt,line join=round] ( 24.56, 90.05) --
	( 27.31, 90.05);

\path[draw=drawColor,line width= 0.6pt,line join=round] ( 24.56,117.60) --
	( 27.31,117.60);
\end{scope}
\begin{scope}
\path[clip] (  0.00,  0.00) rectangle (144.54,130.09);
\definecolor{drawColor}{gray}{0.20}

\path[draw=drawColor,line width= 0.6pt,line join=round] ( 32.39, 27.94) --
	( 32.39, 30.69);

\path[draw=drawColor,line width= 0.6pt,line join=round] ( 57.19, 27.94) --
	( 57.19, 30.69);

\path[draw=drawColor,line width= 0.6pt,line join=round] ( 81.99, 27.94) --
	( 81.99, 30.69);

\path[draw=drawColor,line width= 0.6pt,line join=round] (106.78, 27.94) --
	(106.78, 30.69);

\path[draw=drawColor,line width= 0.6pt,line join=round] (131.58, 27.94) --
	(131.58, 30.69);
\end{scope}
\begin{scope}
\path[clip] (  0.00,  0.00) rectangle (144.54,130.09);
\definecolor{drawColor}{gray}{0.30}

\node[text=drawColor,anchor=base,inner sep=0pt, outer sep=0pt, scale=  0.88] at ( 32.39, 19.68) {0};

\node[text=drawColor,anchor=base,inner sep=0pt, outer sep=0pt, scale=  0.88] at ( 57.19, 19.68) {1};

\node[text=drawColor,anchor=base,inner sep=0pt, outer sep=0pt, scale=  0.88] at ( 81.99, 19.68) {2};

\node[text=drawColor,anchor=base,inner sep=0pt, outer sep=0pt, scale=  0.88] at (106.78, 19.68) {3};

\node[text=drawColor,anchor=base,inner sep=0pt, outer sep=0pt, scale=  0.88] at (131.58, 19.68) {4};
\end{scope}
\begin{scope}
\path[clip] (  0.00,  0.00) rectangle (144.54,130.09);
\definecolor{drawColor}{RGB}{0,0,0}

\node[text=drawColor,anchor=base,inner sep=0pt, outer sep=0pt, scale=  1.10] at ( 83.18,  7.64) {Chunk Size (KB)};
\end{scope}
\begin{scope}
\path[clip] (  0.00,  0.00) rectangle (144.54,130.09);
\definecolor{drawColor}{RGB}{0,0,0}

\node[text=drawColor,rotate= 90.00,anchor=base,inner sep=0pt, outer sep=0pt, scale=  1.10] at ( 13.08, 77.64) {Density};
\end{scope}
\end{tikzpicture}

%% file: fig/csd_gear_variants_legendonly.tex
% Created by tikzDevice version 0.12.6 on 2024-04-19 15:33:44
% !TEX encoding = UTF-8 Unicode
\begin{tikzpicture}[x=1pt,y=1pt]
\definecolor{fillColor}{RGB}{255,255,255}
\path[use as bounding box,fill=fillColor,fill opacity=0.00] (0,0) rectangle (252.94, 36.13);
\begin{scope}
\path[clip] (  0.00,  0.00) rectangle (252.94, 36.13);
\definecolor{drawColor}{RGB}{248,118,109}

\path[draw=drawColor,line width= 0.6pt,line join=round] ( 23.40, 18.07) -- ( 37.28, 18.07);
\end{scope}
\begin{scope}
\path[clip] (  0.00,  0.00) rectangle (252.94, 36.13);
\definecolor{drawColor}{RGB}{124,174,0}

\path[draw=drawColor,line width= 0.6pt,dash pattern=on 2pt off 2pt ,line join=round] ( 81.17, 18.07) -- ( 95.04, 18.07);
\end{scope}
\begin{scope}
\path[clip] (  0.00,  0.00) rectangle (252.94, 36.13);
\definecolor{drawColor}{RGB}{0,191,196}

\path[draw=drawColor,line width= 0.6pt,dash pattern=on 4pt off 2pt ,line join=round] (135.45, 18.07) -- (149.32, 18.07);
\end{scope}
\begin{scope}
\path[clip] (  0.00,  0.00) rectangle (252.94, 36.13);
\definecolor{drawColor}{RGB}{199,124,255}

\path[draw=drawColor,line width= 0.6pt,dash pattern=on 4pt off 4pt ,line join=round] (189.73, 18.07) -- (203.61, 18.07);
\end{scope}
\begin{scope}
\path[clip] (  0.00,  0.00) rectangle (252.94, 36.13);
\definecolor{drawColor}{RGB}{0,0,0}

\node[text=drawColor,anchor=base west,inner sep=0pt, outer sep=0pt, scale=  1.00] at ( 44.51, 14.62) {Vanilla};
\end{scope}
\begin{scope}
\path[clip] (  0.00,  0.00) rectangle (252.94, 36.13);
\definecolor{drawColor}{RGB}{0,0,0}

\node[text=drawColor,anchor=base west,inner sep=0pt, outer sep=0pt, scale=  1.00] at (102.28, 14.62) {NC-1};
\end{scope}
\begin{scope}
\path[clip] (  0.00,  0.00) rectangle (252.94, 36.13);
\definecolor{drawColor}{RGB}{0,0,0}

\node[text=drawColor,anchor=base west,inner sep=0pt, outer sep=0pt, scale=  1.00] at (156.56, 14.62) {NC-2};
\end{scope}
\begin{scope}
\path[clip] (  0.00,  0.00) rectangle (252.94, 36.13);
\definecolor{drawColor}{RGB}{0,0,0}

\node[text=drawColor,anchor=base west,inner sep=0pt, outer sep=0pt, scale=  1.00] at (210.84, 14.62) {NC-3};
\end{scope}
\end{tikzpicture}

%% file: sections/dedup.tex
The degree of achievable deduplication is a key metric in the evaluation of CDC algorithms.
It determines how effectively the algorithm can identify and eliminate redundant data,
thereby minimizing storage requirements.
In this section, we investigate the comparative deduplication performance of the selected algorithms.
Additionally, we explore how the characteristics of the dataset, such as entropy,
influence the deduplication achieved by each algorithm.
We proceed similar to earlier analyses:
We will apply each algorithm to each dataset,
report on the results,
and derive insights into the characteristics of each algorithm and dataset.
As key metric we use the deduplication ratio.
It indicates the ratio of storage space that can be saved due to the elimination of redundant chunks,
as a value in $[0,1]$.
For instance, a deduplication ratio of $0.4$ on a dataset of \qty{1}{GB}
indicates that the same dataset can be represented in \qty{600}{MB} of unique chunks.
Note that we do not consider the overhead of metadata here.
We evaluate the algorithms on the four realistic datasets CODE, WEB, LNX, and PDF.
Evaluations on the RAND dataset are moot,
as no deduplication is possible with realistic chunk sizes.
For the other datasets, we determined optimal window sizes for the \ac{BSW} algorithms (\Cref{fig:dedup_window_sizes}).

%%%%%%%%%%%%%%%%%%%%%%%%%%%%%%%%%%%%%%%%%%%%%%%%%%%%
\subsection{General Overview}

\begin{figure*}[t!]
\centering
\subfloat[LNX]{\scalebox{0.85}{\input{fig/dedup_overview_lnx}}%
    \label{fig:dedup_overview:lnx}}
\hfil
\subfloat[PDF]{\scalebox{0.85}{\input{fig/dedup_overview_pdf}}%
	\label{fig:dedup_overview:pdf}}
\hfil
\subfloat[WEB]{\scalebox{0.85}{\input{fig/dedup_overview_web}}%
    \label{fig:dedup_overview:web}}
\hfil
\subfloat[CODE]{\scalebox{0.85}{\input{fig/dedup_overview_code}}%
    \label{fig:dedup_overview:code}}
 
\vspace{-2em}
\input{fig/dedup_overview_legendonly}
\vspace{-2em}
  \caption{Overview of deduplication ratios. Note the varying scales on the y-axis.}
  \label{fig:dedup_overview}
\end{figure*}
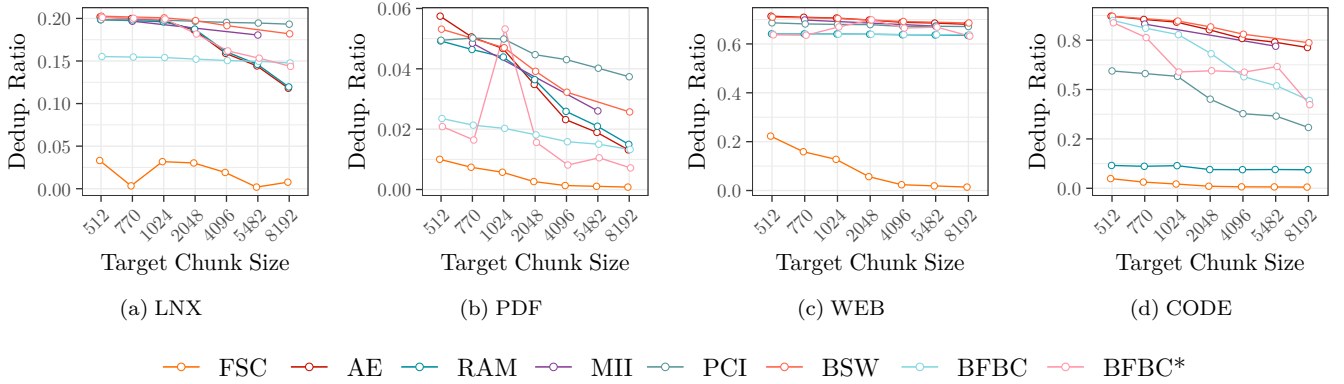

% intro
In \cref{fig:dedup_overview},
we give an overview of the deduplication performance of each chunking algorithm on each dataset and over a range of target chunk sizes.
%We do not display results for QuickCDC in our experiments, as it performed identically to the wrapped algorithm.
We summarize Rabin, Buzhash, and Gear into one category of BSW algorithms.
These algorithms mainly differ in their choice of hash function.
As investigated in \Cref{sec:distribution}, this \emph{does} lead to differences in chunk size distributions.
Interestingly, however, we find only minuscule ($< 0.01$) differences in the deduplication ratios achieved.
%Further, we choose the most favorable window size, as derived by the experiment outlined earlier.
Generally, for all algorithms, the deduplication ratio drops with growing target chunk sizes,
although to varying degrees.
This is expected as smaller chunks have a higher chance of being duplicates.
It is particularly noticeable in the CODE dataset.
We suspect that this is due to the nature of the content:
Minor modifications to source code (which is purely text-based) affect only a few bytes,
while modifications to binary files (a large portion of the files in WEB) potentially affect a longer sequence of bytes, up to the entire file.

The top performers across all settings are the BSW algorithms and AE,
with no considerable differences among each other.
However, there is a slight divergence as the target chunk size increases.
Presumably, AE's deduplication performance lags behind as the window size increases.
We speculate that this difference may become more pronounced for target chunk sizes larger than those tested.
Notably, although PCI occasionally emerges as the top performing (LNX and PDF on high targets),
it is essential to consider this in the context of the mean chunk sizes it generates
(\eg, $<\qty{1}{KiB}$ on $\mu=\qty{8}{KiB}$).
Furthermore, its degradation for larger chunk sizes is less pronounced than in the other algorithms.
RAM also establishes itself among the top performers, with the exception of the CODE dataset.
The problem with RAM on CODE likely stems from the fact that it forms very large chunks
and thus fails to find duplicates (\cf\Cref{sec:distribution}).
Finally, we also observe MII with competitive deduplication results in most scenarios.
BFBC and BFBC* both fall behind the competition.
Despite being among the top performers on CODE on a target of \qty{512}{B},
their performance declines rapidly when run with higher target chunk size configurations.
This is in contrast to results presented in the original publication~\cite{bfbc},
which we discuss in \cref{sec:discussion:bfbc}.

\subsection{Normalized Chunking}

The goal of \ac{NC}, as proposed for FastCDC~\cite{fastcdc},
is the reduction of chunk-size variance.
We investigated this claim in \Cref{sec:distribution}.
The question remains as to how this affects deduplication performance.
Because the same byte sequences at different positions will be subject to inconsistent matching rules,
it is plausible to expect a potentially degrading effect.
We measure the attained deduplication ratio with three levels of normalized chunking applied to Gear,
in comparison to vanilla Gear without NC.
The results are presented in \cref{fig:dedup_gear}.
Surprisingly, even with NC-3, we see only marginal detrimental effects.
The largest difference can be observed at a target of \qty{8}{KiB} on CODE (\cref{fig:dedup_gear_variants:code}),
where NC-2 surpasses Vanilla by two percentage points
While generally, the performance differences are minor,
the implications of this are large since NC can drastically reduce chunk-size variance.

%% file: fig/dedup_overview_lnx.tex
% Created by tikzDevice version 0.12.6 on 2024-07-13 14:07:29
% !TEX encoding = UTF-8 Unicode
\begin{tikzpicture}[x=1pt,y=1pt]
\definecolor{fillColor}{RGB}{255,255,255}
\path[use as bounding box,fill=fillColor,fill opacity=0.00] (0,0) rectangle (144.54,130.09);
\begin{scope}
\path[clip] (  0.00,  0.00) rectangle (144.54,130.09);
\definecolor{drawColor}{RGB}{255,255,255}
\definecolor{fillColor}{RGB}{255,255,255}

\path[draw=drawColor,line width= 0.6pt,line join=round,line cap=round,fill=fillColor] (  0.00,  0.00) rectangle (144.54,130.09);
\end{scope}
\begin{scope}
\path[clip] ( 38.56, 40.85) rectangle (139.04,124.59);
\definecolor{fillColor}{RGB}{255,255,255}

\path[fill=fillColor] ( 38.56, 40.85) rectangle (139.04,124.59);
\definecolor{drawColor}{gray}{0.92}

\path[draw=drawColor,line width= 0.3pt,line join=round] ( 38.56, 53.42) --
	(139.04, 53.42);

\path[draw=drawColor,line width= 0.3pt,line join=round] ( 38.56, 72.38) --
	(139.04, 72.38);

\path[draw=drawColor,line width= 0.3pt,line join=round] ( 38.56, 91.34) --
	(139.04, 91.34);

\path[draw=drawColor,line width= 0.3pt,line join=round] ( 38.56,110.30) --
	(139.04,110.30);

\path[draw=drawColor,line width= 0.6pt,line join=round] ( 38.56, 43.94) --
	(139.04, 43.94);

\path[draw=drawColor,line width= 0.6pt,line join=round] ( 38.56, 62.90) --
	(139.04, 62.90);

\path[draw=drawColor,line width= 0.6pt,line join=round] ( 38.56, 81.86) --
	(139.04, 81.86);

\path[draw=drawColor,line width= 0.6pt,line join=round] ( 38.56,100.82) --
	(139.04,100.82);

\path[draw=drawColor,line width= 0.6pt,line join=round] ( 38.56,119.78) --
	(139.04,119.78);

\path[draw=drawColor,line width= 0.6pt,line join=round] ( 46.93, 40.85) --
	( 46.93,124.59);

\path[draw=drawColor,line width= 0.6pt,line join=round] ( 60.89, 40.85) --
	( 60.89,124.59);

\path[draw=drawColor,line width= 0.6pt,line join=round] ( 74.84, 40.85) --
	( 74.84,124.59);

\path[draw=drawColor,line width= 0.6pt,line join=round] ( 88.80, 40.85) --
	( 88.80,124.59);

\path[draw=drawColor,line width= 0.6pt,line join=round] (102.75, 40.85) --
	(102.75,124.59);

\path[draw=drawColor,line width= 0.6pt,line join=round] (116.71, 40.85) --
	(116.71,124.59);

\path[draw=drawColor,line width= 0.6pt,line join=round] (130.67, 40.85) --
	(130.67,124.59);
\definecolor{drawColor}{RGB}{255,111,0}

\path[draw=drawColor,line width= 0.6pt,line join=round] ( 46.33, 56.51) --
	( 60.29, 45.22) --
	( 74.24, 56.00) --
	( 88.20, 55.41) --
	(102.16, 51.19) --
	(116.11, 44.66) --
	(130.07, 46.85);
\definecolor{drawColor}{RGB}{199,16,0}

\path[draw=drawColor,line width= 0.6pt,line join=round] ( 46.53,120.47) --
	( 60.49,119.73) --
	( 74.44,118.88) --
	( 88.40,114.59) --
	(102.36,104.24) --
	(116.31, 98.62) --
	(130.27, 88.69);
\definecolor{drawColor}{RGB}{0,142,160}

\path[draw=drawColor,line width= 0.6pt,line join=round] ( 46.73,119.20) --
	( 60.69,118.80) --
	( 74.64,118.28) --
	( 88.60,115.02) --
	(102.55,104.98) --
	(116.51, 99.32) --
	(130.47, 89.26);
\definecolor{drawColor}{RGB}{138,65,152}

\path[draw=drawColor,line width= 0.6pt,line join=round] ( 60.89,118.54) --
	(116.71,112.38);
\definecolor{drawColor}{RGB}{90,149,153}

\path[draw=drawColor,line width= 0.6pt,line join=round] ( 46.93,119.21) --
	( 61.08,119.12) --
	( 74.84,119.02) --
	( 88.80,118.49) --
	(102.75,118.01) --
	(116.91,117.73) --
	(130.67,117.22);
\definecolor{drawColor}{RGB}{255,99,72}

\path[draw=drawColor,line width= 0.6pt,line join=round] ( 47.13,120.78) --
	( 75.04,120.05) --
	( 89.00,118.81) --
	(102.95,116.55) --
	(130.87,112.92);
\definecolor{drawColor}{RGB}{132,215,225}

\path[draw=drawColor,line width= 0.6pt,line join=round] ( 47.33,102.80) --
	( 61.28,102.58) --
	( 75.24,102.36) --
	( 89.20,101.65) --
	(103.15,101.02) --
	(117.11,100.67) --
	(131.07, 99.95);
\definecolor{drawColor}{RGB}{255,149,168}

\path[draw=drawColor,line width= 0.6pt,line join=round] ( 47.53,120.20) --
	( 61.48,119.86) --
	( 75.44,119.51) --
	( 89.40,112.85) --
	(103.35,105.32) --
	(117.31,102.07) --
	(131.26, 98.42);
\definecolor{drawColor}{RGB}{255,111,0}

\path[draw=drawColor,line width= 0.4pt,line join=round,line cap=round,fill=fillColor] ( 46.33, 56.51) circle (  1.43);

\path[draw=drawColor,line width= 0.4pt,line join=round,line cap=round,fill=fillColor] ( 74.24, 56.00) circle (  1.43);

\path[draw=drawColor,line width= 0.4pt,line join=round,line cap=round,fill=fillColor] ( 88.20, 55.41) circle (  1.43);

\path[draw=drawColor,line width= 0.4pt,line join=round,line cap=round,fill=fillColor] (102.16, 51.19) circle (  1.43);

\path[draw=drawColor,line width= 0.4pt,line join=round,line cap=round,fill=fillColor] (130.07, 46.85) circle (  1.43);

\path[draw=drawColor,line width= 0.4pt,line join=round,line cap=round,fill=fillColor] ( 60.29, 45.22) circle (  1.43);

\path[draw=drawColor,line width= 0.4pt,line join=round,line cap=round,fill=fillColor] (116.11, 44.66) circle (  1.43);
\definecolor{drawColor}{RGB}{199,16,0}

\path[draw=drawColor,line width= 0.4pt,line join=round,line cap=round,fill=fillColor] ( 46.53,120.47) circle (  1.43);

\path[draw=drawColor,line width= 0.4pt,line join=round,line cap=round,fill=fillColor] ( 74.44,118.88) circle (  1.43);

\path[draw=drawColor,line width= 0.4pt,line join=round,line cap=round,fill=fillColor] ( 88.40,114.59) circle (  1.43);

\path[draw=drawColor,line width= 0.4pt,line join=round,line cap=round,fill=fillColor] (102.36,104.24) circle (  1.43);

\path[draw=drawColor,line width= 0.4pt,line join=round,line cap=round,fill=fillColor] (130.27, 88.69) circle (  1.43);

\path[draw=drawColor,line width= 0.4pt,line join=round,line cap=round,fill=fillColor] ( 60.49,119.73) circle (  1.43);

\path[draw=drawColor,line width= 0.4pt,line join=round,line cap=round,fill=fillColor] (116.31, 98.62) circle (  1.43);
\definecolor{drawColor}{RGB}{0,142,160}

\path[draw=drawColor,line width= 0.4pt,line join=round,line cap=round,fill=fillColor] ( 46.73,119.20) circle (  1.43);

\path[draw=drawColor,line width= 0.4pt,line join=round,line cap=round,fill=fillColor] ( 74.64,118.28) circle (  1.43);

\path[draw=drawColor,line width= 0.4pt,line join=round,line cap=round,fill=fillColor] ( 88.60,115.02) circle (  1.43);

\path[draw=drawColor,line width= 0.4pt,line join=round,line cap=round,fill=fillColor] (102.55,104.98) circle (  1.43);

\path[draw=drawColor,line width= 0.4pt,line join=round,line cap=round,fill=fillColor] (130.47, 89.26) circle (  1.43);

\path[draw=drawColor,line width= 0.4pt,line join=round,line cap=round,fill=fillColor] ( 60.69,118.80) circle (  1.43);

\path[draw=drawColor,line width= 0.4pt,line join=round,line cap=round,fill=fillColor] (116.51, 99.32) circle (  1.43);
\definecolor{drawColor}{RGB}{138,65,152}

\path[draw=drawColor,line width= 0.4pt,line join=round,line cap=round,fill=fillColor] ( 60.89,118.54) circle (  1.43);

\path[draw=drawColor,line width= 0.4pt,line join=round,line cap=round,fill=fillColor] (116.71,112.38) circle (  1.43);
\definecolor{drawColor}{RGB}{90,149,153}

\path[draw=drawColor,line width= 0.4pt,line join=round,line cap=round,fill=fillColor] ( 46.93,119.21) circle (  1.43);

\path[draw=drawColor,line width= 0.4pt,line join=round,line cap=round,fill=fillColor] ( 74.84,119.02) circle (  1.43);

\path[draw=drawColor,line width= 0.4pt,line join=round,line cap=round,fill=fillColor] ( 88.80,118.49) circle (  1.43);

\path[draw=drawColor,line width= 0.4pt,line join=round,line cap=round,fill=fillColor] (102.75,118.01) circle (  1.43);

\path[draw=drawColor,line width= 0.4pt,line join=round,line cap=round,fill=fillColor] (130.67,117.22) circle (  1.43);

\path[draw=drawColor,line width= 0.4pt,line join=round,line cap=round,fill=fillColor] ( 61.08,119.12) circle (  1.43);

\path[draw=drawColor,line width= 0.4pt,line join=round,line cap=round,fill=fillColor] (116.91,117.73) circle (  1.43);
\definecolor{drawColor}{RGB}{255,99,72}

\path[draw=drawColor,line width= 0.4pt,line join=round,line cap=round,fill=fillColor] ( 47.13,120.78) circle (  1.43);

\path[draw=drawColor,line width= 0.4pt,line join=round,line cap=round,fill=fillColor] ( 75.04,120.05) circle (  1.43);

\path[draw=drawColor,line width= 0.4pt,line join=round,line cap=round,fill=fillColor] ( 89.00,118.81) circle (  1.43);

\path[draw=drawColor,line width= 0.4pt,line join=round,line cap=round,fill=fillColor] (102.95,116.55) circle (  1.43);

\path[draw=drawColor,line width= 0.4pt,line join=round,line cap=round,fill=fillColor] (130.87,112.92) circle (  1.43);
\definecolor{drawColor}{RGB}{132,215,225}

\path[draw=drawColor,line width= 0.4pt,line join=round,line cap=round,fill=fillColor] ( 47.33,102.80) circle (  1.43);

\path[draw=drawColor,line width= 0.4pt,line join=round,line cap=round,fill=fillColor] ( 75.24,102.36) circle (  1.43);

\path[draw=drawColor,line width= 0.4pt,line join=round,line cap=round,fill=fillColor] ( 89.20,101.65) circle (  1.43);

\path[draw=drawColor,line width= 0.4pt,line join=round,line cap=round,fill=fillColor] (103.15,101.02) circle (  1.43);

\path[draw=drawColor,line width= 0.4pt,line join=round,line cap=round,fill=fillColor] (131.07, 99.95) circle (  1.43);

\path[draw=drawColor,line width= 0.4pt,line join=round,line cap=round,fill=fillColor] ( 61.28,102.58) circle (  1.43);

\path[draw=drawColor,line width= 0.4pt,line join=round,line cap=round,fill=fillColor] (117.11,100.67) circle (  1.43);
\definecolor{drawColor}{RGB}{255,149,168}

\path[draw=drawColor,line width= 0.4pt,line join=round,line cap=round,fill=fillColor] ( 47.53,120.20) circle (  1.43);

\path[draw=drawColor,line width= 0.4pt,line join=round,line cap=round,fill=fillColor] ( 75.44,119.51) circle (  1.43);

\path[draw=drawColor,line width= 0.4pt,line join=round,line cap=round,fill=fillColor] ( 89.40,112.85) circle (  1.43);

\path[draw=drawColor,line width= 0.4pt,line join=round,line cap=round,fill=fillColor] (103.35,105.32) circle (  1.43);

\path[draw=drawColor,line width= 0.4pt,line join=round,line cap=round,fill=fillColor] (131.26, 98.42) circle (  1.43);

\path[draw=drawColor,line width= 0.4pt,line join=round,line cap=round,fill=fillColor] ( 61.48,119.86) circle (  1.43);

\path[draw=drawColor,line width= 0.4pt,line join=round,line cap=round,fill=fillColor] (117.31,102.07) circle (  1.43);
\definecolor{drawColor}{gray}{0.20}

\path[draw=drawColor,line width= 0.6pt,line join=round,line cap=round] ( 38.56, 40.85) rectangle (139.04,124.59);
\end{scope}
\begin{scope}
\path[clip] (  0.00,  0.00) rectangle (144.54,130.09);
\definecolor{drawColor}{gray}{0.30}

\node[text=drawColor,anchor=base east,inner sep=0pt, outer sep=0pt, scale=  0.88] at ( 33.61, 40.91) {0.00};

\node[text=drawColor,anchor=base east,inner sep=0pt, outer sep=0pt, scale=  0.88] at ( 33.61, 59.87) {0.05};

\node[text=drawColor,anchor=base east,inner sep=0pt, outer sep=0pt, scale=  0.88] at ( 33.61, 78.83) {0.10};

\node[text=drawColor,anchor=base east,inner sep=0pt, outer sep=0pt, scale=  0.88] at ( 33.61, 97.79) {0.15};

\node[text=drawColor,anchor=base east,inner sep=0pt, outer sep=0pt, scale=  0.88] at ( 33.61,116.75) {0.20};
\end{scope}
\begin{scope}
\path[clip] (  0.00,  0.00) rectangle (144.54,130.09);
\definecolor{drawColor}{gray}{0.20}

\path[draw=drawColor,line width= 0.6pt,line join=round] ( 35.81, 43.94) --
	( 38.56, 43.94);

\path[draw=drawColor,line width= 0.6pt,line join=round] ( 35.81, 62.90) --
	( 38.56, 62.90);

\path[draw=drawColor,line width= 0.6pt,line join=round] ( 35.81, 81.86) --
	( 38.56, 81.86);

\path[draw=drawColor,line width= 0.6pt,line join=round] ( 35.81,100.82) --
	( 38.56,100.82);

\path[draw=drawColor,line width= 0.6pt,line join=round] ( 35.81,119.78) --
	( 38.56,119.78);
\end{scope}
\begin{scope}
\path[clip] (  0.00,  0.00) rectangle (144.54,130.09);
\definecolor{drawColor}{gray}{0.20}

\path[draw=drawColor,line width= 0.6pt,line join=round] ( 46.93, 38.10) --
	( 46.93, 40.85);

\path[draw=drawColor,line width= 0.6pt,line join=round] ( 60.89, 38.10) --
	( 60.89, 40.85);

\path[draw=drawColor,line width= 0.6pt,line join=round] ( 74.84, 38.10) --
	( 74.84, 40.85);

\path[draw=drawColor,line width= 0.6pt,line join=round] ( 88.80, 38.10) --
	( 88.80, 40.85);

\path[draw=drawColor,line width= 0.6pt,line join=round] (102.75, 38.10) --
	(102.75, 40.85);

\path[draw=drawColor,line width= 0.6pt,line join=round] (116.71, 38.10) --
	(116.71, 40.85);

\path[draw=drawColor,line width= 0.6pt,line join=round] (130.67, 38.10) --
	(130.67, 40.85);
\end{scope}
\begin{scope}
\path[clip] (  0.00,  0.00) rectangle (144.54,130.09);
\definecolor{drawColor}{gray}{0.30}

\node[text=drawColor,rotate= 45.00,anchor=base east,inner sep=0pt, outer sep=0pt, scale=  0.88] at ( 51.21, 31.62) {512};

\node[text=drawColor,rotate= 45.00,anchor=base east,inner sep=0pt, outer sep=0pt, scale=  0.88] at ( 65.17, 31.62) {770};

\node[text=drawColor,rotate= 45.00,anchor=base east,inner sep=0pt, outer sep=0pt, scale=  0.88] at ( 79.13, 31.62) {1024};

\node[text=drawColor,rotate= 45.00,anchor=base east,inner sep=0pt, outer sep=0pt, scale=  0.88] at ( 93.08, 31.62) {2048};

\node[text=drawColor,rotate= 45.00,anchor=base east,inner sep=0pt, outer sep=0pt, scale=  0.88] at (107.04, 31.62) {4096};

\node[text=drawColor,rotate= 45.00,anchor=base east,inner sep=0pt, outer sep=0pt, scale=  0.88] at (121.00, 31.62) {5482};

\node[text=drawColor,rotate= 45.00,anchor=base east,inner sep=0pt, outer sep=0pt, scale=  0.88] at (134.95, 31.62) {8192};
\end{scope}
\begin{scope}
\path[clip] (  0.00,  0.00) rectangle (144.54,130.09);
\definecolor{drawColor}{RGB}{0,0,0}

\node[text=drawColor,anchor=base,inner sep=0pt, outer sep=0pt, scale=  1.10] at ( 88.80,  7.64) {Target Chunk Size};
\end{scope}
\begin{scope}
\path[clip] (  0.00,  0.00) rectangle (144.54,130.09);
\definecolor{drawColor}{RGB}{0,0,0}

\node[text=drawColor,rotate= 90.00,anchor=base,inner sep=0pt, outer sep=0pt, scale=  1.10] at ( 13.08, 82.72) {Dedup. Ratio};
\end{scope}
\end{tikzpicture}

%% file: fig/dedup_overview_pdf.tex
% Created by tikzDevice version 0.12.6 on 2024-07-13 14:07:30
% !TEX encoding = UTF-8 Unicode
\begin{tikzpicture}[x=1pt,y=1pt]
\definecolor{fillColor}{RGB}{255,255,255}
\path[use as bounding box,fill=fillColor,fill opacity=0.00] (0,0) rectangle (144.54,130.09);
\begin{scope}
\path[clip] (  0.00,  0.00) rectangle (144.54,130.09);
\definecolor{drawColor}{RGB}{255,255,255}
\definecolor{fillColor}{RGB}{255,255,255}

\path[draw=drawColor,line width= 0.6pt,line join=round,line cap=round,fill=fillColor] (  0.00,  0.00) rectangle (144.54,130.09);
\end{scope}
\begin{scope}
\path[clip] ( 38.56, 40.85) rectangle (139.04,124.59);
\definecolor{fillColor}{RGB}{255,255,255}

\path[fill=fillColor] ( 38.56, 40.85) rectangle (139.04,124.59);
\definecolor{drawColor}{gray}{0.92}

\path[draw=drawColor,line width= 0.3pt,line join=round] ( 38.56, 57.02) --
	(139.04, 57.02);

\path[draw=drawColor,line width= 0.3pt,line join=round] ( 38.56, 83.90) --
	(139.04, 83.90);

\path[draw=drawColor,line width= 0.3pt,line join=round] ( 38.56,110.78) --
	(139.04,110.78);

\path[draw=drawColor,line width= 0.6pt,line join=round] ( 38.56, 43.58) --
	(139.04, 43.58);

\path[draw=drawColor,line width= 0.6pt,line join=round] ( 38.56, 70.46) --
	(139.04, 70.46);

\path[draw=drawColor,line width= 0.6pt,line join=round] ( 38.56, 97.34) --
	(139.04, 97.34);

\path[draw=drawColor,line width= 0.6pt,line join=round] ( 38.56,124.22) --
	(139.04,124.22);

\path[draw=drawColor,line width= 0.6pt,line join=round] ( 46.93, 40.85) --
	( 46.93,124.59);

\path[draw=drawColor,line width= 0.6pt,line join=round] ( 60.89, 40.85) --
	( 60.89,124.59);

\path[draw=drawColor,line width= 0.6pt,line join=round] ( 74.84, 40.85) --
	( 74.84,124.59);

\path[draw=drawColor,line width= 0.6pt,line join=round] ( 88.80, 40.85) --
	( 88.80,124.59);

\path[draw=drawColor,line width= 0.6pt,line join=round] (102.75, 40.85) --
	(102.75,124.59);

\path[draw=drawColor,line width= 0.6pt,line join=round] (116.71, 40.85) --
	(116.71,124.59);

\path[draw=drawColor,line width= 0.6pt,line join=round] (130.67, 40.85) --
	(130.67,124.59);
\definecolor{drawColor}{RGB}{255,111,0}

\path[draw=drawColor,line width= 0.6pt,line join=round] ( 46.33, 57.04) --
	( 60.29, 53.50) --
	( 74.24, 51.28) --
	( 88.20, 47.13) --
	(102.16, 45.36) --
	(116.11, 45.01) --
	(130.07, 44.66);
\definecolor{drawColor}{RGB}{199,16,0}

\path[draw=drawColor,line width= 0.6pt,line join=round] ( 46.53,120.78) --
	( 60.49,111.55) --
	( 74.44,106.39) --
	( 88.40, 90.38) --
	(102.36, 74.72) --
	(116.31, 69.02) --
	(130.27, 61.26);
\definecolor{drawColor}{RGB}{0,142,160}

\path[draw=drawColor,line width= 0.6pt,line join=round] ( 46.73,109.74) --
	( 60.69,105.95) --
	( 74.64,102.59) --
	( 88.60, 92.51) --
	(102.55, 78.38) --
	(116.51, 71.77) --
	(130.47, 63.68);
\definecolor{drawColor}{RGB}{138,65,152}

\path[draw=drawColor,line width= 0.6pt,line join=round] ( 60.89,108.74) --
	(116.71, 78.59);
\definecolor{drawColor}{RGB}{90,149,153}

\path[draw=drawColor,line width= 0.6pt,line join=round] ( 46.93,110.06) --
	( 61.08,111.01) --
	( 74.84,110.62) --
	( 88.80,103.64) --
	(102.75,101.43) --
	(116.91, 97.61) --
	(130.67, 93.81);
\definecolor{drawColor}{RGB}{255,99,72}

\path[draw=drawColor,line width= 0.6pt,line join=round] ( 47.13,114.99) --
	( 75.04,106.74) --
	( 89.00, 96.22) --
	(102.95, 86.91) --
	(130.87, 78.14);
\definecolor{drawColor}{RGB}{132,215,225}

\path[draw=drawColor,line width= 0.6pt,line join=round] ( 47.33, 75.19) --
	( 61.28, 72.24) --
	( 75.24, 70.79) --
	( 89.20, 67.93) --
	(103.15, 64.87) --
	(117.11, 63.75) --
	(131.07, 61.43);
\definecolor{drawColor}{RGB}{255,149,168}

\path[draw=drawColor,line width= 0.6pt,line join=round] ( 47.53, 71.60) --
	( 61.48, 65.64) --
	( 75.44,115.07) --
	( 89.40, 64.55) --
	(103.35, 54.50) --
	(117.31, 57.71) --
	(131.26, 53.19);
\definecolor{drawColor}{RGB}{255,111,0}

\path[draw=drawColor,line width= 0.4pt,line join=round,line cap=round,fill=fillColor] ( 46.33, 57.04) circle (  1.43);

\path[draw=drawColor,line width= 0.4pt,line join=round,line cap=round,fill=fillColor] ( 74.24, 51.28) circle (  1.43);

\path[draw=drawColor,line width= 0.4pt,line join=round,line cap=round,fill=fillColor] ( 88.20, 47.13) circle (  1.43);

\path[draw=drawColor,line width= 0.4pt,line join=round,line cap=round,fill=fillColor] (102.16, 45.36) circle (  1.43);

\path[draw=drawColor,line width= 0.4pt,line join=round,line cap=round,fill=fillColor] (130.07, 44.66) circle (  1.43);

\path[draw=drawColor,line width= 0.4pt,line join=round,line cap=round,fill=fillColor] ( 60.29, 53.50) circle (  1.43);

\path[draw=drawColor,line width= 0.4pt,line join=round,line cap=round,fill=fillColor] (116.11, 45.01) circle (  1.43);
\definecolor{drawColor}{RGB}{199,16,0}

\path[draw=drawColor,line width= 0.4pt,line join=round,line cap=round,fill=fillColor] ( 46.53,120.78) circle (  1.43);

\path[draw=drawColor,line width= 0.4pt,line join=round,line cap=round,fill=fillColor] ( 74.44,106.39) circle (  1.43);

\path[draw=drawColor,line width= 0.4pt,line join=round,line cap=round,fill=fillColor] ( 88.40, 90.38) circle (  1.43);

\path[draw=drawColor,line width= 0.4pt,line join=round,line cap=round,fill=fillColor] (102.36, 74.72) circle (  1.43);

\path[draw=drawColor,line width= 0.4pt,line join=round,line cap=round,fill=fillColor] (130.27, 61.26) circle (  1.43);

\path[draw=drawColor,line width= 0.4pt,line join=round,line cap=round,fill=fillColor] ( 60.49,111.55) circle (  1.43);

\path[draw=drawColor,line width= 0.4pt,line join=round,line cap=round,fill=fillColor] (116.31, 69.02) circle (  1.43);
\definecolor{drawColor}{RGB}{0,142,160}

\path[draw=drawColor,line width= 0.4pt,line join=round,line cap=round,fill=fillColor] ( 46.73,109.74) circle (  1.43);

\path[draw=drawColor,line width= 0.4pt,line join=round,line cap=round,fill=fillColor] ( 74.64,102.59) circle (  1.43);

\path[draw=drawColor,line width= 0.4pt,line join=round,line cap=round,fill=fillColor] ( 88.60, 92.51) circle (  1.43);

\path[draw=drawColor,line width= 0.4pt,line join=round,line cap=round,fill=fillColor] (102.55, 78.38) circle (  1.43);

\path[draw=drawColor,line width= 0.4pt,line join=round,line cap=round,fill=fillColor] (130.47, 63.68) circle (  1.43);

\path[draw=drawColor,line width= 0.4pt,line join=round,line cap=round,fill=fillColor] ( 60.69,105.95) circle (  1.43);

\path[draw=drawColor,line width= 0.4pt,line join=round,line cap=round,fill=fillColor] (116.51, 71.77) circle (  1.43);
\definecolor{drawColor}{RGB}{138,65,152}

\path[draw=drawColor,line width= 0.4pt,line join=round,line cap=round,fill=fillColor] ( 60.89,108.74) circle (  1.43);

\path[draw=drawColor,line width= 0.4pt,line join=round,line cap=round,fill=fillColor] (116.71, 78.59) circle (  1.43);
\definecolor{drawColor}{RGB}{90,149,153}

\path[draw=drawColor,line width= 0.4pt,line join=round,line cap=round,fill=fillColor] ( 46.93,110.06) circle (  1.43);

\path[draw=drawColor,line width= 0.4pt,line join=round,line cap=round,fill=fillColor] ( 74.84,110.62) circle (  1.43);

\path[draw=drawColor,line width= 0.4pt,line join=round,line cap=round,fill=fillColor] ( 88.80,103.64) circle (  1.43);

\path[draw=drawColor,line width= 0.4pt,line join=round,line cap=round,fill=fillColor] (102.75,101.43) circle (  1.43);

\path[draw=drawColor,line width= 0.4pt,line join=round,line cap=round,fill=fillColor] (130.67, 93.81) circle (  1.43);

\path[draw=drawColor,line width= 0.4pt,line join=round,line cap=round,fill=fillColor] ( 61.08,111.01) circle (  1.43);

\path[draw=drawColor,line width= 0.4pt,line join=round,line cap=round,fill=fillColor] (116.91, 97.61) circle (  1.43);
\definecolor{drawColor}{RGB}{255,99,72}

\path[draw=drawColor,line width= 0.4pt,line join=round,line cap=round,fill=fillColor] ( 47.13,114.99) circle (  1.43);

\path[draw=drawColor,line width= 0.4pt,line join=round,line cap=round,fill=fillColor] ( 75.04,106.74) circle (  1.43);

\path[draw=drawColor,line width= 0.4pt,line join=round,line cap=round,fill=fillColor] ( 89.00, 96.22) circle (  1.43);

\path[draw=drawColor,line width= 0.4pt,line join=round,line cap=round,fill=fillColor] (102.95, 86.91) circle (  1.43);

\path[draw=drawColor,line width= 0.4pt,line join=round,line cap=round,fill=fillColor] (130.87, 78.14) circle (  1.43);
\definecolor{drawColor}{RGB}{132,215,225}

\path[draw=drawColor,line width= 0.4pt,line join=round,line cap=round,fill=fillColor] ( 47.33, 75.19) circle (  1.43);

\path[draw=drawColor,line width= 0.4pt,line join=round,line cap=round,fill=fillColor] ( 75.24, 70.79) circle (  1.43);

\path[draw=drawColor,line width= 0.4pt,line join=round,line cap=round,fill=fillColor] ( 89.20, 67.93) circle (  1.43);

\path[draw=drawColor,line width= 0.4pt,line join=round,line cap=round,fill=fillColor] (103.15, 64.87) circle (  1.43);

\path[draw=drawColor,line width= 0.4pt,line join=round,line cap=round,fill=fillColor] (131.07, 61.43) circle (  1.43);

\path[draw=drawColor,line width= 0.4pt,line join=round,line cap=round,fill=fillColor] ( 61.28, 72.24) circle (  1.43);

\path[draw=drawColor,line width= 0.4pt,line join=round,line cap=round,fill=fillColor] (117.11, 63.75) circle (  1.43);
\definecolor{drawColor}{RGB}{255,149,168}

\path[draw=drawColor,line width= 0.4pt,line join=round,line cap=round,fill=fillColor] ( 47.53, 71.60) circle (  1.43);

\path[draw=drawColor,line width= 0.4pt,line join=round,line cap=round,fill=fillColor] ( 75.44,115.07) circle (  1.43);

\path[draw=drawColor,line width= 0.4pt,line join=round,line cap=round,fill=fillColor] ( 89.40, 64.55) circle (  1.43);

\path[draw=drawColor,line width= 0.4pt,line join=round,line cap=round,fill=fillColor] (103.35, 54.50) circle (  1.43);

\path[draw=drawColor,line width= 0.4pt,line join=round,line cap=round,fill=fillColor] (131.26, 53.19) circle (  1.43);

\path[draw=drawColor,line width= 0.4pt,line join=round,line cap=round,fill=fillColor] ( 61.48, 65.64) circle (  1.43);

\path[draw=drawColor,line width= 0.4pt,line join=round,line cap=round,fill=fillColor] (117.31, 57.71) circle (  1.43);
\definecolor{drawColor}{gray}{0.20}

\path[draw=drawColor,line width= 0.6pt,line join=round,line cap=round] ( 38.56, 40.85) rectangle (139.04,124.59);
\end{scope}
\begin{scope}
\path[clip] (  0.00,  0.00) rectangle (144.54,130.09);
\definecolor{drawColor}{gray}{0.30}

\node[text=drawColor,anchor=base east,inner sep=0pt, outer sep=0pt, scale=  0.88] at ( 33.61, 40.55) {0.00};

\node[text=drawColor,anchor=base east,inner sep=0pt, outer sep=0pt, scale=  0.88] at ( 33.61, 67.43) {0.02};

\node[text=drawColor,anchor=base east,inner sep=0pt, outer sep=0pt, scale=  0.88] at ( 33.61, 94.31) {0.04};

\node[text=drawColor,anchor=base east,inner sep=0pt, outer sep=0pt, scale=  0.88] at ( 33.61,121.19) {0.06};
\end{scope}
\begin{scope}
\path[clip] (  0.00,  0.00) rectangle (144.54,130.09);
\definecolor{drawColor}{gray}{0.20}

\path[draw=drawColor,line width= 0.6pt,line join=round] ( 35.81, 43.58) --
	( 38.56, 43.58);

\path[draw=drawColor,line width= 0.6pt,line join=round] ( 35.81, 70.46) --
	( 38.56, 70.46);

\path[draw=drawColor,line width= 0.6pt,line join=round] ( 35.81, 97.34) --
	( 38.56, 97.34);

\path[draw=drawColor,line width= 0.6pt,line join=round] ( 35.81,124.22) --
	( 38.56,124.22);
\end{scope}
\begin{scope}
\path[clip] (  0.00,  0.00) rectangle (144.54,130.09);
\definecolor{drawColor}{gray}{0.20}

\path[draw=drawColor,line width= 0.6pt,line join=round] ( 46.93, 38.10) --
	( 46.93, 40.85);

\path[draw=drawColor,line width= 0.6pt,line join=round] ( 60.89, 38.10) --
	( 60.89, 40.85);

\path[draw=drawColor,line width= 0.6pt,line join=round] ( 74.84, 38.10) --
	( 74.84, 40.85);

\path[draw=drawColor,line width= 0.6pt,line join=round] ( 88.80, 38.10) --
	( 88.80, 40.85);

\path[draw=drawColor,line width= 0.6pt,line join=round] (102.75, 38.10) --
	(102.75, 40.85);

\path[draw=drawColor,line width= 0.6pt,line join=round] (116.71, 38.10) --
	(116.71, 40.85);

\path[draw=drawColor,line width= 0.6pt,line join=round] (130.67, 38.10) --
	(130.67, 40.85);
\end{scope}
\begin{scope}
\path[clip] (  0.00,  0.00) rectangle (144.54,130.09);
\definecolor{drawColor}{gray}{0.30}

\node[text=drawColor,rotate= 45.00,anchor=base east,inner sep=0pt, outer sep=0pt, scale=  0.88] at ( 51.21, 31.62) {512};

\node[text=drawColor,rotate= 45.00,anchor=base east,inner sep=0pt, outer sep=0pt, scale=  0.88] at ( 65.17, 31.62) {770};

\node[text=drawColor,rotate= 45.00,anchor=base east,inner sep=0pt, outer sep=0pt, scale=  0.88] at ( 79.13, 31.62) {1024};

\node[text=drawColor,rotate= 45.00,anchor=base east,inner sep=0pt, outer sep=0pt, scale=  0.88] at ( 93.08, 31.62) {2048};

\node[text=drawColor,rotate= 45.00,anchor=base east,inner sep=0pt, outer sep=0pt, scale=  0.88] at (107.04, 31.62) {4096};

\node[text=drawColor,rotate= 45.00,anchor=base east,inner sep=0pt, outer sep=0pt, scale=  0.88] at (121.00, 31.62) {5482};

\node[text=drawColor,rotate= 45.00,anchor=base east,inner sep=0pt, outer sep=0pt, scale=  0.88] at (134.95, 31.62) {8192};
\end{scope}
\begin{scope}
\path[clip] (  0.00,  0.00) rectangle (144.54,130.09);
\definecolor{drawColor}{RGB}{0,0,0}

\node[text=drawColor,anchor=base,inner sep=0pt, outer sep=0pt, scale=  1.10] at ( 88.80,  7.64) {Target Chunk Size};
\end{scope}
\begin{scope}
\path[clip] (  0.00,  0.00) rectangle (144.54,130.09);
\definecolor{drawColor}{RGB}{0,0,0}

\node[text=drawColor,rotate= 90.00,anchor=base,inner sep=0pt, outer sep=0pt, scale=  1.10] at ( 13.08, 82.72) {Dedup. Ratio};
\end{scope}
\end{tikzpicture}

%% file: fig/dedup_overview_web.tex
% Created by tikzDevice version 0.12.6 on 2024-07-13 14:07:31
% !TEX encoding = UTF-8 Unicode
\begin{tikzpicture}[x=1pt,y=1pt]
\definecolor{fillColor}{RGB}{255,255,255}
\path[use as bounding box,fill=fillColor,fill opacity=0.00] (0,0) rectangle (144.54,130.09);
\begin{scope}
\path[clip] (  0.00,  0.00) rectangle (144.54,130.09);
\definecolor{drawColor}{RGB}{255,255,255}
\definecolor{fillColor}{RGB}{255,255,255}

\path[draw=drawColor,line width= 0.6pt,line join=round,line cap=round,fill=fillColor] (  0.00,  0.00) rectangle (144.54,130.09);
\end{scope}
\begin{scope}
\path[clip] ( 34.16, 40.85) rectangle (139.04,124.59);
\definecolor{fillColor}{RGB}{255,255,255}

\path[fill=fillColor] ( 34.16, 40.85) rectangle (139.04,124.59);
\definecolor{drawColor}{gray}{0.92}

\path[draw=drawColor,line width= 0.3pt,line join=round] ( 34.16, 54.02) --
	(139.04, 54.02);

\path[draw=drawColor,line width= 0.3pt,line join=round] ( 34.16, 75.78) --
	(139.04, 75.78);

\path[draw=drawColor,line width= 0.3pt,line join=round] ( 34.16, 97.54) --
	(139.04, 97.54);

\path[draw=drawColor,line width= 0.3pt,line join=round] ( 34.16,119.30) --
	(139.04,119.30);

\path[draw=drawColor,line width= 0.6pt,line join=round] ( 34.16, 43.14) --
	(139.04, 43.14);

\path[draw=drawColor,line width= 0.6pt,line join=round] ( 34.16, 64.90) --
	(139.04, 64.90);

\path[draw=drawColor,line width= 0.6pt,line join=round] ( 34.16, 86.66) --
	(139.04, 86.66);

\path[draw=drawColor,line width= 0.6pt,line join=round] ( 34.16,108.42) --
	(139.04,108.42);

\path[draw=drawColor,line width= 0.6pt,line join=round] ( 42.90, 40.85) --
	( 42.90,124.59);

\path[draw=drawColor,line width= 0.6pt,line join=round] ( 57.46, 40.85) --
	( 57.46,124.59);

\path[draw=drawColor,line width= 0.6pt,line join=round] ( 72.03, 40.85) --
	( 72.03,124.59);

\path[draw=drawColor,line width= 0.6pt,line join=round] ( 86.60, 40.85) --
	( 86.60,124.59);

\path[draw=drawColor,line width= 0.6pt,line join=round] (101.17, 40.85) --
	(101.17,124.59);

\path[draw=drawColor,line width= 0.6pt,line join=round] (115.73, 40.85) --
	(115.73,124.59);

\path[draw=drawColor,line width= 0.6pt,line join=round] (130.30, 40.85) --
	(130.30,124.59);
\definecolor{drawColor}{RGB}{255,111,0}

\path[draw=drawColor,line width= 0.6pt,line join=round] ( 42.27, 67.37) --
	( 56.84, 60.42) --
	( 71.41, 57.01) --
	( 85.97, 49.32) --
	(100.54, 45.72) --
	(115.11, 45.20) --
	(129.68, 44.66);
\definecolor{drawColor}{RGB}{199,16,0}

\path[draw=drawColor,line width= 0.6pt,line join=round] ( 42.48,120.78) --
	( 57.05,120.30) --
	( 71.61,119.96) --
	( 86.18,119.05) --
	(100.75,118.06) --
	(115.32,117.71) --
	(129.88,117.16);
\definecolor{drawColor}{RGB}{0,142,160}

\path[draw=drawColor,line width= 0.6pt,line join=round] ( 42.69,112.90) --
	( 57.26,112.87) --
	( 71.82,112.84) --
	( 86.39,112.73) --
	(100.96,112.53) --
	(115.52,112.42) --
	(130.09,112.20);
\definecolor{drawColor}{RGB}{138,65,152}

\path[draw=drawColor,line width= 0.6pt,line join=round] ( 57.46,119.08) --
	(115.73,116.36);
\definecolor{drawColor}{RGB}{90,149,153}

\path[draw=drawColor,line width= 0.6pt,line join=round] ( 42.90,117.82) --
	( 57.67,117.33) --
	( 72.03,117.12) --
	( 86.60,117.01) --
	(101.17,116.21) --
	(115.94,116.27) --
	(130.30,116.16);
\definecolor{drawColor}{RGB}{255,99,72}

\path[draw=drawColor,line width= 0.6pt,line join=round] ( 43.10,120.35) --
	( 72.24,119.76) --
	( 86.81,119.07) --
	(101.37,118.42) --
	(130.51,117.75);
\definecolor{drawColor}{RGB}{132,215,225}

\path[draw=drawColor,line width= 0.6pt,line join=round] ( 43.31,112.80) --
	( 57.88,112.79) --
	( 72.45,112.76) --
	( 87.01,112.70) --
	(101.58,112.59) --
	(116.15,112.54) --
	(130.72,112.48);
\definecolor{drawColor}{RGB}{255,149,168}

\path[draw=drawColor,line width= 0.6pt,line join=round] ( 43.52,112.47) --
	( 58.09,112.30) --
	( 72.66,116.00) --
	( 87.22,119.15) --
	(101.79,115.49) --
	(116.36,115.89) --
	(130.92,111.88);
\definecolor{drawColor}{RGB}{255,111,0}

\path[draw=drawColor,line width= 0.4pt,line join=round,line cap=round,fill=fillColor] ( 42.27, 67.37) circle (  1.43);

\path[draw=drawColor,line width= 0.4pt,line join=round,line cap=round,fill=fillColor] ( 71.41, 57.01) circle (  1.43);

\path[draw=drawColor,line width= 0.4pt,line join=round,line cap=round,fill=fillColor] ( 85.97, 49.32) circle (  1.43);

\path[draw=drawColor,line width= 0.4pt,line join=round,line cap=round,fill=fillColor] (100.54, 45.72) circle (  1.43);

\path[draw=drawColor,line width= 0.4pt,line join=round,line cap=round,fill=fillColor] (129.68, 44.66) circle (  1.43);

\path[draw=drawColor,line width= 0.4pt,line join=round,line cap=round,fill=fillColor] ( 56.84, 60.42) circle (  1.43);

\path[draw=drawColor,line width= 0.4pt,line join=round,line cap=round,fill=fillColor] (115.11, 45.20) circle (  1.43);
\definecolor{drawColor}{RGB}{199,16,0}

\path[draw=drawColor,line width= 0.4pt,line join=round,line cap=round,fill=fillColor] ( 42.48,120.78) circle (  1.43);

\path[draw=drawColor,line width= 0.4pt,line join=round,line cap=round,fill=fillColor] ( 71.61,119.96) circle (  1.43);

\path[draw=drawColor,line width= 0.4pt,line join=round,line cap=round,fill=fillColor] ( 86.18,119.05) circle (  1.43);

\path[draw=drawColor,line width= 0.4pt,line join=round,line cap=round,fill=fillColor] (100.75,118.06) circle (  1.43);

\path[draw=drawColor,line width= 0.4pt,line join=round,line cap=round,fill=fillColor] (129.88,117.16) circle (  1.43);

\path[draw=drawColor,line width= 0.4pt,line join=round,line cap=round,fill=fillColor] ( 57.05,120.30) circle (  1.43);

\path[draw=drawColor,line width= 0.4pt,line join=round,line cap=round,fill=fillColor] (115.32,117.71) circle (  1.43);
\definecolor{drawColor}{RGB}{0,142,160}

\path[draw=drawColor,line width= 0.4pt,line join=round,line cap=round,fill=fillColor] ( 42.69,112.90) circle (  1.43);

\path[draw=drawColor,line width= 0.4pt,line join=round,line cap=round,fill=fillColor] ( 71.82,112.84) circle (  1.43);

\path[draw=drawColor,line width= 0.4pt,line join=round,line cap=round,fill=fillColor] ( 86.39,112.73) circle (  1.43);

\path[draw=drawColor,line width= 0.4pt,line join=round,line cap=round,fill=fillColor] (100.96,112.53) circle (  1.43);

\path[draw=drawColor,line width= 0.4pt,line join=round,line cap=round,fill=fillColor] (130.09,112.20) circle (  1.43);

\path[draw=drawColor,line width= 0.4pt,line join=round,line cap=round,fill=fillColor] ( 57.26,112.87) circle (  1.43);

\path[draw=drawColor,line width= 0.4pt,line join=round,line cap=round,fill=fillColor] (115.52,112.42) circle (  1.43);
\definecolor{drawColor}{RGB}{138,65,152}

\path[draw=drawColor,line width= 0.4pt,line join=round,line cap=round,fill=fillColor] ( 57.46,119.08) circle (  1.43);

\path[draw=drawColor,line width= 0.4pt,line join=round,line cap=round,fill=fillColor] (115.73,116.36) circle (  1.43);
\definecolor{drawColor}{RGB}{90,149,153}

\path[draw=drawColor,line width= 0.4pt,line join=round,line cap=round,fill=fillColor] ( 42.90,117.82) circle (  1.43);

\path[draw=drawColor,line width= 0.4pt,line join=round,line cap=round,fill=fillColor] ( 72.03,117.12) circle (  1.43);

\path[draw=drawColor,line width= 0.4pt,line join=round,line cap=round,fill=fillColor] ( 86.60,117.01) circle (  1.43);

\path[draw=drawColor,line width= 0.4pt,line join=round,line cap=round,fill=fillColor] (101.17,116.21) circle (  1.43);

\path[draw=drawColor,line width= 0.4pt,line join=round,line cap=round,fill=fillColor] (130.30,116.16) circle (  1.43);

\path[draw=drawColor,line width= 0.4pt,line join=round,line cap=round,fill=fillColor] ( 57.67,117.33) circle (  1.43);

\path[draw=drawColor,line width= 0.4pt,line join=round,line cap=round,fill=fillColor] (115.94,116.27) circle (  1.43);
\definecolor{drawColor}{RGB}{255,99,72}

\path[draw=drawColor,line width= 0.4pt,line join=round,line cap=round,fill=fillColor] ( 43.10,120.35) circle (  1.43);

\path[draw=drawColor,line width= 0.4pt,line join=round,line cap=round,fill=fillColor] ( 72.24,119.76) circle (  1.43);

\path[draw=drawColor,line width= 0.4pt,line join=round,line cap=round,fill=fillColor] ( 86.81,119.07) circle (  1.43);

\path[draw=drawColor,line width= 0.4pt,line join=round,line cap=round,fill=fillColor] (101.37,118.42) circle (  1.43);

\path[draw=drawColor,line width= 0.4pt,line join=round,line cap=round,fill=fillColor] (130.51,117.75) circle (  1.43);
\definecolor{drawColor}{RGB}{132,215,225}

\path[draw=drawColor,line width= 0.4pt,line join=round,line cap=round,fill=fillColor] ( 43.31,112.80) circle (  1.43);

\path[draw=drawColor,line width= 0.4pt,line join=round,line cap=round,fill=fillColor] ( 72.45,112.76) circle (  1.43);

\path[draw=drawColor,line width= 0.4pt,line join=round,line cap=round,fill=fillColor] ( 87.01,112.70) circle (  1.43);

\path[draw=drawColor,line width= 0.4pt,line join=round,line cap=round,fill=fillColor] (101.58,112.59) circle (  1.43);

\path[draw=drawColor,line width= 0.4pt,line join=round,line cap=round,fill=fillColor] (130.72,112.48) circle (  1.43);

\path[draw=drawColor,line width= 0.4pt,line join=round,line cap=round,fill=fillColor] ( 57.88,112.79) circle (  1.43);

\path[draw=drawColor,line width= 0.4pt,line join=round,line cap=round,fill=fillColor] (116.15,112.54) circle (  1.43);
\definecolor{drawColor}{RGB}{255,149,168}

\path[draw=drawColor,line width= 0.4pt,line join=round,line cap=round,fill=fillColor] ( 43.52,112.47) circle (  1.43);

\path[draw=drawColor,line width= 0.4pt,line join=round,line cap=round,fill=fillColor] ( 72.66,116.00) circle (  1.43);

\path[draw=drawColor,line width= 0.4pt,line join=round,line cap=round,fill=fillColor] ( 87.22,119.15) circle (  1.43);

\path[draw=drawColor,line width= 0.4pt,line join=round,line cap=round,fill=fillColor] (101.79,115.49) circle (  1.43);

\path[draw=drawColor,line width= 0.4pt,line join=round,line cap=round,fill=fillColor] (130.92,111.88) circle (  1.43);

\path[draw=drawColor,line width= 0.4pt,line join=round,line cap=round,fill=fillColor] ( 58.09,112.30) circle (  1.43);

\path[draw=drawColor,line width= 0.4pt,line join=round,line cap=round,fill=fillColor] (116.36,115.89) circle (  1.43);
\definecolor{drawColor}{gray}{0.20}

\path[draw=drawColor,line width= 0.6pt,line join=round,line cap=round] ( 34.16, 40.85) rectangle (139.04,124.59);
\end{scope}
\begin{scope}
\path[clip] (  0.00,  0.00) rectangle (144.54,130.09);
\definecolor{drawColor}{gray}{0.30}

\node[text=drawColor,anchor=base east,inner sep=0pt, outer sep=0pt, scale=  0.88] at ( 29.21, 40.11) {0.0};

\node[text=drawColor,anchor=base east,inner sep=0pt, outer sep=0pt, scale=  0.88] at ( 29.21, 61.87) {0.2};

\node[text=drawColor,anchor=base east,inner sep=0pt, outer sep=0pt, scale=  0.88] at ( 29.21, 83.63) {0.4};

\node[text=drawColor,anchor=base east,inner sep=0pt, outer sep=0pt, scale=  0.88] at ( 29.21,105.39) {0.6};
\end{scope}
\begin{scope}
\path[clip] (  0.00,  0.00) rectangle (144.54,130.09);
\definecolor{drawColor}{gray}{0.20}

\path[draw=drawColor,line width= 0.6pt,line join=round] ( 31.41, 43.14) --
	( 34.16, 43.14);

\path[draw=drawColor,line width= 0.6pt,line join=round] ( 31.41, 64.90) --
	( 34.16, 64.90);

\path[draw=drawColor,line width= 0.6pt,line join=round] ( 31.41, 86.66) --
	( 34.16, 86.66);

\path[draw=drawColor,line width= 0.6pt,line join=round] ( 31.41,108.42) --
	( 34.16,108.42);
\end{scope}
\begin{scope}
\path[clip] (  0.00,  0.00) rectangle (144.54,130.09);
\definecolor{drawColor}{gray}{0.20}

\path[draw=drawColor,line width= 0.6pt,line join=round] ( 42.90, 38.10) --
	( 42.90, 40.85);

\path[draw=drawColor,line width= 0.6pt,line join=round] ( 57.46, 38.10) --
	( 57.46, 40.85);

\path[draw=drawColor,line width= 0.6pt,line join=round] ( 72.03, 38.10) --
	( 72.03, 40.85);

\path[draw=drawColor,line width= 0.6pt,line join=round] ( 86.60, 38.10) --
	( 86.60, 40.85);

\path[draw=drawColor,line width= 0.6pt,line join=round] (101.17, 38.10) --
	(101.17, 40.85);

\path[draw=drawColor,line width= 0.6pt,line join=round] (115.73, 38.10) --
	(115.73, 40.85);

\path[draw=drawColor,line width= 0.6pt,line join=round] (130.30, 38.10) --
	(130.30, 40.85);
\end{scope}
\begin{scope}
\path[clip] (  0.00,  0.00) rectangle (144.54,130.09);
\definecolor{drawColor}{gray}{0.30}

\node[text=drawColor,rotate= 45.00,anchor=base east,inner sep=0pt, outer sep=0pt, scale=  0.88] at ( 47.18, 31.62) {512};

\node[text=drawColor,rotate= 45.00,anchor=base east,inner sep=0pt, outer sep=0pt, scale=  0.88] at ( 61.75, 31.62) {770};

\node[text=drawColor,rotate= 45.00,anchor=base east,inner sep=0pt, outer sep=0pt, scale=  0.88] at ( 76.32, 31.62) {1024};

\node[text=drawColor,rotate= 45.00,anchor=base east,inner sep=0pt, outer sep=0pt, scale=  0.88] at ( 90.88, 31.62) {2048};

\node[text=drawColor,rotate= 45.00,anchor=base east,inner sep=0pt, outer sep=0pt, scale=  0.88] at (105.45, 31.62) {4096};

\node[text=drawColor,rotate= 45.00,anchor=base east,inner sep=0pt, outer sep=0pt, scale=  0.88] at (120.02, 31.62) {5482};

\node[text=drawColor,rotate= 45.00,anchor=base east,inner sep=0pt, outer sep=0pt, scale=  0.88] at (134.59, 31.62) {8192};
\end{scope}
\begin{scope}
\path[clip] (  0.00,  0.00) rectangle (144.54,130.09);
\definecolor{drawColor}{RGB}{0,0,0}

\node[text=drawColor,anchor=base,inner sep=0pt, outer sep=0pt, scale=  1.10] at ( 86.60,  7.64) {Target Chunk Size};
\end{scope}
\begin{scope}
\path[clip] (  0.00,  0.00) rectangle (144.54,130.09);
\definecolor{drawColor}{RGB}{0,0,0}

\node[text=drawColor,rotate= 90.00,anchor=base,inner sep=0pt, outer sep=0pt, scale=  1.10] at ( 13.08, 82.72) {Dedup. Ratio};
\end{scope}
\end{tikzpicture}

%% file: fig/dedup_overview_code.tex
% Created by tikzDevice version 0.12.6 on 2024-07-13 14:07:25
% !TEX encoding = UTF-8 Unicode
\begin{tikzpicture}[x=1pt,y=1pt]
\definecolor{fillColor}{RGB}{255,255,255}
\path[use as bounding box,fill=fillColor,fill opacity=0.00] (0,0) rectangle (144.54,130.09);
\begin{scope}
\path[clip] (  0.00,  0.00) rectangle (144.54,130.09);
\definecolor{drawColor}{RGB}{255,255,255}
\definecolor{fillColor}{RGB}{255,255,255}

\path[draw=drawColor,line width= 0.6pt,line join=round,line cap=round,fill=fillColor] (  0.00,  0.00) rectangle (144.54,130.09);
\end{scope}
\begin{scope}
\path[clip] ( 34.16, 40.85) rectangle (139.04,124.59);
\definecolor{fillColor}{RGB}{255,255,255}

\path[fill=fillColor] ( 34.16, 40.85) rectangle (139.04,124.59);
\definecolor{drawColor}{gray}{0.92}

\path[draw=drawColor,line width= 0.3pt,line join=round] ( 34.16, 55.13) --
	(139.04, 55.13);

\path[draw=drawColor,line width= 0.3pt,line join=round] ( 34.16, 77.12) --
	(139.04, 77.12);

\path[draw=drawColor,line width= 0.3pt,line join=round] ( 34.16, 99.11) --
	(139.04, 99.11);

\path[draw=drawColor,line width= 0.3pt,line join=round] ( 34.16,121.10) --
	(139.04,121.10);

\path[draw=drawColor,line width= 0.6pt,line join=round] ( 34.16, 44.13) --
	(139.04, 44.13);

\path[draw=drawColor,line width= 0.6pt,line join=round] ( 34.16, 66.12) --
	(139.04, 66.12);

\path[draw=drawColor,line width= 0.6pt,line join=round] ( 34.16, 88.11) --
	(139.04, 88.11);

\path[draw=drawColor,line width= 0.6pt,line join=round] ( 34.16,110.10) --
	(139.04,110.10);

\path[draw=drawColor,line width= 0.6pt,line join=round] ( 42.90, 40.85) --
	( 42.90,124.59);

\path[draw=drawColor,line width= 0.6pt,line join=round] ( 57.46, 40.85) --
	( 57.46,124.59);

\path[draw=drawColor,line width= 0.6pt,line join=round] ( 72.03, 40.85) --
	( 72.03,124.59);

\path[draw=drawColor,line width= 0.6pt,line join=round] ( 86.60, 40.85) --
	( 86.60,124.59);

\path[draw=drawColor,line width= 0.6pt,line join=round] (101.17, 40.85) --
	(101.17,124.59);

\path[draw=drawColor,line width= 0.6pt,line join=round] (115.73, 40.85) --
	(115.73,124.59);

\path[draw=drawColor,line width= 0.6pt,line join=round] (130.30, 40.85) --
	(130.30,124.59);
\definecolor{drawColor}{RGB}{255,111,0}

\path[draw=drawColor,line width= 0.6pt,line join=round] ( 42.27, 48.45) --
	( 56.84, 46.90) --
	( 71.41, 46.04) --
	( 85.97, 45.05) --
	(100.54, 44.78) --
	(115.11, 44.75) --
	(129.68, 44.66);
\definecolor{drawColor}{RGB}{199,16,0}

\path[draw=drawColor,line width= 0.6pt,line join=round] ( 42.48,120.78) --
	( 57.05,119.33) --
	( 71.61,118.13) --
	( 86.18,114.72) --
	(100.75,110.85) --
	(115.32,109.21) --
	(129.88,106.82);
\definecolor{drawColor}{RGB}{0,142,160}

\path[draw=drawColor,line width= 0.6pt,line join=round] ( 42.69, 54.31) --
	( 57.26, 53.93) --
	( 71.82, 54.21) --
	( 86.39, 52.47) --
	(100.96, 52.42) --
	(115.52, 52.50) --
	(130.09, 52.38);
\definecolor{drawColor}{RGB}{138,65,152}

\path[draw=drawColor,line width= 0.6pt,line join=round] ( 57.46,117.25) --
	(115.73,107.40);
\definecolor{drawColor}{RGB}{90,149,153}

\path[draw=drawColor,line width= 0.6pt,line join=round] ( 42.90, 96.38) --
	( 57.67, 95.22) --
	( 72.03, 93.98) --
	( 86.60, 83.82) --
	(101.17, 77.36) --
	(115.94, 76.28) --
	(130.30, 71.21);
\definecolor{drawColor}{RGB}{255,99,72}

\path[draw=drawColor,line width= 0.6pt,line join=round] ( 43.10,120.72) --
	( 72.24,118.62) --
	( 86.81,116.00) --
	(101.37,112.75) --
	(130.51,108.96);
\definecolor{drawColor}{RGB}{132,215,225}

\path[draw=drawColor,line width= 0.6pt,line join=round] ( 43.31,118.87) --
	( 57.88,115.41) --
	( 72.45,112.54) --
	( 87.01,104.08) --
	(101.58, 93.81) --
	(116.15, 89.76) --
	(130.72, 83.24);
\definecolor{drawColor}{RGB}{255,149,168}

\path[draw=drawColor,line width= 0.6pt,line join=round] ( 43.52,117.85) --
	( 58.09,111.25) --
	( 72.66, 95.91) --
	( 87.22, 96.48) --
	(101.79, 95.87) --
	(116.36, 98.34) --
	(130.92, 81.37);
\definecolor{drawColor}{RGB}{255,111,0}

\path[draw=drawColor,line width= 0.4pt,line join=round,line cap=round,fill=fillColor] ( 42.27, 48.45) circle (  1.43);

\path[draw=drawColor,line width= 0.4pt,line join=round,line cap=round,fill=fillColor] ( 71.41, 46.04) circle (  1.43);

\path[draw=drawColor,line width= 0.4pt,line join=round,line cap=round,fill=fillColor] ( 85.97, 45.05) circle (  1.43);

\path[draw=drawColor,line width= 0.4pt,line join=round,line cap=round,fill=fillColor] (100.54, 44.78) circle (  1.43);

\path[draw=drawColor,line width= 0.4pt,line join=round,line cap=round,fill=fillColor] (129.68, 44.66) circle (  1.43);

\path[draw=drawColor,line width= 0.4pt,line join=round,line cap=round,fill=fillColor] ( 56.84, 46.90) circle (  1.43);

\path[draw=drawColor,line width= 0.4pt,line join=round,line cap=round,fill=fillColor] (115.11, 44.75) circle (  1.43);
\definecolor{drawColor}{RGB}{199,16,0}

\path[draw=drawColor,line width= 0.4pt,line join=round,line cap=round,fill=fillColor] ( 42.48,120.78) circle (  1.43);

\path[draw=drawColor,line width= 0.4pt,line join=round,line cap=round,fill=fillColor] ( 71.61,118.13) circle (  1.43);

\path[draw=drawColor,line width= 0.4pt,line join=round,line cap=round,fill=fillColor] ( 86.18,114.72) circle (  1.43);

\path[draw=drawColor,line width= 0.4pt,line join=round,line cap=round,fill=fillColor] (100.75,110.85) circle (  1.43);

\path[draw=drawColor,line width= 0.4pt,line join=round,line cap=round,fill=fillColor] (129.88,106.82) circle (  1.43);

\path[draw=drawColor,line width= 0.4pt,line join=round,line cap=round,fill=fillColor] ( 57.05,119.33) circle (  1.43);

\path[draw=drawColor,line width= 0.4pt,line join=round,line cap=round,fill=fillColor] (115.32,109.21) circle (  1.43);
\definecolor{drawColor}{RGB}{0,142,160}

\path[draw=drawColor,line width= 0.4pt,line join=round,line cap=round,fill=fillColor] ( 42.69, 54.31) circle (  1.43);

\path[draw=drawColor,line width= 0.4pt,line join=round,line cap=round,fill=fillColor] ( 71.82, 54.21) circle (  1.43);

\path[draw=drawColor,line width= 0.4pt,line join=round,line cap=round,fill=fillColor] ( 86.39, 52.47) circle (  1.43);

\path[draw=drawColor,line width= 0.4pt,line join=round,line cap=round,fill=fillColor] (100.96, 52.42) circle (  1.43);

\path[draw=drawColor,line width= 0.4pt,line join=round,line cap=round,fill=fillColor] (130.09, 52.38) circle (  1.43);

\path[draw=drawColor,line width= 0.4pt,line join=round,line cap=round,fill=fillColor] ( 57.26, 53.93) circle (  1.43);

\path[draw=drawColor,line width= 0.4pt,line join=round,line cap=round,fill=fillColor] (115.52, 52.50) circle (  1.43);
\definecolor{drawColor}{RGB}{138,65,152}

\path[draw=drawColor,line width= 0.4pt,line join=round,line cap=round,fill=fillColor] ( 57.46,117.25) circle (  1.43);

\path[draw=drawColor,line width= 0.4pt,line join=round,line cap=round,fill=fillColor] (115.73,107.40) circle (  1.43);
\definecolor{drawColor}{RGB}{90,149,153}

\path[draw=drawColor,line width= 0.4pt,line join=round,line cap=round,fill=fillColor] ( 42.90, 96.38) circle (  1.43);

\path[draw=drawColor,line width= 0.4pt,line join=round,line cap=round,fill=fillColor] ( 72.03, 93.98) circle (  1.43);

\path[draw=drawColor,line width= 0.4pt,line join=round,line cap=round,fill=fillColor] ( 86.60, 83.82) circle (  1.43);

\path[draw=drawColor,line width= 0.4pt,line join=round,line cap=round,fill=fillColor] (101.17, 77.36) circle (  1.43);

\path[draw=drawColor,line width= 0.4pt,line join=round,line cap=round,fill=fillColor] (130.30, 71.21) circle (  1.43);

\path[draw=drawColor,line width= 0.4pt,line join=round,line cap=round,fill=fillColor] ( 57.67, 95.22) circle (  1.43);

\path[draw=drawColor,line width= 0.4pt,line join=round,line cap=round,fill=fillColor] (115.94, 76.28) circle (  1.43);
\definecolor{drawColor}{RGB}{255,99,72}

\path[draw=drawColor,line width= 0.4pt,line join=round,line cap=round,fill=fillColor] ( 43.10,120.72) circle (  1.43);

\path[draw=drawColor,line width= 0.4pt,line join=round,line cap=round,fill=fillColor] ( 72.24,118.62) circle (  1.43);

\path[draw=drawColor,line width= 0.4pt,line join=round,line cap=round,fill=fillColor] ( 86.81,116.00) circle (  1.43);

\path[draw=drawColor,line width= 0.4pt,line join=round,line cap=round,fill=fillColor] (101.37,112.75) circle (  1.43);

\path[draw=drawColor,line width= 0.4pt,line join=round,line cap=round,fill=fillColor] (130.51,108.96) circle (  1.43);
\definecolor{drawColor}{RGB}{132,215,225}

\path[draw=drawColor,line width= 0.4pt,line join=round,line cap=round,fill=fillColor] ( 43.31,118.87) circle (  1.43);

\path[draw=drawColor,line width= 0.4pt,line join=round,line cap=round,fill=fillColor] ( 72.45,112.54) circle (  1.43);

\path[draw=drawColor,line width= 0.4pt,line join=round,line cap=round,fill=fillColor] ( 87.01,104.08) circle (  1.43);

\path[draw=drawColor,line width= 0.4pt,line join=round,line cap=round,fill=fillColor] (101.58, 93.81) circle (  1.43);

\path[draw=drawColor,line width= 0.4pt,line join=round,line cap=round,fill=fillColor] (130.72, 83.24) circle (  1.43);

\path[draw=drawColor,line width= 0.4pt,line join=round,line cap=round,fill=fillColor] ( 57.88,115.41) circle (  1.43);

\path[draw=drawColor,line width= 0.4pt,line join=round,line cap=round,fill=fillColor] (116.15, 89.76) circle (  1.43);
\definecolor{drawColor}{RGB}{255,149,168}

\path[draw=drawColor,line width= 0.4pt,line join=round,line cap=round,fill=fillColor] ( 43.52,117.85) circle (  1.43);

\path[draw=drawColor,line width= 0.4pt,line join=round,line cap=round,fill=fillColor] ( 72.66, 95.91) circle (  1.43);

\path[draw=drawColor,line width= 0.4pt,line join=round,line cap=round,fill=fillColor] ( 87.22, 96.48) circle (  1.43);

\path[draw=drawColor,line width= 0.4pt,line join=round,line cap=round,fill=fillColor] (101.79, 95.87) circle (  1.43);

\path[draw=drawColor,line width= 0.4pt,line join=round,line cap=round,fill=fillColor] (130.92, 81.37) circle (  1.43);

\path[draw=drawColor,line width= 0.4pt,line join=round,line cap=round,fill=fillColor] ( 58.09,111.25) circle (  1.43);

\path[draw=drawColor,line width= 0.4pt,line join=round,line cap=round,fill=fillColor] (116.36, 98.34) circle (  1.43);
\definecolor{drawColor}{gray}{0.20}

\path[draw=drawColor,line width= 0.6pt,line join=round,line cap=round] ( 34.16, 40.85) rectangle (139.04,124.59);
\end{scope}
\begin{scope}
\path[clip] (  0.00,  0.00) rectangle (144.54,130.09);
\definecolor{drawColor}{gray}{0.30}

\node[text=drawColor,anchor=base east,inner sep=0pt, outer sep=0pt, scale=  0.88] at ( 29.21, 41.10) {0.0};

\node[text=drawColor,anchor=base east,inner sep=0pt, outer sep=0pt, scale=  0.88] at ( 29.21, 63.09) {0.2};

\node[text=drawColor,anchor=base east,inner sep=0pt, outer sep=0pt, scale=  0.88] at ( 29.21, 85.08) {0.5};

\node[text=drawColor,anchor=base east,inner sep=0pt, outer sep=0pt, scale=  0.88] at ( 29.21,107.07) {0.8};
\end{scope}
\begin{scope}
\path[clip] (  0.00,  0.00) rectangle (144.54,130.09);
\definecolor{drawColor}{gray}{0.20}

\path[draw=drawColor,line width= 0.6pt,line join=round] ( 31.41, 44.13) --
	( 34.16, 44.13);

\path[draw=drawColor,line width= 0.6pt,line join=round] ( 31.41, 66.12) --
	( 34.16, 66.12);

\path[draw=drawColor,line width= 0.6pt,line join=round] ( 31.41, 88.11) --
	( 34.16, 88.11);

\path[draw=drawColor,line width= 0.6pt,line join=round] ( 31.41,110.10) --
	( 34.16,110.10);
\end{scope}
\begin{scope}
\path[clip] (  0.00,  0.00) rectangle (144.54,130.09);
\definecolor{drawColor}{gray}{0.20}

\path[draw=drawColor,line width= 0.6pt,line join=round] ( 42.90, 38.10) --
	( 42.90, 40.85);

\path[draw=drawColor,line width= 0.6pt,line join=round] ( 57.46, 38.10) --
	( 57.46, 40.85);

\path[draw=drawColor,line width= 0.6pt,line join=round] ( 72.03, 38.10) --
	( 72.03, 40.85);

\path[draw=drawColor,line width= 0.6pt,line join=round] ( 86.60, 38.10) --
	( 86.60, 40.85);

\path[draw=drawColor,line width= 0.6pt,line join=round] (101.17, 38.10) --
	(101.17, 40.85);

\path[draw=drawColor,line width= 0.6pt,line join=round] (115.73, 38.10) --
	(115.73, 40.85);

\path[draw=drawColor,line width= 0.6pt,line join=round] (130.30, 38.10) --
	(130.30, 40.85);
\end{scope}
\begin{scope}
\path[clip] (  0.00,  0.00) rectangle (144.54,130.09);
\definecolor{drawColor}{gray}{0.30}

\node[text=drawColor,rotate= 45.00,anchor=base east,inner sep=0pt, outer sep=0pt, scale=  0.88] at ( 47.18, 31.62) {512};

\node[text=drawColor,rotate= 45.00,anchor=base east,inner sep=0pt, outer sep=0pt, scale=  0.88] at ( 61.75, 31.62) {770};

\node[text=drawColor,rotate= 45.00,anchor=base east,inner sep=0pt, outer sep=0pt, scale=  0.88] at ( 76.32, 31.62) {1024};

\node[text=drawColor,rotate= 45.00,anchor=base east,inner sep=0pt, outer sep=0pt, scale=  0.88] at ( 90.88, 31.62) {2048};

\node[text=drawColor,rotate= 45.00,anchor=base east,inner sep=0pt, outer sep=0pt, scale=  0.88] at (105.45, 31.62) {4096};

\node[text=drawColor,rotate= 45.00,anchor=base east,inner sep=0pt, outer sep=0pt, scale=  0.88] at (120.02, 31.62) {5482};

\node[text=drawColor,rotate= 45.00,anchor=base east,inner sep=0pt, outer sep=0pt, scale=  0.88] at (134.59, 31.62) {8192};
\end{scope}
\begin{scope}
\path[clip] (  0.00,  0.00) rectangle (144.54,130.09);
\definecolor{drawColor}{RGB}{0,0,0}

\node[text=drawColor,anchor=base,inner sep=0pt, outer sep=0pt, scale=  1.10] at ( 86.60,  7.64) {Target Chunk Size};
\end{scope}
\begin{scope}
\path[clip] (  0.00,  0.00) rectangle (144.54,130.09);
\definecolor{drawColor}{RGB}{0,0,0}

\node[text=drawColor,rotate= 90.00,anchor=base,inner sep=0pt, outer sep=0pt, scale=  1.10] at ( 13.08, 82.72) {Dedup. Ratio};
\end{scope}
\end{tikzpicture}

%% file: fig/dedup_overview_legendonly.tex
% Created by tikzDevice version 0.12.6 on 2024-07-17 20:42:03
% !TEX encoding = UTF-8 Unicode
\begin{tikzpicture}[x=1pt,y=1pt]
\definecolor{fillColor}{RGB}{255,255,255}
\path[use as bounding box,fill=fillColor,fill opacity=0.00] (0,0) rectangle (433.62, 72.27);
\begin{scope}
\path[clip] (  0.00,  0.00) rectangle (433.62, 72.27);
\definecolor{drawColor}{RGB}{255,111,0}

\path[draw=drawColor,line width= 0.6pt,line join=round] ( 27.29, 36.13) -- ( 41.16, 36.13);
\end{scope}
\begin{scope}
\path[clip] (  0.00,  0.00) rectangle (433.62, 72.27);
\definecolor{drawColor}{RGB}{255,111,0}
\definecolor{fillColor}{RGB}{255,255,255}

\path[draw=drawColor,line width= 0.4pt,line join=round,line cap=round,fill=fillColor] ( 34.23, 36.13) circle (  1.43);
\end{scope}
\begin{scope}
\path[clip] (  0.00,  0.00) rectangle (433.62, 72.27);
\definecolor{drawColor}{RGB}{199,16,0}

\path[draw=drawColor,line width= 0.6pt,line join=round] ( 75.71, 36.13) -- ( 89.58, 36.13);
\end{scope}
\begin{scope}
\path[clip] (  0.00,  0.00) rectangle (433.62, 72.27);
\definecolor{drawColor}{RGB}{199,16,0}
\definecolor{fillColor}{RGB}{255,255,255}

\path[draw=drawColor,line width= 0.4pt,line join=round,line cap=round,fill=fillColor] ( 82.65, 36.13) circle (  1.43);
\end{scope}
\begin{scope}
\path[clip] (  0.00,  0.00) rectangle (433.62, 72.27);
\definecolor{drawColor}{RGB}{0,142,160}

\path[draw=drawColor,line width= 0.6pt,line join=round] (117.44, 36.13) -- (131.32, 36.13);
\end{scope}
\begin{scope}
\path[clip] (  0.00,  0.00) rectangle (433.62, 72.27);
\definecolor{drawColor}{RGB}{0,142,160}
\definecolor{fillColor}{RGB}{255,255,255}

\path[draw=drawColor,line width= 0.4pt,line join=round,line cap=round,fill=fillColor] (124.38, 36.13) circle (  1.43);
\end{scope}
\begin{scope}
\path[clip] (  0.00,  0.00) rectangle (433.62, 72.27);
\definecolor{drawColor}{RGB}{138,65,152}

\path[draw=drawColor,line width= 0.6pt,line join=round] (168.09, 36.13) -- (181.97, 36.13);
\end{scope}
\begin{scope}
\path[clip] (  0.00,  0.00) rectangle (433.62, 72.27);
\definecolor{drawColor}{RGB}{138,65,152}
\definecolor{fillColor}{RGB}{255,255,255}

\path[draw=drawColor,line width= 0.4pt,line join=round,line cap=round,fill=fillColor] (175.03, 36.13) circle (  1.43);
\end{scope}
\begin{scope}
\path[clip] (  0.00,  0.00) rectangle (433.62, 72.27);
\definecolor{drawColor}{RGB}{90,149,153}

\path[draw=drawColor,line width= 0.6pt,line join=round] (210.38, 36.13) -- (224.25, 36.13);
\end{scope}
\begin{scope}
\path[clip] (  0.00,  0.00) rectangle (433.62, 72.27);
\definecolor{drawColor}{RGB}{90,149,153}
\definecolor{fillColor}{RGB}{255,255,255}

\path[draw=drawColor,line width= 0.4pt,line join=round,line cap=round,fill=fillColor] (217.32, 36.13) circle (  1.43);
\end{scope}
\begin{scope}
\path[clip] (  0.00,  0.00) rectangle (433.62, 72.27);
\definecolor{drawColor}{RGB}{255,99,72}

\path[draw=drawColor,line width= 0.6pt,line join=round] (255.46, 36.13) -- (269.33, 36.13);
\end{scope}
\begin{scope}
\path[clip] (  0.00,  0.00) rectangle (433.62, 72.27);
\definecolor{drawColor}{RGB}{255,99,72}
\definecolor{fillColor}{RGB}{255,255,255}

\path[draw=drawColor,line width= 0.4pt,line join=round,line cap=round,fill=fillColor] (262.39, 36.13) circle (  1.43);
\end{scope}
\begin{scope}
\path[clip] (  0.00,  0.00) rectangle (433.62, 72.27);
\definecolor{drawColor}{RGB}{132,215,225}

\path[draw=drawColor,line width= 0.6pt,line join=round] (306.67, 36.13) -- (320.54, 36.13);
\end{scope}
\begin{scope}
\path[clip] (  0.00,  0.00) rectangle (433.62, 72.27);
\definecolor{drawColor}{RGB}{132,215,225}
\definecolor{fillColor}{RGB}{255,255,255}

\path[draw=drawColor,line width= 0.4pt,line join=round,line cap=round,fill=fillColor] (313.60, 36.13) circle (  1.43);
\end{scope}
\begin{scope}
\path[clip] (  0.00,  0.00) rectangle (433.62, 72.27);
\definecolor{drawColor}{RGB}{255,149,168}

\path[draw=drawColor,line width= 0.6pt,line join=round] (361.78, 36.13) -- (375.66, 36.13);
\end{scope}
\begin{scope}
\path[clip] (  0.00,  0.00) rectangle (433.62, 72.27);
\definecolor{drawColor}{RGB}{255,149,168}
\definecolor{fillColor}{RGB}{255,255,255}

\path[draw=drawColor,line width= 0.4pt,line join=round,line cap=round,fill=fillColor] (368.72, 36.13) circle (  1.43);
\end{scope}
\begin{scope}
\path[clip] (  0.00,  0.00) rectangle (433.62, 72.27);
\definecolor{drawColor}{RGB}{0,0,0}

\node[text=drawColor,anchor=base west,inner sep=0pt, outer sep=0pt, scale=  1.00] at ( 48.40, 32.69) {FSC};
\end{scope}
\begin{scope}
\path[clip] (  0.00,  0.00) rectangle (433.62, 72.27);
\definecolor{drawColor}{RGB}{0,0,0}

\node[text=drawColor,anchor=base west,inner sep=0pt, outer sep=0pt, scale=  1.00] at ( 96.82, 32.69) {AE};
\end{scope}
\begin{scope}
\path[clip] (  0.00,  0.00) rectangle (433.62, 72.27);
\definecolor{drawColor}{RGB}{0,0,0}

\node[text=drawColor,anchor=base west,inner sep=0pt, outer sep=0pt, scale=  1.00] at (138.55, 32.69) {RAM};
\end{scope}
\begin{scope}
\path[clip] (  0.00,  0.00) rectangle (433.62, 72.27);
\definecolor{drawColor}{RGB}{0,0,0}

\node[text=drawColor,anchor=base west,inner sep=0pt, outer sep=0pt, scale=  1.00] at (189.20, 32.69) {MII};
\end{scope}
\begin{scope}
\path[clip] (  0.00,  0.00) rectangle (433.62, 72.27);
\definecolor{drawColor}{RGB}{0,0,0}

\node[text=drawColor,anchor=base west,inner sep=0pt, outer sep=0pt, scale=  1.00] at (231.49, 32.69) {PCI};
\end{scope}
\begin{scope}
\path[clip] (  0.00,  0.00) rectangle (433.62, 72.27);
\definecolor{drawColor}{RGB}{0,0,0}

\node[text=drawColor,anchor=base west,inner sep=0pt, outer sep=0pt, scale=  1.00] at (276.57, 32.69) {BSW};
\end{scope}
\begin{scope}
\path[clip] (  0.00,  0.00) rectangle (433.62, 72.27);
\definecolor{drawColor}{RGB}{0,0,0}

\node[text=drawColor,anchor=base west,inner sep=0pt, outer sep=0pt, scale=  1.00] at (327.78, 32.69) {BFBC};
\end{scope}
\begin{scope}
\path[clip] (  0.00,  0.00) rectangle (433.62, 72.27);
\definecolor{drawColor}{RGB}{0,0,0}

\node[text=drawColor,anchor=base west,inner sep=0pt, outer sep=0pt, scale=  1.00] at (382.89, 32.69) {BFBC*};
\end{scope}
\end{tikzpicture}

%% file: sections/discussion.tex
Our theoretical analyses and experimental evaluations conducted throughout this study
reproduce existing results but also reveal new insights.
The theoretical analysis uncovered novel aspects and corrects existing formulas, extending the original contributions of the algorithms’ developers.
Through rigorous and impartial experimentation, we identified previously unreported performance characteristics while also validating many of the original claims.
With our discussion, we aim to synthesize those findings, providing a comprehensive overview of the deduplication landscape and also offering practical recommendations based on our findings.

\subsection{Contrasting Results}

While our study confirms many of the claims made by the algorithm developers, we also encountered several discrepancies.
These differences arose from factors such as variations in dataset characteristics, experimental setups, and implementation details.
Thus, this section is dedicated to contrasting our results with the ones from the literature,
and reflecting on their implications.
 
\subsubsection{AE}

As we illuminate in \cref{sec:algorithms:ae}, the formula by which the parameterization in \ac{AE} is derived
does not reflect the actual behavior of the algorithm fully,
but has been disseminated in this form throughout several studies \cite{ae,ae2,ellappan2021dynamic,fastcdc2,jin2023accelerating,ellappan2023smart},
and moreover, the open-source chunking algorithm evaluation platform Destor and DedupBench,
whose authors we have notified about the issue.
According to our own experiments on RAND, the previously established formula renders chunks \qty{13}{\%} smaller than the target when it is set to \qty{512}{B}, 
and \qty{39}{\%} smaller with a target of \qty{8}{KiB}.
In fact, the higher the target, the more pronounced the deviation.
Because smaller chunks are easier to deduplicate, earlier results on the deduplication ratio of \ac{AE} must be taken with a grain of salt.
In the original paper~\cite{ae}, the authors suggest that AE is superior to BSW in terms of deduplication.
While \ac{AE} has shown superior performance in some instances of our experiment, and is competitive in most,
its performance sometimes degrades with higher target chunk sizes.
Our results suggest that the AE algorithm cannot robustly handle high target chunk sizes, compared to, \eg, BSW.
It is likely that this has been overlooked previously due to aforementioned incorrect target chunk size calculations and evaluations on lower targets.
%We argue that this may be another flaw in the algorithm that has been overlooked due to the assumption of an incorrect formula and also testing on lower targets.

\subsubsection{RAM}

We demonstrate the incapability of RAM to deal with low-entropy datasets.
This is especially pronounced on CODE (\cf \cref{tab:csd_means_sd_full}).
The authors of RAM noticed this flaw~\cite{ram}; however, they did not presented the extent of this shortcoming.
In their study, they evaluate RAM on high-entropy datasets that, as we argue, do not represent realistic candidates for data deduplication.
%This is because redundant data chunks are scarce in such datasets.
Indeed, deduplication is only evaluated in the form of \enquote{bytes saved per second},
which counterintuitively conflates throughput with deduplication.
Moreover, the paper suggests chunk-size variances similar to AE, while the experiments on none of our real-world datasets support this statement.
Even with our LNX dataset, which is composed of similar files than their \enquote{Dataset 1}, we cannot confirm comparable results.

\subsubsection{BFBC}\label{sec:discussion:bfbc}

The authors of BFBC~\cite{bfbc} compare their performance to Rabin's in an experiment that uses two source code-based datasets, similar to our CODE dataset.
On both datasets, the deduplication achieved by BFBC outperforms Rabin's, whereas in our experiment, Rabin emerges as the superior algorithm.
We note that their experiment differs from ours in enforcing minimum and maximum chunk size lengths.
That is, they evaluate the algorithms on chunk size ranges instead of targets.
In these experiments, BFBC performed best in the range 128--256\,B.
However, we note that the result they report is similar for the range 128--8192\,B, which is the setting closest to our experiment on target \qty{512}{B}.
In \cref{tab:bfbc_results_vs_ours}, we contrast their results with ours.
Even when considering the best results we achieved using BFBC, it is still outperformed by Rabin.
We postulate that the chunk size limits imposed on Rabin in their work impacted the results negatively.
Since the exact settings chosen for Rabin in their analysis are not disclosed, we are unable to confirm this hypothesis.
Our comprehensive experiments and results ultimately do not support the claims of improved deduplication ratios for BFBC.
In terms of throughput, as well, we find that more efficient \ac{BSW} algorithms, \eg Gear, outperform BFBC.

\begin{table}[tb]
    \centering
	\caption{Comparison of Deduplication Ratio Results Presented in \cite{bfbc} vs. in Our Own Experiments}
    \label{tab:bfbc_results_vs_ours}
	\input{tables/bfbc-results-vs-ours}
\end{table}

\subsection{Summary}% and Recommendations

After an extensive performance evaluation of various algorithms across multiple settings,
we now attempt to give a conclusive summary of the results and findings.
%Furthermore, we strive to give practical advice to system engineers seeking to select an optimal chunking algorithm for their needs.

%To this end, we begin by summarizing our results across four key metrics: 
%deduplication performance (\emph{efficacy}), throughput (\emph{efficiency}), 
%the deviation of the empirical mean to the target chunk size, 
%as well as the produced chunks' variation around the mean.
%
Our research indicates that traditional BSW algorithms are still unbeaten in terms of deduplication.
However, we find that similar performance can also be attained with AE and MII.
RAM proved to be very sensitive \wrt entropy in a dataset:
%In relation to the entropy of a dataset, the algorithm tends to produce very high chunk-size variances.
On datasets with particularly low entropy, such as text-based data (CODE), the algorithm fails to find chunk boundaries, resulting in the generation of pathologically large chunks.
Ultimately, these effects render the slight efficiency gains of RAM over AE negligible.

Only slightly behind AE in terms of throughput, Gear proves to be the fastest BSW algorithm
and simultaneously maintains the healthiest levels of chunk size mean and variance among that group.
Furthermore, with the help of NC, we find that the otherwise high variance in chunk size can be mitigated without impairing the deduplication ratio.
This technique becomes especially powerful when combined with minimum chunk size skipping, which can improve throughput significantly~\cite{quickcdc}.
%This technique becomes especially powerful when combined with minimum chunk size skipping, which can improve throughput by up to \qty{50}{\%}.
%\leo{is this our result? if no, ref. Also, I don't think we talk about minimum chunk size skipping anymore, so we need to introduce it here.}
%\marcel{maybe before the camera-ready, we can reference our QuickCDC paper}

Although the issue of chunk-size variance is of lesser importance in the context of data synchronization, 
which is the intended use case for MII and PCI, 
we were unable to identify any advantages over alternative solutions such as AE or Gear,
which offer comparable or superior deduplication and significantly higher throughput.

Finally, we also find that BFBC and BFBC* offer no real advantage over algorithms such as Gear or AE.
Moreover, whereas the algorithm for BFBC does not consider higher entropy datasets,
both BFBC and BFBC* have no mechanism to ensure consistent chunk sizes.
Furthermore, they add complexity through the initial process of collecting statistics over a dataset.
We note, however, that the optimized divisor selection algorithm in BFBC* successfully rectifies chunk size averages,
and thus offers a real improvement over the original implementation of BFBC.

\begin{table}[h]
    \centering
    
    \begin{threeparttable}
    \caption{Performance Summary of CDC Algorithms Based on Our Experiments}
    \label{table:perf_summary}
    \input{tables/summary}
    \begin{tablenotes}
        % \item[a] using Normalized Chunking
        \item[†] except on the CODE dataset
    \end{tablenotes}
    \end{threeparttable}
\end{table}

These findings are summarized in simplified form in \cref{table:perf_summary}.
A checkmark indicates that the algorithm was among the top performers with regard to the respective metric.
The table reveals AE as the only CDC algorithm competitive on all metrics.
However, this result requires a nuanced interpretation. 
While AE performed admirably within our tested range of target chunk sizes, its deduplication efficacy may degrade with larger chunk sizes more strongly than the more robust BSW algorithms. 
Additionally, our analysis reveals that \ac{NC} significantly reduces Gear’s chunk-size variance, though not to the level achieved by AE, without corrupting deduplication performance.
This reduction also enables the safe skipping of a minimum chunk size, which in turn boosts throughput.
Given these considerations, Gear with NC emerges as a robust and efficient alternative, making it an equally attractive choice for various applications.

%% file: tables/bfbc-results-vs-ours.tex
% \small
\begin{tabularx}{0.75\columnwidth}{lccc}
\toprule
    & \multicolumn{2}{c}{Their Results~\cite{bfbc}} & \multicolumn{1}{c}{Ours} \\
    \cmidrule(lr){2-3}\cmidrule(lr){4-4}
& Dataset 1          & Dataset 2          & CODE      \\
\midrule
BFBC  & \qty{79.8}{\%}     & \qty{96.7}{\%}     & \qty{85.0}{\%} \\
Rabin & \qty{68.2}{\%}     & \qty{92.6}{\%}     & \qty{87.1}{\%} \\
\bottomrule
\end{tabularx}

%% file: tables/summary.tex
% \small
\begin{tabularx}{\columnwidth}{lcccc}
\toprule
    &&&\multicolumn{2}{c}{Chunk Sizes}\\
    \cmidrule(lr){4-5}
    Algorithm & Dedup. & Throughput & Mean & SD\\
\midrule
Rabin, Buzhash & \checkmark & & & \\
Gear & \checkmark & \checkmark & \checkmark & \\
Gear with NC & \checkmark & \checkmark & \checkmark & $\text{\checkmark}$ \\
AE & \checkmark & \checkmark & \checkmark & \checkmark \\
RAM & \checkmark\tnote{\dag} & \checkmark & & \\
PCI & & & & \\
MII & \checkmark & & & \\
BFBC & & \checkmark & & \\
BFBC* & & \checkmark & \checkmark & \\
\bottomrule
\end{tabularx}

%% file: sections/conclusion.tex
In this work, we present a comprehensive and impartial evaluation of state-of-the-art \ac{CDC} algorithms.
Furthermore, we provide an analytical framework, as well as a set of benchmarks for the evaluation of future advancements in CDC.
Our rigorous theoretical analysis and extensive experimental validation yield both reproducible results and novel insights.
%As such, we derive formulas to tune parameters for target chunk sizes in AE, RAM, and MII.
%Moreover, we debunk the formula for AE found in the literature.
Our comparison highlights several limitations and shortcomings that are not apparent from previous studies.
We find that many researchers promote their algorithm under conditions or assumptions that fail to hold up in realistic scenarios,
often relying on biased datasets, misleading metrics, or narrowly defined test cases.
Finally, we recognize Gear with \ac{NC} and \ac{AE} as the most attractive choice for \ac{CDC}, 
despite more recent advancements in algorithm development.
% that are not apparent in the experiments conducted by the respective authors of RAM, MII, PCI, and BFBC.
%Ultimately, our results render the recent developments of MII, PCI, and BFBC ineffectual in comparison with prior techniques.
%Our research indicates that traditional BSW algorithms are still unbeaten in terms of deduplication efficacy.
%Among the BSW algorithms, Gear stands out with low chunk-size variance, and by far the best throughput.
%Gear, enhanced with NC, emerges as particularly robust, making it an optimal choice for most datasets and applications.
%In summary, our theoretical and experimental evaluation demonstrates
%the effectiveness and constraints of the application of CDC algorithms for data reduction. 
We believe that our findings and methodologies will significantly contribute to the optimization of storage and bandwidth efficiency in cloud computing infrastructures.

%% file: sections/appendix.tex
\section*{Appendix A: Chunking Algorithms}\label{app:algos}
\addcontentsline{toc}{section}{Appendix A: Implementation of Chunking Algorithms}

Here, we include the pseudocode to the algorithms outlined in \Cref{sec:algorithms}.

\begin{algorithm}
	\caption{$\textproc{BSW}(d, l)$}
	\label{alg:bsw}
	{
	% \small
	\input{algorithms/bsw.tex}
	}
\end{algorithm}
\begin{algorithm}
	\caption{$\textproc{AE}(d, l)$}
	\label{alg:ae}
	{%\small
		\input{algorithms/ae.tex}
	}
\end{algorithm}
\begin{algorithm}
	\caption{$\textproc{RAM}(d, l)$}
	\label{alg:ram}
	{
 % \small
		\input{algorithms/ram.tex}
	}
\end{algorithm}
\begin{algorithm}
	\caption{$\textproc{MII}(d, l)$}
	\label{alg:mii}
	{%\small
		\input{algorithms/mii.tex}
	}
\end{algorithm}
\begin{algorithm}
	\caption{$\textproc{PCI}(d, l)$}
	\label{alg:pciOpt}
	{%\small
		\input{algorithms/pci_opt.tex}
	}
\end{algorithm}
\begin{algorithm}
	\caption{$\textproc{BFBC}(d, l)$}
	\label{alg:bfbc}
	{%\small
		\input{algorithms/bfbc.tex}
	}
\end{algorithm}
\begin{algorithm}
	\input{algorithms/bfbc-params.tex}
	\label{alg:bfbcParams}
\end{algorithm}

\onecolumn
\clearpage
\begin{landscape}
    
\section*{Appendix B: Extended Results}\label{app:results}
\addcontentsline{toc}{section}{Appendix B: Extended Results}

\input{tab/csd_means_sd_full}

\vspace{-3.62em}
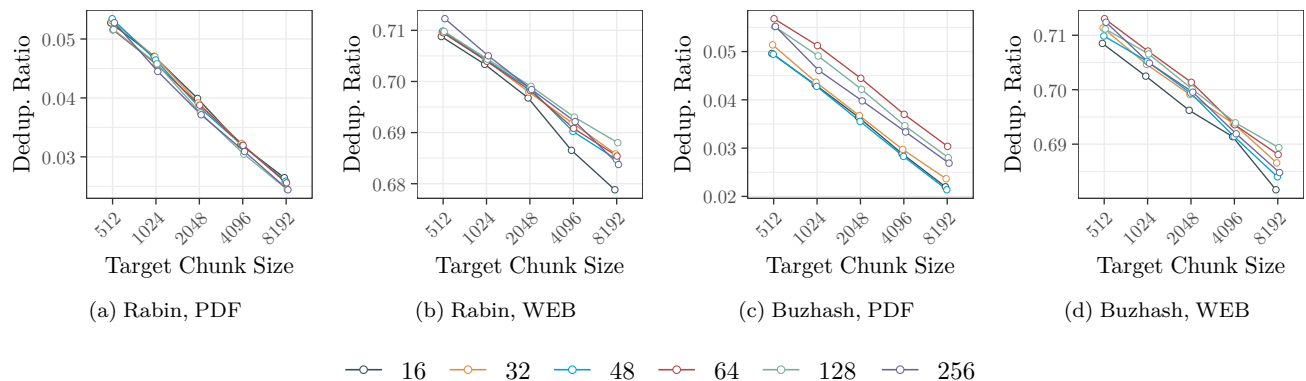
\begin{figure}[h]
          \centering
    		\subfloat[Rabin, PDF]{\scalebox{0.85}{\input{fig/dedup_rabin_window_sizes_pdf}}%
        		\label{fig:dedup_rabin_window_sizes_pdf}}
    		\subfloat[Rabin, WEB]{\scalebox{0.85}{\input{fig/dedup_rabin_window_sizes_web}}%
        		\label{fig:dedup_rabin_window_sizes_web}}
    		\subfloat[Buzhash, PDF]{\scalebox{0.85}{\input{fig/dedup_buzhash_window_sizes_pdf}}%
        		\label{fig:dedup_buzhash_window_sizes_pdf}}
		    \subfloat[Buzhash, WEB]{\scalebox{0.85}{\input{fig/dedup_buzhash_window_sizes_web}}%
        		\label{fig:dedup_buzhash_window_sizes_web}}
			\vspace{-2em}
			\input{fig/dedup_window_sizes_legendonly}
			\vspace{-3em}
			\caption{Deduplication ratio of Rabin and Buzhash on different window sizes.}
			\label{fig:dedup_window_sizes}
\end{figure}
          
\vspace{-2.4em}

\begin{figure}[h]
			\centering
			\subfloat[LNX]{\scalebox{0.85}{\input{fig/dedup_gear_variants_lnx}}%
			    \label{fig:dedup_gear_variants:lnx}}
			\subfloat[PDF]{\scalebox{0.85}{\input{fig/dedup_gear_variants_pdf}}%
			    \label{fig:dedup_gear_variants:pdf}}
			\subfloat[WEB]{\scalebox{0.85}{\input{fig/dedup_gear_variants_web}}%
			    \label{fig:dedup_gear_variants:web}}\subfloat[CODE]{\scalebox{0.85}{\input{fig/dedup_gear_variants_code}}%
	    	\label{fig:dedup_gear_variants:code}}
			\vspace{-2em}
			\input{fig/dedup_gear_variants_legendonly}
			\vspace{-3em}
			\caption{Deduplication performance of Gear with different levels of NC, as well as without (Vanilla).}
			\label{fig:dedup_gear}
\end{figure}
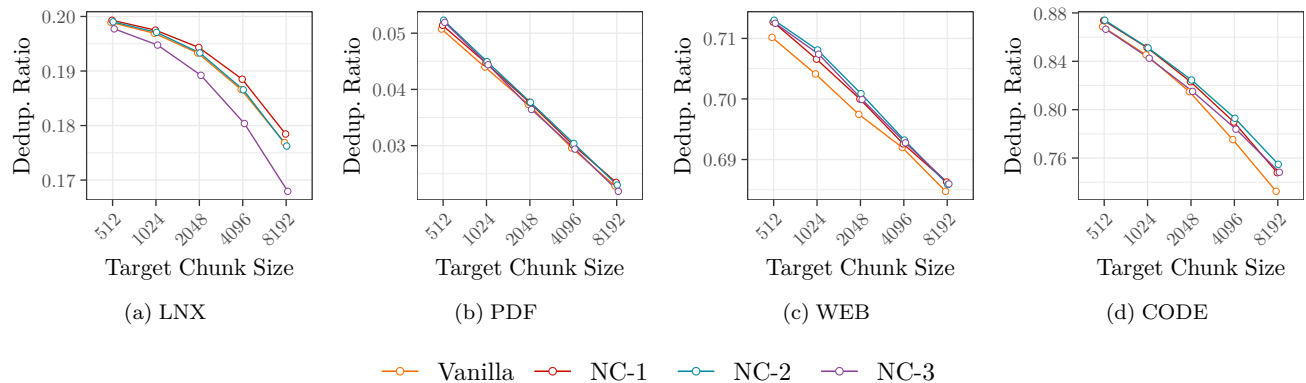

\end{landscape}

%% file: algorithms/bsw.tex
\hspace*{\algorithmicindent} \textbf{Input:} Data stream $d$, data length $l$\\
\hspace*{\algorithmicindent} \textbf{Predefined:} Window size $w$, bitmask length $b$
\begin{algorithmic}[1]
	\For{$i \leftarrow w$ to $l$}
		\State $f \gets H(d[i-w],\ldots,d[i])$
		\If{$f \; \& \; (2^b - 1) = 0$}
			\State \Return{$i$}
		\EndIf
	\EndFor
\end{algorithmic}

%% file: algorithms/ae.tex
\hspace*{\algorithmicindent} \textbf{Input:} Data stream $d$, data length $l$ \\
\hspace*{\algorithmicindent} \textbf{Predefined:} Horizon length $h$
\begin{algorithmic}[1]
	\State $x_\text{val} \gets 0$
	\State $x_\text{pos} \gets 0$
	\For{$i \leftarrow 1$ to $l$}
    	\If{$d[i] \leq x_\text{val}$}
			\If{$i = x_\text{pos}+h$}
				\State \Return{$i$}
			\EndIf
		\Else
			\State $x_\text{val} \gets d[i]$
			\State $x_\text{pos} \gets i$
		\EndIf
	\EndFor
\end{algorithmic}

%% file: algorithms/ram.tex
\hspace*{\algorithmicindent} \textbf{Input:} Data stream $d$, data length $l$\\
\hspace*{\algorithmicindent} \textbf{Predefined:} Horizon length $h$
\begin{algorithmic}[1]
	\State $x \gets 0$
	\For{$i \leftarrow 1$ to $l$}
		\If{$i \leq h$}
			\If{$d[i] > x$}
				\State $x \gets d[i]$
			\EndIf
		\Else
			\If{$d[i] \geq x$}
				\State \Return{$i$}
			\EndIf
		\EndIf
	\EndFor
\end{algorithmic}

%% file: algorithms/mii.tex
\hspace*{\algorithmicindent} \textbf{Input:} Data stream $d$, data length $l$\\
\hspace*{\algorithmicindent} \textbf{Predefined:} Window size $w$ 
\begin{algorithmic}[1]
	\State $c \gets 0$
	\For{$i \leftarrow 2$ to $l$}
    	\If{$d[i] > d[i-1]$}
			\State $c \gets c+1$
			\If{$c=w$}
				\State \Return{$i$}
			\EndIf
		\Else
			\State $c\gets 0$
		\EndIf
	\EndFor
\end{algorithmic}

%% file: algorithms/pci_opt.tex
\hspace*{\algorithmicindent} \textbf{Input:} Data stream $d$, data length $l$\\
\hspace*{\algorithmicindent} \textbf{Predefined:} Window size $w$, threshold $\theta$ 
\begin{algorithmic}[1]
	\State $v \gets (0,0,\ldots,0), \;\vert v \vert=w$ \Comment{Current window.}
	\State $p \gets 0$ \Comment{Popcount in $v$.}
	\For{$i \leftarrow 1$ to $l$}
		\State $p \gets p - \textproc{Popcount}(v[i\bmod w]) + \textproc{Popcount}(d[i])$
		\State $v[i\bmod w] \gets d[i]$
    	\If{$i \geq w$ \textbf{and} $p \geq \theta$}
			\State \Return{$i$}
		\EndIf
	\EndFor
\end{algorithmic}

%% file: algorithms/bfbc.tex
\hspace*{\algorithmicindent} \textbf{Input:} Data stream $d$, data length $l$\\
\hspace*{\algorithmicindent} \textbf{Predefined:} Divisors $D$, minimum chunk length $\lambda_\text{min}$
\begin{algorithmic}[1]
	\For{$i \leftarrow 1$ to $l$}
    	\If{$i > \lambda_\text{min}$}
    		\ForEach{$(b_0, b_1) \in D$}
    			\If{$(d[i-1],\,d[i]) = (b_0, b_1)$}
					\State \Return{$i$}
				\EndIf
			\EndFor
		\EndIf
	\EndFor
\end{algorithmic}

%% file: algorithms/bfbc-params.tex
\begin{small}
\caption{$\textproc{DetermineBfbcDivisors}(F,\mu,l)$}
\hspace*{\algorithmicindent} \textbf{Input:} Frequencies of top-frequent byte pairs $F$, target chunk size $\mu$, file length $l$ \\\hspace*{\algorithmicindent} \textbf{Output:} Set of divisors as indices of $F$
\begin{algorithmic}[1]
	\State $D \gets \{\}$
	\For{$i \leftarrow 1$ to $|F|$}
		\If{$D = \emptyset$}
	    	\If{$\mu(\{i\}) \geq \mu$}
    			\State $D \gets D \cup \{i\}$
			\EndIf
		\Else
			\If{$|\mu-\mu(D \cup \{i\})| < |\mu-\mu(D)|$}
    			\State $D \gets D \cup \{i\}$
			\EndIf
		\EndIf
	\EndFor	
	\State \Return $D$
\end{algorithmic}
\end{small}

%% file: tab/csd_means_sd_full.tex
\definecolor{tempcolor1}{rgb}{1,0.96,0.94} 
\definecolor{tempcolor2}{rgb}{0.99,0.88,0.82} 
\definecolor{tempcolor3}{rgb}{0.4,0,0.05} 
\definecolor{tempcolor4}{rgb}{0.98,0.41,0.29} 
\definecolor{tempcolor5}{rgb}{0.98,0.73,0.63} 
\definecolor{tempcolor6}{rgb}{0.98,0.57,0.45} 
\definecolor{tempcolor7}{rgb}{0.93,0.23,0.17} 
\definecolor{tempcolor8}{rgb}{0.64,0.06,0.08} 
\definecolor{tempcolor9}{rgb}{0.79,0.09,0.11} 
\definecolor{black}{rgb}{0.0, 0.0, 0.0} 
\definecolor{black}{rgb}{0.0, 0.0, 0.0} 
\definecolor{black}{rgb}{0.0, 0.0, 0.0} 
\begin{center}
\color{black}
{\footnotesize%
\begin{longtable}{lcrrrrrrrrrrrrrrrrrrrrrrr}
\caption{Means and Standard Deviation of Chunks Produced by Chunking Algorithms on Different Target Chunk Sizes and Datasets %
%Cells are left blank where the target chunk size was not applicable.
}\label{tab:csd_means_sd_full}\\
\hline
\multirow{2}{*}{Algorithm}&&\multicolumn{2}{c}{512\,B}&&\multicolumn{2}{c}{737\,B}&&\multicolumn{2}{c}{1024\,B}&&\multicolumn{2}{c}{2048\,B}&&\multicolumn{2}{c}{4096\,B}&&\multicolumn{2}{c}{5152\,B}&&\multicolumn{2}{c}{8192\,B}\\ 
\cline{3-4}\cline{6-7}\cline{9-10}\cline{12-13}\cline{15-16}\cline{18-19}\cline{21-22}
&&$\bar{cs}$&$s$&&$\bar{cs}$&$s$&&$\bar{cs}$&$s$&&$\bar{cs}$&$s$&&$\bar{cs}$&$s$&&$\bar{cs}$&$s$&&$\bar{cs}$&$s$\\ 
\hline\noalign{\smallskip}

\textbf{RAND}&&&&&&&&&&&&&&&&&&&&&&&&\\ 
Rabin&&\cellcolor{tempcolor1}542&\cellcolor{tempcolor6}511&&&&&\cellcolor{tempcolor1}1055&\cellcolor{tempcolor6}1024&&\cellcolor{tempcolor1}2078&\cellcolor{tempcolor4}\color{white}2047&&\cellcolor{tempcolor1}4127&\cellcolor{tempcolor4}\color{white}4091&&&&&\cellcolor{tempcolor1}8220&\cellcolor{tempcolor4}\color{white}8179\\ 
Buzhash&&\cellcolor{tempcolor2}576&\cellcolor{tempcolor6}512&&&&&\cellcolor{tempcolor2}1089&\cellcolor{tempcolor6}1026&&\cellcolor{tempcolor1}2115&\cellcolor{tempcolor4}\color{white}2053&&\cellcolor{tempcolor1}4166&\cellcolor{tempcolor4}\color{white}4101&&&&&\cellcolor{tempcolor1}8267&\cellcolor{tempcolor4}\color{white}8213\\ 
Gear&&\cellcolor{tempcolor1}512&\cellcolor{tempcolor6}510&&&&&\cellcolor{tempcolor1}1024&\cellcolor{tempcolor4}\color{white}1023&&\cellcolor{tempcolor1}2048&\cellcolor{tempcolor4}\color{white}2047&&\cellcolor{tempcolor1}4098&\cellcolor{tempcolor4}\color{white}4096&&&&&\cellcolor{tempcolor1}8184&\cellcolor{tempcolor4}\color{white}8175\\ 
Gear NC-1&&\cellcolor{tempcolor2}558&\cellcolor{tempcolor5}339&&\cellcolor{tempcolor5}993&\cellcolor{tempcolor5}613&&\cellcolor{tempcolor2}1116&\cellcolor{tempcolor5}680&&\cellcolor{tempcolor2}2233&\cellcolor{tempcolor5}1362&&\cellcolor{tempcolor2}4464&\cellcolor{tempcolor5}2725&&\cellcolor{tempcolor2}5045&\cellcolor{tempcolor5}3134&&\cellcolor{tempcolor2}8928&\cellcolor{tempcolor5}5451\\ 
Gear NC-2&&\cellcolor{tempcolor2}553&\cellcolor{tempcolor2}210&&\cellcolor{tempcolor2}913&\cellcolor{tempcolor2}349&&\cellcolor{tempcolor2}1105&\cellcolor{tempcolor2}420&&\cellcolor{tempcolor2}2209&\cellcolor{tempcolor2}841&&\cellcolor{tempcolor2}4420&\cellcolor{tempcolor2}1685&&\cellcolor{tempcolor1}5391&\cellcolor{tempcolor2}2140&&\cellcolor{tempcolor2}8842&\cellcolor{tempcolor2}3372\\ 
Gear NC-3&&\cellcolor{tempcolor1}538&\cellcolor{tempcolor1}130&&\cellcolor{tempcolor2}852&\cellcolor{tempcolor1}203&&\cellcolor{tempcolor1}1076&\cellcolor{tempcolor1}261&&\cellcolor{tempcolor1}2150&\cellcolor{tempcolor2}522&&\cellcolor{tempcolor1}4300&\cellcolor{tempcolor2}1047&&\cellcolor{tempcolor1}5479&\cellcolor{tempcolor2}1432&&\cellcolor{tempcolor1}8603&\cellcolor{tempcolor2}2093\\ 
AE&&\cellcolor{tempcolor1}512&\cellcolor{tempcolor1}136&&\cellcolor{tempcolor1}769&\cellcolor{tempcolor1}179&&\cellcolor{tempcolor1}1024&\cellcolor{tempcolor1}209&&\cellcolor{tempcolor1}2048&\cellcolor{tempcolor1}252&&\cellcolor{tempcolor1}4095&\cellcolor{tempcolor1}255&&\cellcolor{tempcolor1}5480&\cellcolor{tempcolor1}255&&\cellcolor{tempcolor1}8191&\cellcolor{tempcolor1}255\\ 
RAM&&\cellcolor{tempcolor2}544&\cellcolor{tempcolor2}234&&\cellcolor{tempcolor1}785&\cellcolor{tempcolor1}247&&\cellcolor{tempcolor1}1030&\cellcolor{tempcolor1}252&&\cellcolor{tempcolor1}2048&\cellcolor{tempcolor1}255&&\cellcolor{tempcolor1}4096&\cellcolor{tempcolor1}255&&\cellcolor{tempcolor1}5482&\cellcolor{tempcolor1}255&&\cellcolor{tempcolor1}8192&\cellcolor{tempcolor1}255\\ 
PCI&&\cellcolor{tempcolor1}506&\cellcolor{tempcolor6}465&&\cellcolor{tempcolor1}761&\cellcolor{tempcolor6}731&&\cellcolor{tempcolor1}1017&\cellcolor{tempcolor6}991&&\cellcolor{tempcolor1}2085&\cellcolor{tempcolor4}\color{white}2036&&\cellcolor{tempcolor1}4292&\cellcolor{tempcolor4}\color{white}4255&&\cellcolor{tempcolor1}5388&\cellcolor{tempcolor4}\color{white}5341&&\cellcolor{tempcolor1}8690&\cellcolor{tempcolor4}\color{white}8642\\ 
MII&&&&&\cellcolor{tempcolor2}887&\cellcolor{tempcolor6}882&&&&&&&&&&&\cellcolor{tempcolor2}6238&\cellcolor{tempcolor4}\color{white}6228&&&\\ 
BFBC&&\cellcolor{tempcolor3}\color{white}22019&\cellcolor{tempcolor6}21604&&\cellcolor{tempcolor3}\color{white}22280&\cellcolor{tempcolor6}21604&&\cellcolor{tempcolor3}\color{white}22535&\cellcolor{tempcolor6}21603&&\cellcolor{tempcolor3}\color{white}23555&\cellcolor{tempcolor6}21602&&\cellcolor{tempcolor3}\color{white}25595&\cellcolor{tempcolor6}21598&&\cellcolor{tempcolor3}\color{white}26983&\cellcolor{tempcolor6}21596&&\cellcolor{tempcolor3}\color{white}29703&\cellcolor{tempcolor6}21594\\ 
BFBC*&&\cellcolor{tempcolor1}509&\cellcolor{tempcolor6}507&&\cellcolor{tempcolor1}765&\cellcolor{tempcolor6}764&&\cellcolor{tempcolor1}1016&\cellcolor{tempcolor4}\color{white}1015&&\cellcolor{tempcolor1}2031&\cellcolor{tempcolor4}\color{white}2028&&\cellcolor{tempcolor1}4062&\cellcolor{tempcolor4}\color{white}4062&&\cellcolor{tempcolor1}5414&\cellcolor{tempcolor4}\color{white}5413&&\cellcolor{tempcolor1}8121&\cellcolor{tempcolor4}\color{white}8128\\ 
\hline\noalign{\smallskip}
\textbf{LNX}&&&&&&&&&&&&&&&&&&&&&&&&\\ 
Rabin&&\cellcolor{tempcolor1}507&\cellcolor{tempcolor6}512&&&&&\cellcolor{tempcolor2}923&\cellcolor{tempcolor4}\color{white}1018&&\cellcolor{tempcolor5}1617&\cellcolor{tempcolor7}\color{white}1996&&\cellcolor{tempcolor6}2628&\cellcolor{tempcolor9}\color{white}3810&&&&&\cellcolor{tempcolor4}\color{white}3842&\cellcolor{tempcolor8}\color{white}6897\\ 
Buzhash&&\cellcolor{tempcolor2}556&\cellcolor{tempcolor6}512&&&&&\cellcolor{tempcolor1}1016&\cellcolor{tempcolor4}\color{white}1022&&\cellcolor{tempcolor2}1848&\cellcolor{tempcolor4}\color{white}2027&&\cellcolor{tempcolor5}3235&\cellcolor{tempcolor7}\color{white}3980&&&&&\cellcolor{tempcolor6}5247&\cellcolor{tempcolor9}\color{white}7581\\ 
Gear&&\cellcolor{tempcolor1}514&\cellcolor{tempcolor3}\color{white}1235&&&&&\cellcolor{tempcolor1}1029&\cellcolor{tempcolor8}\color{white}1902&&\cellcolor{tempcolor1}2058&\cellcolor{tempcolor9}\color{white}3075&&\cellcolor{tempcolor1}4115&\cellcolor{tempcolor7}\color{white}5288&&&&&\cellcolor{tempcolor1}8223&\cellcolor{tempcolor7}\color{white}9512\\ 
Gear NC-1&&\cellcolor{tempcolor2}560&\cellcolor{tempcolor3}\color{white}1214&&\cellcolor{tempcolor5}997&\cellcolor{tempcolor8}\color{white}1680&&\cellcolor{tempcolor2}1121&\cellcolor{tempcolor9}\color{white}1792&&\cellcolor{tempcolor2}2242&\cellcolor{tempcolor7}\color{white}2727&&\cellcolor{tempcolor2}4482&\cellcolor{tempcolor4}\color{white}4339&&\cellcolor{tempcolor2}5063&\cellcolor{tempcolor4}\color{white}4761&&\cellcolor{tempcolor2}8972&\cellcolor{tempcolor6}7338\\ 
Gear NC-2&&\cellcolor{tempcolor2}555&\cellcolor{tempcolor3}\color{white}1178&&\cellcolor{tempcolor5}917&\cellcolor{tempcolor9}\color{white}1531&&\cellcolor{tempcolor2}1109&\cellcolor{tempcolor9}\color{white}1691&&\cellcolor{tempcolor2}2218&\cellcolor{tempcolor4}\color{white}2477&&\cellcolor{tempcolor2}4438&\cellcolor{tempcolor6}3719&&\cellcolor{tempcolor1}5412&\cellcolor{tempcolor6}4238&&\cellcolor{tempcolor2}8873&\cellcolor{tempcolor6}5808\\ 
Gear NC-3&&\cellcolor{tempcolor1}540&\cellcolor{tempcolor3}\color{white}1154&&\cellcolor{tempcolor2}855&\cellcolor{tempcolor8}\color{white}1452&&\cellcolor{tempcolor1}1080&\cellcolor{tempcolor9}\color{white}1636&&\cellcolor{tempcolor1}2157&\cellcolor{tempcolor4}\color{white}2342&&\cellcolor{tempcolor1}4315&\cellcolor{tempcolor6}3410&&\cellcolor{tempcolor1}5496&\cellcolor{tempcolor6}3928&&\cellcolor{tempcolor1}8631&\cellcolor{tempcolor5}5063\\ 
AE&&\cellcolor{tempcolor1}505&\cellcolor{tempcolor1}138&&\cellcolor{tempcolor1}761&\cellcolor{tempcolor1}184&&\cellcolor{tempcolor1}1019&\cellcolor{tempcolor1}219&&\cellcolor{tempcolor1}2053&\cellcolor{tempcolor1}287&&\cellcolor{tempcolor1}4112&\cellcolor{tempcolor1}354&&\cellcolor{tempcolor1}5501&\cellcolor{tempcolor1}397&&\cellcolor{tempcolor1}8217&\cellcolor{tempcolor1}455\\ 
RAM&&\cellcolor{tempcolor2}544&\cellcolor{tempcolor3}\color{white}1154&&\cellcolor{tempcolor1}788&\cellcolor{tempcolor8}\color{white}1383&&\cellcolor{tempcolor1}1037&\cellcolor{tempcolor9}\color{white}1589&&\cellcolor{tempcolor1}2067&\cellcolor{tempcolor4}\color{white}2212&&\cellcolor{tempcolor1}4133&\cellcolor{tempcolor4}\color{white}3788&&\cellcolor{tempcolor1}5527&\cellcolor{tempcolor6}4365&&\cellcolor{tempcolor1}8252&\cellcolor{tempcolor5}5319\\ 
PCI&&\cellcolor{tempcolor4}\color{white}238&\cellcolor{tempcolor3}\color{white}2236&&\cellcolor{tempcolor7}\color{white}272&\cellcolor{tempcolor3}\color{white}2413&&\cellcolor{tempcolor9}\color{white}310&\cellcolor{tempcolor3}\color{white}2593&&\cellcolor{tempcolor9}\color{white}452&\cellcolor{tempcolor3}\color{white}3211&&\cellcolor{tempcolor8}\color{white}583&\cellcolor{tempcolor3}\color{white}3783&&\cellcolor{tempcolor8}\color{white}631&\cellcolor{tempcolor3}\color{white}4009&&\cellcolor{tempcolor8}\color{white}740&\cellcolor{tempcolor3}\color{white}4528\\ 
MII&&&&&\cellcolor{tempcolor5}920&\cellcolor{tempcolor3}\color{white}1909&&&&&&&&&&&\cellcolor{tempcolor2}6239&\cellcolor{tempcolor9}\color{white}9107&&&\\ 
BFBC&&\cellcolor{tempcolor3}\color{white}15625&\cellcolor{tempcolor3}\color{white}33037&&\cellcolor{tempcolor3}\color{white}18427&\cellcolor{tempcolor3}\color{white}35154&&\cellcolor{tempcolor3}\color{white}20115&\cellcolor{tempcolor8}\color{white}36270&&\cellcolor{tempcolor3}\color{white}23905&\cellcolor{tempcolor9}\color{white}38418&&\cellcolor{tempcolor3}\color{white}28185&\cellcolor{tempcolor9}\color{white}40406&&\cellcolor{tempcolor3}\color{white}30447&\cellcolor{tempcolor7}\color{white}41325&&\cellcolor{tempcolor3}\color{white}34288&\cellcolor{tempcolor7}\color{white}42727\\ 
BFBC*&&\cellcolor{tempcolor2}579&\cellcolor{tempcolor3}\color{white}1464&&\cellcolor{tempcolor2}895&\cellcolor{tempcolor3}\color{white}2077&&\cellcolor{tempcolor2}1195&\cellcolor{tempcolor3}\color{white}2720&&\cellcolor{tempcolor5}2448&\cellcolor{tempcolor3}\color{white}6127&&\cellcolor{tempcolor5}4961&\cellcolor{tempcolor3}\color{white}14980&&\cellcolor{tempcolor2}6474&\cellcolor{tempcolor3}\color{white}21447&&\cellcolor{tempcolor1}7794&\cellcolor{tempcolor3}\color{white}28928\\ 
\hline\noalign{\smallskip}
\textbf{PDF}&&&&&&&&&&&&&&&&&&&&&&&&\\ 
Rabin&&\cellcolor{tempcolor1}516&\cellcolor{tempcolor7}\color{white}658&&&&&\cellcolor{tempcolor2}957&\cellcolor{tempcolor7}\color{white}1213&&\cellcolor{tempcolor2}1728&\cellcolor{tempcolor7}\color{white}2281&&\cellcolor{tempcolor5}2880&\cellcolor{tempcolor9}\color{white}4304&&&&&\cellcolor{tempcolor4}\color{white}4341&\cellcolor{tempcolor8}\color{white}7778\\ 
Buzhash&&\cellcolor{tempcolor1}539&\cellcolor{tempcolor4}\color{white}577&&&&&\cellcolor{tempcolor2}946&\cellcolor{tempcolor4}\color{white}1080&&\cellcolor{tempcolor5}1612&\cellcolor{tempcolor7}\color{white}2038&&\cellcolor{tempcolor6}2555&\cellcolor{tempcolor9}\color{white}3768&&&&&\cellcolor{tempcolor4}\color{white}3645&\cellcolor{tempcolor8}\color{white}6652\\ 
Gear&&\cellcolor{tempcolor1}508&\cellcolor{tempcolor7}\color{white}667&&&&&\cellcolor{tempcolor1}1018&\cellcolor{tempcolor7}\color{white}1271&&\cellcolor{tempcolor1}1971&\cellcolor{tempcolor7}\color{white}2360&&\cellcolor{tempcolor2}3838&\cellcolor{tempcolor7}\color{white}4620&&&&&\cellcolor{tempcolor1}8352&\cellcolor{tempcolor4}\color{white}9399\\ 
Gear NC-1&&\cellcolor{tempcolor2}559&\cellcolor{tempcolor6}535&&\cellcolor{tempcolor5}985&\cellcolor{tempcolor6}866&&\cellcolor{tempcolor2}1103&\cellcolor{tempcolor6}939&&\cellcolor{tempcolor1}2172&\cellcolor{tempcolor6}1766&&\cellcolor{tempcolor2}4506&\cellcolor{tempcolor6}3240&&\cellcolor{tempcolor2}5088&\cellcolor{tempcolor6}3644&&\cellcolor{tempcolor2}8964&\cellcolor{tempcolor6}6351\\ 
Gear NC-2&&\cellcolor{tempcolor2}553&\cellcolor{tempcolor5}433&&\cellcolor{tempcolor2}911&\cellcolor{tempcolor5}634&&\cellcolor{tempcolor2}1096&\cellcolor{tempcolor5}718&&\cellcolor{tempcolor2}2232&\cellcolor{tempcolor5}1225&&\cellcolor{tempcolor2}4450&\cellcolor{tempcolor5}2269&&\cellcolor{tempcolor1}5416&\cellcolor{tempcolor5}2721&&\cellcolor{tempcolor2}8823&\cellcolor{tempcolor5}4181\\ 
Gear NC-3&&\cellcolor{tempcolor1}538&\cellcolor{tempcolor5}336&&\cellcolor{tempcolor2}859&\cellcolor{tempcolor5}508&&\cellcolor{tempcolor1}1084&\cellcolor{tempcolor2}584&&\cellcolor{tempcolor1}2165&\cellcolor{tempcolor2}946&&\cellcolor{tempcolor1}4314&\cellcolor{tempcolor2}1616&&\cellcolor{tempcolor1}5484&\cellcolor{tempcolor2}2002&&\cellcolor{tempcolor1}8575&\cellcolor{tempcolor2}3017\\ 
AE&&\cellcolor{tempcolor1}507&\cellcolor{tempcolor1}146&&\cellcolor{tempcolor1}772&\cellcolor{tempcolor1}202&&\cellcolor{tempcolor1}1039&\cellcolor{tempcolor1}247&&\cellcolor{tempcolor1}2109&\cellcolor{tempcolor1}372&&\cellcolor{tempcolor1}4209&\cellcolor{tempcolor1}568&&\cellcolor{tempcolor1}5628&\cellcolor{tempcolor1}743&&\cellcolor{tempcolor1}8399&\cellcolor{tempcolor1}1047\\ 
RAM&&\cellcolor{tempcolor5}618&\cellcolor{tempcolor3}\color{white}4188&&\cellcolor{tempcolor2}890&\cellcolor{tempcolor3}\color{white}5052&&\cellcolor{tempcolor2}1166&\cellcolor{tempcolor3}\color{white}5779&&\cellcolor{tempcolor2}2283&\cellcolor{tempcolor3}\color{white}8091&&\cellcolor{tempcolor2}4486&\cellcolor{tempcolor3}\color{white}11300&&\cellcolor{tempcolor2}5972&\cellcolor{tempcolor3}\color{white}13043&&\cellcolor{tempcolor2}8855&\cellcolor{tempcolor8}\color{white}15834\\ 
PCI&&\cellcolor{tempcolor4}\color{white}250&\cellcolor{tempcolor3}\color{white}2486&&\cellcolor{tempcolor7}\color{white}270&\cellcolor{tempcolor3}\color{white}2628&&\cellcolor{tempcolor9}\color{white}297&\cellcolor{tempcolor3}\color{white}2780&&\cellcolor{tempcolor9}\color{white}445&\cellcolor{tempcolor3}\color{white}3636&&\cellcolor{tempcolor8}\color{white}530&\cellcolor{tempcolor3}\color{white}4012&&\cellcolor{tempcolor8}\color{white}602&\cellcolor{tempcolor3}\color{white}4394&&\cellcolor{tempcolor8}\color{white}700&\cellcolor{tempcolor3}\color{white}4862\\ 
MII&&&&&\cellcolor{tempcolor2}685&\cellcolor{tempcolor3}\color{white}3343&&&&&&&&&&&\cellcolor{tempcolor7}\color{white}2397&\cellcolor{tempcolor3}\color{white}8825&&&\\ 
BFBC&&\cellcolor{tempcolor3}\color{white}2445&\cellcolor{tempcolor3}\color{white}8026&&\cellcolor{tempcolor3}\color{white}3447&\cellcolor{tempcolor3}\color{white}9364&&\cellcolor{tempcolor3}\color{white}4253&\cellcolor{tempcolor3}\color{white}10252&&\cellcolor{tempcolor3}\color{white}6652&\cellcolor{tempcolor8}\color{white}12286&&\cellcolor{tempcolor3}\color{white}10116&\cellcolor{tempcolor9}\color{white}14309&&\cellcolor{tempcolor3}\color{white}12068&\cellcolor{tempcolor7}\color{white}15171&&\cellcolor{tempcolor8}\color{white}15803&\cellcolor{tempcolor4}\color{white}16520\\ 
BFBC*&&\cellcolor{tempcolor1}481&\cellcolor{tempcolor3}\color{white}6907&&\cellcolor{tempcolor5}997&\cellcolor{tempcolor3}\color{white}10484&&\cellcolor{tempcolor1}1025&\cellcolor{tempcolor3}\color{white}9552&&\cellcolor{tempcolor1}2051&\cellcolor{tempcolor3}\color{white}14521&&\cellcolor{tempcolor1}4106&\cellcolor{tempcolor3}\color{white}31292&&\cellcolor{tempcolor1}5482&\cellcolor{tempcolor3}\color{white}25065&&\cellcolor{tempcolor1}8242&\cellcolor{tempcolor3}\color{white}40550\\ 
\hline\noalign{\smallskip}
\textbf{WEB}&&&&&&&&&&&&&&&&&&&&&&&&\\ 
Rabin&&\cellcolor{tempcolor2}451&\cellcolor{tempcolor4}\color{white}516&&&&&\cellcolor{tempcolor5}744&\cellcolor{tempcolor7}\color{white}998&&\cellcolor{tempcolor4}\color{white}1113&\cellcolor{tempcolor8}\color{white}1834&&\cellcolor{tempcolor7}\color{white}1511&\cellcolor{tempcolor3}\color{white}3184&&&&&\cellcolor{tempcolor9}\color{white}1832&\cellcolor{tempcolor3}\color{white}5120\\ 
Buzhash&&\cellcolor{tempcolor2}547&\cellcolor{tempcolor6}520&&&&&\cellcolor{tempcolor1}969&\cellcolor{tempcolor4}\color{white}1018&&\cellcolor{tempcolor5}1648&\cellcolor{tempcolor7}\color{white}1992&&\cellcolor{tempcolor6}2660&\cellcolor{tempcolor9}\color{white}3839&&&&&\cellcolor{tempcolor4}\color{white}3793&\cellcolor{tempcolor8}\color{white}6652\\ 
Gear&&\cellcolor{tempcolor1}516&\cellcolor{tempcolor4}\color{white}550&&&&&\cellcolor{tempcolor1}1046&\cellcolor{tempcolor4}\color{white}1111&&\cellcolor{tempcolor1}2075&\cellcolor{tempcolor4}\color{white}2184&&\cellcolor{tempcolor1}4123&\cellcolor{tempcolor4}\color{white}4267&&&&&\cellcolor{tempcolor1}8166&\cellcolor{tempcolor4}\color{white}8497\\ 
Gear NC-1&&\cellcolor{tempcolor2}564&\cellcolor{tempcolor5}356&&\cellcolor{tempcolor5}1010&\cellcolor{tempcolor5}665&&\cellcolor{tempcolor2}1129&\cellcolor{tempcolor5}728&&\cellcolor{tempcolor2}2264&\cellcolor{tempcolor5}1461&&\cellcolor{tempcolor2}4514&\cellcolor{tempcolor5}2889&&\cellcolor{tempcolor2}5068&\cellcolor{tempcolor5}3290&&\cellcolor{tempcolor2}9055&\cellcolor{tempcolor5}5710\\ 
Gear NC-2&&\cellcolor{tempcolor2}557&\cellcolor{tempcolor2}218&&\cellcolor{tempcolor5}922&\cellcolor{tempcolor2}368&&\cellcolor{tempcolor2}1112&\cellcolor{tempcolor2}436&&\cellcolor{tempcolor2}2229&\cellcolor{tempcolor2}889&&\cellcolor{tempcolor2}4529&\cellcolor{tempcolor2}1788&&\cellcolor{tempcolor1}5474&\cellcolor{tempcolor5}2266&&\cellcolor{tempcolor2}8952&\cellcolor{tempcolor5}3595\\ 
Gear NC-3&&\cellcolor{tempcolor1}539&\cellcolor{tempcolor1}135&&\cellcolor{tempcolor2}856&\cellcolor{tempcolor1}212&&\cellcolor{tempcolor1}1078&\cellcolor{tempcolor1}269&&\cellcolor{tempcolor1}2166&\cellcolor{tempcolor2}531&&\cellcolor{tempcolor1}4340&\cellcolor{tempcolor2}1109&&\cellcolor{tempcolor1}5525&\cellcolor{tempcolor2}1482&&\cellcolor{tempcolor1}8672&\cellcolor{tempcolor2}2211\\ 
AE&&\cellcolor{tempcolor1}490&\cellcolor{tempcolor1}137&&\cellcolor{tempcolor1}744&\cellcolor{tempcolor1}209&&\cellcolor{tempcolor1}1002&\cellcolor{tempcolor1}253&&\cellcolor{tempcolor1}2063&\cellcolor{tempcolor1}380&&\cellcolor{tempcolor1}4164&\cellcolor{tempcolor1}554&&\cellcolor{tempcolor1}5576&\cellcolor{tempcolor1}662&&\cellcolor{tempcolor1}8339&\cellcolor{tempcolor1}990\\ 
RAM&&\cellcolor{tempcolor2}601&\cellcolor{tempcolor3}\color{white}12033&&\cellcolor{tempcolor2}882&\cellcolor{tempcolor3}\color{white}14679&&\cellcolor{tempcolor2}1166&\cellcolor{tempcolor3}\color{white}17054&&\cellcolor{tempcolor2}2332&\cellcolor{tempcolor3}\color{white}24339&&\cellcolor{tempcolor2}4650&\cellcolor{tempcolor3}\color{white}34631&&\cellcolor{tempcolor2}6213&\cellcolor{tempcolor3}\color{white}39897&&\cellcolor{tempcolor2}9245&\cellcolor{tempcolor3}\color{white}48800\\ 
PCI&&\cellcolor{tempcolor1}495&\cellcolor{tempcolor3}\color{white}3321&&\cellcolor{tempcolor5}598&\cellcolor{tempcolor3}\color{white}3796&&\cellcolor{tempcolor6}674&\cellcolor{tempcolor3}\color{white}4255&&\cellcolor{tempcolor4}\color{white}1090&\cellcolor{tempcolor3}\color{white}6270&&\cellcolor{tempcolor7}\color{white}1426&\cellcolor{tempcolor3}\color{white}11024&&\cellcolor{tempcolor9}\color{white}1704&\cellcolor{tempcolor3}\color{white}11118&&\cellcolor{tempcolor9}\color{white}2043&\cellcolor{tempcolor3}\color{white}13215\\ 
MII&&&&&\cellcolor{tempcolor2}856&\cellcolor{tempcolor3}\color{white}1981&&&&&&&&&&&\cellcolor{tempcolor2}5018&\cellcolor{tempcolor3}\color{white}11503&&&\\ 
BFBC&&\cellcolor{tempcolor3}\color{white}1256&\cellcolor{tempcolor3}\color{white}18359&&\cellcolor{tempcolor3}\color{white}1872&\cellcolor{tempcolor3}\color{white}22389&&\cellcolor{tempcolor3}\color{white}2422&\cellcolor{tempcolor3}\color{white}25448&&\cellcolor{tempcolor3}\color{white}4357&\cellcolor{tempcolor3}\color{white}34050&&\cellcolor{tempcolor8}\color{white}7558&\cellcolor{tempcolor3}\color{white}44714&&\cellcolor{tempcolor9}\color{white}9516&\cellcolor{tempcolor3}\color{white}50090&&\cellcolor{tempcolor7}\color{white}13125&\cellcolor{tempcolor3}\color{white}58682\\ 
BFBC*&&\cellcolor{tempcolor2}453&\cellcolor{tempcolor3}\color{white}11319&&\cellcolor{tempcolor1}763&\cellcolor{tempcolor3}\color{white}13940&&\cellcolor{tempcolor8}\color{white}1875&\cellcolor{tempcolor3}\color{white}13780&&\cellcolor{tempcolor1}2026&\cellcolor{tempcolor3}\color{white}12155&&\cellcolor{tempcolor1}4107&\cellcolor{tempcolor3}\color{white}24116&&\cellcolor{tempcolor1}5492&\cellcolor{tempcolor3}\color{white}29886&&\cellcolor{tempcolor1}8195&\cellcolor{tempcolor3}\color{white}52117\\ 
\hline\noalign{\smallskip}
\textbf{CODE}&&&&&&&&&&&&&&&&&&&&&&&&\\ 
Rabin&&\cellcolor{tempcolor1}542&\cellcolor{tempcolor3}\color{white}3006&&&&&\cellcolor{tempcolor1}1051&\cellcolor{tempcolor3}\color{white}4281&&\cellcolor{tempcolor1}2018&\cellcolor{tempcolor3}\color{white}6210&&\cellcolor{tempcolor2}3741&\cellcolor{tempcolor3}\color{white}9208&&&&&\cellcolor{tempcolor5}6447&\cellcolor{tempcolor3}\color{white}14125\\ 
Buzhash&&\cellcolor{tempcolor2}574&\cellcolor{tempcolor3}\color{white}3084&&&&&\cellcolor{tempcolor1}1041&\cellcolor{tempcolor3}\color{white}4229&&\cellcolor{tempcolor1}1943&\cellcolor{tempcolor3}\color{white}6014&&\cellcolor{tempcolor2}3406&\cellcolor{tempcolor3}\color{white}8638&&&&&\cellcolor{tempcolor6}5454&\cellcolor{tempcolor3}\color{white}12580\\ 
Gear&&\cellcolor{tempcolor1}519&\cellcolor{tempcolor3}\color{white}2951&&&&&\cellcolor{tempcolor1}1037&\cellcolor{tempcolor3}\color{white}4261&&\cellcolor{tempcolor1}2062&\cellcolor{tempcolor3}\color{white}6278&&\cellcolor{tempcolor1}4249&\cellcolor{tempcolor3}\color{white}9906&&&&&\cellcolor{tempcolor1}8385&\cellcolor{tempcolor3}\color{white}16303\\ 
Gear NC-1&&\cellcolor{tempcolor2}568&\cellcolor{tempcolor3}\color{white}3045&&\cellcolor{tempcolor6}1021&\cellcolor{tempcolor3}\color{white}4118&&\cellcolor{tempcolor2}1138&\cellcolor{tempcolor3}\color{white}4352&&\cellcolor{tempcolor2}2286&\cellcolor{tempcolor3}\color{white}6278&&\cellcolor{tempcolor2}4568&\cellcolor{tempcolor3}\color{white}9201&&\cellcolor{tempcolor2}5108&\cellcolor{tempcolor3}\color{white}9809&&\cellcolor{tempcolor2}9247&\cellcolor{tempcolor9}\color{white}14227\\ 
Gear NC-2&&\cellcolor{tempcolor2}558&\cellcolor{tempcolor3}\color{white}3004&&\cellcolor{tempcolor5}934&\cellcolor{tempcolor3}\color{white}3895&&\cellcolor{tempcolor2}1123&\cellcolor{tempcolor3}\color{white}4273&&\cellcolor{tempcolor2}2252&\cellcolor{tempcolor3}\color{white}6097&&\cellcolor{tempcolor2}4495&\cellcolor{tempcolor3}\color{white}8738&&\cellcolor{tempcolor1}5430&\cellcolor{tempcolor8}\color{white}9671&&\cellcolor{tempcolor2}9146&\cellcolor{tempcolor9}\color{white}12810\\ 
Gear NC-3&&\cellcolor{tempcolor1}540&\cellcolor{tempcolor1}159&&\cellcolor{tempcolor2}859&\cellcolor{tempcolor3}\color{white}3723&&\cellcolor{tempcolor1}1083&\cellcolor{tempcolor3}\color{white}4180&&\cellcolor{tempcolor1}2171&\cellcolor{tempcolor3}\color{white}5934&&\cellcolor{tempcolor2}4387&\cellcolor{tempcolor3}\color{white}8475&&\cellcolor{tempcolor1}5565&\cellcolor{tempcolor8}\color{white}9584&&\cellcolor{tempcolor2}8727&\cellcolor{tempcolor9}\color{white}12114\\ 
AE&&\cellcolor{tempcolor1}484&\cellcolor{tempcolor1}145&&\cellcolor{tempcolor1}758&\cellcolor{tempcolor1}217&&\cellcolor{tempcolor1}1043&\cellcolor{tempcolor1}289&&\cellcolor{tempcolor2}2242&\cellcolor{tempcolor2}570&&\cellcolor{tempcolor2}4647&\cellcolor{tempcolor2}1136&&\cellcolor{tempcolor2}6268&\cellcolor{tempcolor2}1529&&\cellcolor{tempcolor2}9472&\cellcolor{tempcolor2}2327\\ 
RAM&&\cellcolor{tempcolor3}\color{white}15989&\cellcolor{tempcolor3}\color{white}759876&&\cellcolor{tempcolor3}\color{white}26865&\cellcolor{tempcolor3}\color{white}975111&&\cellcolor{tempcolor3}\color{white}33879&\cellcolor{tempcolor3}\color{white}1070323&&\cellcolor{tempcolor3}\color{white}66846&\cellcolor{tempcolor3}\color{white}1733430&&\cellcolor{tempcolor3}\color{white}129537&\cellcolor{tempcolor3}\color{white}2407519&&\cellcolor{tempcolor3}\color{white}169158&\cellcolor{tempcolor3}\color{white}2758360&&\cellcolor{tempcolor3}\color{white}250739&\cellcolor{tempcolor3}\color{white}3364372\\ 
PCI&&\cellcolor{tempcolor3}\color{white}4358&\cellcolor{tempcolor3}\color{white}47128&&\cellcolor{tempcolor3}\color{white}5124&\cellcolor{tempcolor3}\color{white}52722&&\cellcolor{tempcolor3}\color{white}5978&\cellcolor{tempcolor3}\color{white}58003&&\cellcolor{tempcolor3}\color{white}12021&\cellcolor{tempcolor3}\color{white}115074&&\cellcolor{tempcolor3}\color{white}17925&\cellcolor{tempcolor3}\color{white}173403&&\cellcolor{tempcolor3}\color{white}21797&\cellcolor{tempcolor3}\color{white}186508&&\cellcolor{tempcolor3}\color{white}26463&\cellcolor{tempcolor3}\color{white}299234\\ 
MII&&&&&\cellcolor{tempcolor5}563&\cellcolor{tempcolor3}\color{white}3917&&&&&&&&&&&\cellcolor{tempcolor7}\color{white}2334&\cellcolor{tempcolor3}\color{white}13300&&&\\ 
BFBC&&\cellcolor{tempcolor2}458&\cellcolor{tempcolor3}\color{white}3046&&\cellcolor{tempcolor1}734&\cellcolor{tempcolor3}\color{white}3853&&\cellcolor{tempcolor1}1002&\cellcolor{tempcolor3}\color{white}4501&&\cellcolor{tempcolor1}2078&\cellcolor{tempcolor3}\color{white}6466&&\cellcolor{tempcolor1}4207&\cellcolor{tempcolor3}\color{white}9167&&\cellcolor{tempcolor1}5642&\cellcolor{tempcolor3}\color{white}10609&&\cellcolor{tempcolor1}8435&\cellcolor{tempcolor9}\color{white}12932\\ 
BFBC*&&\cellcolor{tempcolor1}506&\cellcolor{tempcolor3}\color{white}7706&&\cellcolor{tempcolor1}764&\cellcolor{tempcolor3}\color{white}16540&&\cellcolor{tempcolor1}1019&\cellcolor{tempcolor3}\color{white}32168&&\cellcolor{tempcolor6}2824&\cellcolor{tempcolor3}\color{white}24637&&\cellcolor{tempcolor1}4084&\cellcolor{tempcolor3}\color{white}48854&&\cellcolor{tempcolor1}5473&\cellcolor{tempcolor3}\color{white}56841&&\cellcolor{tempcolor1}8189&\cellcolor{tempcolor3}\color{white}127743\\ 
\hline
\end{longtable}}
\end{center}
\color{black}

%% file: fig/dedup_rabin_window_sizes_pdf.tex
% Created by tikzDevice version 0.12.6 on 2024-04-19 17:33:43
% !TEX encoding = UTF-8 Unicode
\begin{tikzpicture}[x=1pt,y=1pt]
\definecolor{fillColor}{RGB}{255,255,255}
\path[use as bounding box,fill=fillColor,fill opacity=0.00] (0,0) rectangle (144.54,130.09);
\begin{scope}
\path[clip] (  0.00,  0.00) rectangle (144.54,130.09);
\definecolor{drawColor}{RGB}{255,255,255}
\definecolor{fillColor}{RGB}{255,255,255}

\path[draw=drawColor,line width= 0.6pt,line join=round,line cap=round,fill=fillColor] (  0.00,  0.00) rectangle (144.54,130.09);
\end{scope}
\begin{scope}
\path[clip] ( 38.56, 40.85) rectangle (139.04,124.59);
\definecolor{fillColor}{RGB}{255,255,255}

\path[fill=fillColor] ( 38.56, 40.85) rectangle (139.04,124.59);
\definecolor{drawColor}{gray}{0.92}

\path[draw=drawColor,line width= 0.3pt,line join=round] ( 38.56, 46.20) --
	(139.04, 46.20);

\path[draw=drawColor,line width= 0.3pt,line join=round] ( 38.56, 72.42) --
	(139.04, 72.42);

\path[draw=drawColor,line width= 0.3pt,line join=round] ( 38.56, 98.64) --
	(139.04, 98.64);

\path[draw=drawColor,line width= 0.6pt,line join=round] ( 38.56, 59.31) --
	(139.04, 59.31);

\path[draw=drawColor,line width= 0.6pt,line join=round] ( 38.56, 85.53) --
	(139.04, 85.53);

\path[draw=drawColor,line width= 0.6pt,line join=round] ( 38.56,111.75) --
	(139.04,111.75);

\path[draw=drawColor,line width= 0.6pt,line join=round] ( 50.15, 40.85) --
	( 50.15,124.59);

\path[draw=drawColor,line width= 0.6pt,line join=round] ( 69.47, 40.85) --
	( 69.47,124.59);

\path[draw=drawColor,line width= 0.6pt,line join=round] ( 88.80, 40.85) --
	( 88.80,124.59);

\path[draw=drawColor,line width= 0.6pt,line join=round] (108.12, 40.85) --
	(108.12,124.59);

\path[draw=drawColor,line width= 0.6pt,line join=round] (127.45, 40.85) --
	(127.45,124.59);
\definecolor{drawColor}{RGB}{55,78,85}

\path[draw=drawColor,line width= 0.6pt,line join=round] ( 49.34,118.93) --
	( 68.67,104.01) --
	( 87.99, 85.31) --
	(107.32, 64.73) --
	(126.64, 50.02);
\definecolor{drawColor}{RGB}{223,143,68}

\path[draw=drawColor,line width= 0.6pt,line join=round] ( 49.67,119.95) --
	( 68.99,103.87) --
	( 88.31, 83.33) --
	(107.64, 65.18) --
	(126.96, 48.07);
\definecolor{drawColor}{RGB}{0,161,213}

\path[draw=drawColor,line width= 0.6pt,line join=round] ( 49.99,120.78) --
	( 69.31,102.43) --
	( 88.64, 81.47) --
	(107.96, 64.38) --
	(127.28, 48.48);
\definecolor{drawColor}{RGB}{178,71,69}

\path[draw=drawColor,line width= 0.6pt,line join=round] ( 50.31,115.84) --
	( 69.63,100.53) --
	( 88.96, 82.31) --
	(108.28, 64.40) --
	(127.61, 47.75);
\definecolor{drawColor}{RGB}{121,175,151}

\path[draw=drawColor,line width= 0.6pt,line join=round] ( 50.63,115.94) --
	( 69.96,100.68) --
	( 89.28, 78.86) --
	(108.60, 60.52) --
	(127.93, 44.80);
\definecolor{drawColor}{RGB}{106,101,153}

\path[draw=drawColor,line width= 0.6pt,line join=round] ( 50.95,118.98) --
	( 70.28, 97.32) --
	( 89.60, 78.02) --
	(108.93, 61.66) --
	(128.25, 44.66);
\definecolor{drawColor}{RGB}{55,78,85}

\path[draw=drawColor,line width= 0.4pt,line join=round,line cap=round,fill=fillColor] ( 49.34,118.93) circle (  1.43);

\path[draw=drawColor,line width= 0.4pt,line join=round,line cap=round,fill=fillColor] ( 68.67,104.01) circle (  1.43);

\path[draw=drawColor,line width= 0.4pt,line join=round,line cap=round,fill=fillColor] ( 87.99, 85.31) circle (  1.43);

\path[draw=drawColor,line width= 0.4pt,line join=round,line cap=round,fill=fillColor] (107.32, 64.73) circle (  1.43);

\path[draw=drawColor,line width= 0.4pt,line join=round,line cap=round,fill=fillColor] (126.64, 50.02) circle (  1.43);
\definecolor{drawColor}{RGB}{223,143,68}

\path[draw=drawColor,line width= 0.4pt,line join=round,line cap=round,fill=fillColor] ( 49.67,119.95) circle (  1.43);

\path[draw=drawColor,line width= 0.4pt,line join=round,line cap=round,fill=fillColor] ( 68.99,103.87) circle (  1.43);

\path[draw=drawColor,line width= 0.4pt,line join=round,line cap=round,fill=fillColor] ( 88.31, 83.33) circle (  1.43);

\path[draw=drawColor,line width= 0.4pt,line join=round,line cap=round,fill=fillColor] (107.64, 65.18) circle (  1.43);

\path[draw=drawColor,line width= 0.4pt,line join=round,line cap=round,fill=fillColor] (126.96, 48.07) circle (  1.43);
\definecolor{drawColor}{RGB}{0,161,213}

\path[draw=drawColor,line width= 0.4pt,line join=round,line cap=round,fill=fillColor] ( 49.99,120.78) circle (  1.43);

\path[draw=drawColor,line width= 0.4pt,line join=round,line cap=round,fill=fillColor] ( 69.31,102.43) circle (  1.43);

\path[draw=drawColor,line width= 0.4pt,line join=round,line cap=round,fill=fillColor] ( 88.64, 81.47) circle (  1.43);

\path[draw=drawColor,line width= 0.4pt,line join=round,line cap=round,fill=fillColor] (107.96, 64.38) circle (  1.43);

\path[draw=drawColor,line width= 0.4pt,line join=round,line cap=round,fill=fillColor] (127.28, 48.48) circle (  1.43);
\definecolor{drawColor}{RGB}{178,71,69}

\path[draw=drawColor,line width= 0.4pt,line join=round,line cap=round,fill=fillColor] ( 50.31,115.84) circle (  1.43);

\path[draw=drawColor,line width= 0.4pt,line join=round,line cap=round,fill=fillColor] ( 69.63,100.53) circle (  1.43);

\path[draw=drawColor,line width= 0.4pt,line join=round,line cap=round,fill=fillColor] ( 88.96, 82.31) circle (  1.43);

\path[draw=drawColor,line width= 0.4pt,line join=round,line cap=round,fill=fillColor] (108.28, 64.40) circle (  1.43);

\path[draw=drawColor,line width= 0.4pt,line join=round,line cap=round,fill=fillColor] (127.61, 47.75) circle (  1.43);
\definecolor{drawColor}{RGB}{121,175,151}

\path[draw=drawColor,line width= 0.4pt,line join=round,line cap=round,fill=fillColor] ( 50.63,115.94) circle (  1.43);

\path[draw=drawColor,line width= 0.4pt,line join=round,line cap=round,fill=fillColor] ( 69.96,100.68) circle (  1.43);

\path[draw=drawColor,line width= 0.4pt,line join=round,line cap=round,fill=fillColor] ( 89.28, 78.86) circle (  1.43);

\path[draw=drawColor,line width= 0.4pt,line join=round,line cap=round,fill=fillColor] (108.60, 60.52) circle (  1.43);

\path[draw=drawColor,line width= 0.4pt,line join=round,line cap=round,fill=fillColor] (127.93, 44.80) circle (  1.43);
\definecolor{drawColor}{RGB}{106,101,153}

\path[draw=drawColor,line width= 0.4pt,line join=round,line cap=round,fill=fillColor] ( 50.95,118.98) circle (  1.43);

\path[draw=drawColor,line width= 0.4pt,line join=round,line cap=round,fill=fillColor] ( 70.28, 97.32) circle (  1.43);

\path[draw=drawColor,line width= 0.4pt,line join=round,line cap=round,fill=fillColor] ( 89.60, 78.02) circle (  1.43);

\path[draw=drawColor,line width= 0.4pt,line join=round,line cap=round,fill=fillColor] (108.93, 61.66) circle (  1.43);

\path[draw=drawColor,line width= 0.4pt,line join=round,line cap=round,fill=fillColor] (128.25, 44.66) circle (  1.43);
\definecolor{drawColor}{gray}{0.20}

\path[draw=drawColor,line width= 0.6pt,line join=round,line cap=round] ( 38.56, 40.85) rectangle (139.04,124.59);
\end{scope}
\begin{scope}
\path[clip] (  0.00,  0.00) rectangle (144.54,130.09);
\definecolor{drawColor}{gray}{0.30}

\node[text=drawColor,anchor=base east,inner sep=0pt, outer sep=0pt, scale=  0.88] at ( 33.61, 56.28) {0.03};

\node[text=drawColor,anchor=base east,inner sep=0pt, outer sep=0pt, scale=  0.88] at ( 33.61, 82.50) {0.04};

\node[text=drawColor,anchor=base east,inner sep=0pt, outer sep=0pt, scale=  0.88] at ( 33.61,108.72) {0.05};
\end{scope}
\begin{scope}
\path[clip] (  0.00,  0.00) rectangle (144.54,130.09);
\definecolor{drawColor}{gray}{0.20}

\path[draw=drawColor,line width= 0.6pt,line join=round] ( 35.81, 59.31) --
	( 38.56, 59.31);

\path[draw=drawColor,line width= 0.6pt,line join=round] ( 35.81, 85.53) --
	( 38.56, 85.53);

\path[draw=drawColor,line width= 0.6pt,line join=round] ( 35.81,111.75) --
	( 38.56,111.75);
\end{scope}
\begin{scope}
\path[clip] (  0.00,  0.00) rectangle (144.54,130.09);
\definecolor{drawColor}{gray}{0.20}

\path[draw=drawColor,line width= 0.6pt,line join=round] ( 50.15, 38.10) --
	( 50.15, 40.85);

\path[draw=drawColor,line width= 0.6pt,line join=round] ( 69.47, 38.10) --
	( 69.47, 40.85);

\path[draw=drawColor,line width= 0.6pt,line join=round] ( 88.80, 38.10) --
	( 88.80, 40.85);

\path[draw=drawColor,line width= 0.6pt,line join=round] (108.12, 38.10) --
	(108.12, 40.85);

\path[draw=drawColor,line width= 0.6pt,line join=round] (127.45, 38.10) --
	(127.45, 40.85);
\end{scope}
\begin{scope}
\path[clip] (  0.00,  0.00) rectangle (144.54,130.09);
\definecolor{drawColor}{gray}{0.30}

\node[text=drawColor,rotate= 45.00,anchor=base east,inner sep=0pt, outer sep=0pt, scale=  0.88] at ( 54.44, 31.62) {512};

\node[text=drawColor,rotate= 45.00,anchor=base east,inner sep=0pt, outer sep=0pt, scale=  0.88] at ( 73.76, 31.62) {1024};

\node[text=drawColor,rotate= 45.00,anchor=base east,inner sep=0pt, outer sep=0pt, scale=  0.88] at ( 93.08, 31.62) {2048};

\node[text=drawColor,rotate= 45.00,anchor=base east,inner sep=0pt, outer sep=0pt, scale=  0.88] at (112.41, 31.62) {4096};

\node[text=drawColor,rotate= 45.00,anchor=base east,inner sep=0pt, outer sep=0pt, scale=  0.88] at (131.73, 31.62) {8192};
\end{scope}
\begin{scope}
\path[clip] (  0.00,  0.00) rectangle (144.54,130.09);
\definecolor{drawColor}{RGB}{0,0,0}

\node[text=drawColor,anchor=base,inner sep=0pt, outer sep=0pt, scale=  1.10] at ( 88.80,  7.64) {Target Chunk Size};
\end{scope}
\begin{scope}
\path[clip] (  0.00,  0.00) rectangle (144.54,130.09);
\definecolor{drawColor}{RGB}{0,0,0}

\node[text=drawColor,rotate= 90.00,anchor=base,inner sep=0pt, outer sep=0pt, scale=  1.10] at ( 13.08, 82.72) {Dedup. Ratio};
\end{scope}
\end{tikzpicture}

%% file: fig/dedup_rabin_window_sizes_web.tex
% Created by tikzDevice version 0.12.6 on 2024-04-19 17:33:45
% !TEX encoding = UTF-8 Unicode
\begin{tikzpicture}[x=1pt,y=1pt]
\definecolor{fillColor}{RGB}{255,255,255}
\path[use as bounding box,fill=fillColor,fill opacity=0.00] (0,0) rectangle (144.54,130.09);
\begin{scope}
\path[clip] (  0.00,  0.00) rectangle (144.54,130.09);
\definecolor{drawColor}{RGB}{255,255,255}
\definecolor{fillColor}{RGB}{255,255,255}

\path[draw=drawColor,line width= 0.6pt,line join=round,line cap=round,fill=fillColor] (  0.00,  0.00) rectangle (144.54,130.09);
\end{scope}
\begin{scope}
\path[clip] ( 38.56, 40.85) rectangle (139.04,124.59);
\definecolor{fillColor}{RGB}{255,255,255}

\path[fill=fillColor] ( 38.56, 40.85) rectangle (139.04,124.59);
\definecolor{drawColor}{gray}{0.92}

\path[draw=drawColor,line width= 0.3pt,line join=round] ( 38.56, 58.73) --
	(139.04, 58.73);

\path[draw=drawColor,line width= 0.3pt,line join=round] ( 38.56, 81.48) --
	(139.04, 81.48);

\path[draw=drawColor,line width= 0.3pt,line join=round] ( 38.56,104.23) --
	(139.04,104.23);

\path[draw=drawColor,line width= 0.6pt,line join=round] ( 38.56, 47.35) --
	(139.04, 47.35);

\path[draw=drawColor,line width= 0.6pt,line join=round] ( 38.56, 70.10) --
	(139.04, 70.10);

\path[draw=drawColor,line width= 0.6pt,line join=round] ( 38.56, 92.85) --
	(139.04, 92.85);

\path[draw=drawColor,line width= 0.6pt,line join=round] ( 38.56,115.60) --
	(139.04,115.60);

\path[draw=drawColor,line width= 0.6pt,line join=round] ( 50.15, 40.85) --
	( 50.15,124.59);

\path[draw=drawColor,line width= 0.6pt,line join=round] ( 69.47, 40.85) --
	( 69.47,124.59);

\path[draw=drawColor,line width= 0.6pt,line join=round] ( 88.80, 40.85) --
	( 88.80,124.59);

\path[draw=drawColor,line width= 0.6pt,line join=round] (108.12, 40.85) --
	(108.12,124.59);

\path[draw=drawColor,line width= 0.6pt,line join=round] (127.45, 40.85) --
	(127.45,124.59);
\definecolor{drawColor}{RGB}{55,78,85}

\path[draw=drawColor,line width= 0.6pt,line join=round] ( 49.34,112.85) --
	( 68.67,100.49) --
	( 87.99, 85.53) --
	(107.32, 62.21) --
	(126.64, 44.66);
\definecolor{drawColor}{RGB}{223,143,68}

\path[draw=drawColor,line width= 0.6pt,line join=round] ( 49.67,114.77) --
	( 68.99,102.54) --
	( 88.31, 88.12) --
	(107.64, 74.45) --
	(126.96, 60.60);
\definecolor{drawColor}{RGB}{0,161,213}

\path[draw=drawColor,line width= 0.6pt,line join=round] ( 49.99,115.28) --
	( 69.31,102.43) --
	( 88.64, 89.93) --
	(107.96, 70.64) --
	(127.28, 58.21);
\definecolor{drawColor}{RGB}{178,71,69}

\path[draw=drawColor,line width= 0.6pt,line join=round] ( 50.31,114.50) --
	( 69.63,101.83) --
	( 88.96, 88.81) --
	(108.28, 71.96) --
	(127.61, 59.66);
\definecolor{drawColor}{RGB}{121,175,151}

\path[draw=drawColor,line width= 0.6pt,line join=round] ( 50.63,115.10) --
	( 69.96,102.75) --
	( 89.28, 90.68) --
	(108.60, 76.97) --
	(127.93, 65.61);
\definecolor{drawColor}{RGB}{106,101,153}

\path[draw=drawColor,line width= 0.6pt,line join=round] ( 50.95,120.78) --
	( 70.28,104.26) --
	( 89.60, 89.25) --
	(108.93, 74.97) --
	(128.25, 55.90);
\definecolor{drawColor}{RGB}{55,78,85}

\path[draw=drawColor,line width= 0.4pt,line join=round,line cap=round,fill=fillColor] ( 49.34,112.85) circle (  1.43);

\path[draw=drawColor,line width= 0.4pt,line join=round,line cap=round,fill=fillColor] ( 68.67,100.49) circle (  1.43);

\path[draw=drawColor,line width= 0.4pt,line join=round,line cap=round,fill=fillColor] ( 87.99, 85.53) circle (  1.43);

\path[draw=drawColor,line width= 0.4pt,line join=round,line cap=round,fill=fillColor] (107.32, 62.21) circle (  1.43);

\path[draw=drawColor,line width= 0.4pt,line join=round,line cap=round,fill=fillColor] (126.64, 44.66) circle (  1.43);
\definecolor{drawColor}{RGB}{223,143,68}

\path[draw=drawColor,line width= 0.4pt,line join=round,line cap=round,fill=fillColor] ( 49.67,114.77) circle (  1.43);

\path[draw=drawColor,line width= 0.4pt,line join=round,line cap=round,fill=fillColor] ( 68.99,102.54) circle (  1.43);

\path[draw=drawColor,line width= 0.4pt,line join=round,line cap=round,fill=fillColor] ( 88.31, 88.12) circle (  1.43);

\path[draw=drawColor,line width= 0.4pt,line join=round,line cap=round,fill=fillColor] (107.64, 74.45) circle (  1.43);

\path[draw=drawColor,line width= 0.4pt,line join=round,line cap=round,fill=fillColor] (126.96, 60.60) circle (  1.43);
\definecolor{drawColor}{RGB}{0,161,213}

\path[draw=drawColor,line width= 0.4pt,line join=round,line cap=round,fill=fillColor] ( 49.99,115.28) circle (  1.43);

\path[draw=drawColor,line width= 0.4pt,line join=round,line cap=round,fill=fillColor] ( 69.31,102.43) circle (  1.43);

\path[draw=drawColor,line width= 0.4pt,line join=round,line cap=round,fill=fillColor] ( 88.64, 89.93) circle (  1.43);

\path[draw=drawColor,line width= 0.4pt,line join=round,line cap=round,fill=fillColor] (107.96, 70.64) circle (  1.43);

\path[draw=drawColor,line width= 0.4pt,line join=round,line cap=round,fill=fillColor] (127.28, 58.21) circle (  1.43);
\definecolor{drawColor}{RGB}{178,71,69}

\path[draw=drawColor,line width= 0.4pt,line join=round,line cap=round,fill=fillColor] ( 50.31,114.50) circle (  1.43);

\path[draw=drawColor,line width= 0.4pt,line join=round,line cap=round,fill=fillColor] ( 69.63,101.83) circle (  1.43);

\path[draw=drawColor,line width= 0.4pt,line join=round,line cap=round,fill=fillColor] ( 88.96, 88.81) circle (  1.43);

\path[draw=drawColor,line width= 0.4pt,line join=round,line cap=round,fill=fillColor] (108.28, 71.96) circle (  1.43);

\path[draw=drawColor,line width= 0.4pt,line join=round,line cap=round,fill=fillColor] (127.61, 59.66) circle (  1.43);
\definecolor{drawColor}{RGB}{121,175,151}

\path[draw=drawColor,line width= 0.4pt,line join=round,line cap=round,fill=fillColor] ( 50.63,115.10) circle (  1.43);

\path[draw=drawColor,line width= 0.4pt,line join=round,line cap=round,fill=fillColor] ( 69.96,102.75) circle (  1.43);

\path[draw=drawColor,line width= 0.4pt,line join=round,line cap=round,fill=fillColor] ( 89.28, 90.68) circle (  1.43);

\path[draw=drawColor,line width= 0.4pt,line join=round,line cap=round,fill=fillColor] (108.60, 76.97) circle (  1.43);

\path[draw=drawColor,line width= 0.4pt,line join=round,line cap=round,fill=fillColor] (127.93, 65.61) circle (  1.43);
\definecolor{drawColor}{RGB}{106,101,153}

\path[draw=drawColor,line width= 0.4pt,line join=round,line cap=round,fill=fillColor] ( 50.95,120.78) circle (  1.43);

\path[draw=drawColor,line width= 0.4pt,line join=round,line cap=round,fill=fillColor] ( 70.28,104.26) circle (  1.43);

\path[draw=drawColor,line width= 0.4pt,line join=round,line cap=round,fill=fillColor] ( 89.60, 89.25) circle (  1.43);

\path[draw=drawColor,line width= 0.4pt,line join=round,line cap=round,fill=fillColor] (108.93, 74.97) circle (  1.43);

\path[draw=drawColor,line width= 0.4pt,line join=round,line cap=round,fill=fillColor] (128.25, 55.90) circle (  1.43);
\definecolor{drawColor}{gray}{0.20}

\path[draw=drawColor,line width= 0.6pt,line join=round,line cap=round] ( 38.56, 40.85) rectangle (139.04,124.59);
\end{scope}
\begin{scope}
\path[clip] (  0.00,  0.00) rectangle (144.54,130.09);
\definecolor{drawColor}{gray}{0.30}

\node[text=drawColor,anchor=base east,inner sep=0pt, outer sep=0pt, scale=  0.88] at ( 33.61, 44.32) {0.68};

\node[text=drawColor,anchor=base east,inner sep=0pt, outer sep=0pt, scale=  0.88] at ( 33.61, 67.07) {0.69};

\node[text=drawColor,anchor=base east,inner sep=0pt, outer sep=0pt, scale=  0.88] at ( 33.61, 89.82) {0.70};

\node[text=drawColor,anchor=base east,inner sep=0pt, outer sep=0pt, scale=  0.88] at ( 33.61,112.57) {0.71};
\end{scope}
\begin{scope}
\path[clip] (  0.00,  0.00) rectangle (144.54,130.09);
\definecolor{drawColor}{gray}{0.20}

\path[draw=drawColor,line width= 0.6pt,line join=round] ( 35.81, 47.35) --
	( 38.56, 47.35);

\path[draw=drawColor,line width= 0.6pt,line join=round] ( 35.81, 70.10) --
	( 38.56, 70.10);

\path[draw=drawColor,line width= 0.6pt,line join=round] ( 35.81, 92.85) --
	( 38.56, 92.85);

\path[draw=drawColor,line width= 0.6pt,line join=round] ( 35.81,115.60) --
	( 38.56,115.60);
\end{scope}
\begin{scope}
\path[clip] (  0.00,  0.00) rectangle (144.54,130.09);
\definecolor{drawColor}{gray}{0.20}

\path[draw=drawColor,line width= 0.6pt,line join=round] ( 50.15, 38.10) --
	( 50.15, 40.85);

\path[draw=drawColor,line width= 0.6pt,line join=round] ( 69.47, 38.10) --
	( 69.47, 40.85);

\path[draw=drawColor,line width= 0.6pt,line join=round] ( 88.80, 38.10) --
	( 88.80, 40.85);

\path[draw=drawColor,line width= 0.6pt,line join=round] (108.12, 38.10) --
	(108.12, 40.85);

\path[draw=drawColor,line width= 0.6pt,line join=round] (127.45, 38.10) --
	(127.45, 40.85);
\end{scope}
\begin{scope}
\path[clip] (  0.00,  0.00) rectangle (144.54,130.09);
\definecolor{drawColor}{gray}{0.30}

\node[text=drawColor,rotate= 45.00,anchor=base east,inner sep=0pt, outer sep=0pt, scale=  0.88] at ( 54.44, 31.62) {512};

\node[text=drawColor,rotate= 45.00,anchor=base east,inner sep=0pt, outer sep=0pt, scale=  0.88] at ( 73.76, 31.62) {1024};

\node[text=drawColor,rotate= 45.00,anchor=base east,inner sep=0pt, outer sep=0pt, scale=  0.88] at ( 93.08, 31.62) {2048};

\node[text=drawColor,rotate= 45.00,anchor=base east,inner sep=0pt, outer sep=0pt, scale=  0.88] at (112.41, 31.62) {4096};

\node[text=drawColor,rotate= 45.00,anchor=base east,inner sep=0pt, outer sep=0pt, scale=  0.88] at (131.73, 31.62) {8192};
\end{scope}
\begin{scope}
\path[clip] (  0.00,  0.00) rectangle (144.54,130.09);
\definecolor{drawColor}{RGB}{0,0,0}

\node[text=drawColor,anchor=base,inner sep=0pt, outer sep=0pt, scale=  1.10] at ( 88.80,  7.64) {Target Chunk Size};
\end{scope}
\begin{scope}
\path[clip] (  0.00,  0.00) rectangle (144.54,130.09);
\definecolor{drawColor}{RGB}{0,0,0}

\node[text=drawColor,rotate= 90.00,anchor=base,inner sep=0pt, outer sep=0pt, scale=  1.10] at ( 13.08, 82.72) {Dedup. Ratio};
\end{scope}
\end{tikzpicture}

%% file: fig/dedup_buzhash_window_sizes_pdf.tex
% Created by tikzDevice version 0.12.6 on 2024-04-19 17:33:44
% !TEX encoding = UTF-8 Unicode
\begin{tikzpicture}[x=1pt,y=1pt]
\definecolor{fillColor}{RGB}{255,255,255}
\path[use as bounding box,fill=fillColor,fill opacity=0.00] (0,0) rectangle (144.54,130.09);
\begin{scope}
\path[clip] (  0.00,  0.00) rectangle (144.54,130.09);
\definecolor{drawColor}{RGB}{255,255,255}
\definecolor{fillColor}{RGB}{255,255,255}

\path[draw=drawColor,line width= 0.6pt,line join=round,line cap=round,fill=fillColor] (  0.00,  0.00) rectangle (144.54,130.09);
\end{scope}
\begin{scope}
\path[clip] ( 38.56, 40.85) rectangle (139.04,124.59);
\definecolor{fillColor}{RGB}{255,255,255}

\path[fill=fillColor] ( 38.56, 40.85) rectangle (139.04,124.59);
\definecolor{drawColor}{gray}{0.92}

\path[draw=drawColor,line width= 0.3pt,line join=round] ( 38.56, 52.45) --
	(139.04, 52.45);

\path[draw=drawColor,line width= 0.3pt,line join=round] ( 38.56, 73.93) --
	(139.04, 73.93);

\path[draw=drawColor,line width= 0.3pt,line join=round] ( 38.56, 95.42) --
	(139.04, 95.42);

\path[draw=drawColor,line width= 0.3pt,line join=round] ( 38.56,116.91) --
	(139.04,116.91);

\path[draw=drawColor,line width= 0.6pt,line join=round] ( 38.56, 41.70) --
	(139.04, 41.70);

\path[draw=drawColor,line width= 0.6pt,line join=round] ( 38.56, 63.19) --
	(139.04, 63.19);

\path[draw=drawColor,line width= 0.6pt,line join=round] ( 38.56, 84.68) --
	(139.04, 84.68);

\path[draw=drawColor,line width= 0.6pt,line join=round] ( 38.56,106.16) --
	(139.04,106.16);

\path[draw=drawColor,line width= 0.6pt,line join=round] ( 50.15, 40.85) --
	( 50.15,124.59);

\path[draw=drawColor,line width= 0.6pt,line join=round] ( 69.47, 40.85) --
	( 69.47,124.59);

\path[draw=drawColor,line width= 0.6pt,line join=round] ( 88.80, 40.85) --
	( 88.80,124.59);

\path[draw=drawColor,line width= 0.6pt,line join=round] (108.12, 40.85) --
	(108.12,124.59);

\path[draw=drawColor,line width= 0.6pt,line join=round] (127.45, 40.85) --
	(127.45,124.59);
\definecolor{drawColor}{RGB}{55,78,85}

\path[draw=drawColor,line width= 0.6pt,line join=round] ( 49.34,105.29) --
	( 68.67, 91.43) --
	( 87.99, 76.91) --
	(107.32, 60.76) --
	(126.64, 45.98);
\definecolor{drawColor}{RGB}{223,143,68}

\path[draw=drawColor,line width= 0.6pt,line join=round] ( 49.67,109.15) --
	( 68.99, 92.53) --
	( 88.31, 77.65) --
	(107.64, 62.60) --
	(126.96, 49.56);
\definecolor{drawColor}{RGB}{0,161,213}

\path[draw=drawColor,line width= 0.6pt,line join=round] ( 49.99,104.96) --
	( 69.31, 90.65) --
	( 88.64, 75.08) --
	(107.96, 59.51) --
	(127.28, 44.66);
\definecolor{drawColor}{RGB}{178,71,69}

\path[draw=drawColor,line width= 0.6pt,line join=round] ( 50.31,120.78) --
	( 69.63,108.80) --
	( 88.96, 94.26) --
	(108.28, 78.22) --
	(127.61, 63.98);
\definecolor{drawColor}{RGB}{121,175,151}

\path[draw=drawColor,line width= 0.6pt,line join=round] ( 50.63,117.25) --
	( 69.96,104.22) --
	( 89.28, 89.34) --
	(108.60, 73.14) --
	(127.93, 58.95);
\definecolor{drawColor}{RGB}{106,101,153}

\path[draw=drawColor,line width= 0.6pt,line join=round] ( 50.95,117.35) --
	( 70.28, 97.73) --
	( 89.60, 84.21) --
	(108.93, 70.36) --
	(128.25, 56.49);
\definecolor{drawColor}{RGB}{55,78,85}

\path[draw=drawColor,line width= 0.4pt,line join=round,line cap=round,fill=fillColor] ( 49.34,105.29) circle (  1.43);

\path[draw=drawColor,line width= 0.4pt,line join=round,line cap=round,fill=fillColor] ( 68.67, 91.43) circle (  1.43);

\path[draw=drawColor,line width= 0.4pt,line join=round,line cap=round,fill=fillColor] ( 87.99, 76.91) circle (  1.43);

\path[draw=drawColor,line width= 0.4pt,line join=round,line cap=round,fill=fillColor] (107.32, 60.76) circle (  1.43);

\path[draw=drawColor,line width= 0.4pt,line join=round,line cap=round,fill=fillColor] (126.64, 45.98) circle (  1.43);
\definecolor{drawColor}{RGB}{223,143,68}

\path[draw=drawColor,line width= 0.4pt,line join=round,line cap=round,fill=fillColor] ( 49.67,109.15) circle (  1.43);

\path[draw=drawColor,line width= 0.4pt,line join=round,line cap=round,fill=fillColor] ( 68.99, 92.53) circle (  1.43);

\path[draw=drawColor,line width= 0.4pt,line join=round,line cap=round,fill=fillColor] ( 88.31, 77.65) circle (  1.43);

\path[draw=drawColor,line width= 0.4pt,line join=round,line cap=round,fill=fillColor] (107.64, 62.60) circle (  1.43);

\path[draw=drawColor,line width= 0.4pt,line join=round,line cap=round,fill=fillColor] (126.96, 49.56) circle (  1.43);
\definecolor{drawColor}{RGB}{0,161,213}

\path[draw=drawColor,line width= 0.4pt,line join=round,line cap=round,fill=fillColor] ( 49.99,104.96) circle (  1.43);

\path[draw=drawColor,line width= 0.4pt,line join=round,line cap=round,fill=fillColor] ( 69.31, 90.65) circle (  1.43);

\path[draw=drawColor,line width= 0.4pt,line join=round,line cap=round,fill=fillColor] ( 88.64, 75.08) circle (  1.43);

\path[draw=drawColor,line width= 0.4pt,line join=round,line cap=round,fill=fillColor] (107.96, 59.51) circle (  1.43);

\path[draw=drawColor,line width= 0.4pt,line join=round,line cap=round,fill=fillColor] (127.28, 44.66) circle (  1.43);
\definecolor{drawColor}{RGB}{178,71,69}

\path[draw=drawColor,line width= 0.4pt,line join=round,line cap=round,fill=fillColor] ( 50.31,120.78) circle (  1.43);

\path[draw=drawColor,line width= 0.4pt,line join=round,line cap=round,fill=fillColor] ( 69.63,108.80) circle (  1.43);

\path[draw=drawColor,line width= 0.4pt,line join=round,line cap=round,fill=fillColor] ( 88.96, 94.26) circle (  1.43);

\path[draw=drawColor,line width= 0.4pt,line join=round,line cap=round,fill=fillColor] (108.28, 78.22) circle (  1.43);

\path[draw=drawColor,line width= 0.4pt,line join=round,line cap=round,fill=fillColor] (127.61, 63.98) circle (  1.43);
\definecolor{drawColor}{RGB}{121,175,151}

\path[draw=drawColor,line width= 0.4pt,line join=round,line cap=round,fill=fillColor] ( 50.63,117.25) circle (  1.43);

\path[draw=drawColor,line width= 0.4pt,line join=round,line cap=round,fill=fillColor] ( 69.96,104.22) circle (  1.43);

\path[draw=drawColor,line width= 0.4pt,line join=round,line cap=round,fill=fillColor] ( 89.28, 89.34) circle (  1.43);

\path[draw=drawColor,line width= 0.4pt,line join=round,line cap=round,fill=fillColor] (108.60, 73.14) circle (  1.43);

\path[draw=drawColor,line width= 0.4pt,line join=round,line cap=round,fill=fillColor] (127.93, 58.95) circle (  1.43);
\definecolor{drawColor}{RGB}{106,101,153}

\path[draw=drawColor,line width= 0.4pt,line join=round,line cap=round,fill=fillColor] ( 50.95,117.35) circle (  1.43);

\path[draw=drawColor,line width= 0.4pt,line join=round,line cap=round,fill=fillColor] ( 70.28, 97.73) circle (  1.43);

\path[draw=drawColor,line width= 0.4pt,line join=round,line cap=round,fill=fillColor] ( 89.60, 84.21) circle (  1.43);

\path[draw=drawColor,line width= 0.4pt,line join=round,line cap=round,fill=fillColor] (108.93, 70.36) circle (  1.43);

\path[draw=drawColor,line width= 0.4pt,line join=round,line cap=round,fill=fillColor] (128.25, 56.49) circle (  1.43);
\definecolor{drawColor}{gray}{0.20}

\path[draw=drawColor,line width= 0.6pt,line join=round,line cap=round] ( 38.56, 40.85) rectangle (139.04,124.59);
\end{scope}
\begin{scope}
\path[clip] (  0.00,  0.00) rectangle (144.54,130.09);
\definecolor{drawColor}{gray}{0.30}

\node[text=drawColor,anchor=base east,inner sep=0pt, outer sep=0pt, scale=  0.88] at ( 33.61, 38.67) {0.02};

\node[text=drawColor,anchor=base east,inner sep=0pt, outer sep=0pt, scale=  0.88] at ( 33.61, 60.16) {0.03};

\node[text=drawColor,anchor=base east,inner sep=0pt, outer sep=0pt, scale=  0.88] at ( 33.61, 81.65) {0.04};

\node[text=drawColor,anchor=base east,inner sep=0pt, outer sep=0pt, scale=  0.88] at ( 33.61,103.13) {0.05};
\end{scope}
\begin{scope}
\path[clip] (  0.00,  0.00) rectangle (144.54,130.09);
\definecolor{drawColor}{gray}{0.20}

\path[draw=drawColor,line width= 0.6pt,line join=round] ( 35.81, 41.70) --
	( 38.56, 41.70);

\path[draw=drawColor,line width= 0.6pt,line join=round] ( 35.81, 63.19) --
	( 38.56, 63.19);

\path[draw=drawColor,line width= 0.6pt,line join=round] ( 35.81, 84.68) --
	( 38.56, 84.68);

\path[draw=drawColor,line width= 0.6pt,line join=round] ( 35.81,106.16) --
	( 38.56,106.16);
\end{scope}
\begin{scope}
\path[clip] (  0.00,  0.00) rectangle (144.54,130.09);
\definecolor{drawColor}{gray}{0.20}

\path[draw=drawColor,line width= 0.6pt,line join=round] ( 50.15, 38.10) --
	( 50.15, 40.85);

\path[draw=drawColor,line width= 0.6pt,line join=round] ( 69.47, 38.10) --
	( 69.47, 40.85);

\path[draw=drawColor,line width= 0.6pt,line join=round] ( 88.80, 38.10) --
	( 88.80, 40.85);

\path[draw=drawColor,line width= 0.6pt,line join=round] (108.12, 38.10) --
	(108.12, 40.85);

\path[draw=drawColor,line width= 0.6pt,line join=round] (127.45, 38.10) --
	(127.45, 40.85);
\end{scope}
\begin{scope}
\path[clip] (  0.00,  0.00) rectangle (144.54,130.09);
\definecolor{drawColor}{gray}{0.30}

\node[text=drawColor,rotate= 45.00,anchor=base east,inner sep=0pt, outer sep=0pt, scale=  0.88] at ( 54.44, 31.62) {512};

\node[text=drawColor,rotate= 45.00,anchor=base east,inner sep=0pt, outer sep=0pt, scale=  0.88] at ( 73.76, 31.62) {1024};

\node[text=drawColor,rotate= 45.00,anchor=base east,inner sep=0pt, outer sep=0pt, scale=  0.88] at ( 93.08, 31.62) {2048};

\node[text=drawColor,rotate= 45.00,anchor=base east,inner sep=0pt, outer sep=0pt, scale=  0.88] at (112.41, 31.62) {4096};

\node[text=drawColor,rotate= 45.00,anchor=base east,inner sep=0pt, outer sep=0pt, scale=  0.88] at (131.73, 31.62) {8192};
\end{scope}
\begin{scope}
\path[clip] (  0.00,  0.00) rectangle (144.54,130.09);
\definecolor{drawColor}{RGB}{0,0,0}

\node[text=drawColor,anchor=base,inner sep=0pt, outer sep=0pt, scale=  1.10] at ( 88.80,  7.64) {Target Chunk Size};
\end{scope}
\begin{scope}
\path[clip] (  0.00,  0.00) rectangle (144.54,130.09);
\definecolor{drawColor}{RGB}{0,0,0}

\node[text=drawColor,rotate= 90.00,anchor=base,inner sep=0pt, outer sep=0pt, scale=  1.10] at ( 13.08, 82.72) {Dedup. Ratio};
\end{scope}
\end{tikzpicture}

%% file: fig/dedup_buzhash_window_sizes_web.tex
% Created by tikzDevice version 0.12.6 on 2024-04-19 17:33:45
% !TEX encoding = UTF-8 Unicode
\begin{tikzpicture}[x=1pt,y=1pt]
\definecolor{fillColor}{RGB}{255,255,255}
\path[use as bounding box,fill=fillColor,fill opacity=0.00] (0,0) rectangle (144.54,130.09);
\begin{scope}
\path[clip] (  0.00,  0.00) rectangle (144.54,130.09);
\definecolor{drawColor}{RGB}{255,255,255}
\definecolor{fillColor}{RGB}{255,255,255}

\path[draw=drawColor,line width= 0.6pt,line join=round,line cap=round,fill=fillColor] (  0.00,  0.00) rectangle (144.54,130.09);
\end{scope}
\begin{scope}
\path[clip] ( 38.56, 40.85) rectangle (139.04,124.59);
\definecolor{fillColor}{RGB}{255,255,255}

\path[fill=fillColor] ( 38.56, 40.85) rectangle (139.04,124.59);
\definecolor{drawColor}{gray}{0.92}

\path[draw=drawColor,line width= 0.3pt,line join=round] ( 38.56, 52.83) --
	(139.04, 52.83);

\path[draw=drawColor,line width= 0.3pt,line join=round] ( 38.56, 77.06) --
	(139.04, 77.06);

\path[draw=drawColor,line width= 0.3pt,line join=round] ( 38.56,101.30) --
	(139.04,101.30);

\path[draw=drawColor,line width= 0.6pt,line join=round] ( 38.56, 64.95) --
	(139.04, 64.95);

\path[draw=drawColor,line width= 0.6pt,line join=round] ( 38.56, 89.18) --
	(139.04, 89.18);

\path[draw=drawColor,line width= 0.6pt,line join=round] ( 38.56,113.42) --
	(139.04,113.42);

\path[draw=drawColor,line width= 0.6pt,line join=round] ( 50.15, 40.85) --
	( 50.15,124.59);

\path[draw=drawColor,line width= 0.6pt,line join=round] ( 69.47, 40.85) --
	( 69.47,124.59);

\path[draw=drawColor,line width= 0.6pt,line join=round] ( 88.80, 40.85) --
	( 88.80,124.59);

\path[draw=drawColor,line width= 0.6pt,line join=round] (108.12, 40.85) --
	(108.12,124.59);

\path[draw=drawColor,line width= 0.6pt,line join=round] (127.45, 40.85) --
	(127.45,124.59);
\definecolor{drawColor}{RGB}{55,78,85}

\path[draw=drawColor,line width= 0.6pt,line join=round] ( 49.34,109.79) --
	( 68.67, 95.22) --
	( 87.99, 80.03) --
	(107.32, 68.29) --
	(126.64, 44.66);
\definecolor{drawColor}{RGB}{223,143,68}

\path[draw=drawColor,line width= 0.6pt,line join=round] ( 49.67,116.64) --
	( 68.99,100.47) --
	( 88.31, 87.04) --
	(107.64, 74.17) --
	(126.96, 56.63);
\definecolor{drawColor}{RGB}{0,161,213}

\path[draw=drawColor,line width= 0.6pt,line join=round] ( 49.99,113.09) --
	( 69.31,101.80) --
	( 88.64, 87.66) --
	(107.96, 68.40) --
	(127.28, 50.41);
\definecolor{drawColor}{RGB}{178,71,69}

\path[draw=drawColor,line width= 0.6pt,line join=round] ( 50.31,120.78) --
	( 69.63,106.38) --
	( 88.96, 92.48) --
	(108.28, 73.60) --
	(127.61, 60.31);
\definecolor{drawColor}{RGB}{121,175,151}

\path[draw=drawColor,line width= 0.6pt,line join=round] ( 50.63,116.20) --
	( 69.96,105.09) --
	( 89.28, 89.49) --
	(108.60, 74.44) --
	(127.93, 63.40);
\definecolor{drawColor}{RGB}{106,101,153}

\path[draw=drawColor,line width= 0.6pt,line join=round] ( 50.95,119.19) --
	( 70.28,100.99) --
	( 89.60, 88.09) --
	(108.93, 69.56) --
	(128.25, 52.31);
\definecolor{drawColor}{RGB}{55,78,85}

\path[draw=drawColor,line width= 0.4pt,line join=round,line cap=round,fill=fillColor] ( 49.34,109.79) circle (  1.43);

\path[draw=drawColor,line width= 0.4pt,line join=round,line cap=round,fill=fillColor] ( 68.67, 95.22) circle (  1.43);

\path[draw=drawColor,line width= 0.4pt,line join=round,line cap=round,fill=fillColor] ( 87.99, 80.03) circle (  1.43);

\path[draw=drawColor,line width= 0.4pt,line join=round,line cap=round,fill=fillColor] (107.32, 68.29) circle (  1.43);

\path[draw=drawColor,line width= 0.4pt,line join=round,line cap=round,fill=fillColor] (126.64, 44.66) circle (  1.43);
\definecolor{drawColor}{RGB}{223,143,68}

\path[draw=drawColor,line width= 0.4pt,line join=round,line cap=round,fill=fillColor] ( 49.67,116.64) circle (  1.43);

\path[draw=drawColor,line width= 0.4pt,line join=round,line cap=round,fill=fillColor] ( 68.99,100.47) circle (  1.43);

\path[draw=drawColor,line width= 0.4pt,line join=round,line cap=round,fill=fillColor] ( 88.31, 87.04) circle (  1.43);

\path[draw=drawColor,line width= 0.4pt,line join=round,line cap=round,fill=fillColor] (107.64, 74.17) circle (  1.43);

\path[draw=drawColor,line width= 0.4pt,line join=round,line cap=round,fill=fillColor] (126.96, 56.63) circle (  1.43);
\definecolor{drawColor}{RGB}{0,161,213}

\path[draw=drawColor,line width= 0.4pt,line join=round,line cap=round,fill=fillColor] ( 49.99,113.09) circle (  1.43);

\path[draw=drawColor,line width= 0.4pt,line join=round,line cap=round,fill=fillColor] ( 69.31,101.80) circle (  1.43);

\path[draw=drawColor,line width= 0.4pt,line join=round,line cap=round,fill=fillColor] ( 88.64, 87.66) circle (  1.43);

\path[draw=drawColor,line width= 0.4pt,line join=round,line cap=round,fill=fillColor] (107.96, 68.40) circle (  1.43);

\path[draw=drawColor,line width= 0.4pt,line join=round,line cap=round,fill=fillColor] (127.28, 50.41) circle (  1.43);
\definecolor{drawColor}{RGB}{178,71,69}

\path[draw=drawColor,line width= 0.4pt,line join=round,line cap=round,fill=fillColor] ( 50.31,120.78) circle (  1.43);

\path[draw=drawColor,line width= 0.4pt,line join=round,line cap=round,fill=fillColor] ( 69.63,106.38) circle (  1.43);

\path[draw=drawColor,line width= 0.4pt,line join=round,line cap=round,fill=fillColor] ( 88.96, 92.48) circle (  1.43);

\path[draw=drawColor,line width= 0.4pt,line join=round,line cap=round,fill=fillColor] (108.28, 73.60) circle (  1.43);

\path[draw=drawColor,line width= 0.4pt,line join=round,line cap=round,fill=fillColor] (127.61, 60.31) circle (  1.43);
\definecolor{drawColor}{RGB}{121,175,151}

\path[draw=drawColor,line width= 0.4pt,line join=round,line cap=round,fill=fillColor] ( 50.63,116.20) circle (  1.43);

\path[draw=drawColor,line width= 0.4pt,line join=round,line cap=round,fill=fillColor] ( 69.96,105.09) circle (  1.43);

\path[draw=drawColor,line width= 0.4pt,line join=round,line cap=round,fill=fillColor] ( 89.28, 89.49) circle (  1.43);

\path[draw=drawColor,line width= 0.4pt,line join=round,line cap=round,fill=fillColor] (108.60, 74.44) circle (  1.43);

\path[draw=drawColor,line width= 0.4pt,line join=round,line cap=round,fill=fillColor] (127.93, 63.40) circle (  1.43);
\definecolor{drawColor}{RGB}{106,101,153}

\path[draw=drawColor,line width= 0.4pt,line join=round,line cap=round,fill=fillColor] ( 50.95,119.19) circle (  1.43);

\path[draw=drawColor,line width= 0.4pt,line join=round,line cap=round,fill=fillColor] ( 70.28,100.99) circle (  1.43);

\path[draw=drawColor,line width= 0.4pt,line join=round,line cap=round,fill=fillColor] ( 89.60, 88.09) circle (  1.43);

\path[draw=drawColor,line width= 0.4pt,line join=round,line cap=round,fill=fillColor] (108.93, 69.56) circle (  1.43);

\path[draw=drawColor,line width= 0.4pt,line join=round,line cap=round,fill=fillColor] (128.25, 52.31) circle (  1.43);
\definecolor{drawColor}{gray}{0.20}

\path[draw=drawColor,line width= 0.6pt,line join=round,line cap=round] ( 38.56, 40.85) rectangle (139.04,124.59);
\end{scope}
\begin{scope}
\path[clip] (  0.00,  0.00) rectangle (144.54,130.09);
\definecolor{drawColor}{gray}{0.30}

\node[text=drawColor,anchor=base east,inner sep=0pt, outer sep=0pt, scale=  0.88] at ( 33.61, 61.91) {0.69};

\node[text=drawColor,anchor=base east,inner sep=0pt, outer sep=0pt, scale=  0.88] at ( 33.61, 86.15) {0.70};

\node[text=drawColor,anchor=base east,inner sep=0pt, outer sep=0pt, scale=  0.88] at ( 33.61,110.39) {0.71};
\end{scope}
\begin{scope}
\path[clip] (  0.00,  0.00) rectangle (144.54,130.09);
\definecolor{drawColor}{gray}{0.20}

\path[draw=drawColor,line width= 0.6pt,line join=round] ( 35.81, 64.95) --
	( 38.56, 64.95);

\path[draw=drawColor,line width= 0.6pt,line join=round] ( 35.81, 89.18) --
	( 38.56, 89.18);

\path[draw=drawColor,line width= 0.6pt,line join=round] ( 35.81,113.42) --
	( 38.56,113.42);
\end{scope}
\begin{scope}
\path[clip] (  0.00,  0.00) rectangle (144.54,130.09);
\definecolor{drawColor}{gray}{0.20}

\path[draw=drawColor,line width= 0.6pt,line join=round] ( 50.15, 38.10) --
	( 50.15, 40.85);

\path[draw=drawColor,line width= 0.6pt,line join=round] ( 69.47, 38.10) --
	( 69.47, 40.85);

\path[draw=drawColor,line width= 0.6pt,line join=round] ( 88.80, 38.10) --
	( 88.80, 40.85);

\path[draw=drawColor,line width= 0.6pt,line join=round] (108.12, 38.10) --
	(108.12, 40.85);

\path[draw=drawColor,line width= 0.6pt,line join=round] (127.45, 38.10) --
	(127.45, 40.85);
\end{scope}
\begin{scope}
\path[clip] (  0.00,  0.00) rectangle (144.54,130.09);
\definecolor{drawColor}{gray}{0.30}

\node[text=drawColor,rotate= 45.00,anchor=base east,inner sep=0pt, outer sep=0pt, scale=  0.88] at ( 54.44, 31.62) {512};

\node[text=drawColor,rotate= 45.00,anchor=base east,inner sep=0pt, outer sep=0pt, scale=  0.88] at ( 73.76, 31.62) {1024};

\node[text=drawColor,rotate= 45.00,anchor=base east,inner sep=0pt, outer sep=0pt, scale=  0.88] at ( 93.08, 31.62) {2048};

\node[text=drawColor,rotate= 45.00,anchor=base east,inner sep=0pt, outer sep=0pt, scale=  0.88] at (112.41, 31.62) {4096};

\node[text=drawColor,rotate= 45.00,anchor=base east,inner sep=0pt, outer sep=0pt, scale=  0.88] at (131.73, 31.62) {8192};
\end{scope}
\begin{scope}
\path[clip] (  0.00,  0.00) rectangle (144.54,130.09);
\definecolor{drawColor}{RGB}{0,0,0}

\node[text=drawColor,anchor=base,inner sep=0pt, outer sep=0pt, scale=  1.10] at ( 88.80,  7.64) {Target Chunk Size};
\end{scope}
\begin{scope}
\path[clip] (  0.00,  0.00) rectangle (144.54,130.09);
\definecolor{drawColor}{RGB}{0,0,0}

\node[text=drawColor,rotate= 90.00,anchor=base,inner sep=0pt, outer sep=0pt, scale=  1.10] at ( 13.08, 82.72) {Dedup. Ratio};
\end{scope}
\end{tikzpicture}

%% file: fig/dedup_window_sizes_legendonly.tex
% Created by tikzDevice version 0.12.6 on 2024-07-17 22:33:06
% !TEX encoding = UTF-8 Unicode
\begin{tikzpicture}[x=1pt,y=1pt]
\definecolor{fillColor}{RGB}{255,255,255}
\path[use as bounding box,fill=fillColor,fill opacity=0.00] (0,0) rectangle (252.94, 72.27);
\begin{scope}
\path[clip] (  0.00,  0.00) rectangle (252.94, 72.27);
\definecolor{drawColor}{RGB}{55,78,85}

\path[draw=drawColor,line width= 0.6pt,line join=round] (  9.61, 36.13) -- ( 23.48, 36.13);
\end{scope}
\begin{scope}
\path[clip] (  0.00,  0.00) rectangle (252.94, 72.27);
\definecolor{drawColor}{RGB}{55,78,85}
\definecolor{fillColor}{RGB}{255,255,255}

\path[draw=drawColor,line width= 0.4pt,line join=round,line cap=round,fill=fillColor] ( 16.54, 36.13) circle (  1.43);
\end{scope}
\begin{scope}
\path[clip] (  0.00,  0.00) rectangle (252.94, 72.27);
\definecolor{drawColor}{RGB}{223,143,68}

\path[draw=drawColor,line width= 0.6pt,line join=round] ( 49.11, 36.13) -- ( 62.99, 36.13);
\end{scope}
\begin{scope}
\path[clip] (  0.00,  0.00) rectangle (252.94, 72.27);
\definecolor{drawColor}{RGB}{223,143,68}
\definecolor{fillColor}{RGB}{255,255,255}

\path[draw=drawColor,line width= 0.4pt,line join=round,line cap=round,fill=fillColor] ( 56.05, 36.13) circle (  1.43);
\end{scope}
\begin{scope}
\path[clip] (  0.00,  0.00) rectangle (252.94, 72.27);
\definecolor{drawColor}{RGB}{0,161,213}

\path[draw=drawColor,line width= 0.6pt,line join=round] ( 88.62, 36.13) -- (102.50, 36.13);
\end{scope}
\begin{scope}
\path[clip] (  0.00,  0.00) rectangle (252.94, 72.27);
\definecolor{drawColor}{RGB}{0,161,213}
\definecolor{fillColor}{RGB}{255,255,255}

\path[draw=drawColor,line width= 0.4pt,line join=round,line cap=round,fill=fillColor] ( 95.56, 36.13) circle (  1.43);
\end{scope}
\begin{scope}
\path[clip] (  0.00,  0.00) rectangle (252.94, 72.27);
\definecolor{drawColor}{RGB}{178,71,69}

\path[draw=drawColor,line width= 0.6pt,line join=round] (128.13, 36.13) -- (142.00, 36.13);
\end{scope}
\begin{scope}
\path[clip] (  0.00,  0.00) rectangle (252.94, 72.27);
\definecolor{drawColor}{RGB}{178,71,69}
\definecolor{fillColor}{RGB}{255,255,255}

\path[draw=drawColor,line width= 0.4pt,line join=round,line cap=round,fill=fillColor] (135.06, 36.13) circle (  1.43);
\end{scope}
\begin{scope}
\path[clip] (  0.00,  0.00) rectangle (252.94, 72.27);
\definecolor{drawColor}{RGB}{121,175,151}

\path[draw=drawColor,line width= 0.6pt,line join=round] (167.63, 36.13) -- (181.51, 36.13);
\end{scope}
\begin{scope}
\path[clip] (  0.00,  0.00) rectangle (252.94, 72.27);
\definecolor{drawColor}{RGB}{121,175,151}
\definecolor{fillColor}{RGB}{255,255,255}

\path[draw=drawColor,line width= 0.4pt,line join=round,line cap=round,fill=fillColor] (174.57, 36.13) circle (  1.43);
\end{scope}
\begin{scope}
\path[clip] (  0.00,  0.00) rectangle (252.94, 72.27);
\definecolor{drawColor}{RGB}{106,101,153}

\path[draw=drawColor,line width= 0.6pt,line join=round] (212.72, 36.13) -- (226.60, 36.13);
\end{scope}
\begin{scope}
\path[clip] (  0.00,  0.00) rectangle (252.94, 72.27);
\definecolor{drawColor}{RGB}{106,101,153}
\definecolor{fillColor}{RGB}{255,255,255}

\path[draw=drawColor,line width= 0.4pt,line join=round,line cap=round,fill=fillColor] (219.66, 36.13) circle (  1.43);
\end{scope}
\begin{scope}
\path[clip] (  0.00,  0.00) rectangle (252.94, 72.27);
\definecolor{drawColor}{RGB}{0,0,0}

\node[text=drawColor,anchor=base west,inner sep=0pt, outer sep=0pt, scale=  1.00] at ( 30.72, 32.69) {16};
\end{scope}
\begin{scope}
\path[clip] (  0.00,  0.00) rectangle (252.94, 72.27);
\definecolor{drawColor}{RGB}{0,0,0}

\node[text=drawColor,anchor=base west,inner sep=0pt, outer sep=0pt, scale=  1.00] at ( 70.22, 32.69) {32};
\end{scope}
\begin{scope}
\path[clip] (  0.00,  0.00) rectangle (252.94, 72.27);
\definecolor{drawColor}{RGB}{0,0,0}

\node[text=drawColor,anchor=base west,inner sep=0pt, outer sep=0pt, scale=  1.00] at (109.73, 32.69) {48};
\end{scope}
\begin{scope}
\path[clip] (  0.00,  0.00) rectangle (252.94, 72.27);
\definecolor{drawColor}{RGB}{0,0,0}

\node[text=drawColor,anchor=base west,inner sep=0pt, outer sep=0pt, scale=  1.00] at (149.24, 32.69) {64};
\end{scope}
\begin{scope}
\path[clip] (  0.00,  0.00) rectangle (252.94, 72.27);
\definecolor{drawColor}{RGB}{0,0,0}

\node[text=drawColor,anchor=base west,inner sep=0pt, outer sep=0pt, scale=  1.00] at (188.74, 32.69) {128};
\end{scope}
\begin{scope}
\path[clip] (  0.00,  0.00) rectangle (252.94, 72.27);
\definecolor{drawColor}{RGB}{0,0,0}

\node[text=drawColor,anchor=base west,inner sep=0pt, outer sep=0pt, scale=  1.00] at (233.83, 32.69) {256};
\end{scope}
\end{tikzpicture}

%% file: fig/dedup_gear_variants_lnx.tex
% Created by tikzDevice version 0.12.6 on 2024-04-19 16:30:57
% !TEX encoding = UTF-8 Unicode
\begin{tikzpicture}[x=1pt,y=1pt]
\definecolor{fillColor}{RGB}{255,255,255}
\path[use as bounding box,fill=fillColor,fill opacity=0.00] (0,0) rectangle (144.54,130.09);
\begin{scope}
\path[clip] (  0.00,  0.00) rectangle (144.54,130.09);
\definecolor{drawColor}{RGB}{255,255,255}
\definecolor{fillColor}{RGB}{255,255,255}

\path[draw=drawColor,line width= 0.6pt,line join=round,line cap=round,fill=fillColor] (  0.00,  0.00) rectangle (144.54,130.09);
\end{scope}
\begin{scope}
\path[clip] ( 38.56, 40.85) rectangle (139.04,124.59);
\definecolor{fillColor}{RGB}{255,255,255}

\path[fill=fillColor] ( 38.56, 40.85) rectangle (139.04,124.59);
\definecolor{drawColor}{gray}{0.92}

\path[draw=drawColor,line width= 0.3pt,line join=round] ( 38.56, 61.79) --
	(139.04, 61.79);

\path[draw=drawColor,line width= 0.3pt,line join=round] ( 38.56, 86.06) --
	(139.04, 86.06);

\path[draw=drawColor,line width= 0.3pt,line join=round] ( 38.56,110.32) --
	(139.04,110.32);

\path[draw=drawColor,line width= 0.6pt,line join=round] ( 38.56, 49.66) --
	(139.04, 49.66);

\path[draw=drawColor,line width= 0.6pt,line join=round] ( 38.56, 73.93) --
	(139.04, 73.93);

\path[draw=drawColor,line width= 0.6pt,line join=round] ( 38.56, 98.19) --
	(139.04, 98.19);

\path[draw=drawColor,line width= 0.6pt,line join=round] ( 38.56,122.46) --
	(139.04,122.46);

\path[draw=drawColor,line width= 0.6pt,line join=round] ( 50.15, 40.85) --
	( 50.15,124.59);

\path[draw=drawColor,line width= 0.6pt,line join=round] ( 69.47, 40.85) --
	( 69.47,124.59);

\path[draw=drawColor,line width= 0.6pt,line join=round] ( 88.80, 40.85) --
	( 88.80,124.59);

\path[draw=drawColor,line width= 0.6pt,line join=round] (108.12, 40.85) --
	(108.12,124.59);

\path[draw=drawColor,line width= 0.6pt,line join=round] (127.45, 40.85) --
	(127.45,124.59);
\definecolor{drawColor}{RGB}{255,111,0}

\path[draw=drawColor,line width= 0.6pt,line join=round] ( 49.42,119.84) --
	( 68.75,114.98) --
	( 88.07,106.21) --
	(107.40, 89.96) --
	(126.72, 66.37);
\definecolor{drawColor}{RGB}{199,16,0}

\path[draw=drawColor,line width= 0.6pt,line join=round] ( 49.91,120.78) --
	( 69.23,116.38) --
	( 88.56,108.68) --
	(107.88, 94.54) --
	(127.20, 70.16);
\definecolor{drawColor}{RGB}{0,142,160}

\path[draw=drawColor,line width= 0.6pt,line join=round] ( 50.39,120.07) --
	( 69.72,115.38) --
	( 89.04,106.24) --
	(108.36, 89.84) --
	(127.69, 64.81);
\definecolor{drawColor}{RGB}{138,65,152}

\path[draw=drawColor,line width= 0.6pt,line join=round] ( 50.87,116.95) --
	( 70.20,109.74) --
	( 89.52, 96.26) --
	(108.85, 74.83) --
	(128.17, 44.66);
\definecolor{drawColor}{RGB}{255,111,0}

\path[draw=drawColor,line width= 0.4pt,line join=round,line cap=round,fill=fillColor] ( 49.42,119.84) circle (  1.43);

\path[draw=drawColor,line width= 0.4pt,line join=round,line cap=round,fill=fillColor] ( 68.75,114.98) circle (  1.43);

\path[draw=drawColor,line width= 0.4pt,line join=round,line cap=round,fill=fillColor] ( 88.07,106.21) circle (  1.43);

\path[draw=drawColor,line width= 0.4pt,line join=round,line cap=round,fill=fillColor] (107.40, 89.96) circle (  1.43);

\path[draw=drawColor,line width= 0.4pt,line join=round,line cap=round,fill=fillColor] (126.72, 66.37) circle (  1.43);
\definecolor{drawColor}{RGB}{199,16,0}

\path[draw=drawColor,line width= 0.4pt,line join=round,line cap=round,fill=fillColor] ( 49.91,120.78) circle (  1.43);

\path[draw=drawColor,line width= 0.4pt,line join=round,line cap=round,fill=fillColor] ( 69.23,116.38) circle (  1.43);

\path[draw=drawColor,line width= 0.4pt,line join=round,line cap=round,fill=fillColor] ( 88.56,108.68) circle (  1.43);

\path[draw=drawColor,line width= 0.4pt,line join=round,line cap=round,fill=fillColor] (107.88, 94.54) circle (  1.43);

\path[draw=drawColor,line width= 0.4pt,line join=round,line cap=round,fill=fillColor] (127.20, 70.16) circle (  1.43);
\definecolor{drawColor}{RGB}{0,142,160}

\path[draw=drawColor,line width= 0.4pt,line join=round,line cap=round,fill=fillColor] ( 50.39,120.07) circle (  1.43);

\path[draw=drawColor,line width= 0.4pt,line join=round,line cap=round,fill=fillColor] ( 69.72,115.38) circle (  1.43);

\path[draw=drawColor,line width= 0.4pt,line join=round,line cap=round,fill=fillColor] ( 89.04,106.24) circle (  1.43);

\path[draw=drawColor,line width= 0.4pt,line join=round,line cap=round,fill=fillColor] (108.36, 89.84) circle (  1.43);

\path[draw=drawColor,line width= 0.4pt,line join=round,line cap=round,fill=fillColor] (127.69, 64.81) circle (  1.43);
\definecolor{drawColor}{RGB}{138,65,152}

\path[draw=drawColor,line width= 0.4pt,line join=round,line cap=round,fill=fillColor] ( 50.87,116.95) circle (  1.43);

\path[draw=drawColor,line width= 0.4pt,line join=round,line cap=round,fill=fillColor] ( 70.20,109.74) circle (  1.43);

\path[draw=drawColor,line width= 0.4pt,line join=round,line cap=round,fill=fillColor] ( 89.52, 96.26) circle (  1.43);

\path[draw=drawColor,line width= 0.4pt,line join=round,line cap=round,fill=fillColor] (108.85, 74.83) circle (  1.43);

\path[draw=drawColor,line width= 0.4pt,line join=round,line cap=round,fill=fillColor] (128.17, 44.66) circle (  1.43);
\definecolor{drawColor}{gray}{0.20}

\path[draw=drawColor,line width= 0.6pt,line join=round,line cap=round] ( 38.56, 40.85) rectangle (139.04,124.59);
\end{scope}
\begin{scope}
\path[clip] (  0.00,  0.00) rectangle (144.54,130.09);
\definecolor{drawColor}{gray}{0.30}

\node[text=drawColor,anchor=base east,inner sep=0pt, outer sep=0pt, scale=  0.88] at ( 33.61, 46.63) {0.17};

\node[text=drawColor,anchor=base east,inner sep=0pt, outer sep=0pt, scale=  0.88] at ( 33.61, 70.90) {0.18};

\node[text=drawColor,anchor=base east,inner sep=0pt, outer sep=0pt, scale=  0.88] at ( 33.61, 95.16) {0.19};

\node[text=drawColor,anchor=base east,inner sep=0pt, outer sep=0pt, scale=  0.88] at ( 33.61,119.43) {0.20};
\end{scope}
\begin{scope}
\path[clip] (  0.00,  0.00) rectangle (144.54,130.09);
\definecolor{drawColor}{gray}{0.20}

\path[draw=drawColor,line width= 0.6pt,line join=round] ( 35.81, 49.66) --
	( 38.56, 49.66);

\path[draw=drawColor,line width= 0.6pt,line join=round] ( 35.81, 73.93) --
	( 38.56, 73.93);

\path[draw=drawColor,line width= 0.6pt,line join=round] ( 35.81, 98.19) --
	( 38.56, 98.19);

\path[draw=drawColor,line width= 0.6pt,line join=round] ( 35.81,122.46) --
	( 38.56,122.46);
\end{scope}
\begin{scope}
\path[clip] (  0.00,  0.00) rectangle (144.54,130.09);
\definecolor{drawColor}{gray}{0.20}

\path[draw=drawColor,line width= 0.6pt,line join=round] ( 50.15, 38.10) --
	( 50.15, 40.85);

\path[draw=drawColor,line width= 0.6pt,line join=round] ( 69.47, 38.10) --
	( 69.47, 40.85);

\path[draw=drawColor,line width= 0.6pt,line join=round] ( 88.80, 38.10) --
	( 88.80, 40.85);

\path[draw=drawColor,line width= 0.6pt,line join=round] (108.12, 38.10) --
	(108.12, 40.85);

\path[draw=drawColor,line width= 0.6pt,line join=round] (127.45, 38.10) --
	(127.45, 40.85);
\end{scope}
\begin{scope}
\path[clip] (  0.00,  0.00) rectangle (144.54,130.09);
\definecolor{drawColor}{gray}{0.30}

\node[text=drawColor,rotate= 45.00,anchor=base east,inner sep=0pt, outer sep=0pt, scale=  0.88] at ( 54.44, 31.62) {512};

\node[text=drawColor,rotate= 45.00,anchor=base east,inner sep=0pt, outer sep=0pt, scale=  0.88] at ( 73.76, 31.62) {1024};

\node[text=drawColor,rotate= 45.00,anchor=base east,inner sep=0pt, outer sep=0pt, scale=  0.88] at ( 93.08, 31.62) {2048};

\node[text=drawColor,rotate= 45.00,anchor=base east,inner sep=0pt, outer sep=0pt, scale=  0.88] at (112.41, 31.62) {4096};

\node[text=drawColor,rotate= 45.00,anchor=base east,inner sep=0pt, outer sep=0pt, scale=  0.88] at (131.73, 31.62) {8192};
\end{scope}
\begin{scope}
\path[clip] (  0.00,  0.00) rectangle (144.54,130.09);
\definecolor{drawColor}{RGB}{0,0,0}

\node[text=drawColor,anchor=base,inner sep=0pt, outer sep=0pt, scale=  1.10] at ( 88.80,  7.64) {Target Chunk Size};
\end{scope}
\begin{scope}
\path[clip] (  0.00,  0.00) rectangle (144.54,130.09);
\definecolor{drawColor}{RGB}{0,0,0}

\node[text=drawColor,rotate= 90.00,anchor=base,inner sep=0pt, outer sep=0pt, scale=  1.10] at ( 13.08, 82.72) {Dedup. Ratio};
\end{scope}
\end{tikzpicture}

%% file: fig/dedup_gear_variants_pdf.tex
% Created by tikzDevice version 0.12.6 on 2024-04-19 16:30:58
% !TEX encoding = UTF-8 Unicode
\begin{tikzpicture}[x=1pt,y=1pt]
\definecolor{fillColor}{RGB}{255,255,255}
\path[use as bounding box,fill=fillColor,fill opacity=0.00] (0,0) rectangle (144.54,130.09);
\begin{scope}
\path[clip] (  0.00,  0.00) rectangle (144.54,130.09);
\definecolor{drawColor}{RGB}{255,255,255}
\definecolor{fillColor}{RGB}{255,255,255}

\path[draw=drawColor,line width= 0.6pt,line join=round,line cap=round,fill=fillColor] (  0.00,  0.00) rectangle (144.54,130.09);
\end{scope}
\begin{scope}
\path[clip] ( 38.56, 40.85) rectangle (139.04,124.59);
\definecolor{fillColor}{RGB}{255,255,255}

\path[fill=fillColor] ( 38.56, 40.85) rectangle (139.04,124.59);
\definecolor{drawColor}{gray}{0.92}

\path[draw=drawColor,line width= 0.3pt,line join=round] ( 38.56, 52.47) --
	(139.04, 52.47);

\path[draw=drawColor,line width= 0.3pt,line join=round] ( 38.56, 77.49) --
	(139.04, 77.49);

\path[draw=drawColor,line width= 0.3pt,line join=round] ( 38.56,102.52) --
	(139.04,102.52);

\path[draw=drawColor,line width= 0.6pt,line join=round] ( 38.56, 64.98) --
	(139.04, 64.98);

\path[draw=drawColor,line width= 0.6pt,line join=round] ( 38.56, 90.01) --
	(139.04, 90.01);

\path[draw=drawColor,line width= 0.6pt,line join=round] ( 38.56,115.03) --
	(139.04,115.03);

\path[draw=drawColor,line width= 0.6pt,line join=round] ( 50.15, 40.85) --
	( 50.15,124.59);

\path[draw=drawColor,line width= 0.6pt,line join=round] ( 69.47, 40.85) --
	( 69.47,124.59);

\path[draw=drawColor,line width= 0.6pt,line join=round] ( 88.80, 40.85) --
	( 88.80,124.59);

\path[draw=drawColor,line width= 0.6pt,line join=round] (108.12, 40.85) --
	(108.12,124.59);

\path[draw=drawColor,line width= 0.6pt,line join=round] (127.45, 40.85) --
	(127.45,124.59);
\definecolor{drawColor}{RGB}{255,111,0}

\path[draw=drawColor,line width= 0.6pt,line join=round] ( 49.42,116.87) --
	( 68.75, 99.95) --
	( 88.07, 83.37) --
	(107.40, 63.98) --
	(126.72, 47.06);
\definecolor{drawColor}{RGB}{199,16,0}

\path[draw=drawColor,line width= 0.6pt,line join=round] ( 49.91,118.53) --
	( 69.23,101.75) --
	( 88.56, 84.28) --
	(107.88, 65.40) --
	(127.20, 48.63);
\definecolor{drawColor}{RGB}{0,142,160}

\path[draw=drawColor,line width= 0.6pt,line join=round] ( 50.39,120.78) --
	( 69.72,102.37) --
	( 89.04, 84.28) --
	(108.36, 65.94) --
	(127.69, 47.44);
\definecolor{drawColor}{RGB}{138,65,152}

\path[draw=drawColor,line width= 0.6pt,line join=round] ( 50.87,119.82) --
	( 70.20,100.99) --
	( 89.52, 81.09) --
	(108.85, 63.45) --
	(128.17, 44.66);
\definecolor{drawColor}{RGB}{255,111,0}

\path[draw=drawColor,line width= 0.4pt,line join=round,line cap=round,fill=fillColor] ( 49.42,116.87) circle (  1.43);

\path[draw=drawColor,line width= 0.4pt,line join=round,line cap=round,fill=fillColor] ( 68.75, 99.95) circle (  1.43);

\path[draw=drawColor,line width= 0.4pt,line join=round,line cap=round,fill=fillColor] ( 88.07, 83.37) circle (  1.43);

\path[draw=drawColor,line width= 0.4pt,line join=round,line cap=round,fill=fillColor] (107.40, 63.98) circle (  1.43);

\path[draw=drawColor,line width= 0.4pt,line join=round,line cap=round,fill=fillColor] (126.72, 47.06) circle (  1.43);
\definecolor{drawColor}{RGB}{199,16,0}

\path[draw=drawColor,line width= 0.4pt,line join=round,line cap=round,fill=fillColor] ( 49.91,118.53) circle (  1.43);

\path[draw=drawColor,line width= 0.4pt,line join=round,line cap=round,fill=fillColor] ( 69.23,101.75) circle (  1.43);

\path[draw=drawColor,line width= 0.4pt,line join=round,line cap=round,fill=fillColor] ( 88.56, 84.28) circle (  1.43);

\path[draw=drawColor,line width= 0.4pt,line join=round,line cap=round,fill=fillColor] (107.88, 65.40) circle (  1.43);

\path[draw=drawColor,line width= 0.4pt,line join=round,line cap=round,fill=fillColor] (127.20, 48.63) circle (  1.43);
\definecolor{drawColor}{RGB}{0,142,160}

\path[draw=drawColor,line width= 0.4pt,line join=round,line cap=round,fill=fillColor] ( 50.39,120.78) circle (  1.43);

\path[draw=drawColor,line width= 0.4pt,line join=round,line cap=round,fill=fillColor] ( 69.72,102.37) circle (  1.43);

\path[draw=drawColor,line width= 0.4pt,line join=round,line cap=round,fill=fillColor] ( 89.04, 84.28) circle (  1.43);

\path[draw=drawColor,line width= 0.4pt,line join=round,line cap=round,fill=fillColor] (108.36, 65.94) circle (  1.43);

\path[draw=drawColor,line width= 0.4pt,line join=round,line cap=round,fill=fillColor] (127.69, 47.44) circle (  1.43);
\definecolor{drawColor}{RGB}{138,65,152}

\path[draw=drawColor,line width= 0.4pt,line join=round,line cap=round,fill=fillColor] ( 50.87,119.82) circle (  1.43);

\path[draw=drawColor,line width= 0.4pt,line join=round,line cap=round,fill=fillColor] ( 70.20,100.99) circle (  1.43);

\path[draw=drawColor,line width= 0.4pt,line join=round,line cap=round,fill=fillColor] ( 89.52, 81.09) circle (  1.43);

\path[draw=drawColor,line width= 0.4pt,line join=round,line cap=round,fill=fillColor] (108.85, 63.45) circle (  1.43);

\path[draw=drawColor,line width= 0.4pt,line join=round,line cap=round,fill=fillColor] (128.17, 44.66) circle (  1.43);
\definecolor{drawColor}{gray}{0.20}

\path[draw=drawColor,line width= 0.6pt,line join=round,line cap=round] ( 38.56, 40.85) rectangle (139.04,124.59);
\end{scope}
\begin{scope}
\path[clip] (  0.00,  0.00) rectangle (144.54,130.09);
\definecolor{drawColor}{gray}{0.30}

\node[text=drawColor,anchor=base east,inner sep=0pt, outer sep=0pt, scale=  0.88] at ( 33.61, 61.95) {0.03};

\node[text=drawColor,anchor=base east,inner sep=0pt, outer sep=0pt, scale=  0.88] at ( 33.61, 86.97) {0.04};

\node[text=drawColor,anchor=base east,inner sep=0pt, outer sep=0pt, scale=  0.88] at ( 33.61,112.00) {0.05};
\end{scope}
\begin{scope}
\path[clip] (  0.00,  0.00) rectangle (144.54,130.09);
\definecolor{drawColor}{gray}{0.20}

\path[draw=drawColor,line width= 0.6pt,line join=round] ( 35.81, 64.98) --
	( 38.56, 64.98);

\path[draw=drawColor,line width= 0.6pt,line join=round] ( 35.81, 90.01) --
	( 38.56, 90.01);

\path[draw=drawColor,line width= 0.6pt,line join=round] ( 35.81,115.03) --
	( 38.56,115.03);
\end{scope}
\begin{scope}
\path[clip] (  0.00,  0.00) rectangle (144.54,130.09);
\definecolor{drawColor}{gray}{0.20}

\path[draw=drawColor,line width= 0.6pt,line join=round] ( 50.15, 38.10) --
	( 50.15, 40.85);

\path[draw=drawColor,line width= 0.6pt,line join=round] ( 69.47, 38.10) --
	( 69.47, 40.85);

\path[draw=drawColor,line width= 0.6pt,line join=round] ( 88.80, 38.10) --
	( 88.80, 40.85);

\path[draw=drawColor,line width= 0.6pt,line join=round] (108.12, 38.10) --
	(108.12, 40.85);

\path[draw=drawColor,line width= 0.6pt,line join=round] (127.45, 38.10) --
	(127.45, 40.85);
\end{scope}
\begin{scope}
\path[clip] (  0.00,  0.00) rectangle (144.54,130.09);
\definecolor{drawColor}{gray}{0.30}

\node[text=drawColor,rotate= 45.00,anchor=base east,inner sep=0pt, outer sep=0pt, scale=  0.88] at ( 54.44, 31.62) {512};

\node[text=drawColor,rotate= 45.00,anchor=base east,inner sep=0pt, outer sep=0pt, scale=  0.88] at ( 73.76, 31.62) {1024};

\node[text=drawColor,rotate= 45.00,anchor=base east,inner sep=0pt, outer sep=0pt, scale=  0.88] at ( 93.08, 31.62) {2048};

\node[text=drawColor,rotate= 45.00,anchor=base east,inner sep=0pt, outer sep=0pt, scale=  0.88] at (112.41, 31.62) {4096};

\node[text=drawColor,rotate= 45.00,anchor=base east,inner sep=0pt, outer sep=0pt, scale=  0.88] at (131.73, 31.62) {8192};
\end{scope}
\begin{scope}
\path[clip] (  0.00,  0.00) rectangle (144.54,130.09);
\definecolor{drawColor}{RGB}{0,0,0}

\node[text=drawColor,anchor=base,inner sep=0pt, outer sep=0pt, scale=  1.10] at ( 88.80,  7.64) {Target Chunk Size};
\end{scope}
\begin{scope}
\path[clip] (  0.00,  0.00) rectangle (144.54,130.09);
\definecolor{drawColor}{RGB}{0,0,0}

\node[text=drawColor,rotate= 90.00,anchor=base,inner sep=0pt, outer sep=0pt, scale=  1.10] at ( 13.08, 82.72) {Dedup. Ratio};
\end{scope}
\end{tikzpicture}

%% file: fig/dedup_gear_variants_web.tex
% Created by tikzDevice version 0.12.6 on 2024-04-19 16:50:25
% !TEX encoding = UTF-8 Unicode
\begin{tikzpicture}[x=1pt,y=1pt]
\definecolor{fillColor}{RGB}{255,255,255}
\path[use as bounding box,fill=fillColor,fill opacity=0.00] (0,0) rectangle (144.54,130.09);
\begin{scope}
\path[clip] (  0.00,  0.00) rectangle (144.54,130.09);
\definecolor{drawColor}{RGB}{255,255,255}
\definecolor{fillColor}{RGB}{255,255,255}

\path[draw=drawColor,line width= 0.6pt,line join=round,line cap=round,fill=fillColor] (  0.00,  0.00) rectangle (144.54,130.09);
\end{scope}
\begin{scope}
\path[clip] ( 38.56, 40.85) rectangle (139.04,124.59);
\definecolor{fillColor}{RGB}{255,255,255}

\path[fill=fillColor] ( 38.56, 40.85) rectangle (139.04,124.59);
\definecolor{drawColor}{gray}{0.92}

\path[draw=drawColor,line width= 0.3pt,line join=round] ( 38.56, 45.34) --
	(139.04, 45.34);

\path[draw=drawColor,line width= 0.3pt,line join=round] ( 38.56, 72.28) --
	(139.04, 72.28);

\path[draw=drawColor,line width= 0.3pt,line join=round] ( 38.56, 99.23) --
	(139.04, 99.23);

\path[draw=drawColor,line width= 0.6pt,line join=round] ( 38.56, 58.81) --
	(139.04, 58.81);

\path[draw=drawColor,line width= 0.6pt,line join=round] ( 38.56, 85.76) --
	(139.04, 85.76);

\path[draw=drawColor,line width= 0.6pt,line join=round] ( 38.56,112.70) --
	(139.04,112.70);

\path[draw=drawColor,line width= 0.6pt,line join=round] ( 50.15, 40.85) --
	( 50.15,124.59);

\path[draw=drawColor,line width= 0.6pt,line join=round] ( 69.47, 40.85) --
	( 69.47,124.59);

\path[draw=drawColor,line width= 0.6pt,line join=round] ( 88.80, 40.85) --
	( 88.80,124.59);

\path[draw=drawColor,line width= 0.6pt,line join=round] (108.12, 40.85) --
	(108.12,124.59);

\path[draw=drawColor,line width= 0.6pt,line join=round] (127.45, 40.85) --
	(127.45,124.59);
\definecolor{drawColor}{RGB}{255,111,0}

\path[draw=drawColor,line width= 0.6pt,line join=round] ( 49.42,113.18) --
	( 68.75, 96.89) --
	( 88.07, 78.89) --
	(107.40, 64.17) --
	(126.72, 44.66);
\definecolor{drawColor}{RGB}{199,16,0}

\path[draw=drawColor,line width= 0.6pt,line join=round] ( 49.91,120.00) --
	( 69.23,103.47) --
	( 88.56, 85.71) --
	(107.88, 65.76) --
	(127.20, 48.79);
\definecolor{drawColor}{RGB}{0,142,160}

\path[draw=drawColor,line width= 0.6pt,line join=round] ( 50.39,120.78) --
	( 69.72,107.54) --
	( 89.04, 88.13) --
	(108.36, 67.46) --
	(127.69, 47.64);
\definecolor{drawColor}{RGB}{138,65,152}

\path[draw=drawColor,line width= 0.6pt,line join=round] ( 50.87,119.43) --
	( 70.20,105.68) --
	( 89.52, 85.52) --
	(108.85, 66.24) --
	(128.17, 47.95);
\definecolor{drawColor}{RGB}{255,111,0}

\path[draw=drawColor,line width= 0.4pt,line join=round,line cap=round,fill=fillColor] ( 49.42,113.18) circle (  1.43);

\path[draw=drawColor,line width= 0.4pt,line join=round,line cap=round,fill=fillColor] ( 68.75, 96.89) circle (  1.43);

\path[draw=drawColor,line width= 0.4pt,line join=round,line cap=round,fill=fillColor] ( 88.07, 78.89) circle (  1.43);

\path[draw=drawColor,line width= 0.4pt,line join=round,line cap=round,fill=fillColor] (107.40, 64.17) circle (  1.43);

\path[draw=drawColor,line width= 0.4pt,line join=round,line cap=round,fill=fillColor] (126.72, 44.66) circle (  1.43);
\definecolor{drawColor}{RGB}{199,16,0}

\path[draw=drawColor,line width= 0.4pt,line join=round,line cap=round,fill=fillColor] ( 49.91,120.00) circle (  1.43);

\path[draw=drawColor,line width= 0.4pt,line join=round,line cap=round,fill=fillColor] ( 69.23,103.47) circle (  1.43);

\path[draw=drawColor,line width= 0.4pt,line join=round,line cap=round,fill=fillColor] ( 88.56, 85.71) circle (  1.43);

\path[draw=drawColor,line width= 0.4pt,line join=round,line cap=round,fill=fillColor] (107.88, 65.76) circle (  1.43);

\path[draw=drawColor,line width= 0.4pt,line join=round,line cap=round,fill=fillColor] (127.20, 48.79) circle (  1.43);
\definecolor{drawColor}{RGB}{0,142,160}

\path[draw=drawColor,line width= 0.4pt,line join=round,line cap=round,fill=fillColor] ( 50.39,120.78) circle (  1.43);

\path[draw=drawColor,line width= 0.4pt,line join=round,line cap=round,fill=fillColor] ( 69.72,107.54) circle (  1.43);

\path[draw=drawColor,line width= 0.4pt,line join=round,line cap=round,fill=fillColor] ( 89.04, 88.13) circle (  1.43);

\path[draw=drawColor,line width= 0.4pt,line join=round,line cap=round,fill=fillColor] (108.36, 67.46) circle (  1.43);

\path[draw=drawColor,line width= 0.4pt,line join=round,line cap=round,fill=fillColor] (127.69, 47.64) circle (  1.43);
\definecolor{drawColor}{RGB}{138,65,152}

\path[draw=drawColor,line width= 0.4pt,line join=round,line cap=round,fill=fillColor] ( 50.87,119.43) circle (  1.43);

\path[draw=drawColor,line width= 0.4pt,line join=round,line cap=round,fill=fillColor] ( 70.20,105.68) circle (  1.43);

\path[draw=drawColor,line width= 0.4pt,line join=round,line cap=round,fill=fillColor] ( 89.52, 85.52) circle (  1.43);

\path[draw=drawColor,line width= 0.4pt,line join=round,line cap=round,fill=fillColor] (108.85, 66.24) circle (  1.43);

\path[draw=drawColor,line width= 0.4pt,line join=round,line cap=round,fill=fillColor] (128.17, 47.95) circle (  1.43);
\definecolor{drawColor}{gray}{0.20}

\path[draw=drawColor,line width= 0.6pt,line join=round,line cap=round] ( 38.56, 40.85) rectangle (139.04,124.59);
\end{scope}
\begin{scope}
\path[clip] (  0.00,  0.00) rectangle (144.54,130.09);
\definecolor{drawColor}{gray}{0.30}

\node[text=drawColor,anchor=base east,inner sep=0pt, outer sep=0pt, scale=  0.88] at ( 33.61, 55.78) {0.69};

\node[text=drawColor,anchor=base east,inner sep=0pt, outer sep=0pt, scale=  0.88] at ( 33.61, 82.73) {0.70};

\node[text=drawColor,anchor=base east,inner sep=0pt, outer sep=0pt, scale=  0.88] at ( 33.61,109.67) {0.71};
\end{scope}
\begin{scope}
\path[clip] (  0.00,  0.00) rectangle (144.54,130.09);
\definecolor{drawColor}{gray}{0.20}

\path[draw=drawColor,line width= 0.6pt,line join=round] ( 35.81, 58.81) --
	( 38.56, 58.81);

\path[draw=drawColor,line width= 0.6pt,line join=round] ( 35.81, 85.76) --
	( 38.56, 85.76);

\path[draw=drawColor,line width= 0.6pt,line join=round] ( 35.81,112.70) --
	( 38.56,112.70);
\end{scope}
\begin{scope}
\path[clip] (  0.00,  0.00) rectangle (144.54,130.09);
\definecolor{drawColor}{gray}{0.20}

\path[draw=drawColor,line width= 0.6pt,line join=round] ( 50.15, 38.10) --
	( 50.15, 40.85);

\path[draw=drawColor,line width= 0.6pt,line join=round] ( 69.47, 38.10) --
	( 69.47, 40.85);

\path[draw=drawColor,line width= 0.6pt,line join=round] ( 88.80, 38.10) --
	( 88.80, 40.85);

\path[draw=drawColor,line width= 0.6pt,line join=round] (108.12, 38.10) --
	(108.12, 40.85);

\path[draw=drawColor,line width= 0.6pt,line join=round] (127.45, 38.10) --
	(127.45, 40.85);
\end{scope}
\begin{scope}
\path[clip] (  0.00,  0.00) rectangle (144.54,130.09);
\definecolor{drawColor}{gray}{0.30}

\node[text=drawColor,rotate= 45.00,anchor=base east,inner sep=0pt, outer sep=0pt, scale=  0.88] at ( 54.44, 31.62) {512};

\node[text=drawColor,rotate= 45.00,anchor=base east,inner sep=0pt, outer sep=0pt, scale=  0.88] at ( 73.76, 31.62) {1024};

\node[text=drawColor,rotate= 45.00,anchor=base east,inner sep=0pt, outer sep=0pt, scale=  0.88] at ( 93.08, 31.62) {2048};

\node[text=drawColor,rotate= 45.00,anchor=base east,inner sep=0pt, outer sep=0pt, scale=  0.88] at (112.41, 31.62) {4096};

\node[text=drawColor,rotate= 45.00,anchor=base east,inner sep=0pt, outer sep=0pt, scale=  0.88] at (131.73, 31.62) {8192};
\end{scope}
\begin{scope}
\path[clip] (  0.00,  0.00) rectangle (144.54,130.09);
\definecolor{drawColor}{RGB}{0,0,0}

\node[text=drawColor,anchor=base,inner sep=0pt, outer sep=0pt, scale=  1.10] at ( 88.80,  7.64) {Target Chunk Size};
\end{scope}
\begin{scope}
\path[clip] (  0.00,  0.00) rectangle (144.54,130.09);
\definecolor{drawColor}{RGB}{0,0,0}

\node[text=drawColor,rotate= 90.00,anchor=base,inner sep=0pt, outer sep=0pt, scale=  1.10] at ( 13.08, 82.72) {Dedup. Ratio};
\end{scope}
\end{tikzpicture}

%% file: fig/dedup_gear_variants_code.tex
% Created by tikzDevice version 0.12.6 on 2024-04-19 16:30:55
% !TEX encoding = UTF-8 Unicode
\begin{tikzpicture}[x=1pt,y=1pt]
\definecolor{fillColor}{RGB}{255,255,255}
\path[use as bounding box,fill=fillColor,fill opacity=0.00] (0,0) rectangle (144.54,130.09);
\begin{scope}
\path[clip] (  0.00,  0.00) rectangle (144.54,130.09);
\definecolor{drawColor}{RGB}{255,255,255}
\definecolor{fillColor}{RGB}{255,255,255}

\path[draw=drawColor,line width= 0.6pt,line join=round,line cap=round,fill=fillColor] (  0.00,  0.00) rectangle (144.54,130.09);
\end{scope}
\begin{scope}
\path[clip] ( 38.56, 40.85) rectangle (139.04,124.59);
\definecolor{fillColor}{RGB}{255,255,255}

\path[fill=fillColor] ( 38.56, 40.85) rectangle (139.04,124.59);
\definecolor{drawColor}{gray}{0.92}

\path[draw=drawColor,line width= 0.3pt,line join=round] ( 38.56, 48.70) --
	(139.04, 48.70);

\path[draw=drawColor,line width= 0.3pt,line join=round] ( 38.56, 70.23) --
	(139.04, 70.23);

\path[draw=drawColor,line width= 0.3pt,line join=round] ( 38.56, 91.75) --
	(139.04, 91.75);

\path[draw=drawColor,line width= 0.3pt,line join=round] ( 38.56,113.28) --
	(139.04,113.28);

\path[draw=drawColor,line width= 0.6pt,line join=round] ( 38.56, 59.47) --
	(139.04, 59.47);

\path[draw=drawColor,line width= 0.6pt,line join=round] ( 38.56, 80.99) --
	(139.04, 80.99);

\path[draw=drawColor,line width= 0.6pt,line join=round] ( 38.56,102.52) --
	(139.04,102.52);

\path[draw=drawColor,line width= 0.6pt,line join=round] ( 38.56,124.04) --
	(139.04,124.04);

\path[draw=drawColor,line width= 0.6pt,line join=round] ( 50.15, 40.85) --
	( 50.15,124.59);

\path[draw=drawColor,line width= 0.6pt,line join=round] ( 69.47, 40.85) --
	( 69.47,124.59);

\path[draw=drawColor,line width= 0.6pt,line join=round] ( 88.80, 40.85) --
	( 88.80,124.59);

\path[draw=drawColor,line width= 0.6pt,line join=round] (108.12, 40.85) --
	(108.12,124.59);

\path[draw=drawColor,line width= 0.6pt,line join=round] (127.45, 40.85) --
	(127.45,124.59);
\definecolor{drawColor}{RGB}{255,111,0}

\path[draw=drawColor,line width= 0.6pt,line join=round] ( 49.42,117.94) --
	( 68.75,105.40) --
	( 88.07, 89.01) --
	(107.40, 67.69) --
	(126.72, 44.66);
\definecolor{drawColor}{RGB}{199,16,0}

\path[draw=drawColor,line width= 0.6pt,line join=round] ( 49.91,120.67) --
	( 69.23,108.63) --
	( 88.56, 93.44) --
	(107.88, 75.49) --
	(127.20, 52.96);
\definecolor{drawColor}{RGB}{0,142,160}

\path[draw=drawColor,line width= 0.6pt,line join=round] ( 50.39,120.78) --
	( 69.72,108.47) --
	( 89.04, 94.18) --
	(108.36, 77.15) --
	(127.69, 56.67);
\definecolor{drawColor}{RGB}{138,65,152}

\path[draw=drawColor,line width= 0.6pt,line join=round] ( 50.87,116.78) --
	( 70.20,103.88) --
	( 89.52, 89.10) --
	(108.85, 72.32) --
	(128.17, 53.15);
\definecolor{drawColor}{RGB}{255,111,0}

\path[draw=drawColor,line width= 0.4pt,line join=round,line cap=round,fill=fillColor] ( 49.42,117.94) circle (  1.43);

\path[draw=drawColor,line width= 0.4pt,line join=round,line cap=round,fill=fillColor] ( 68.75,105.40) circle (  1.43);

\path[draw=drawColor,line width= 0.4pt,line join=round,line cap=round,fill=fillColor] ( 88.07, 89.01) circle (  1.43);

\path[draw=drawColor,line width= 0.4pt,line join=round,line cap=round,fill=fillColor] (107.40, 67.69) circle (  1.43);

\path[draw=drawColor,line width= 0.4pt,line join=round,line cap=round,fill=fillColor] (126.72, 44.66) circle (  1.43);
\definecolor{drawColor}{RGB}{199,16,0}

\path[draw=drawColor,line width= 0.4pt,line join=round,line cap=round,fill=fillColor] ( 49.91,120.67) circle (  1.43);

\path[draw=drawColor,line width= 0.4pt,line join=round,line cap=round,fill=fillColor] ( 69.23,108.63) circle (  1.43);

\path[draw=drawColor,line width= 0.4pt,line join=round,line cap=round,fill=fillColor] ( 88.56, 93.44) circle (  1.43);

\path[draw=drawColor,line width= 0.4pt,line join=round,line cap=round,fill=fillColor] (107.88, 75.49) circle (  1.43);

\path[draw=drawColor,line width= 0.4pt,line join=round,line cap=round,fill=fillColor] (127.20, 52.96) circle (  1.43);
\definecolor{drawColor}{RGB}{0,142,160}

\path[draw=drawColor,line width= 0.4pt,line join=round,line cap=round,fill=fillColor] ( 50.39,120.78) circle (  1.43);

\path[draw=drawColor,line width= 0.4pt,line join=round,line cap=round,fill=fillColor] ( 69.72,108.47) circle (  1.43);

\path[draw=drawColor,line width= 0.4pt,line join=round,line cap=round,fill=fillColor] ( 89.04, 94.18) circle (  1.43);

\path[draw=drawColor,line width= 0.4pt,line join=round,line cap=round,fill=fillColor] (108.36, 77.15) circle (  1.43);

\path[draw=drawColor,line width= 0.4pt,line join=round,line cap=round,fill=fillColor] (127.69, 56.67) circle (  1.43);
\definecolor{drawColor}{RGB}{138,65,152}

\path[draw=drawColor,line width= 0.4pt,line join=round,line cap=round,fill=fillColor] ( 50.87,116.78) circle (  1.43);

\path[draw=drawColor,line width= 0.4pt,line join=round,line cap=round,fill=fillColor] ( 70.20,103.88) circle (  1.43);

\path[draw=drawColor,line width= 0.4pt,line join=round,line cap=round,fill=fillColor] ( 89.52, 89.10) circle (  1.43);

\path[draw=drawColor,line width= 0.4pt,line join=round,line cap=round,fill=fillColor] (108.85, 72.32) circle (  1.43);

\path[draw=drawColor,line width= 0.4pt,line join=round,line cap=round,fill=fillColor] (128.17, 53.15) circle (  1.43);
\definecolor{drawColor}{gray}{0.20}

\path[draw=drawColor,line width= 0.6pt,line join=round,line cap=round] ( 38.56, 40.85) rectangle (139.04,124.59);
\end{scope}
\begin{scope}
\path[clip] (  0.00,  0.00) rectangle (144.54,130.09);
\definecolor{drawColor}{gray}{0.30}

\node[text=drawColor,anchor=base east,inner sep=0pt, outer sep=0pt, scale=  0.88] at ( 33.61, 56.44) {0.76};

\node[text=drawColor,anchor=base east,inner sep=0pt, outer sep=0pt, scale=  0.88] at ( 33.61, 77.96) {0.80};

\node[text=drawColor,anchor=base east,inner sep=0pt, outer sep=0pt, scale=  0.88] at ( 33.61, 99.49) {0.84};

\node[text=drawColor,anchor=base east,inner sep=0pt, outer sep=0pt, scale=  0.88] at ( 33.61,121.01) {0.88};
\end{scope}
\begin{scope}
\path[clip] (  0.00,  0.00) rectangle (144.54,130.09);
\definecolor{drawColor}{gray}{0.20}

\path[draw=drawColor,line width= 0.6pt,line join=round] ( 35.81, 59.47) --
	( 38.56, 59.47);

\path[draw=drawColor,line width= 0.6pt,line join=round] ( 35.81, 80.99) --
	( 38.56, 80.99);

\path[draw=drawColor,line width= 0.6pt,line join=round] ( 35.81,102.52) --
	( 38.56,102.52);

\path[draw=drawColor,line width= 0.6pt,line join=round] ( 35.81,124.04) --
	( 38.56,124.04);
\end{scope}
\begin{scope}
\path[clip] (  0.00,  0.00) rectangle (144.54,130.09);
\definecolor{drawColor}{gray}{0.20}

\path[draw=drawColor,line width= 0.6pt,line join=round] ( 50.15, 38.10) --
	( 50.15, 40.85);

\path[draw=drawColor,line width= 0.6pt,line join=round] ( 69.47, 38.10) --
	( 69.47, 40.85);

\path[draw=drawColor,line width= 0.6pt,line join=round] ( 88.80, 38.10) --
	( 88.80, 40.85);

\path[draw=drawColor,line width= 0.6pt,line join=round] (108.12, 38.10) --
	(108.12, 40.85);

\path[draw=drawColor,line width= 0.6pt,line join=round] (127.45, 38.10) --
	(127.45, 40.85);
\end{scope}
\begin{scope}
\path[clip] (  0.00,  0.00) rectangle (144.54,130.09);
\definecolor{drawColor}{gray}{0.30}

\node[text=drawColor,rotate= 45.00,anchor=base east,inner sep=0pt, outer sep=0pt, scale=  0.88] at ( 54.44, 31.62) {512};

\node[text=drawColor,rotate= 45.00,anchor=base east,inner sep=0pt, outer sep=0pt, scale=  0.88] at ( 73.76, 31.62) {1024};

\node[text=drawColor,rotate= 45.00,anchor=base east,inner sep=0pt, outer sep=0pt, scale=  0.88] at ( 93.08, 31.62) {2048};

\node[text=drawColor,rotate= 45.00,anchor=base east,inner sep=0pt, outer sep=0pt, scale=  0.88] at (112.41, 31.62) {4096};

\node[text=drawColor,rotate= 45.00,anchor=base east,inner sep=0pt, outer sep=0pt, scale=  0.88] at (131.73, 31.62) {8192};
\end{scope}
\begin{scope}
\path[clip] (  0.00,  0.00) rectangle (144.54,130.09);
\definecolor{drawColor}{RGB}{0,0,0}

\node[text=drawColor,anchor=base,inner sep=0pt, outer sep=0pt, scale=  1.10] at ( 88.80,  7.64) {Target Chunk Size};
\end{scope}
\begin{scope}
\path[clip] (  0.00,  0.00) rectangle (144.54,130.09);
\definecolor{drawColor}{RGB}{0,0,0}

\node[text=drawColor,rotate= 90.00,anchor=base,inner sep=0pt, outer sep=0pt, scale=  1.10] at ( 13.08, 82.72) {Dedup. Ratio};
\end{scope}
\end{tikzpicture}

%% file: fig/dedup_gear_variants_legendonly.tex
% Created by tikzDevice version 0.12.6 on 2024-04-19 16:30:59
% !TEX encoding = UTF-8 Unicode
\begin{tikzpicture}[x=1pt,y=1pt]
\definecolor{fillColor}{RGB}{255,255,255}
\path[use as bounding box,fill=fillColor,fill opacity=0.00] (0,0) rectangle (252.94, 72.27);
\begin{scope}
\path[clip] (  0.00,  0.00) rectangle (252.94, 72.27);
\definecolor{drawColor}{RGB}{255,111,0}

\path[draw=drawColor,line width= 0.6pt,line join=round] ( 23.40, 36.13) -- ( 37.28, 36.13);
\end{scope}
\begin{scope}
\path[clip] (  0.00,  0.00) rectangle (252.94, 72.27);
\definecolor{drawColor}{RGB}{255,111,0}
\definecolor{fillColor}{RGB}{255,255,255}

\path[draw=drawColor,line width= 0.4pt,line join=round,line cap=round,fill=fillColor] ( 30.34, 36.13) circle (  1.43);
\end{scope}
\begin{scope}
\path[clip] (  0.00,  0.00) rectangle (252.94, 72.27);
\definecolor{drawColor}{RGB}{199,16,0}

\path[draw=drawColor,line width= 0.6pt,line join=round] ( 81.17, 36.13) -- ( 95.04, 36.13);
\end{scope}
\begin{scope}
\path[clip] (  0.00,  0.00) rectangle (252.94, 72.27);
\definecolor{drawColor}{RGB}{199,16,0}
\definecolor{fillColor}{RGB}{255,255,255}

\path[draw=drawColor,line width= 0.4pt,line join=round,line cap=round,fill=fillColor] ( 88.10, 36.13) circle (  1.43);
\end{scope}
\begin{scope}
\path[clip] (  0.00,  0.00) rectangle (252.94, 72.27);
\definecolor{drawColor}{RGB}{0,142,160}

\path[draw=drawColor,line width= 0.6pt,line join=round] (135.45, 36.13) -- (149.32, 36.13);
\end{scope}
\begin{scope}
\path[clip] (  0.00,  0.00) rectangle (252.94, 72.27);
\definecolor{drawColor}{RGB}{0,142,160}
\definecolor{fillColor}{RGB}{255,255,255}

\path[draw=drawColor,line width= 0.4pt,line join=round,line cap=round,fill=fillColor] (142.39, 36.13) circle (  1.43);
\end{scope}
\begin{scope}
\path[clip] (  0.00,  0.00) rectangle (252.94, 72.27);
\definecolor{drawColor}{RGB}{138,65,152}

\path[draw=drawColor,line width= 0.6pt,line join=round] (189.73, 36.13) -- (203.61, 36.13);
\end{scope}
\begin{scope}
\path[clip] (  0.00,  0.00) rectangle (252.94, 72.27);
\definecolor{drawColor}{RGB}{138,65,152}
\definecolor{fillColor}{RGB}{255,255,255}

\path[draw=drawColor,line width= 0.4pt,line join=round,line cap=round,fill=fillColor] (196.67, 36.13) circle (  1.43);
\end{scope}
\begin{scope}
\path[clip] (  0.00,  0.00) rectangle (252.94, 72.27);
\definecolor{drawColor}{RGB}{0,0,0}

\node[text=drawColor,anchor=base west,inner sep=0pt, outer sep=0pt, scale=  1.00] at ( 44.51, 32.69) {Vanilla};
\end{scope}
\begin{scope}
\path[clip] (  0.00,  0.00) rectangle (252.94, 72.27);
\definecolor{drawColor}{RGB}{0,0,0}

\node[text=drawColor,anchor=base west,inner sep=0pt, outer sep=0pt, scale=  1.00] at (102.28, 32.69) {NC-1};
\end{scope}
\begin{scope}
\path[clip] (  0.00,  0.00) rectangle (252.94, 72.27);
\definecolor{drawColor}{RGB}{0,0,0}

\node[text=drawColor,anchor=base west,inner sep=0pt, outer sep=0pt, scale=  1.00] at (156.56, 32.69) {NC-2};
\end{scope}
\begin{scope}
\path[clip] (  0.00,  0.00) rectangle (252.94, 72.27);
\definecolor{drawColor}{RGB}{0,0,0}

\node[text=drawColor,anchor=base west,inner sep=0pt, outer sep=0pt, scale=  1.00] at (210.84, 32.69) {NC-3};
\end{scope}
\end{tikzpicture}